FACOLTÀ DI SCIENZE
MATEMATICHE FISICHE
E NATURALI

SAPIENZA
UNIVERSITÀ DI ROMA

Corso di Laurea Specialistica in Astronomia e Astrofisica

# The search for exomoons and the characterization of exoplanet atmospheres

*Relatore interno : dott. Alessandro Melchiorri*

*Relatore esterno : dott.ssa Giovanna Tinetti*

*Candidato: Giammarco Campanella*

# The search for exomoons and the characterization of exoplanet atmospheres

*Giammarco Campanella*

*Dipartimento di Fisica*

*Università degli studi di Roma "La Sapienza"*

*Associate at Department of Physics & Astronomy*

*University College London*

*A thesis submitted for the MSc Degree in*

*Astronomy and Astrophysics*

*September 4th, 2009*

Università degli Studi di Roma "La Sapienza"

Abstract

# THE SEARCH FOR EXOMOONS AND THE CHARACTERIZATION OF EXOPLANET ATMOSPHERES

by Giammarco Campanella

Since planets were first discovered outside our own Solar System in 1992 (around a pulsar) and in 1995 (around a main sequence star), extrasolar planet studies have become one of the most dynamic research fields in astronomy. Our knowledge of extrasolar planets has grown exponentially, from our understanding of their formation and evolution to the development of different methods to detect them.

Now that more than 370 exoplanets have been discovered, focus has moved from finding planets to characterise these alien worlds. As well as detecting the atmospheres of these exoplanets, part of the characterisation process undoubtedly involves the search for extrasolar moons.

The structure of the thesis is as follows. In Chapter 1 an historical background is provided and some general aspects about ongoing situation in the research field of extrasolar planets are shown.

In Chapter 2, various detection techniques such as radial velocity, microlensing, astrometry, circumstellar disks, pulsar timing and magnetospheric emission are described. A special emphasis is given to the transit photometry technique and to the two already operational transit space missions, CoRoT and Kepler.

A review on the current situation of exoplanet characterization is presented in Chapter 3. We focus on the characterization of transiting planets orbiting very close to their parent star since for them we can already probe their atmospheric constituents. By contrast, the second part of the Chapter is dedicated to the search for extraterrestrial life, both within and beyond the Solar System. The characteristics of the Habitable Zone and the markers for the presence of life (biosignatures) are detailed.

In Chapter 4 we describe the primary transit observations of the hot Jupiter HD 209458b we obtained at 3.6, 4.5, 5.8 and 8.0 μm using the Infrared Array Camera on the Spitzer Space Telescope. We detail the procedures we adopted to correct for the systematic trends present in IRAC data. The lightcurves were fitted, taking into account limb darkening effects, using Markov Chain Monte Carlo and prayer-bead Monte Carlo techniques. We obtained the following depth measurements: at 3.6 μm, 1.469±0.013 % and 1.448±0.013 %; at 4.5 μm, 1.478±0.017 %; at 5.8 μm, 1.549±0.015 % and at 8.0 μm 1.535±0.011 %. Our results indicate the presence of water in the planetary atmosphere.

Chapter 5 is dedicated to the search for exomoons, we review some of the proposed detection techniques and introduce a model for the TTV and TDV signals which permits not only the identification of exomoons but also the derivation of some of their characteristics. We find that these techniques could easily detect Earth-mass exomoons with current instruments.

Finally, in Chapter 6 the detectability of a habitable-zone exomoon around various configurations of exoplanetary systems with the Kepler Mission or photometry of approximately equal quality is investigated. We calculate both the predicted transit timing signal amplitudes and the estimated uncertainty on such measurements in order to calculate the confidence in detecting such bodies across a broad spectrum of orbital arrangements. The effects of photon noise, stellar variability and instrument noise are all accounted for in the analysis. We validate our methodology by simulating synthetic lightcurves and we find that habitable-zone exomoons down to $0.2M_{\oplus}$ may be detected and $\sim$ 25,000 stars could be surveyed for habitable-zone exomoons within Kepler 's field-of-view.

Università degli Studi di Roma "La Sapienza"

Riassunto

# THE SEARCH FOR EXOMOONS AND THE CHARACTERIZATION OF EXOPLANET ATMOSPHERES

di Giammarco Campanella

Partendo da quando vennero scoperti per la prima volta alcuni pianeti esterni al Sistema Solare nel 1992 (attorno ad una pulsar) e nel 1995 (attorno ad una stella di Sequenza Principale), gli studi inerenti ai pianeti extrasolari sono divenuti uno dei campi di ricerca astronomica più dinamici. Negli anni, la nostra conoscenza dei pianeti extrasolari è cresciuta immensamente: si va dalla comprensione dei loro meccanismi di formazione ed evoluzione, allo sviluppo di diversi metodi per rivelarli.

Ora che più di 370 esopianeti sono stati scoperti, l'interesse si è spostato dal trovare i pianeti al caratterizzare questi mondi alieni. Oltre ad investigare le atmosfere di questi pianeti, parte del processo di caratterizzazione riguarda indubbiamente la ricerca delle esolune, i satelliti naturali che orbitano attorno ai pianeti extrasolari.

La struttura della tesi è la seguente. Nel Capitolo 1 viene fornito un quadro storico relativo all'argomento oggetto della tesi e viene presentata la situazione attuale del campo di ricerca dei pianeti extrasolari.

Nel Capitolo 2 vengono descritti alcuni metodi utilizzati per l'individuazione degli esopianeti: la velocità radiale, le microlenti gravitazionali, l'astrometria, i dischi circumstellari, le variazioni degli intervalli di emissioni di una Pulsar e le emissioni magnetosferiche. Inoltre, si illustra con una speciale attenzione la tecnica della fotometria di transito e si presentano le due missioni spaziali già operative nell'ambito: CoRoT e Kepler.

Nel Capitolo 3 viene tracciata la situazione attuale della caratterizzazione degli esopianeti. Si dedica una particolare attenzione a quei pianeti che transitano davanti alla propria stella e che si trovano molto vicino ad essa perché per loro è già possibile sondare la loro costituzione atmosferica. La seconda parte del Capitolo è invece dedicata alla ricerca della vita extraterrestre, riferita sia all'interno che all'esterno del Sistema Solare. Tra l'altro poi vengono illustrate le singolarità della zona abitabile e le caratteristiche spettrali riconducibili proprio all'attività biologica.

Il Capitolo 4 è incentrato sulla descrizione di un progetto di ricerca inerente all'osservazione a 3.6, 4.5, 5.8 e 8.0 μm di alcuni transiti di HD 209458b rilevati utilizzando la camera ad infrarossi IRAC del Telescopio Spaziale Spitzer. Vengono presentate in dettaglio le procedure adottate per correggere le tendenze sistematiche presenti nei dati di IRAC. Le curve di luce sono state quindi fittate, tenendo conto dell'oscuramento al bordo, con le tecniche Monte Carlo "Markov Chain" e "Prayer-Bead". Si sono ottenute le seguenti misure di profondità di transito: a 3.6 μm, 1.469±0.013 % e 1.448±0.013 %; a 4.5 μm, 1.478±0.017 %; a 5.8 μm, 1.549±0.015 % e a 8.0 μm 1.535±0.011 %. I nostri risultati indicano la presenza di acqua nell'atmosfera del pianeta.

Il Capitolo 5 è dedicato alla ricerca delle esolune, vengono esaminate prima alcune tecniche di rivelamento proposte e poi si introduce un modello che descrive le variazioni temporali e di durata del transito (TTV/TDV). Questo modello, oltre a permettere l'identificazione delle esolune, dà la possibilità di derivare alcune delle sue caratteristiche. Si trova che con queste tecniche si potrebbe già rivelare con la strumentazione attualmente disponibile una esoluna di massa pari a quella terrestre.

Infine, nel Capitolo 6 viene investigata la possibilità di rivelare esolune in orbita nella zona abitabile di un ampio numero di configurazioni di sistemi esoplanetari tramite Kepler o fotometria di qualità approssimativamente uguale. Abbiamo calcolato le ampiezze previste dei segnali TTV/TDV e anche le incertezze stimate su queste misure. Cosi è stato possibile calcolare la confidenza che si può raggiungere nel rilevare questi corpi, spaziando un ampio spettro di assetti orbitali. Durante questa analisi sono stati considerati tutti gli effetti derivanti dal rumore fotonico, la variabilità stellare e il rumore strumentale. Abbiamo validato la nostra metodologia simulando curve di luce sintetiche e siamo arrivati alla conclusione che possono essere rivelate esolune nella zona abitabile fino ad un minimo di $0.2M_{\oplus}$ e che $\sim$ 25000 stelle nel campo di vista di Kepler offrono le condizioni per rilevare esolune in orbita nella zona abitabile.

# CONTENTS

## LIST OF FIGURES

*Number* *Page*

LIST OF TABLES

ACKNOWLEDGMENTS/RINGRAZIAMENTI

Wow, è stato un semestre di preparazione Tesi davvero intenso (sotto tutti i punti di vista) in viaggio da Roma all'Inghilterra, alla Puglia, alla Sicilia, alla Scozia e con quanti traslochi ed impacchettamenti vari. Il mio ringraziamento deve andare quindi a tutti coloro che mi hanno accompagnato durante questo denso periodo.

La mia prima menzione và naturalmente ai miei familiari: Emilio, Wilma che non mi ha visto per 7 mesi, Cristiana, Nonna Maria, gli zii e i cugini... scusate per il tempo che non vi ho potuto dedicare e per essermi perso il matrimonio di Mary; in particolare il mio ricordo và a Nonna Rosa e ai suoi viaggi in Brasile che forse hanno ispirato il mio interesse per l'astronomia.

Poi un ringraziamento speciale al Mazza e al Direttore Domaschio e a tutti gli amici di funny cooperation e i compagni d'interazioni simpatiche... we'll keep in touch in a funny way, sempre.

Inoltre devo ringraziare Mara e Nino per quanto hanno fatto per me a Londra, e poi Graziella e Franco in particolare per il pranzo Pasquale in vero stile curdunnese che mi ha fatto sentire come a casa.

Next, I should say thanks to David, Jean-Philippe, Steve, all my mates in London... we had a really good time, and the guys of the UCL Astrophysics Group for the fantastic football matches in Regent's Park. A special, very special thanks to James for the "exquisite" Korean and Indian lunches.

And thanks to my pals at the STFC Summer School, in particular for the unplanned bonfire and for the "personal comunications" exchanged in the pubs.

Per concludere, il mio pensiero va a te Roberta, per avermi sopportato in tutti quei giorni in cui non potevo dedicarmi abbastanza a te... e grazie alla tua famiglia che si è appioppata in Agosto un ospite che passava la maggior parte del suo tempo in studio a scrivere la Tesi.

Ma manca qualcuno? Ah si, Giovanna! Altre 170 pagine di tesi non basterebbero per ribadire quanto ti sono grato per tutto quello che hai fatto per me in questi 6 mesi, festa di compleanno inclusa... qualche giorno e poi incomincia l'attacco al prossimo pianeta!

*Giammarco*

*"There are infinite worlds both like and unlike this world of ours...We must believe that in all worlds there are living creatures and plants and other things we see in this world."*
*Epicurus (c. 300 BCE)*

*"Do there exist many worlds, or is there but a single world? This is one of the most noble and exalted questions in the study of Nature."*
*Albertus Magnus (d. 1280)*

*"To consider the Earth as the only populated world in infinite space is as absurd as to assert that in an entire field of millet, only one grain will grow."*
*Metrodorus of Chios (4th century B.C.)*

*Chapter 1*

# FOREWORD

"Are there other worlds and other beings?" these profound questions have always been part of our history and culture. The greatest figures of Classical Civilisation, from Leucippus (5th century BC) to Epicurus (341 BC - 270 BC), and renowned philosophers, from Giordano Bruno (1548-1600) to Immanuel Kant (1724-1804), supposed we cannot be alone in the Universe. These eminent thinkers were following an ancient philosophical and theological tradition, but their ideas were not based on any experimental or observational evidence [1].

Proof of the existence of other worlds had to wait until Galileo turned his telescope to the night sky 400 years ago. Galileo was the first to truly see the planets and moons in our Solar System as other worlds. Yet it took until the end of the $20^{th}$ Century before we developed telescopes and spacecraft to view –up close-the planets, their moons, and the persistent debris from which they have formed. All of these hold deep secrets of the Earth's origins and likely the beginning of life itself.

It is ironic that what is arguably the most compelling subject in astronomy-the search for other worlds and other life beyond our Solar System-emerges only now, in the $21^{st}$ Century. Four centuries of discovery have brought us a remarkable understanding of the birth and evolution of stars, the history of galaxies, and even cosmology-the development of the entire universe. Not for a lack of imagination or motivation, but simply for the want of technology, our oldest and deepest questions, the ones most relevant to our own origins and fate, have remained beyond our grasp for thousands of years.

About fifty years ago, a great scientist and storyteller like Isaac Asimov estimated that about a billion Earth-like planets may exist just in our galaxy. Even reducing this number a hundred times, as suggested by less optimistic predictions, it remains extremely large. Nevertheless, in our galaxy there is a huge number of stars, around 200 billion, and probably a lot of them host a planetary system. It was only in 1988 that we recorded the first results: some observations of Campbell, Walker and Yang [162] suggested the presence of planetary masses around some near stars. However, Gordon et al. [164] were extremely cautious as planets in orbit was just one of the possible interpretations of the data. Not many took in consideration their studies.

One Year Later, David W. Latham (Harvard Smithsonian Center for Astrophysics) et al. [163] found strong evidences of what could have been a planet around a star, called HD 114762. Since Latham's planet had a mass at least 10 times the one of Jupiter, it was hypothesized that actually it was a Brown dwarf, so also this news did not have a big impact.

In 1992, Alexander Wolszczan (Pennsylvania State University) and Dale A. Frail (National Radio Astronomy Observatory) used a very precise timing method and discovered Earth-mass planets in orbit around PSR 1257+12, as pulsar planets, they surprised many astronomers who expected to find planets only around main sequence stars. Not many believed that a Solar System planet-like was discovered; anyway this was the first clue suggesting that planetary formation is an ordinary process.

Then, in 1995 during a conference in Florence, two Swiss astronomers, Michel Mayor and Didier Queloz (Observatory of Geneva), astonished the entire world by announcing the first detection of a planet orbiting a sun-like star, 51 Pegasi [56]. After its discovery, many teams confirmed the planet's existence and obtained more observations of its properties. It was found that the planet orbits the star in 4.2 Earth days, and is much closer to it than Mercury is to our Sun, yet has a minimum mass about half that of Jupiter (about 150 times that of the Earth). At the time, the presence of a gas giant planet so close to its star was not compatible with theories of planet formation and this 'hot Jupiter' was considered an anomaly; actually it was not even believed that this body, named "51 Peg b", could survive there. Computational models suggested that it is sufficiently massive that its thick atmosphere is not blown away by the star's solar wind [165].

The discovery of this strange and unpredicted object gave a boost to this revolutionary astronomical branch: the study of planetary systems beyond our own.

Nowadays, the search for exoplanets is imposing itself like one of the most interesting field of astronomy. As a matter of fact, we can count more than 80 ground-based and 20 space-based ongoing programs and future projects. More than 370 exoplanets have been discovered with detection rates escalating [3].

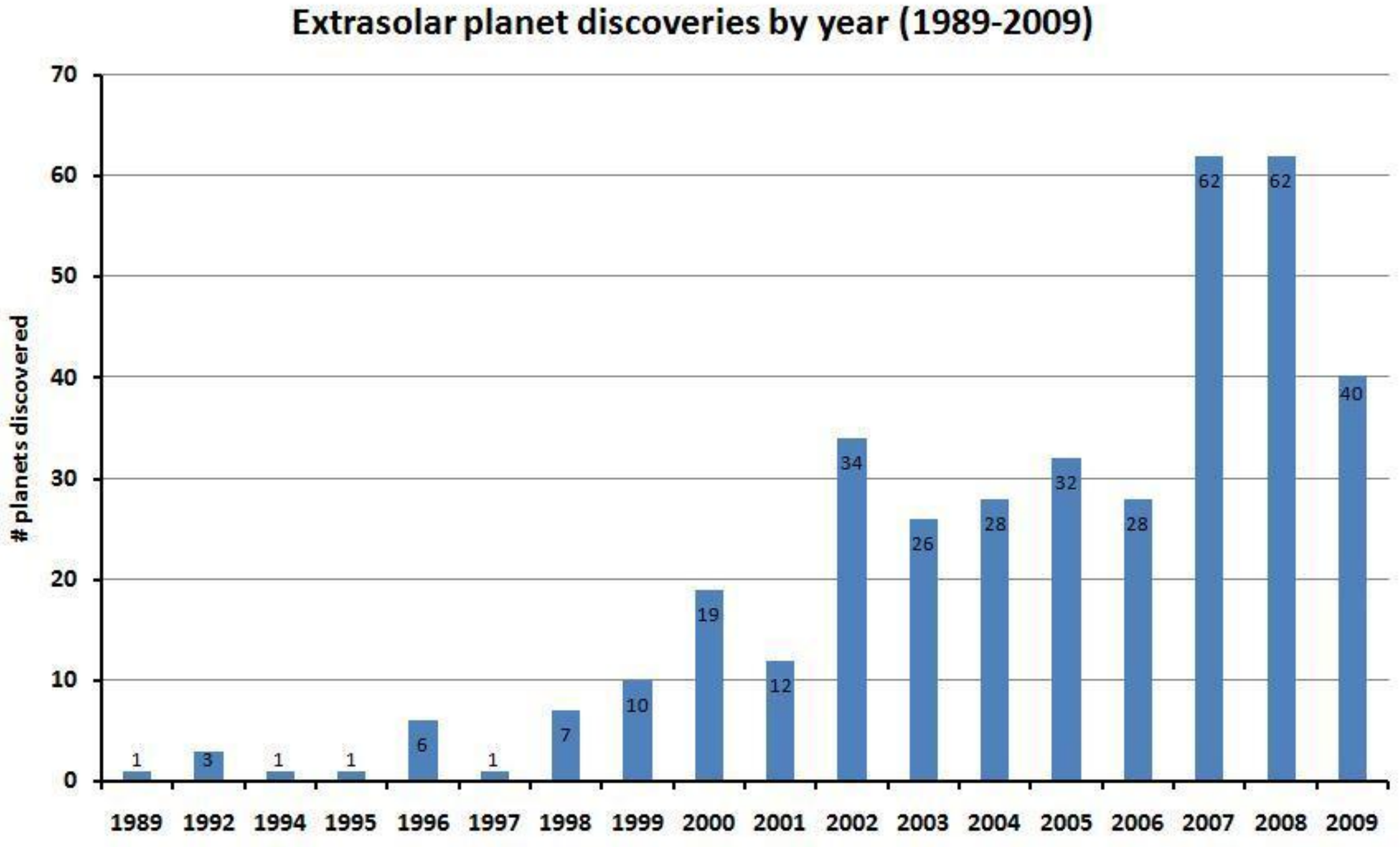

Figure 1.1: Number of extrasolar planet discoveries per year as of August 2009.

A decade ago, a handful of giant planets, had been discovered by the radial velocity method – a revolutionary step. Today Jovian-mass planets and considerably smaller ones (down to 1.9 Earth-masses in the case of GJ 581e [122]) are known by the hundreds, and even planetary systems similar to our own have been found. Gravitational lensing has been used to find a true analog to the Solar System, somewhat scaled down, consisting of both a Jupiter-like and Saturn-like planet on circular orbits a few astronomical units from their star [41]. Coronagraphic and related high-contrast direct-imaging techniques have produced images of nine planetary systems [3]. Precision photometry has led to the discovery of planets transiting their parent stars. The transit geometry allows direct detection of a planet's emergent light in some cases, and characterization of its atmosphere. New methods of finding and characterizing planets -astrometry and optical/infrared imaging-are developing rapidly. Application of these methods, which will greatly increase not just the inventory of planets but also begin to characterize them, is now limited more by resources than technology.

Having entered the phase of extrasolar planets characterization, we can constrain their horizontal and vertical temperature profiles and estimate the contribution of clouds and hazes [4]. Most importantly it is possible to use the wavelength dependence of this extinction to identify key chemical components in the planet's atmosphere. This achievement opens up enormous possibilities in terms of exoplanet characterization. In particular, in this thesis the probing of the atmosphere of HD 209458b in primary transit in the four IRAC bands at 3.6, 4.5, 5.8 and 8 μm will be presented [5]. We find our photometry data to be consistent with the presence of water, while the limitations imposed by the accuracy of our data prevent us from commenting on the possible additional presence of methane, CO and/or $CO_2$.

As well as detecting the atmospheres of exoplanets, part of the characterization process undoubtedly involves the search for extrasolar moons. In this work, we explore the motivations for undergoing such a search, review some of the proposed detection techniques and we investigate the detectability of a habitable-zone exomoon around various configurations for exoplanetary systems with the *Kepler Mission* or photometry of approximately equal quality. We find that Saturn-like planets offer the best opportunity for detecting a single large satellite and that habitable exomoons down to 0.2 $M_\oplus$ should be detectable with the expected performance of *Kepler* [6].

*Chapter 2*

# EXTRASOLAR PLANET DETECTION METHODS

Extrasolar planets are incredibly difficult to detect. This is because any planet is an extremely faint light source compared to its parent star. For this reason, only few extrasolar planets have been observed directly [3].

Instead, indirect methods have to be resorted to find extrasolar planets. Here we present seven different indirect techniques that are currently used:

1. Transit Photometry

2. Radial Velocity

3. Microlensing

4. Astrometry

5. Circumstellar Disks

6. Pulsar Timing

7. Magnetospheric Emission

Many of these methods rely on the measure of the interaction between the exoplanet and its parent star. By observing changes in the parent star, the existence of the planet can be deduced. Since the changes become larger as the planet becomes more massive, it is always easier to detect Jovian planets rather than terrestrial ones.

Table 2.1: Basic quantities for planets [7].

| | **Sun** | **Jupiter** | **Earth** | **HD 209458b** |
|---|---|---|---|---|
| **Mass (kg)** | $1.99 \cdot 10^{30}$ | $1.9 \cdot 10^{27}$ | $5.98 \cdot 10^{24}$ | $1.31 \cdot 10^{27}$ |
| **$M_V$ (mag)** | 4.85 | 25.5 | 27.8 | - |
| **Radius (km)** | 696000 | 71474 | 6378 | 94346 |
| **P (days)** | - | 4329 | 365 | 3.52 |
| **Semimajor Axis (AU)** | - | 5.2 | 1 | 0.045 |
| **RV semiamplitude of reflex motion (m/s)** | - | 12.5 | 0.09 | 86.52 |
| **Projected semimajor axis at 10pc (mas)** | - | 520 | 100 | 4.5 |
| **Contrast** $(L_\odot/L)$ | 1 | $1.82 \cdot 10^{8}$ | $1.5 \cdot 10^{9}$ | - |
| **Transit lightcurve depth (%)** | - | 1.01 | 0.0084 | 1.7 |

## 2.1 Radial Velocity

Radial velocity (RV) method detects planets by measuring a parent star's periodic line of sight velocity change due to its orbit around a common centre of mass, using the Doppler shift of the star's spectral lines.

Beginning with the detection of a planet half the mass of Jupiter around 51 Pegasi, by Meyor & Queloz in 1995, the Doppler spectroscopy technique has been the most successful so far in finding extrasolar planets. In fact, this method is responsible for the initial detection of over 80% of the exoplanets known today.

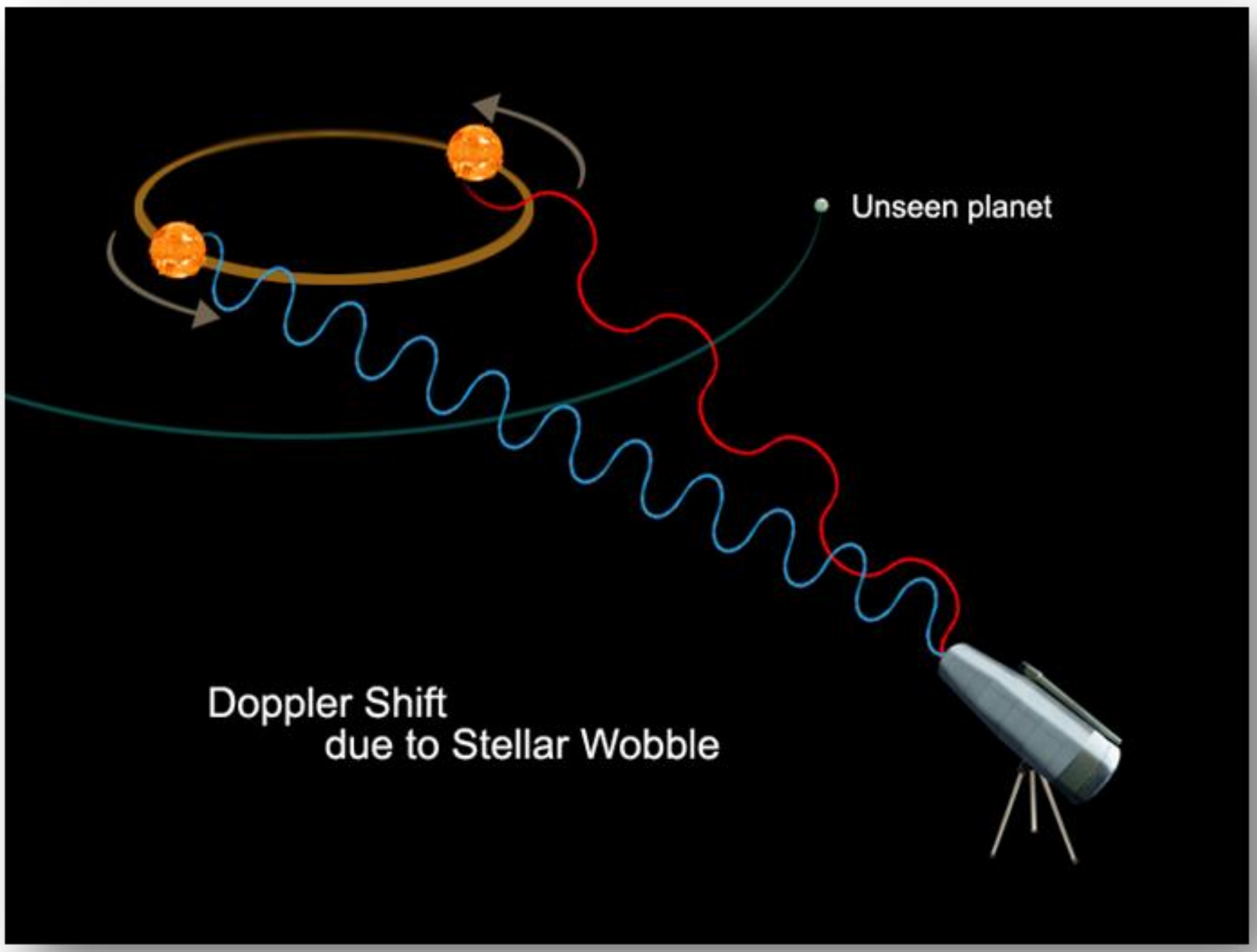

Figure 2.1: Schematic view of the wobble of a star due to an orbiting planet as observed from Earth. The star moves around the barycentre of the planetary system and its spectrum appears blue-shifted as it approaches the observer and red-shifted when it moves away.

The semi-amplitude of the stellar radial velocity signal K resulting from an exoplanetary system is given by

$$K = 28.4 \frac{M_p \sin i}{(M_* + M_p)^{2/3}} P^{-1/3} \ \mathrm{m \cdot s^{-1}}, \tag{2.1}$$

where $M_p$ is in Jupiter masses, $M_*$ is in solar masses, P is in years, and the orbit is assumed to be circular. From this equation, we see immediately that radial velocity detection gives the largest signal for massive planets in short period orbits. For instance, 51 Peg has a 4.2-day orbital period and $M_p \sin i$ of about half of a Jupiter mass. Its velocity semi-amplitude is K = 59 m $s^{-1}$, which is significantly larger than Mayor & Queloz's velocity measurement precision of 13 m $s^{-1}$. In contrast, the radial velocity signal of the Sun due to Jupiter's orbit is only about 13 m $s^{-1}$. We note also that the radial velocity technique is sensitive to $M_p \sin i$, rather than $M_p$. Thus, RVs allow only to measure a lower limit to the mass of a planet, unless the inclination $i$ of the system can be determined by some independent means (e. g. the transit method).

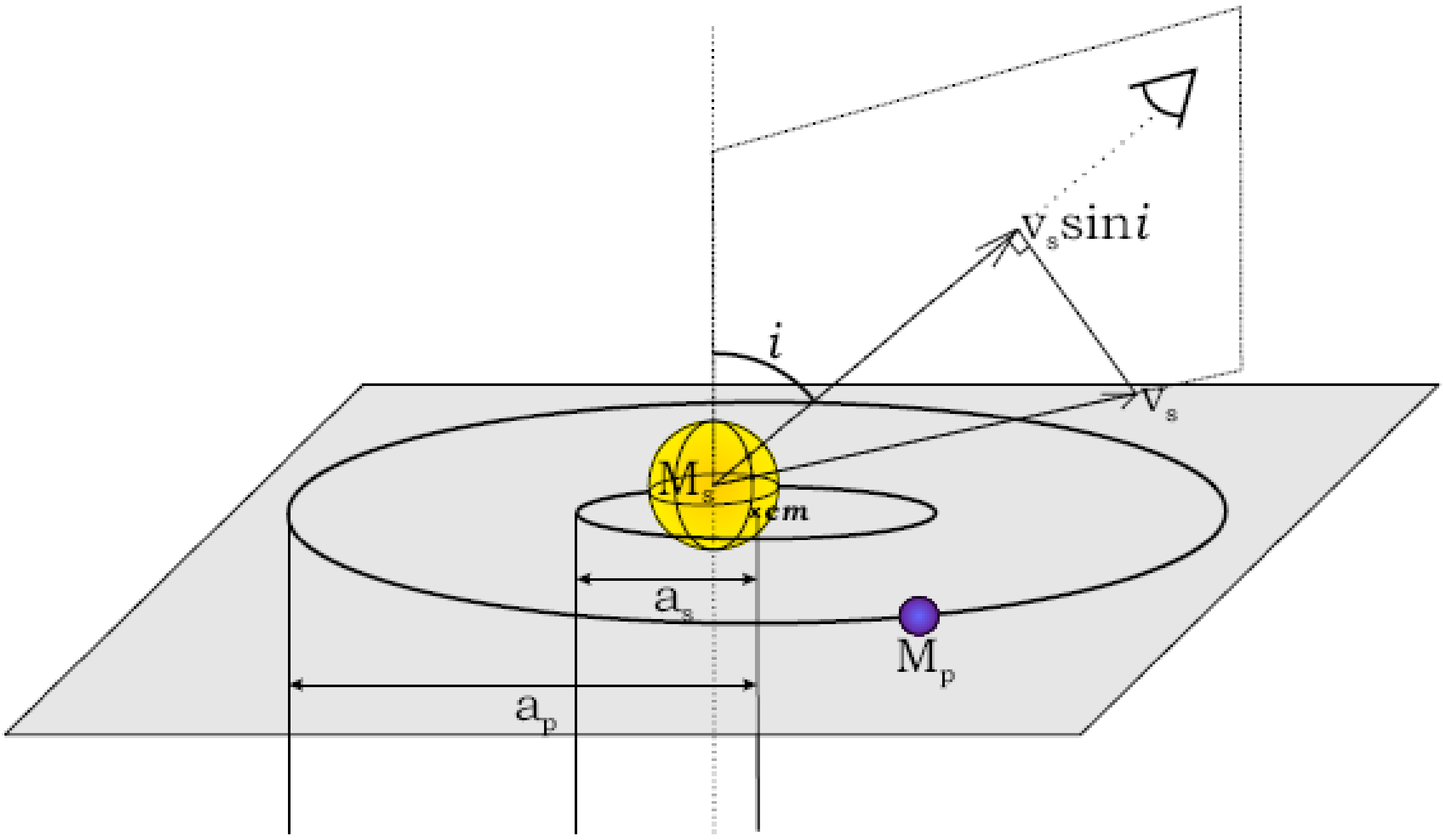

Figure 2.2: Schematic view Orbital parameters of a planet-star system. The star s and the planet p are in circular orbit around the centre of mass *cm* of the system. The orbital radii are $a_s$ for the star and $a_p$ for the planet, these are plotted along the orbital plane. The angle $i$ between the normal to the orbital plane and the line of sight determines the orbital inclination angle. The radial velocity $V_s$ of the star as measured along the line of sight (from the upper right in the diagram) depends on the sine of the orbital inclination angle [8].

Radial velocity measurement precision has continually improved from the initial ~15 m $s^{-1}$ regime [9] to the current state-of-the-art of 1 m $s^{-1}$ or better. The attainment of very high measurement precision has mostly been a matter of controlling systematic measurement errors. The use of a gas absorption cell or the continual illumination with a separate reference spectrum measures intrinsic spectrograph drifts. Enclosing the spectrograph in a vacuum and stabilizing the temperature to mK tolerances can result in the achievement of residuals to planetary orbit fits as low as 10-20 cm $s^{-1}$. This will be the case of HARPS-NEF, a high-resolution optical spectrograph born from the collaboration between Harvard and the Geneva Observatory that will be operational in 2010 on the 4.2-m Herschel Telescope (La Palma, Canary Islands) for follow-up studies of transit candidates from the *Kepler* mission [2].

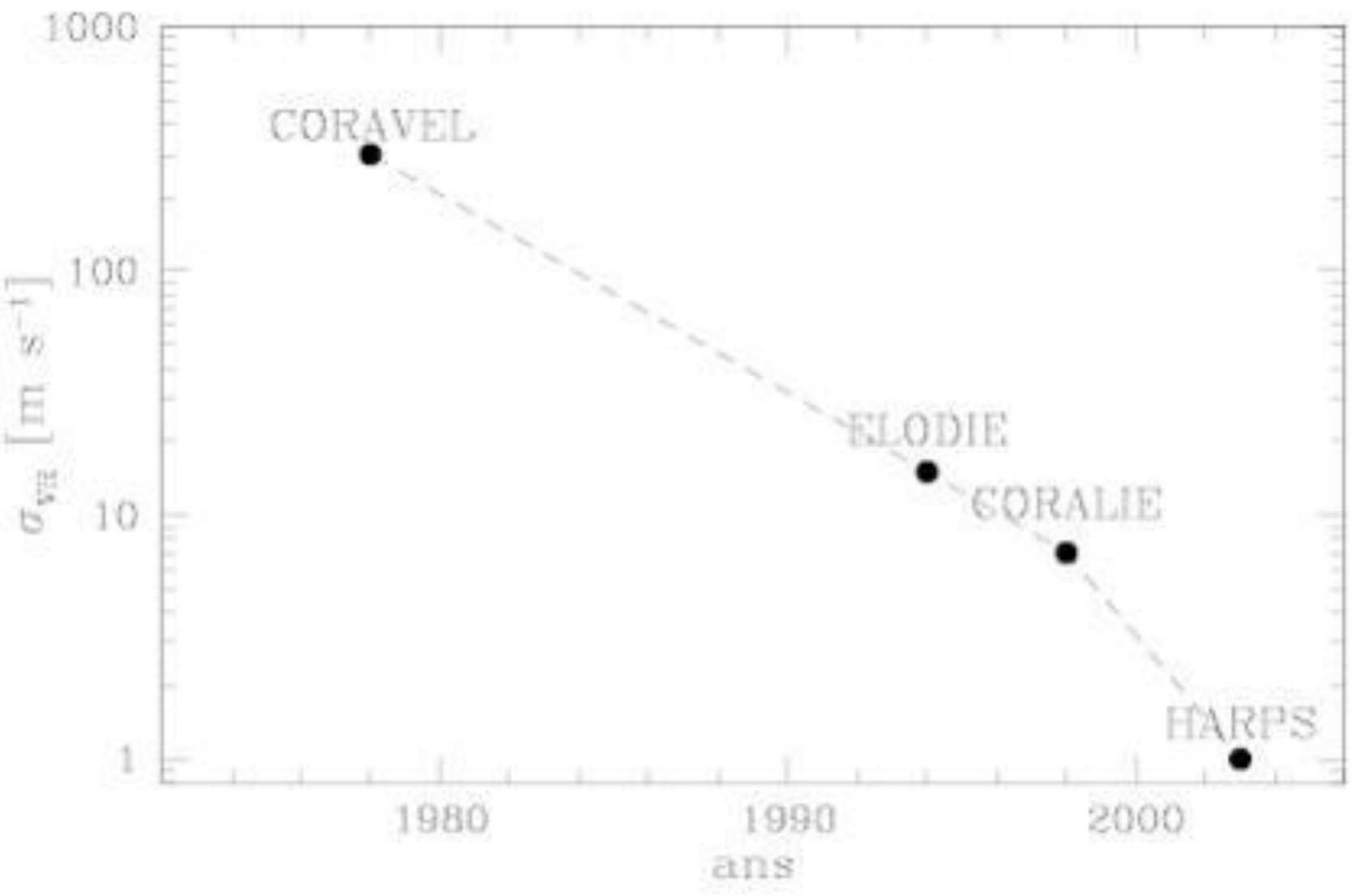

Figure 2.3: Development of Doppler techniques during last decades.

Reaching this level of measurement precision opens up new possibilities for planet detection. While the radial velocity signal of an Earth-mass planet in a 1-AU orbit around a solar-mass star is only 9 cm $s^{-1}$ (Table 2.1), we do have the measurement precision to detect habitable Earths around low-mass M stars, and to detect Super Earths (planets of a few Earth masses) in short-period orbits around solar-mass stars. These Super Earths are objects in the transition region between the masses of ice-giants like Uranus and Neptune and the masses of the rocky terrestrial planets of our inner Solar System. An extremely wide variety of compositions of such planets is possible, with mixtures of various fractions of metals (iron and nickel), silicates, "ices" (water, methane, ammonia, etc.) and $H_2$-He. The detection of these objects, and the measurement of their mass and radius (for those that undergo transits), will open up exciting new areas of planetary astrophysics.

The main factors currently limiting the precision of the RVs, in addition to photon noise, are stellar noise, the wavelength calibration, telescope guiding, stability in the illumination of the spectrograph, and detector-related effects. The only solution to photon noise is more photons. This may come from either larger telescopes equipped with high-precision spectrographs, or more telescope time on existing facilities permitting longer exposures without compromising the size of the samples of stars surveyed for planets. Indications are that the remedy to the problem of stellar noise may be similar. Longer exposures or binning over suitable timescales (again implying access to more telescope time) can reduce astrophysical jitter to some extent by averaging out those intrinsic variations (see, e.g., [10]).

Planets often occur in multi-planet systems. The first of these found was the system around υ And. Intensive follow-up observations often reveal the presence of additional planets in the system. Many times, these planets will be gravitationally interacting with each other. The study of such interactions helps us to understand the dynamics of planetary system formation and evolution.

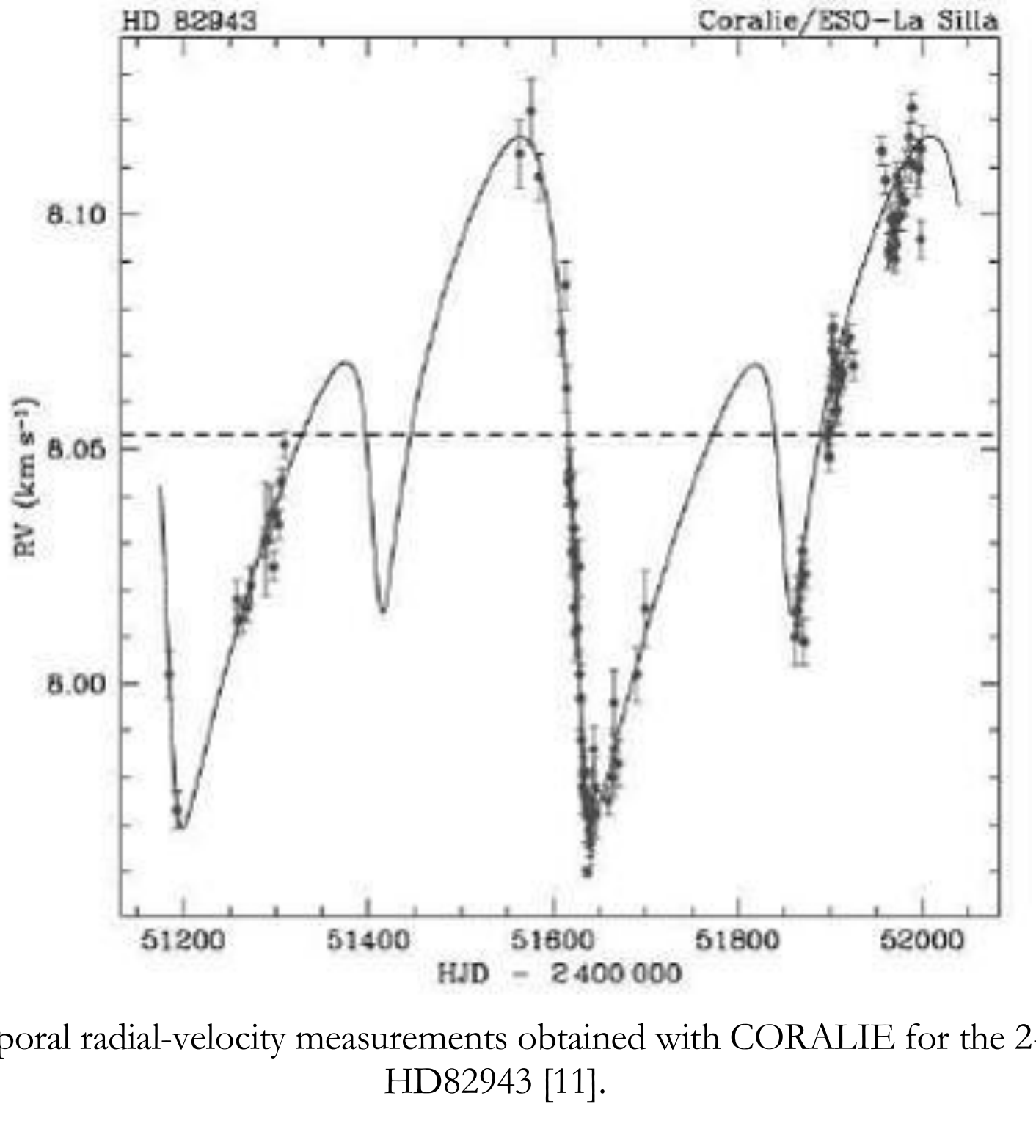

Figure 2.4: Temporal radial-velocity measurements obtained with CORALIE for the 2-planet systems HD82943 [11].

Doppler studies have revealed that approximately 10% of all F, G, and K stars have at least one planet in the mass range 0.3−10 $M_{Jup}$ with periods between 2 and 2000 days [12]. Extrapolating to longer periods, it is estimated that 17–19% of stars have a gas giant within 20 AU. We have also learned that gas giant formation is much more efficient around stars with super-solar metallicity (e.g. [13]).

## 2.2 Transits

The transit method met its first success in 1999 with the transit observation of the RV planet HD 209458b ([14], [15] and Fig. 2.5, left). It then became popular for two reasons: 1) the detection of a planetary transit around a bright star requires a telescope as small as 20-cm in diameter, and many transit surveys were initiated after this first success; 2) studies of a planetary system seen edge-on ($i \approx 90°$) is much richer than for systems at any inclination, because the radius and the mass are directly measured [16].

After the current success of ground-based transit searches like OGLE, major transit discoveries are expected from space, or already obtained (Fig. 2.5, right). Several NASA and ESA missions have been dedicated to exoplanet transits in large part or in total. The Kepler mission aims to determine the frequency of occurrence of Earth analogs, by monitoring ~100,000 solar-type stars (a field in Cygnus) for four years [17]. The CoRoT mission is similarly monitoring numerous solar-type stars to determine the frequency of occurrence of Super Earths, and to find and characterize exoplanets ranging in size from giant planets, down to Super Earths. CoRoT has already identified new giant transiting planets (e.g., [18]), and the CoRoT team is currently working to extend the detections to much smaller planets [2].

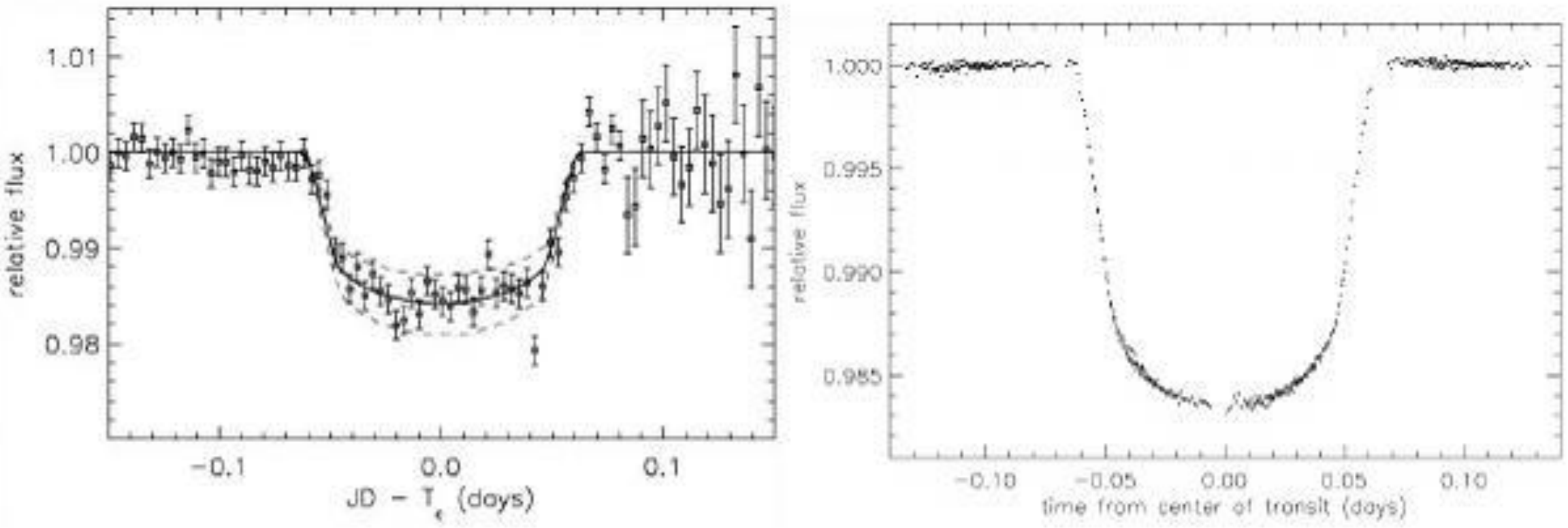

Figure 2.5: Left: the light curve of HD 209458, showing the first observed planetary transit [14]. Right: the same system observed by HST/STIS [19].

### 2.2.1 Principles

The transit method consists in detecting the shallow dip in a stellar light curve when a planet crosses the line of sight towards its host star during its revolution. It thus requires an almost perfect alignment between the observer, the planet, and the star. The transit appears periodically, with a period equal to the revolution period of the planet. The probability $P_{tr}$ for a planetary system to show a transit is a direct relation to the star radius R and the semi-major axis a: $P_{tr} = R/a$.

Applying this equation to the solar system, it follows that the Earth in front of the Sun has a 1/214 probability to produce a transit for a distant observer; for Jupiter the value is 1/1100. With the discovery of the "hot Jupiter" class of planets, the prospects of detecting exoplanets by looking for transits improved dramatically. For 51 Peg b, the transit probability is about 10% and transit events recur every ~ 4 days [16].

A planetary system observed edge-on, as in the case of transiting systems, also offers the geometry for the strongest radial velocity (RV) signal to be observed; moreover, the "sin *i*" indetermination for the mass disappears, so studying the planet transit light curve we can determine its physical parameters: radius, orbital inclination *i*, density, surface gravity, orbital distance *a*.

### 2.2.2 Measured parameters

The planetary transit measured in the stellar light curve is mainly described by three parameters: **depth**, **duration**, **shape**. Depending on the latitude of the transit on the stellar disk, the transit light curve will be U-shaped (central occultation) or V-shaped (grazing occultation). Quantitatively, the related parameter is the duration of the ingress and egress (alternatively, the duration of the flat bottom of the transit) [16].

Let us calculate these parameters in the simplified case of a circular orbit and a stellar disc of uniform brightness. The sketch of a planetary transit is given in Figure 2.6. Ingress (resp. egress) is defined as the phase from contact 1 (resp. 3) to contact 2 (resp. 4). The "flat bottom" corresponds to phases 2 to 3.

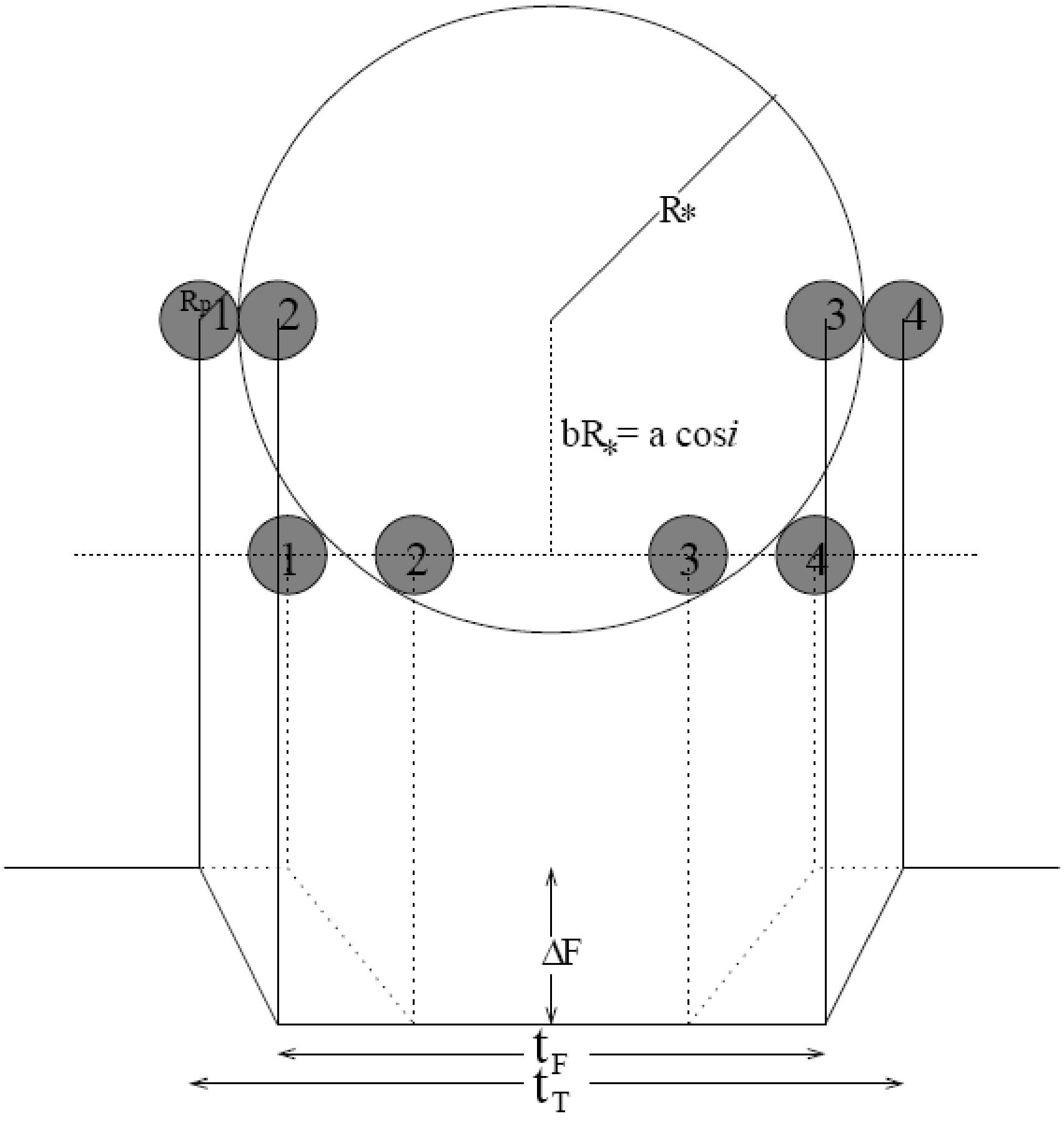

Figure 2.6: Definition of transit light-curve observables. Two schematic light curves are shown on the bottom (solid and dotted lines), and the corresponding geometry of the star and planet is shown on the top. Indicated on the solid light curve are the transit depth ΔF, the total transit duration $t_T$, and the transit duration between ingress and egress $t_F$ (i.e., the "flat part " of the transit light curve when the planet is fully superimposed on the parent star). The planet is shown from first to fourth contact. Also defined are $R_*$, $R_p$, and impact parameter b corresponding to orbital inclination *i* [20].

*Transit depth.* The depth of the transit is related to the star and planet radii (R and r respectively):

$$\Delta F = \frac{F_{off} - F_{on}}{F_{off}} = \left(\frac{r}{R}\right)^2 \tag{2.2}$$

$F_{off}$ is the observed stellar flux out of transit, and $F_{on}$ is the observed flux during transit. This formula neglects the phenomenon known as limb darkening, i.e. the fact that stars appear slightly brighter in their centre than near the edge. Taking limb darkening into account makes the transits slightly deeper than $(r/R)^2$, and gives the light curve a more rounded shape (see Fig 2.7 and par. 2.2.3).

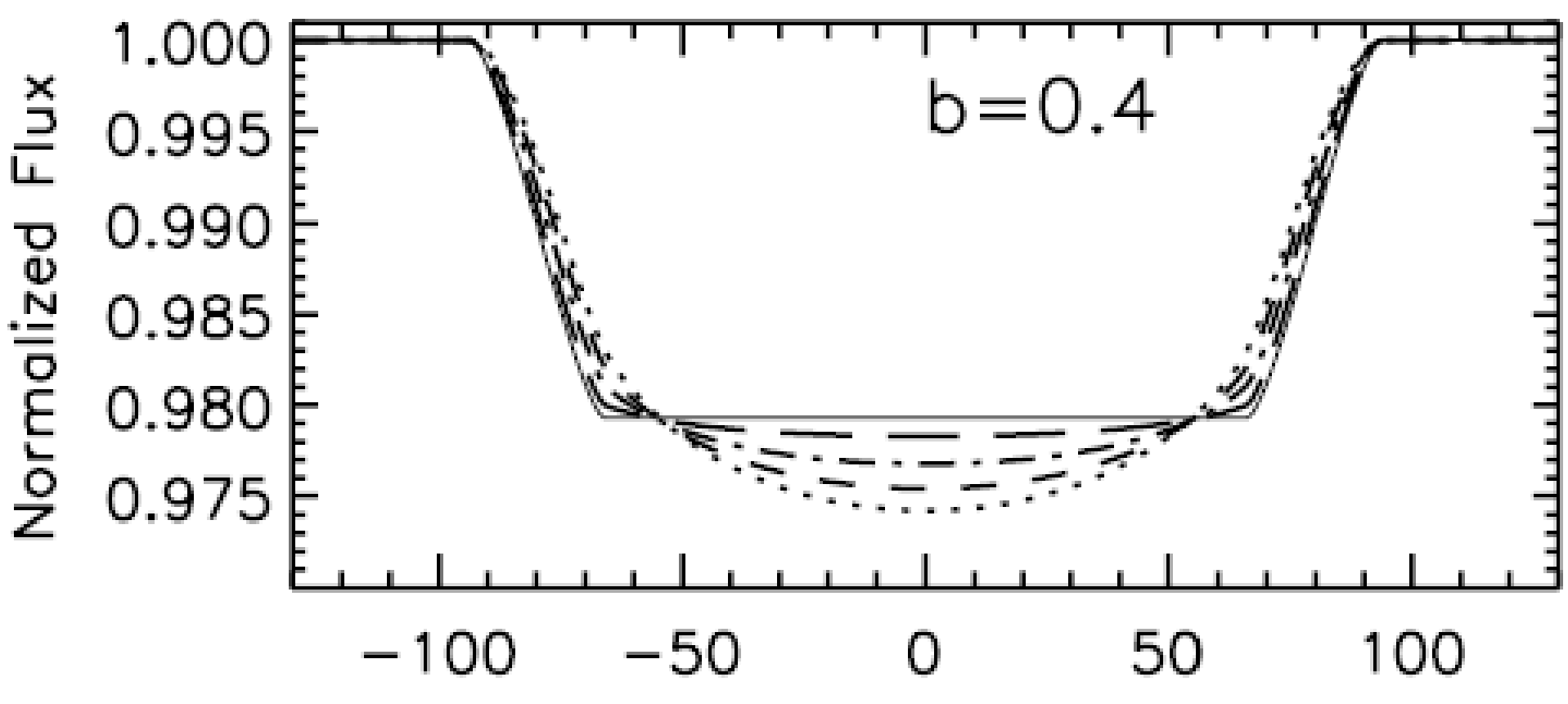

Figure 2.7: Solar limb-darkening dependence of a planet transit light curve. The limb-darkening parameters are from [21]. In these theoretical light curves the planet has $R_p = 1.4\ R_J$ and a = 0.05 AU and the star has $R_* = R_\odot$ and $M_* = M_\odot$. The solid curve shows a transit light curve with limb darkening neglected. The other planet transit light curves have solar limb darkening at wavelengths 3, 0.8, 0.55 and 0.45 μm [20].

*Transit duration.* The total duration of the transit, for a circular orbit, is related to the orbital parameters and to the star radius [20]:

$$d \approx \frac{PR}{\pi a}\sqrt{\left(1+\frac{r}{R}\right)^2 - \left(\frac{a}{R}\cos i\right)^2} \qquad (2.3)$$

Parameter P is the orbital period, in the same unit as d, and a is the orbital radius in the same unit as R and r; i is the inclination of the orbit.

The expression $b = \frac{a}{R}\cos i$ is the impact parameter and represents the projected distance of the planet's centre to the star's equator in units of star's radius. More user-friendly equations of (2.3) are (as expressed against P or *a*):

$$d \simeq 13.0\sqrt{(1-b^2)}\frac{R}{M^{1/2}}a^{1/2} d \simeq 1.8\sqrt{(1-b^2)}\frac{R}{M^{1/3}}P^{1/3}$$

(2.4)

where M is the star's mass, the mass of the planet being negligible. Contrarily to Eq. 2.3, d is here in hours, P in days, a in astronomical unit, and R and M are both in solar units.

*Ingress duration*. Another temporal parameter of the transit is the duration of the ingress or egress:

$$t \approx d\frac{r}{R}\sqrt{1-b^2} \tag{2.5}$$

One may also consider the transit shape by deriving the ratio of the durations of the flat bottom ($t_F$) over the total transit (d):

$$\left(\frac{t_F}{d}\right)^2 = \frac{(1-\frac{r}{R})^2 - (\frac{a}{R}\cos i)^2}{(1+\frac{r}{R})^2 - (\frac{a}{R}\cos i)^2} \tag{2.6}$$

The three equations describing a planetary transit (depth, total duration and ingress duration), can be used to constrain four unknown parameters of the system: r, R, M, b. The star's mass and radius may be independently constrained by other observations, specifically high-resolution spectroscopy, as well as with stellar evolution models. For instance, for low-mass stars, $M \propto R$ is a fairly good approximation.

### 2.2.3 Limb darkening

Light curves taken in a bandpass blueward of ~1 μm for solar-like stars show noticeable limb darkening. Thus, including limb-darkening is important for computing accurate eclipse lightcurves. The general effect of limb darkening is to (1) change the depth of the light curve ΔF as a function of impact parameter (making the light curve deeper for most values of impact parameter), (2) make the flat bottom rounder (and hence the "flat part " shorter, thus reducing $t_F$), and (3) blur the boundary between ingress/egress and the flat bottom [20]. These effects are shown in Figure 2.7.

It is possible to incorporate limb darkening into the mathematical description of a planet transit light curve by parameterizing the transit shape by the slope of the ingress/egress instead of by $t_F$ and $t_T$. The ingress/egress slope is mainly dependent on the time it takes the planet to cross the stellar limb but is also affected by limb darkening. The transit depth ΔF is also dependent on limb darkening and orbital inclination.

The limb-darkening of main-sequence stars is represented by functions of $\mu = \cos\theta$, where $\theta$ is the angle between the normal to the stellar surface and the line of sight to the observer (Figure 2.8). Claret [34] has found that the most accurate limb-darkening functions are the quadratic law in μ and the "nonlinear" law which is a Taylor series to fourth order in $\mu^{1/2}$ (see Appendix B); the latter conserves flux to better than 0.05%. Mandel & Agol [35] have computed analytic functions for transit lightcurves for the quadratic and nonlinear limb-darkening laws which are very used by the community.

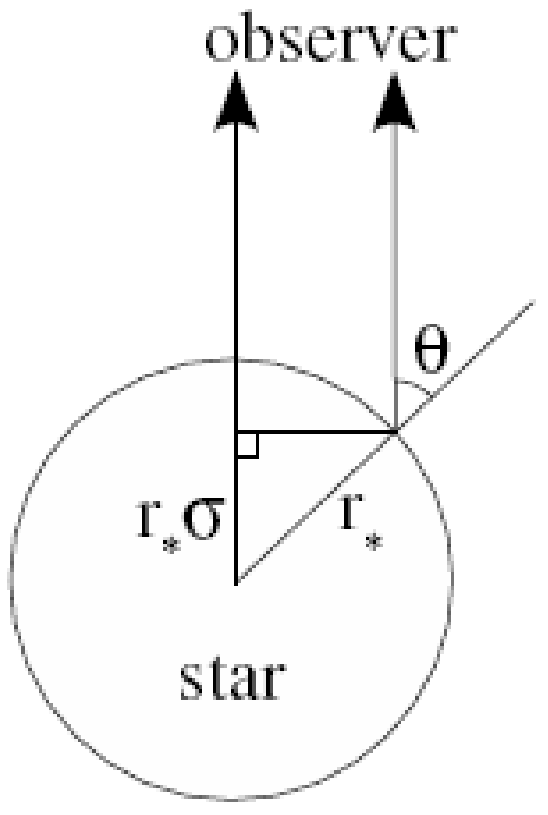

Figure 2.8: Geometry of limb-darkening. Star is seen edge on, with the observer off the top of the page. The star has radius $r_*$ and θ is defined as the angle between the observer and the normal to the stellar surface [35].

### *2.2.4 Elliptical orbits*

Due to the higher probability of detecting transits of short-period planets, which orbits are rapidly circularized by tidal effects, it is usually sufficient to consider the transit formalism for circular orbits, as given above. However, as more and more eccentric transiting planets are discovered, the need for a more general set of equations has become increasingly exigent. Here we present the model of Tingley & Sackett [22] which has been also developed by Kipping [23].

In this case, the transit duration (and shape) depends on the planet position on its orbits (hereafter, the phase angle φ, with respect to the periastron of the ellipse). Figure 2.9 shows how the transit duration evolves with phase angle and orbital eccentricity.

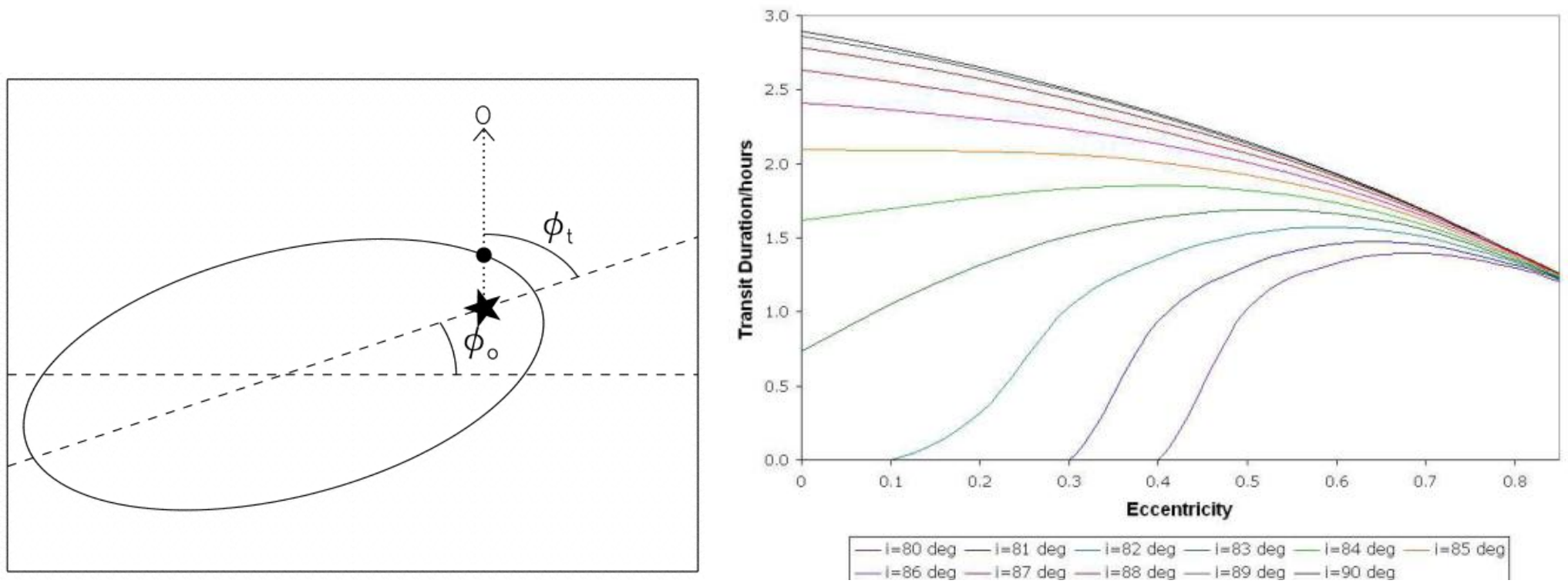

Figure 2.9: Left: orbital geometry showing the phase angle [22]. Right: calculation for d using Kipping's model for various e and i. A convergence occurs for very high e, making an accurate determination of i more difficult [23].

The transit duration in an elliptical orbit with eccentricity e is [22]:

$$d = 2\sqrt{\frac{1-(\rho\cos i)^2}{(R+r)^2}}(R+r)\frac{\sqrt{1-e^2}}{1+e\cos\phi}\left(\frac{P}{2\pi GM}\right)^{1/3} \tag{2.7}$$

where G is the gravitational constant and ϱ is the star-planet distance at the time of the transit, corresponding to a phase angle φ.

Another effect of an eccentricity greater than zero is to change the timing of the secondary eclipse. If $t_1 - t_2$ is the time interval between the primary and secondary eclipses, then we get [24]:

$$\frac{\pi}{2P}(t_1 - t_2 - P/2) \simeq e\cos\omega \tag{2.8}$$

The exact timing of a secondary eclipse may thus be derived if the orbit is precisely known, or alternatively, the observation of the secondary eclipse may provide an accurate estimate of the orbital eccentricity (as in [24]).

The most eccentric orbit discovered for a transiting planet is the one of HD 80606b which has also the longest period (111 days). Its transit egress has also been observed from University of London Observatory, London and has allowed a better constrain of its orbital parameters [149].

### 2.2.5 Measurement requirements

The principal requirement of transit measurements is high photometric and spectrophotometric precision, on relatively bright stars. This need for precision derives from the fact that transit measurements are made in the combined light of the star plus planet. Hence the planet signal is greatly diluted by the stellar photons, and the measurement precision must be as high as possible. Moreover, the time scale of the photometric noise is crucial. Since transits are typically a few hours in duration, the precision of individual measurements should improve as the inverse square-root of the measurement time, for times exceeding several hours. Since a photometric baseline is required before and after transit, a reasonable time scale for the required stability is ~20 hours. Since both visible and IR measurements are important for transits, the instrumentation should be designed to reach the fundamental limits determined by the stellar photon noise (primarily important in the visible) and the background noise caused by the thermal emission of the instrument, telescope, and zodiacal dust (primarily important in the IR).

For more extended ("around-the-orbit") science, the phenomena being measured have time scales of several days, so ~20 day stability is needed. Since there is also scientific motivation to study much longer-term changes on exoplanets, due for example to seasonal effects for transiting planets in longer period orbits, then eventually there may be motivation to extend the stability time to ~20 months.

Ground-based photometry has made great strides, and is achieving sub-milli-magnitude precision in favorable cases [25]. However, for most exoplanet characterization science, space-borne observations will be necessary. The best location for a space-borne transit mission is heliocentric orbit, or placement at a Lagrangian point. Although significant transit science can be done from near-Earth orbit, those orbits have two principal limitations: 1) long uninterrupted observing times, such as are needed for exoplanet around-the-orbit observations, are generally not possible from a near-Earth location, and 2) proximity to the Earth results in time-variable scattered light and thermal radiation that can interfere with achieving the necessary precision [2].

Anyhow it is important to set up suitable strategies of data analysis to discard false alarm detections, that can be caused by stellar effects like flares, spots, coronal effects or intrinsic stellar variation, as well as photometric binaries with grazing eclipses or whose image is blended with another constant star. For the case of ground based observations, attention must also be paid to atmospheric effects like air mass, absorption bands, seeing and scintillation. But transits may be caused by binary stars or brown dwarfs instead of exoplanets. All these are motivations for spectroscopic follow up observations, in order to confirm the real detection of a transiting planet [7].

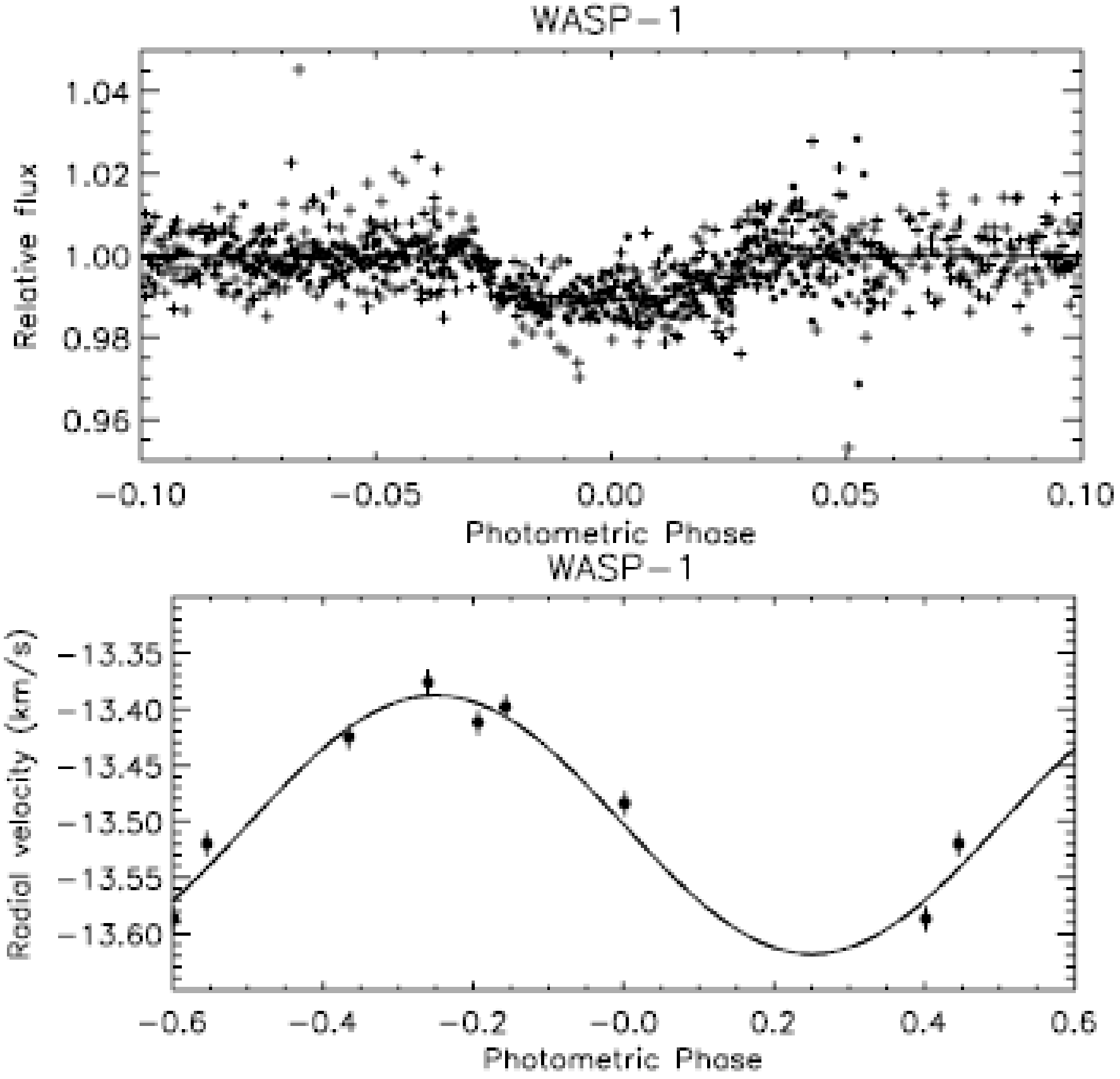

Figure 2.10: The phased light curve and radial-velocity curve of WASP-1b [26].

### *2.2.6 Transit missions and surveys*

New transit candidates are currently searched for in dedicated ground based photometric surveys and are the targets of space missions. Several projects have been successful in detection of exoplanets, among the others we remember:

- SuperWASP (Wide Angle Search for Planets) with 18 discoveries (personal communication) thanks to two robotic observatories, one in the Canaries and the other in South Africa.

- HATNet (Hungarian-made Automated Telescope Network) with 13 detections made by six small fully-automated telescopes in Arizona and Hawaii.

- OGLE (Optical Gravitational Lensing Experiment) has discovered 8 transiting planets using a 1.3 m telescope in Las Campanas, Chile.

- The XO project with 5 planets found with a network of telescopes led by the XO observatory in the Hawaii.

- TrES (Trans-atlantic Exoplanet Survey) with 4 discoveries made with a network of three small-aperture telescopes located in California, Arizona and Canary Islands.

Another extension of transit searches from the ground could be carried out in Antarctica, where the duty cycle is larger and skies potentially transparent and stable for high-accuracy photometry [27].

Here we describe the two space mission already operational: the CoRoT satellite and the Kepler mission.

- *CoRoT*

CoRoT (COnvection ROtation and planetary Transits) is a space mission led by the French Space Agency (CNES) in conjunction with the European Space Agency (ESA), Austria, Belgium, Germany, Spain and Brazil.

The probe hosts a 27 cm diameter off-axis afocal telescope and a focal box with an array of four CCD detectors protected against radiation by aluminum shielding 10mm thick. The satellite has a launch mass of 630 kg, is 4.10 m long and 1.98 m in diameter. It is powered by two solar wings which are rotated toward the Sun every 14 days. Moreover, it has a payload mass of 300 kg, the energy necessary to its operations is of 530 W, data transmission rate at ground is of 1.5 Gbits per day while it has a mass memory capacity of 2 Gbit [30].

The launch took place on 27 December 2006 when a Russian Soyuz 2-1b rocket lifted off from Baikonur Cosmodrome, Kazakistan. At the moment, CoRoT is moving into a circular polar orbit with an altitude of 896 km, this kind of orbit allows the space telescope to observe the same field of view continuously for 150 days. Zone observed are at less than 10° from the perpendicular to the orbit meaning there are no Earth occultations. Mission flight operations were originally scheduled to end 2.5 years from launch but to date flight operations have been extended to January 2010.

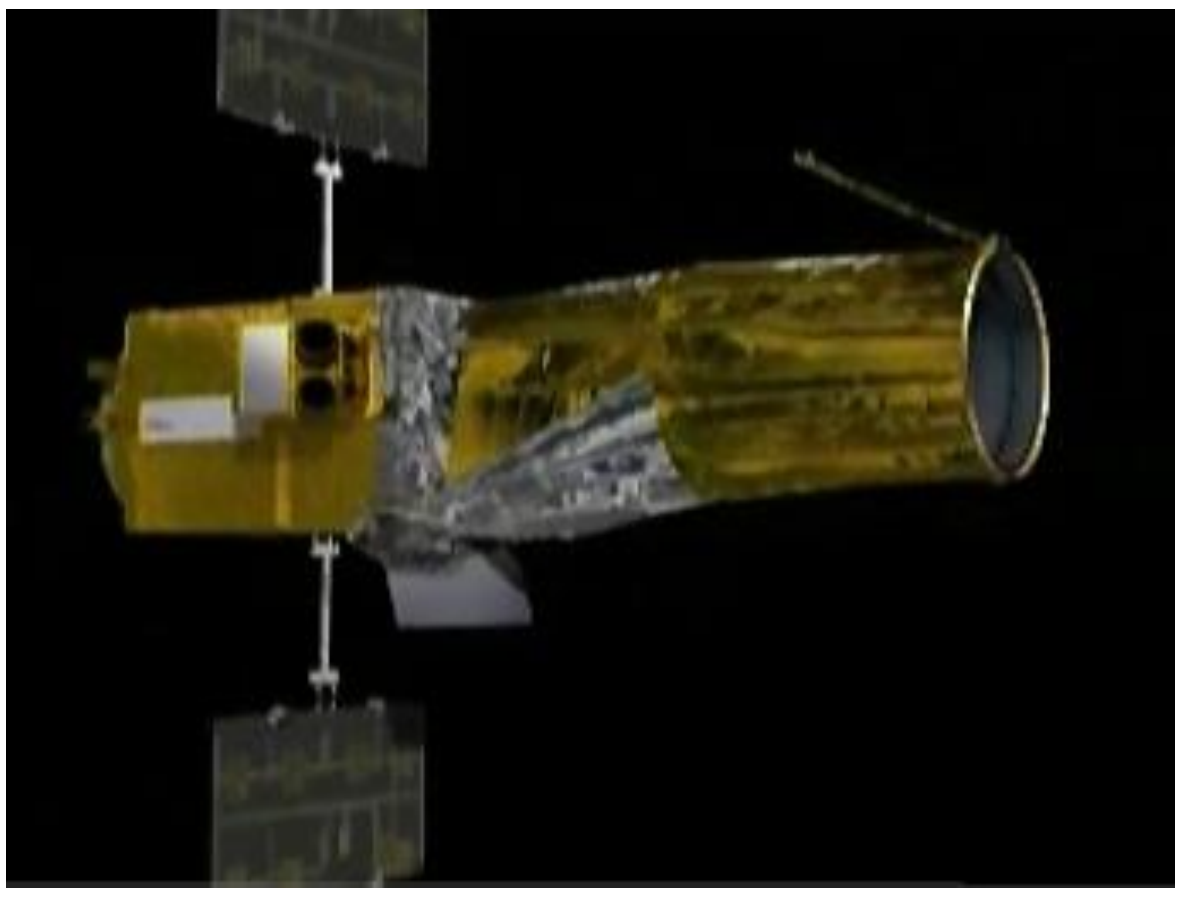

Figure 2.11: A view of the satellite.

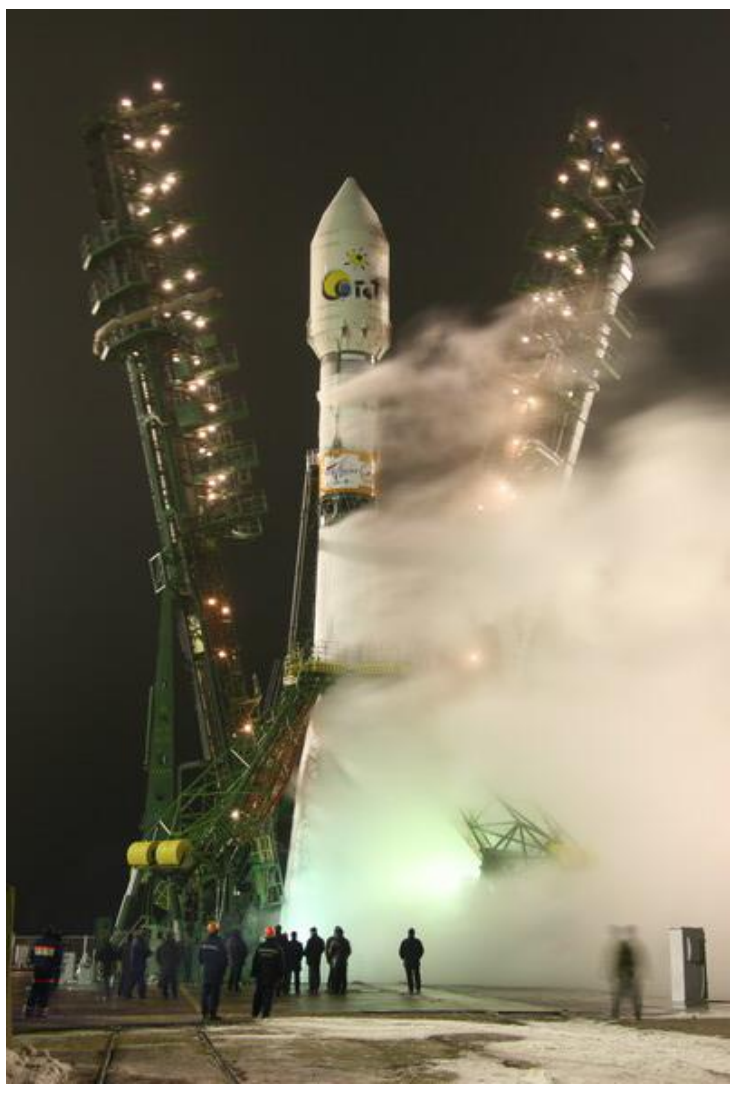

Figure 2.12: The launch.

Commissioning phase began on 18 January 2007 and the first scientific observations started on 3 February 2007. Since then, CoRoT has been conducting a photometric high-precision campaign to search for extrasolar planets with short orbital periods, particularly those of large terrestrial size, and to perform asteroseismology by measuring solar-like oscillations in stars.

For the extrasolar planet study, in each field of view 12000 stars with brightness between 11 and 16 mag are observed in the R band. This is because brighter stars (mag < 11) would saturate the CCDs while the fainter ones would not provide data with a sufficient resolution. At least 2-3 transits have to be detected by CoRoT in order to be sure of the discovery of a new exoplanet (only for giant planets a single transit can be enough) so almost all the detectable planets can have an orbital period shorter than 75 days.

During the northern summer the telescope observes in an area around Serpens Cauda, toward the galactic bulge, and during the winter it observes in Monoceros, towards the anticentre of the Milky Way. Fields of view have been chosen so that sunlight cannot interfere with measurements. During the remaining 30 days between the two main observation periods, CoRoT is pointed to 5 other patches of sky.

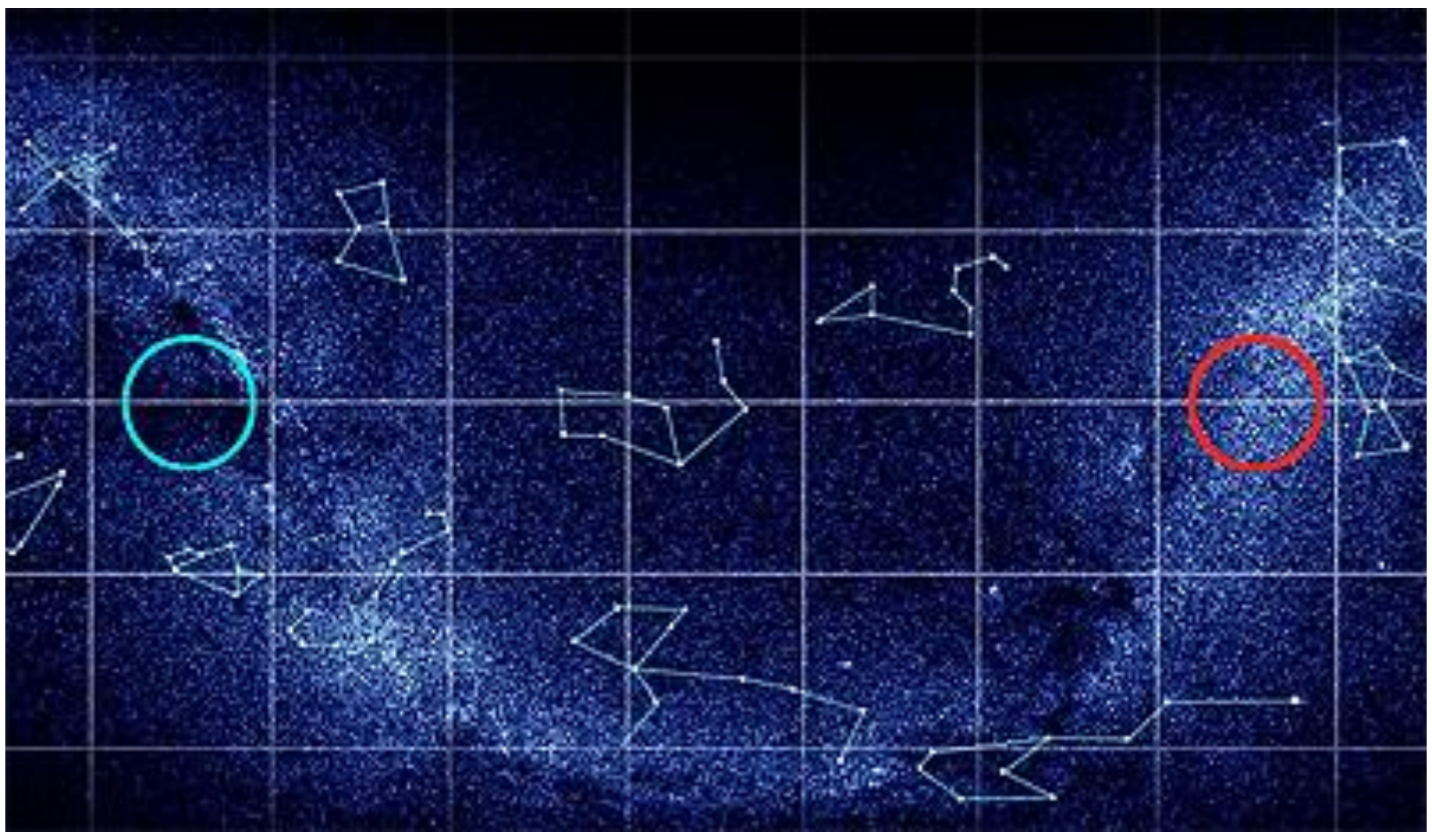

Figure 2.13: The two main fields of view of CoRoT.

Data already collected on ground shows that lightcurves with a precision of a part over 20,000 are obtainable. When many transits are observed, precision will be higher and as a consequence, small planets down to the size of Earth will be in the grasp of COROT [166].

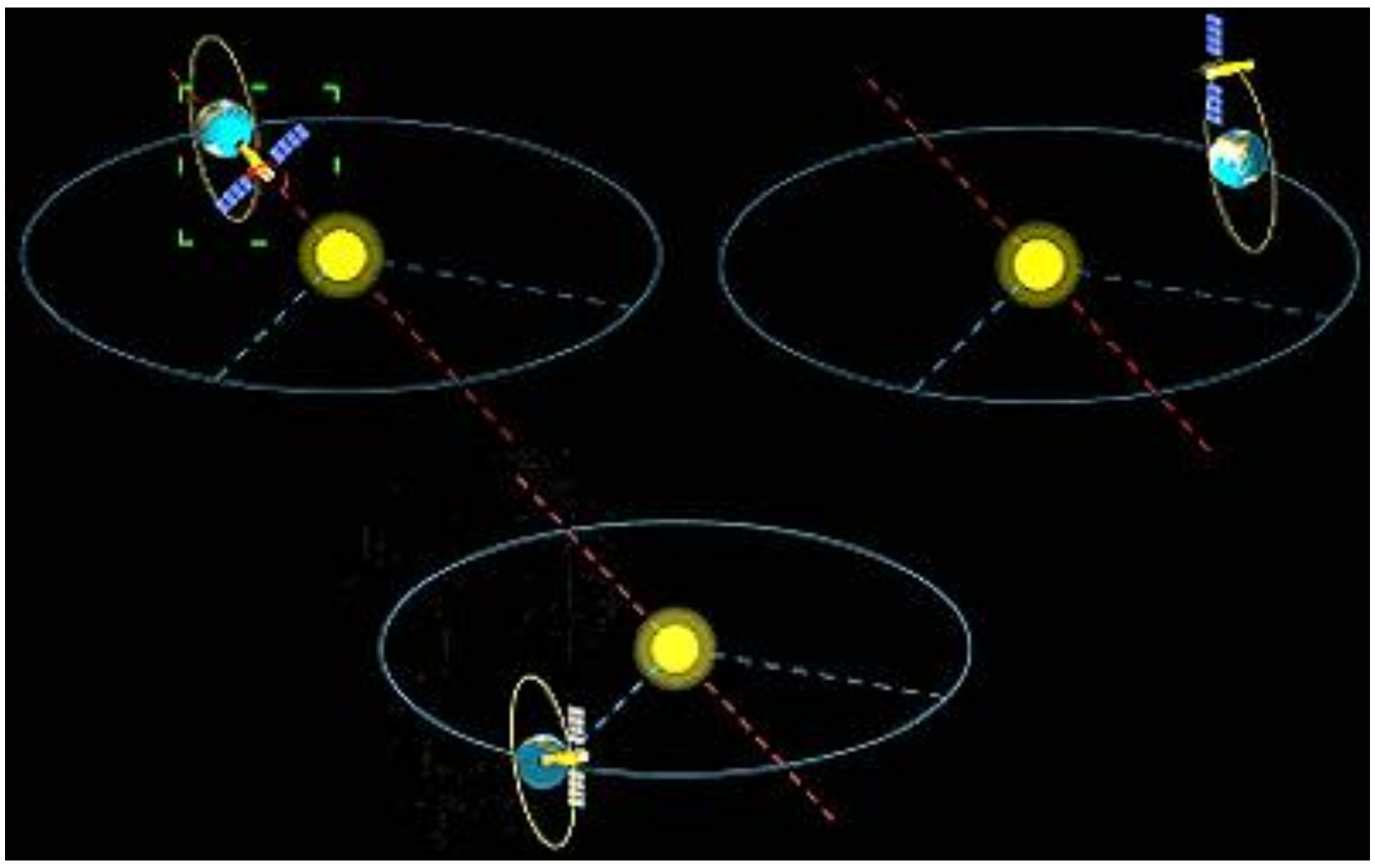

Figure 2.14: The orbit of CoRoT. When the sun begins to interfere with the measurements CoRoT rotates of 180° and starts observing the next area.

To date, CoRoT team has announced the detection of 7 exoplanets. Identification rate is constrained by the fact transit candidates need to be confirmed by radial-velocity follow-up observations (see par. 2.2.5). Moreover, these surveys enable the determination of the mass of the planet (see par. 2.2.1).

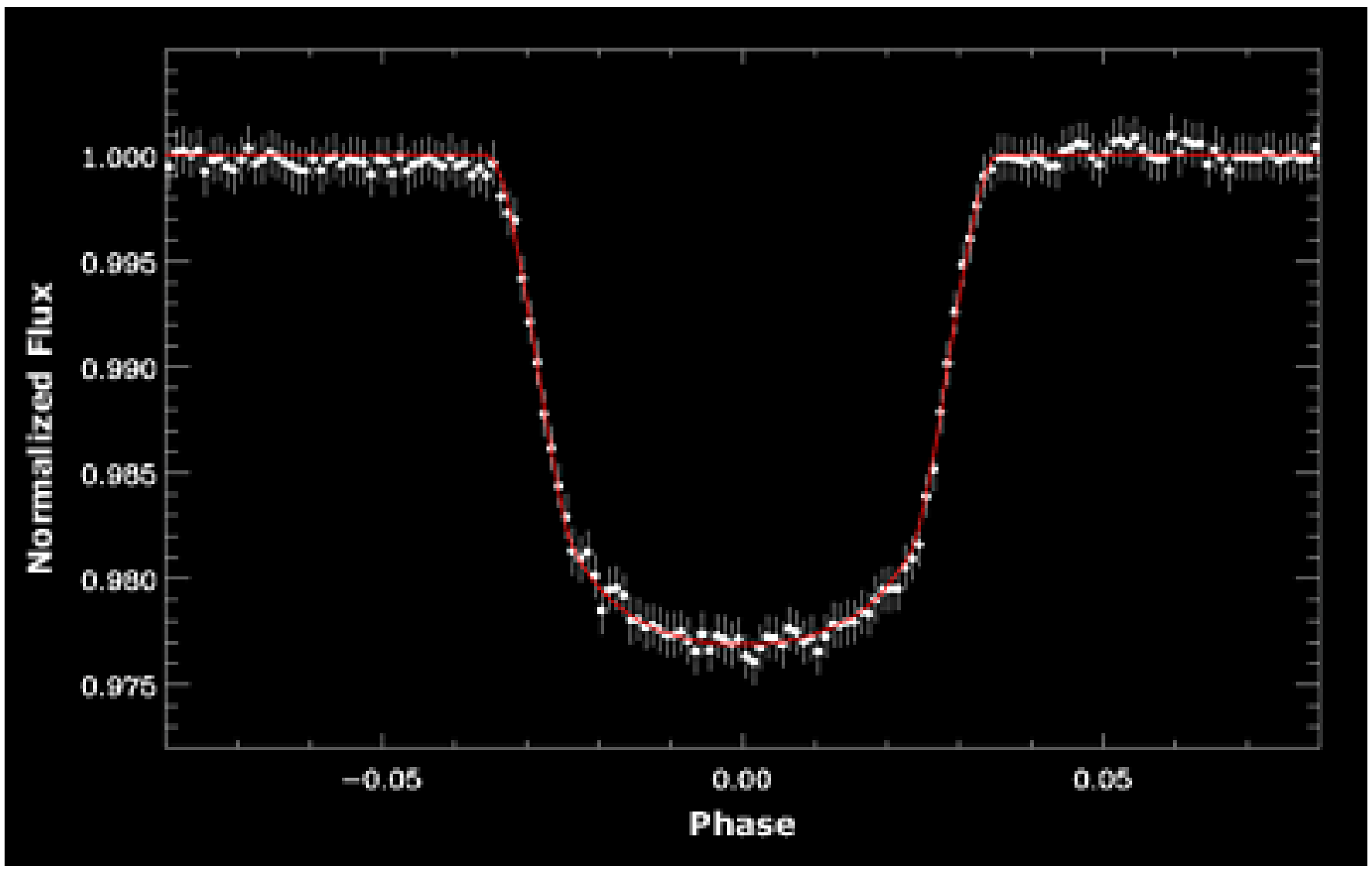

Figure 2.15: The transit light curve of CoRoT-1 b.

- *Kepler*

Kepler is a Discovery-class mission designed to explore the structure and diversity of planetary systems by surveying a large sample of stars.

The primary goal of Kepler is to survey our region of the Milky Way galaxy to discover hundreds of Earth-size or larger planets in or near the habitable zone of solar-like stars and get statistical data on the population and distribution of such planets.

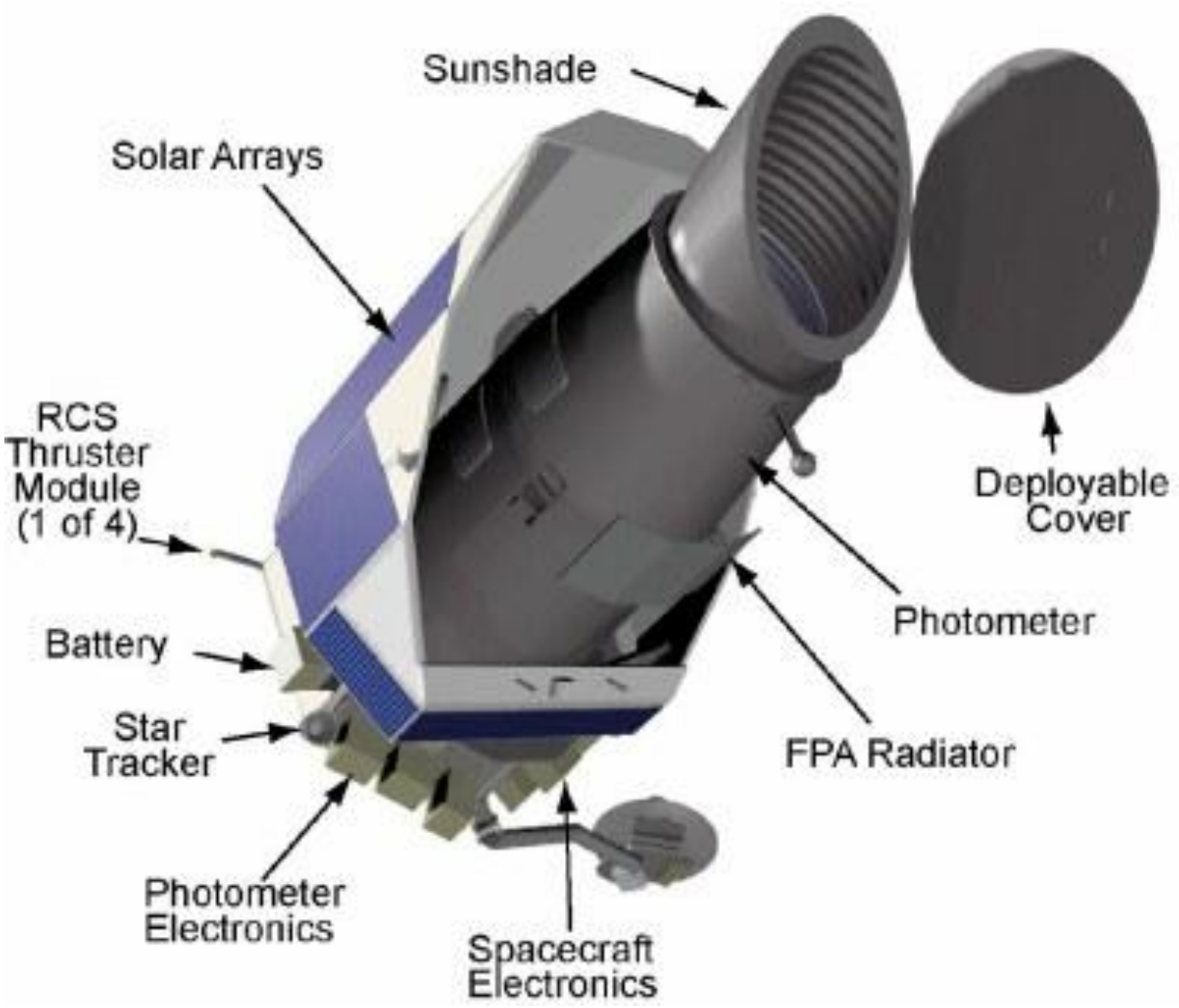

Figure 2.16: Kepler spacecraft.

The satellite hosts a photometer which has a 0.95-meter aperture Schmidt type telescope with a 1.4-meter primary mirror. This aperture is sufficient to reduce the Poisson noise to the level required to obtain a 4σ detection for a single transit from an Earth-size planet transiting a 12$^{th}$ magnitude G2 dwarf with a 6.5 hour transit [36]. Kepler's photometer has a very wide field of view for an astronomical telescope of about 15 degrees in diameter. The total mass at launch is 1052.4 kilograms consisting of 562.7-kilograms for the spacecraft, 478.0-kilograms for the photometer, and 11.7 kilograms of hydrazine propellant. The overall size is about 2.7 meters in diameter and 4.7 meters high. The array is cooled by heat pipes connected to an external radiator. Data from the CCDs are extracted every six seconds to limit saturation and added on board to form a 15-minute sum for each pixel. Data for a subset of target stars can be measured at a cadence of once per minute. This option will be exercised for detecting changes in transit timing due to the presence of multiple planets or moons [36]. Only the pixels of interest from each of the target stars are stored and telemetered to the ground. In fact, the Kepler spacecraft conducts its own partial analysis on board and only transmits scientific data deemed necessary to the mission in order to conserve bandwidth. NASA contacts the spacecraft using the X band communication link twice a week for command and status updates. Scientific data is downloaded once a month using the $K_a$ band link at a maximum data transfer rate of 4.33 Mb/s. Every three months, the spacecraft is also rotated of 90° about the optical axis to maintain the maximum exposure on the solar array (10.2 $m^2$ of collecting surface area which can produce over 1100 W) and to ensure the spacecraft's radiator is pointing towards deep space. The mission's life-cycle cost is estimated at US$600 million, including funding for 3.5 years of operation [31].

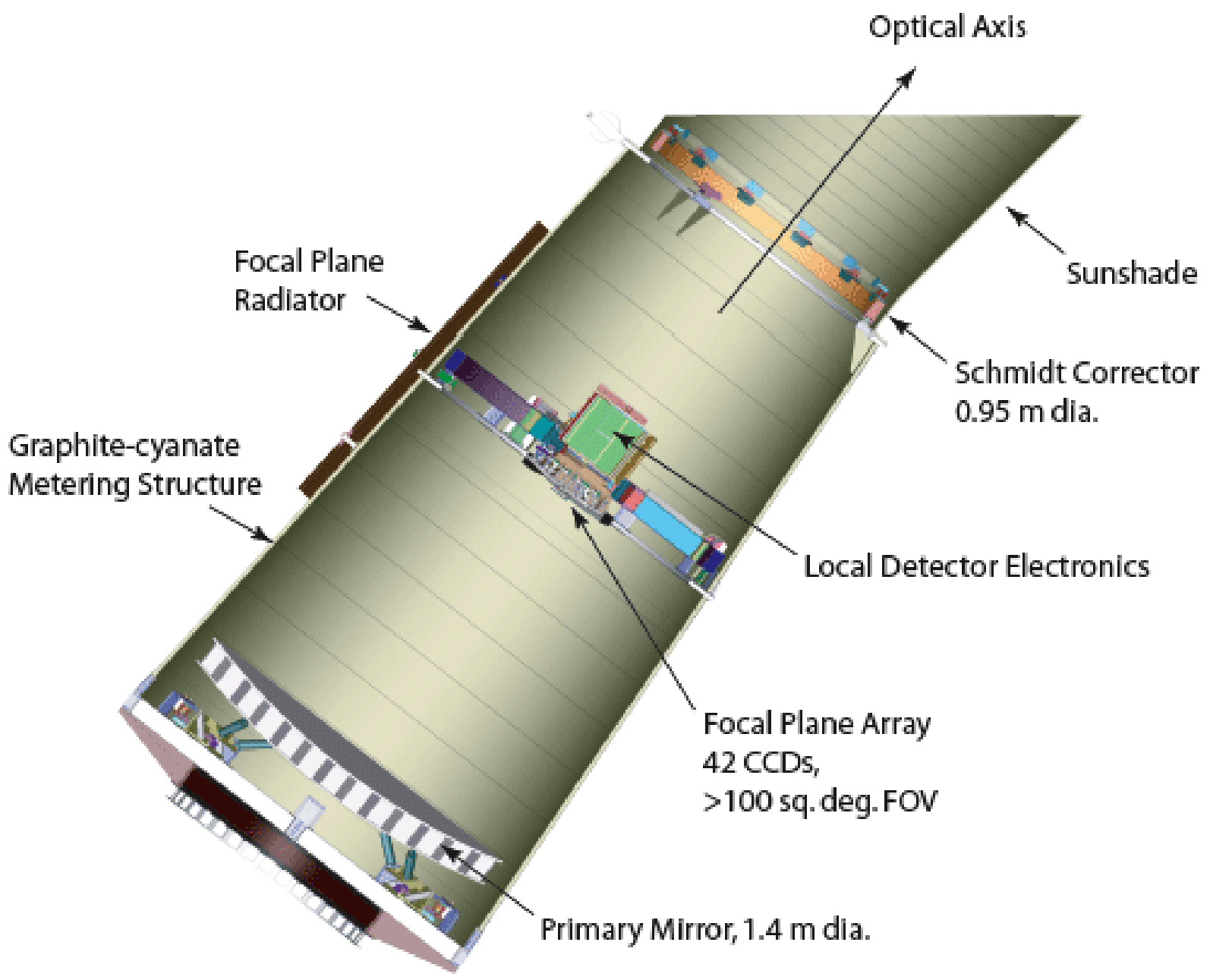

Figure 2.17: Kepler Photometer.

The observatory was launched on 7 March 2009 aboard a Delta II rocket from Cape Canaveral Air Force Station, Florida. Solid propellant has been utilized for the nine strap-on rocket motors and for the third stage; kerosene and liquid oxygen for the first stage, hydrazine and nitrogen tetroxide for the second stage.

Kepler is in an Earth-trailing solar orbit (a = 1.013 AU, P = 371 d) so that Earth does not occlude the stars which are observed continuously and the photometer is not influenced by stray light from Earth. This orbit avoids gravitational perturbations and torques inherent in an Earth orbit, allowing for a more stable viewing platform. The mission is planned to last 3.5 years, with the potential to extend it an additional 2.5 years.

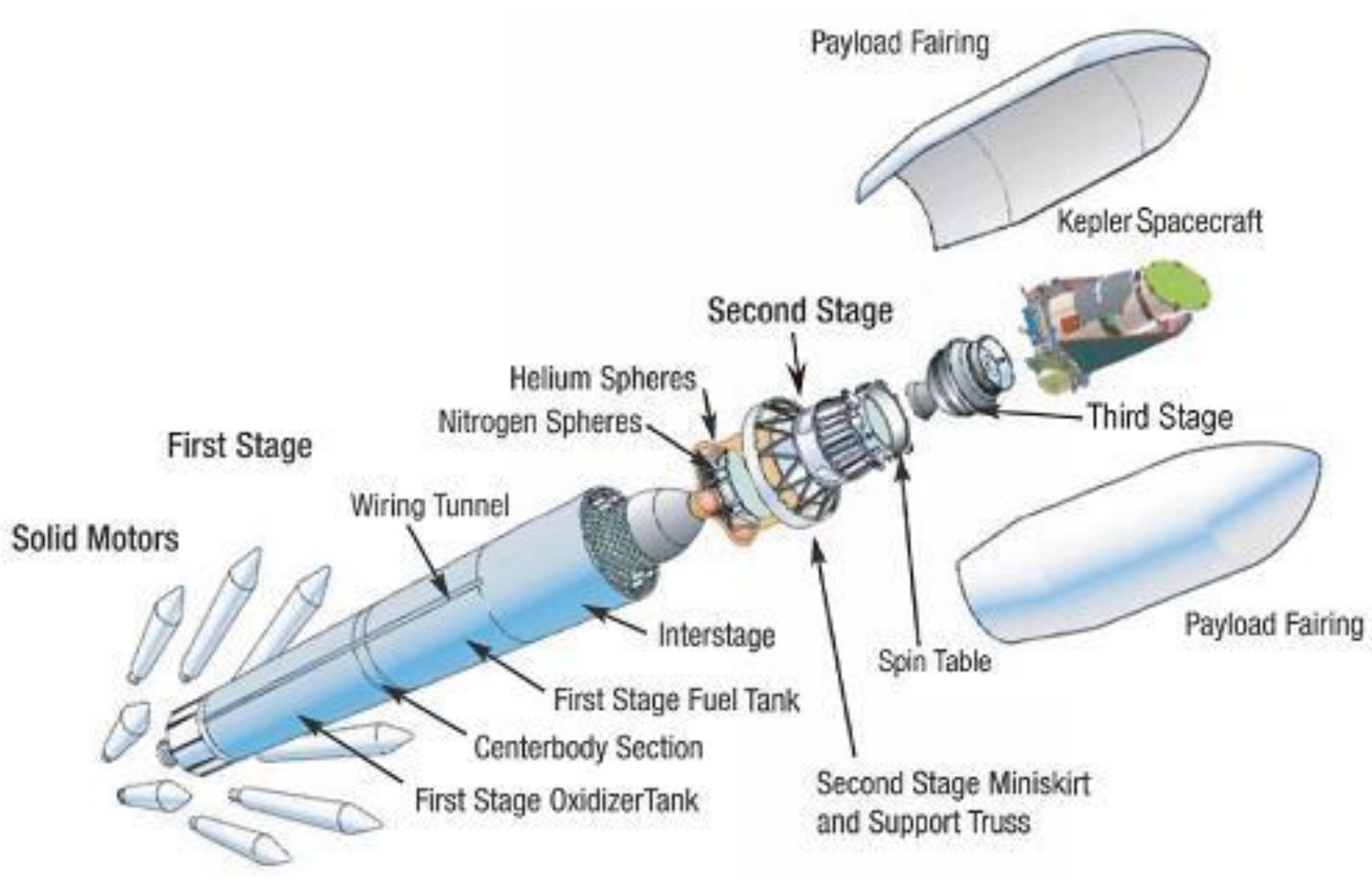

Figure 2.18: Delta Launch Vehicle with Kepler Spacecraft.

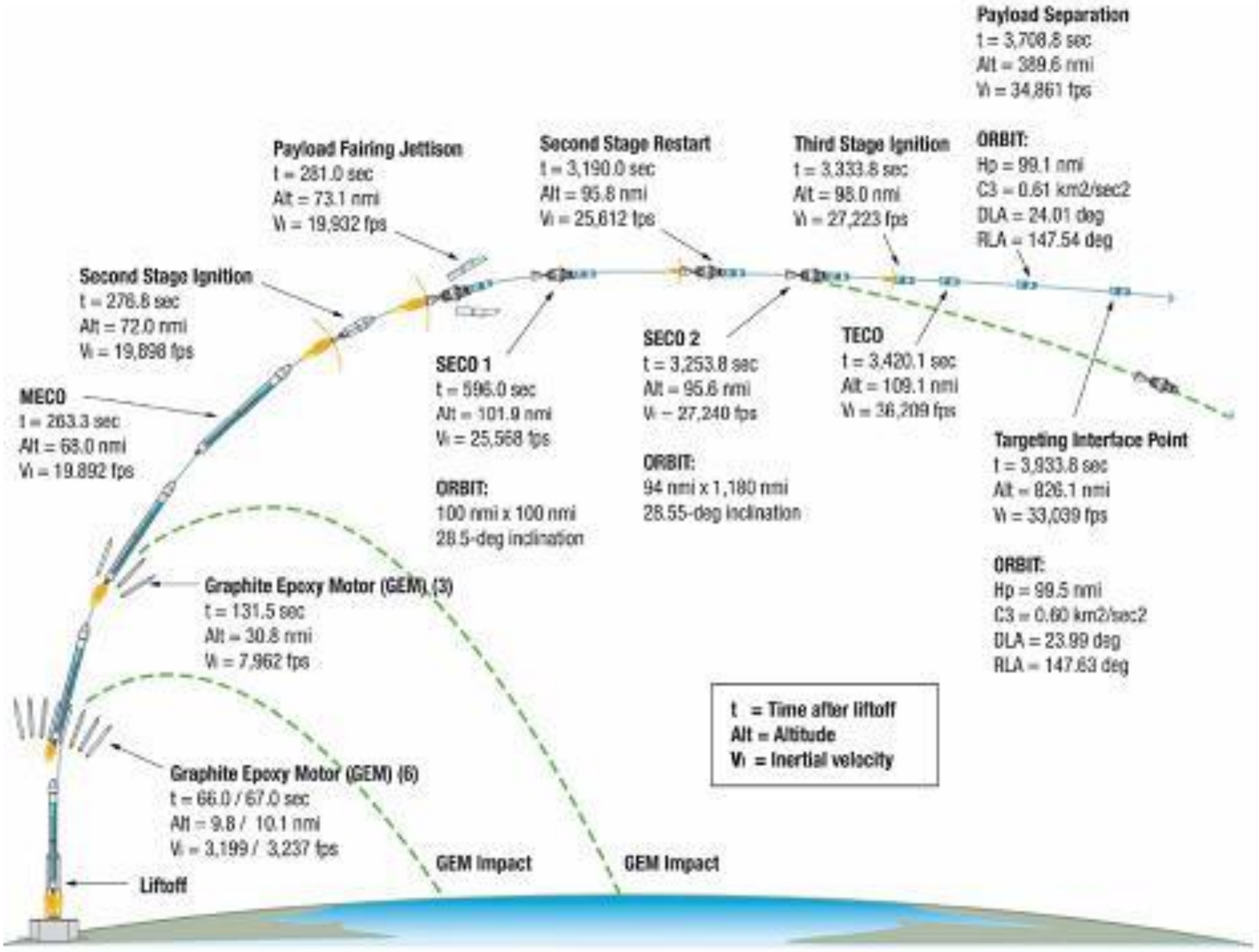

Figure 2.19: Kepler Launch Profile [31].

Commissioning phase lasted for 60 days after launch. Major activities during this period included initial acquisition of the spacecraft's signal and confirmation of a valid radio link with the ground, confirmation that the spacecraft is generating its own electrical power via its solar panels and has returned telemetry recorded during launch. This phase also included jettisoning of the photometer dust cover, checkout and calibration of the photometer, and fine-tuning of the spacecraft's guidance system. On 12 May 2009 Kepler began the scientific observations, while on 19 June 2009 the spacecraft sent its first science data to Earth.

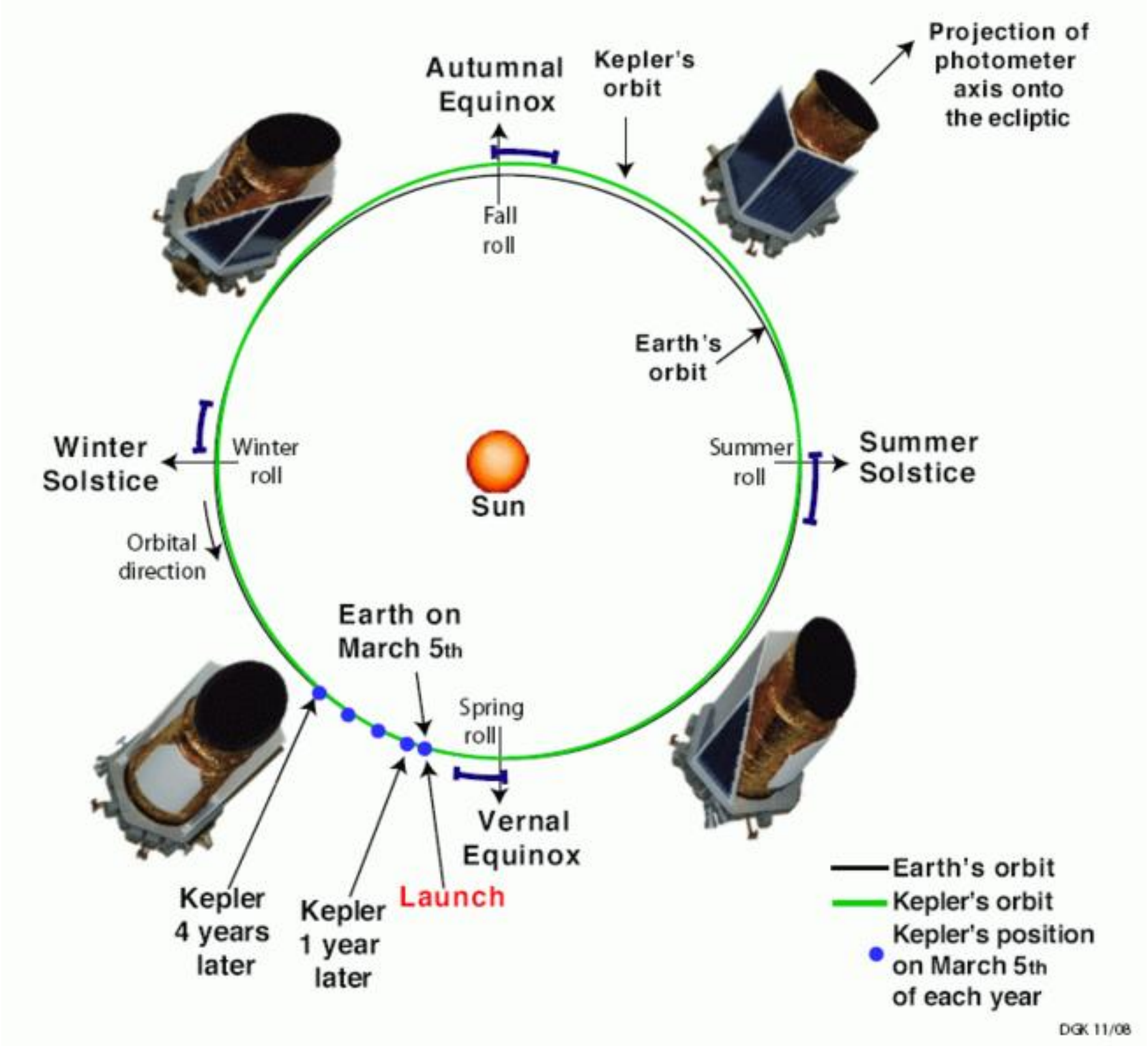

Figure 2.20: Kepler Orbit.

Kepler is designed to continuously monitor the brightness of more than $10^5$ F through M dwarfs (of magnitude from $9^{th}$ to $15^{th}$ and of low variability) in the Milky Way galaxy. Kepler must measure at least three transits that show that the orbital period repeats to a precision of at least 10 ppm and that shows at least a 7σ detection to classify a signal as a valid planetary candidate. A detection threshold of 7σ is required to avoid false positives due to random noise. Fig. 2.21 shows the minimum size planet required to produce an 8σ detection versus the amplitude of the stellar variability assuming that the frequency distribution of the stellar noise is the same as that of the Sun [36].

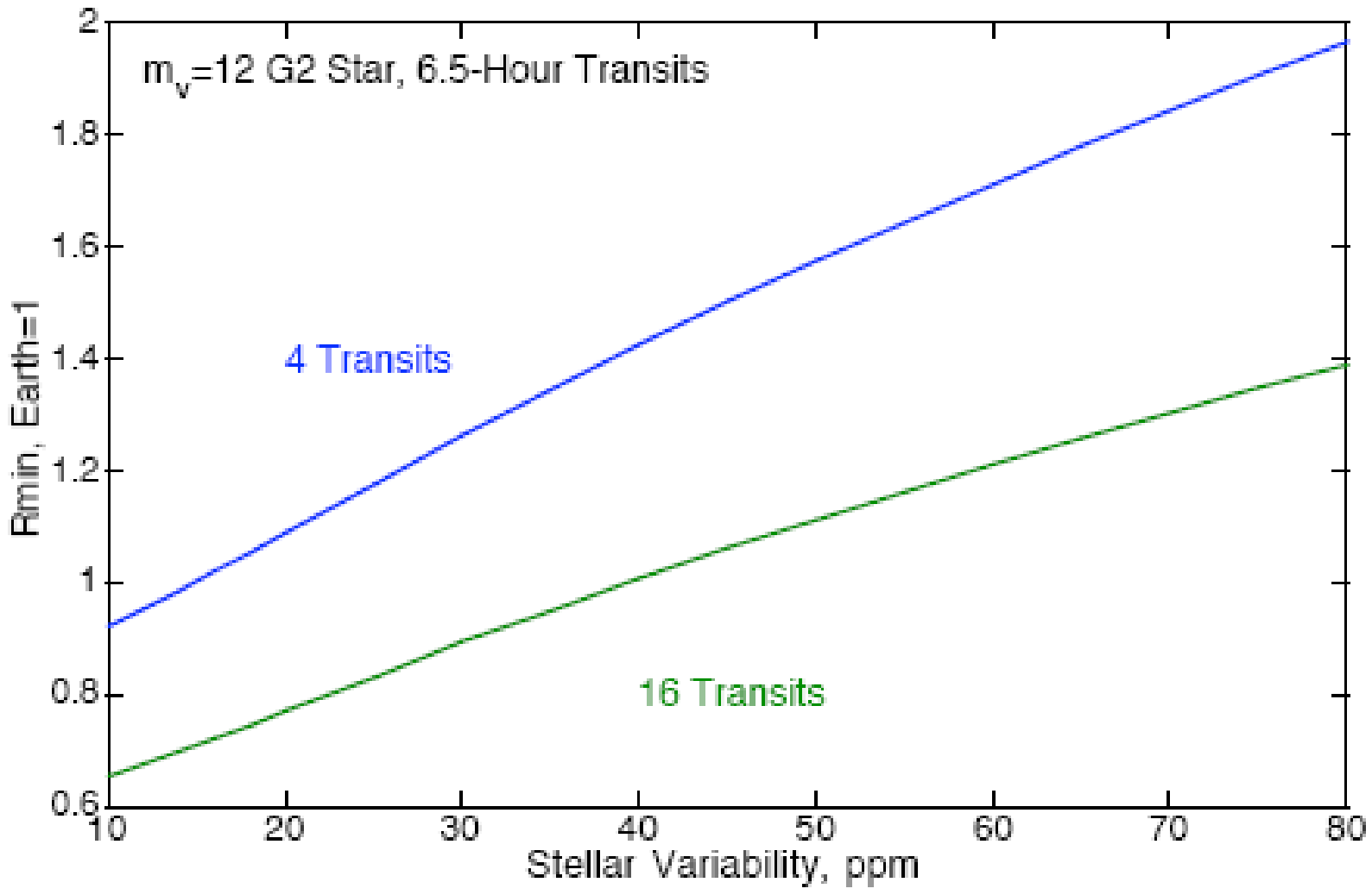

Figure 2.21: Effect of increased stellar variability on the minimum size planet that can be detected with 8σ [36].

Moreover, since larger planets give a signal that is easier to check, the first reported results are expected to be larger Jupiter sized planets in tight orbits. These could be reported after only a few months of operation. Smaller planets, and planets further from their sun will take longer, and discovering planets comparable to Earth is expected to take three years or longer. On 6 August 2009 it was already announced the detection of the previously known exoplanet HAT-P-7b. Measurements were so precise that it was able to detect a smooth rise and fall of the light caused by the changing phases of the planet. The photometric precision reached was in accordance with the expectations [99].

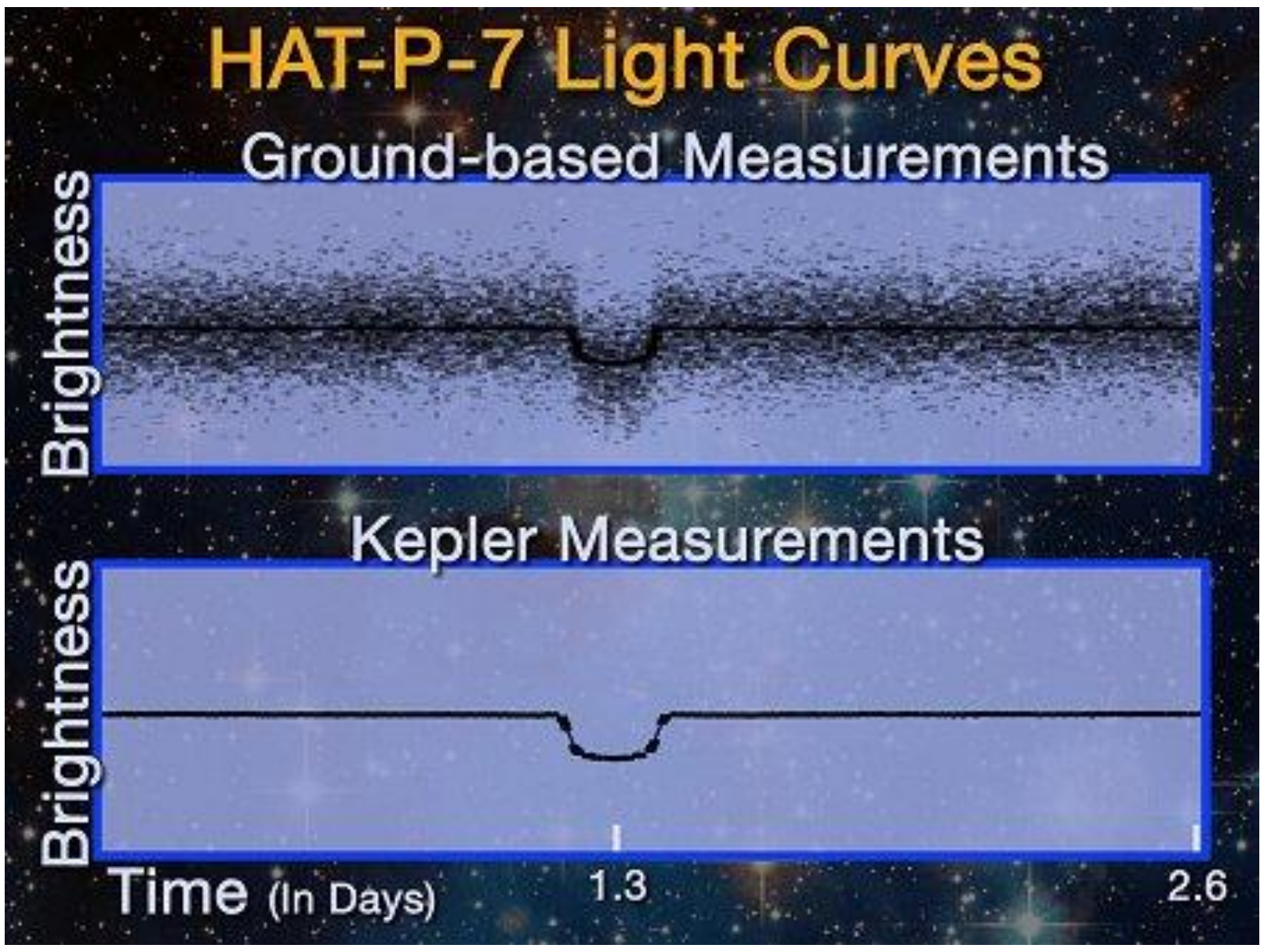

Figure 2.22: A comparison of ground-based and space-based light curves for hot exoplanet HAT-P-7b [99].

The photometer points to a field in the northern constellations of Cygnus, Lyra and Draco, which is well out of the ecliptic plane, so that sunlight never enters the photometer as the spacecraft orbits the Sun. Cygnus is also a good choice to observe because it will never be obscured by Kuiper belt objects or the asteroid belt. An additional benefit of that choice is that Kepler is pointing in the direction of the Solar System's motion around the centre of the galaxy. Thus, the stars which are observed by Kepler are roughly the same distance from the galaxy centre as the Solar System, and also close to the galactic plane. This fact is important if position in the galaxy is related to habitability.

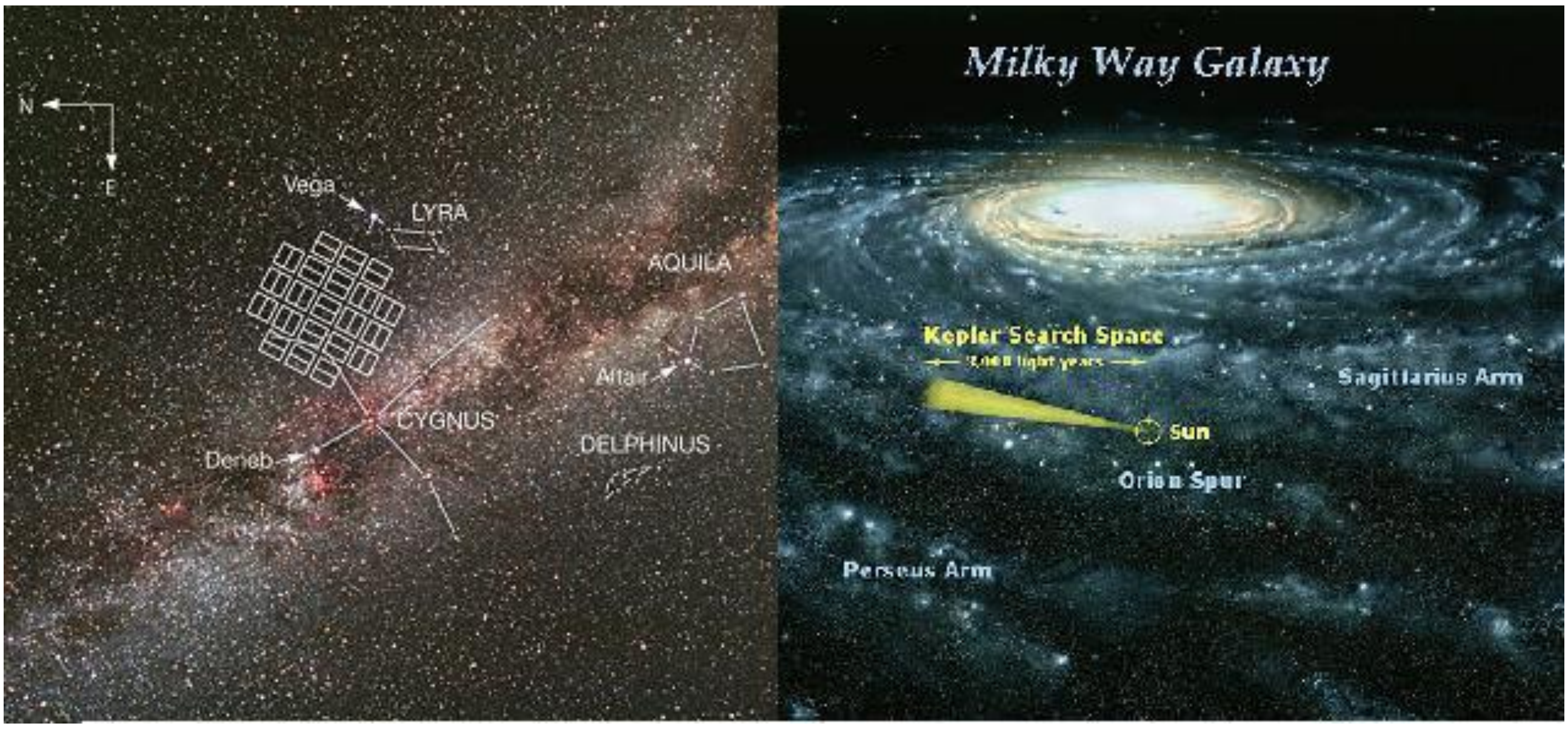

Figure 2.23: Left: The Kepler Field of View. Right: Kepler Mission search in context of the Milky Way.

Kepler has specifically developed technology that will likely be very beneficial to future transit missions. For transit measurements, mathematical techniques to perform high-precision data analysis (e.g., decorrelation techniques), and techniques to execute observations in an optimal manner, carry an importance on a par with, or exceeding, the development of new hardware. The Kepler mission has pioneered measurement techniques to obtain very high precision CCD photometry. These techniques include differential spatial and temporal photometry, and decorrelation methods to compensate for the effects of image motion. Kepler implemented a photometry testbed and demonstrated a photometric precision of $10^{-5}$ for CCD photometry in the laboratory prior to designing and building flight hardware [32].

Although Kepler looks for transiting exoplanets out to considerable distance, about 1000 parsecs, it has the limitation of looking towards only one direction in the galaxy, along the Orion arm. It is possible that terrestrial exoplanet statistics could be somewhat different in other regions of the galaxy, and some very interesting individual planets could be discovered by looking toward those regions.

We are most interested in finding planets relatively nearby to our Solar System because the consequent greater observed brightness of the star and planet permit more detailed characterization to be achieved, so an all-sky survey, albeit less deep than Kepler, is needed. Kepler observes stars in a field just over 100 square degrees in size (about 1/400 of the sky). An all-sky survey merely out to magnitude 12 should find at least eight times as many planets as Kepler does, and the ones found would, on average, be four times closer than the Kepler planets. The recent inference that Super Earths are common around solar-type and lower main-sequence stars [33] adds greatly to the motivation for an all-sky survey. An all-sky survey would find the closest transiting Super Earths or transiting Earth analogs. These results imply that there will be a large number of transiting Super Earths orbiting nearby solar-type stars, and the closest example is likely to be the planet that can be characterized to the highest signal-to-noise ratio [2].

## 2.3 Microlensing

Microlensing is the only known method capable of discovering planets at truly great distances from the Earth. Whereas spectroscopy searches for planets in our immediate galactic neighborhood, up to 100 light years from Earth, and photometry can potentially detect planets at a distance of hundreds of light-years, microlensing can find planets orbiting stars near the centre of the galaxy, thousands of light-years away [168].

Microlensing's disproportionately large contribution stems from its ability to probe regions of parameter space that are currently inaccessible to other methods. In particular, microlensing is most sensitive to planets in the cold, outer regions of systems beyond the snow line, the point in the protoplanetary disk exterior to which the temperature is less than the condensation temperature of water (Lecar et al. [167];

Kennedy & Kenyon [38]). Giant planets are thought to form in the region immediately beyond this line, where the surface density of solids is highest and thus planet formation is most efficient. It is also thought that this region is likely to be the ultimate source of the water for habitable planets. In addition, microlensing is sensitive to very low-mass planets, potentially down to the mass of Mars. Finally, microlensing is the only method capable of detecting old, free-floating planets, hypothesized to be a common by-product of planet formation and evolution [2].

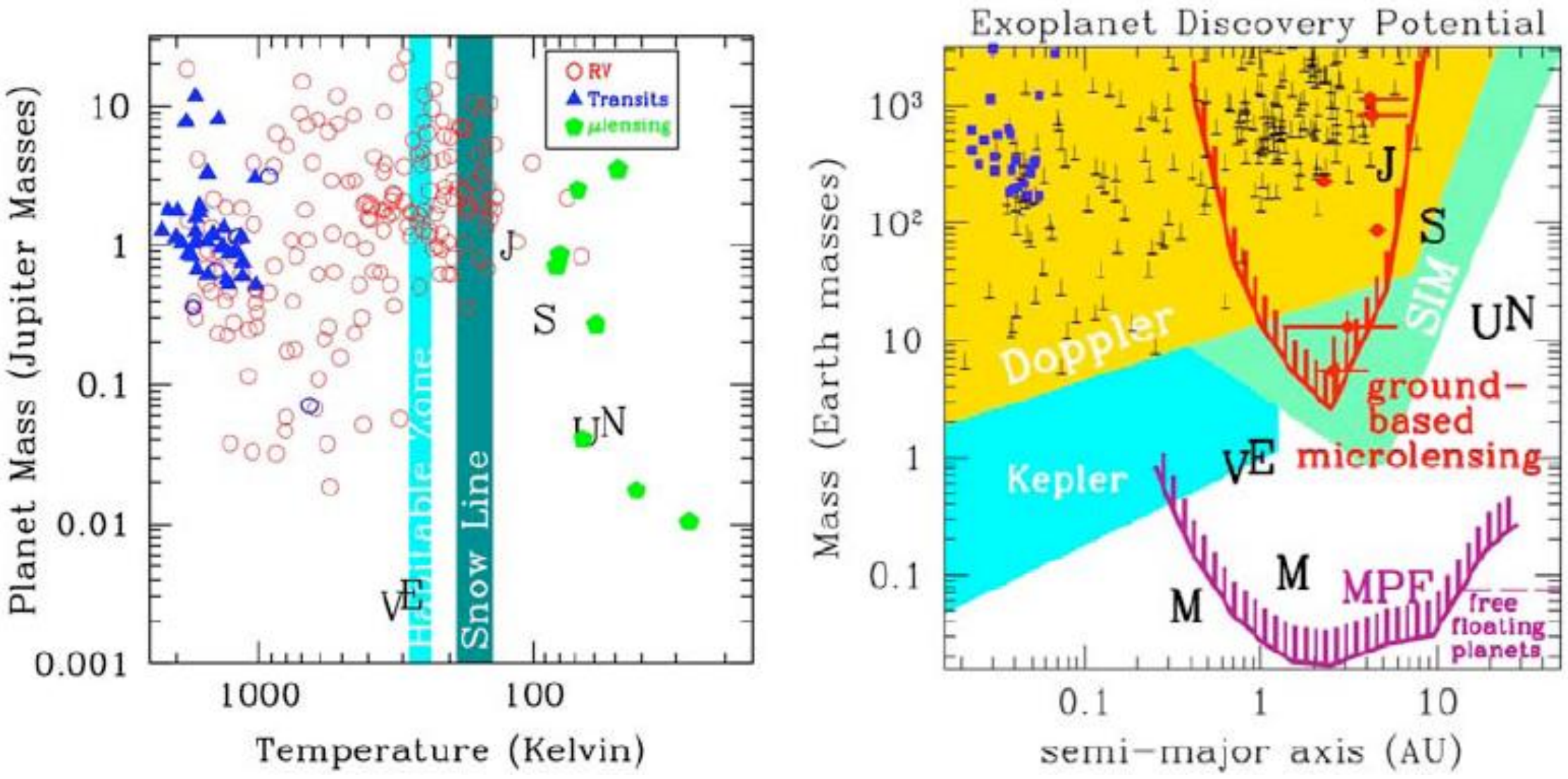

Figure 2.24: Left: Exoplanets detected via transits (triangles), RV (circles), and microlensing (pentagons), as a function of their mass and equilibrium temperature. Blue circles are planets found via RV that were subsequently found to be transiting. Also shown are the approximate locations of the habitable zone (e.g., Kasting et al. [37]) and the snow line (e.g., Kennedy and Kenyon [38]). Right: Groundbased (red) and space-based (purple) microlensing surveys are sensitive to planets above the curve in the mass vs. semi-major axis plane. The gold, green and cyan regions indicate the sensitivities of radial velocity surveys, SIM and *Kepler*, respectively. The location of our Solar System's planets and many exoplanets are indicated, with microlensing discoveries in red [2].

Microlensing is a remarkable astronomical effect, predicted by Einstein's General Theory of Relativity. According to Einstein, when the light emanating from a star passes very close to another star on its way to an observer on Earth, the gravity of the intermediary star will slightly bend the light rays from the source star, causing the two stars to appear farther apart than they normally would. This effect was used by Sir

Arthur Eddington in 1919 to provide the first empirical evidence for General Relativity.

If the source star is positioned not just close to the intermediary star when seen from Earth, but precisely behind it, this effect is doubled. Light rays from the source star pass on both sides of the intermediary, or "lensing" star. Since both light streams are bent by the lensing star's gravity, the source star should appear doubled from Earth -- once on each side of the lensing star (Fig. 2.25 3). In reality, even the most powerful Earth-bound telescope cannot resolve the separate images of the source star and the lensing star between them, seeing instead a single giant disk of light, known as the "Einstein disk," where a star had previously been. The Einstein radius is the angular radius of the Einstein ring in the event of perfect alignment. It depends on the lens mass M, the distance of the lens $d_L$, and the distance of the source $d_S$:

$$\theta_E = \sqrt{\frac{4GM}{c^2}\frac{d_S - d_L}{d_S d_L}} \quad \text{(in radians)} \tag{2.9}$$

For M equal to the mass of the Sun, $d_L$ = 4 kpc and $d_S$ = 8 kpc (typical for a Bulge microlensing event), the Einstein radius is 1 mas.

The resulting effect is a sudden dramatic increase in the brightness of the lensing star, by as much as 1,000 times. This typically lasts for a few weeks or months before the source star moves out of alignment with the lensing star and the brightness subsides. If a planet is positioned close enough to the lensing star so that it crosses one of the two light streams emanating from the source star, the planet's own gravity bends the light stream and temporarily produces a third image of the source star (Fig. 2.25 4). When measured from Earth, this effect appears as a temporary spike of brightness, lasting several hours to several days, superimposed upon the regular pattern of the microlensing event. For planet hunters, such spikes are the telltale signs of the presence of a planet. Furthermore, the precise characteristics of the microlensing light-curve, its intensity and length, tell a great deal about the planet itself. Its total mass, its orbit, and its period can all be deduced with a high degree of accuracy and probability from the microlensing event [1].

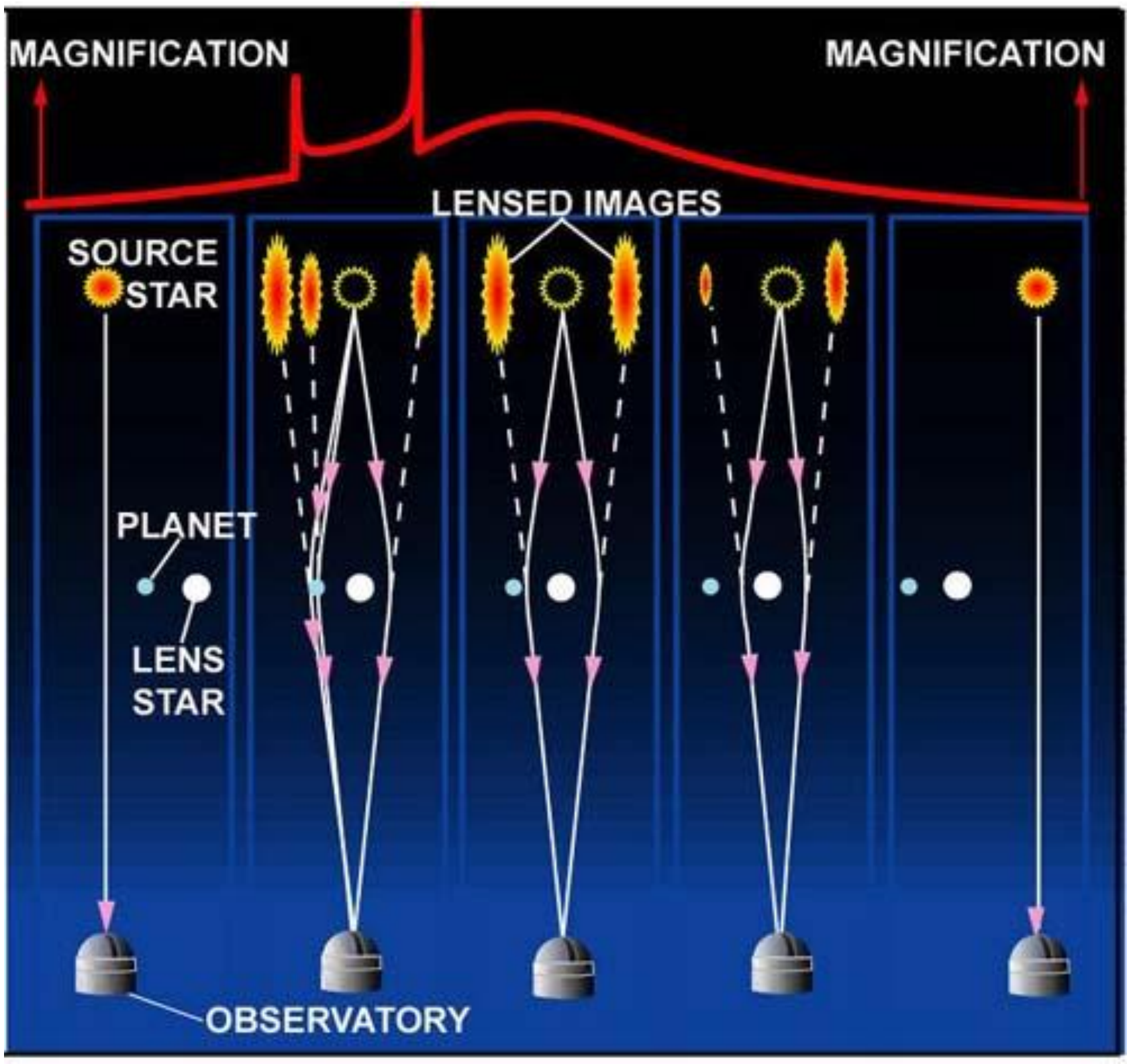

Figure 2.25: Planet detection through Microlensing. From right to left, the lensing star (white) moves in front of the source star (yellow) doubling its image and creating a microlensing event. In the fourth image from the right the planet adds its own microlensing effect, creating the two characteristic spikes in the light curve. Credit: OGLE

Microlensing is capable of finding the furthest and the smallest planets of any currently available method for detecting extrasolar planets. Anyway, these planets will never be observed again. This is because microlensing is dependent on rare and random events - the passage of one star precisely in front of another, as seen from Earth. This makes the discovery of planets by this method both difficult and unpredictable and forces to observe stellar zones with many sources, like our Galactic Bulge and the Large Magellanic Cloud, providing a high, not (strongly) biased, sample of stars. As a result, despite years of intense observations, to date only thirteen planets have been detected with microlensing [168] (out of which 9 have been published [3]).

Table 2.2: The Planets detected by microlensing [3].

| Planet | M ($M_{Jup}$) | R ($R_{Jup}$) | P (days) | a (AU) | e | i | Discov. |
|---|---|---|---|---|---|---|---|
| OGLE235-MOA53 b | 2.6 | - | - | 5.1 | - | - | 2004 |
| OGLE-05-071L b | 3.5 | - | ~ 3600 | 3.6 | - | - | 2005 |
| OGLE-05-169L b | 0.04 | - | 3300 | 2.8 | - | - | 2005 |
| OGLE-05-390L b | 0.017 | - | 3500 | 2.1 | - | - | 2005 |
| MOA-2007-BLG-192-L b | 0.01 | - | - | 0.62 | - | - | 2008 |
| MOA-2007-BLG-400-L b | 0.9 | - | - | 0.85 | - | - | 2008 |
| OGLE-06-109L b | 0.71 | - | 1825 | 2.3 | - | - | 2008 |
| OGLE-06-109L c | 0.27 | - | 5100 | 4.6 | 0.11 | 59 | 2008 |
| MOA-2008-BLG-310-L b | 0.23 | - | - | 1.25 | - | - | 2009 |

The detection of two cool, "Super Earth" planets among the first four planets suggests that these planets are common (Beaulieu et al. [39]; Gould et al. [40]). The detection of a Jupiter/Saturn analogue also suggests that Solar System analogues are probably not rare (Gaudi et al. [41]). The detection of a low-mass planetary companion to a brown-dwarf star suggests that such objects can form planetary systems similar to those around solar-type main-sequence stars (Bennett et al. [42]).

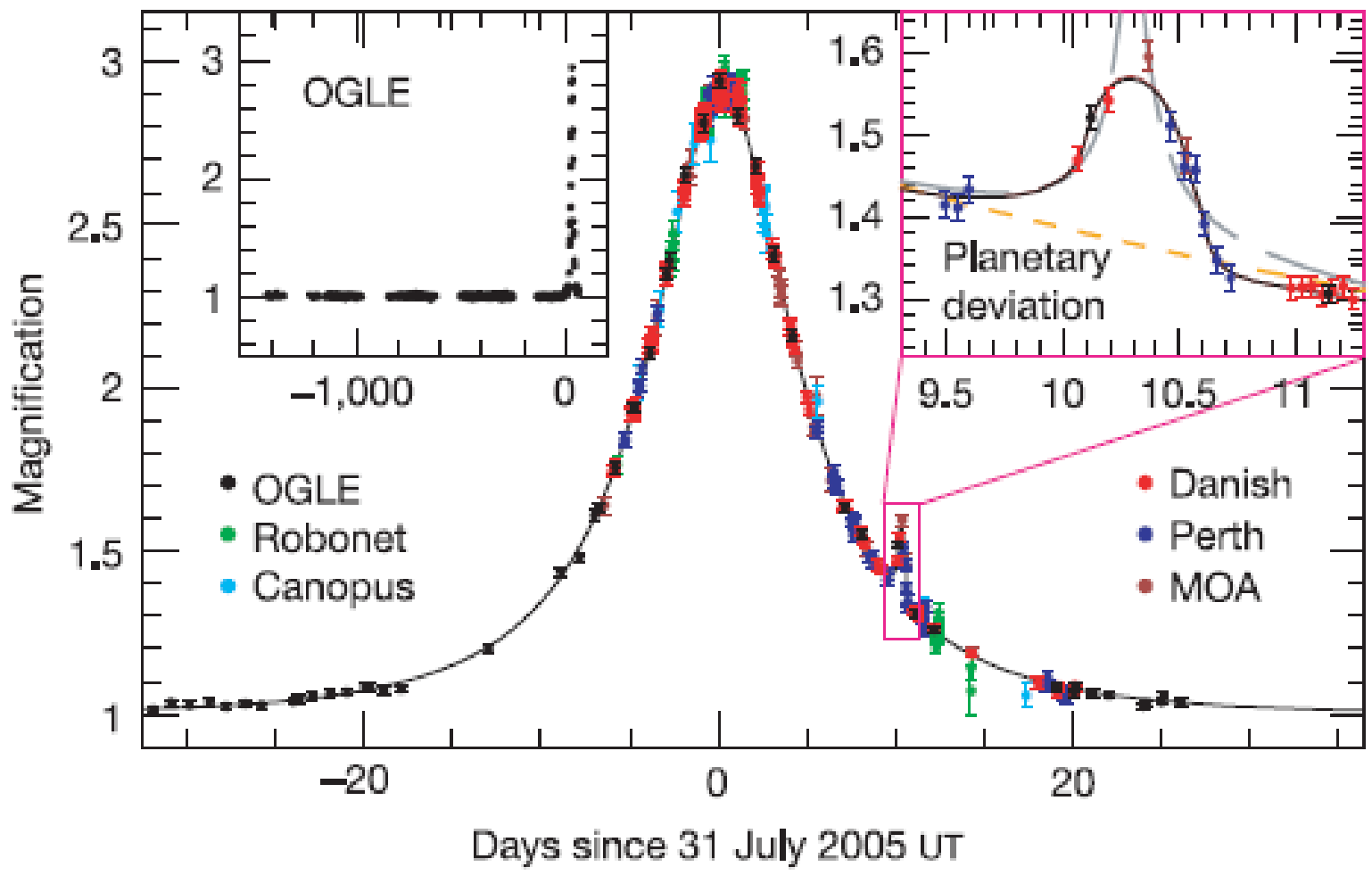

Figure 2.26: The observed light curve of the OGLE-2005-BLG-390 microlensing event and best-fit model plotted as a function of time [39].

Observations are usually performed using networks of robotic telescopes. The most active programs are the NASA/National Science Foundation-funded OGLE and MOA (Microlensing Observations in Astrophysics). MOA is a collaborative project between researchers in New Zealand and Japan. Observations are conducted at New Zealand's Mt. John Observatory using a 1.8 m reflector telescope built for the project.

OGLE (see also par. 2.2.6) is a Polish astronomical project based on observations performed in Las Campanas. Every night the "Warsaw" telescope is pointed toward the same dense field of 100 million stars in the vicinity of the galactic bulge, while the telescope's complex CCD cameras note any change in brightness of any point in the star-field. Every year OGLE detects around 500 microlensing events, but planet detections are extremely rare.

Whenever OGLE detects a microlensing event, it contacts a network of telescopes that specialize in searching for signs of the presence of a planet. The networks, known as PLANET (Probing Lensing Anomalies) and Robonet, include 1 and 2 meter telescopes in La Silla (Chile), Hobart (Tasmania, Australia), Perth (Australia), Boyden (South Africa), Sutherland (Australia), La Palma (Spain), and Haleakala (Hawaii). Together, the telescopes are able to continuously cover each microlensing event, providing an accurate light curve and indicating whether a planet is present or not.

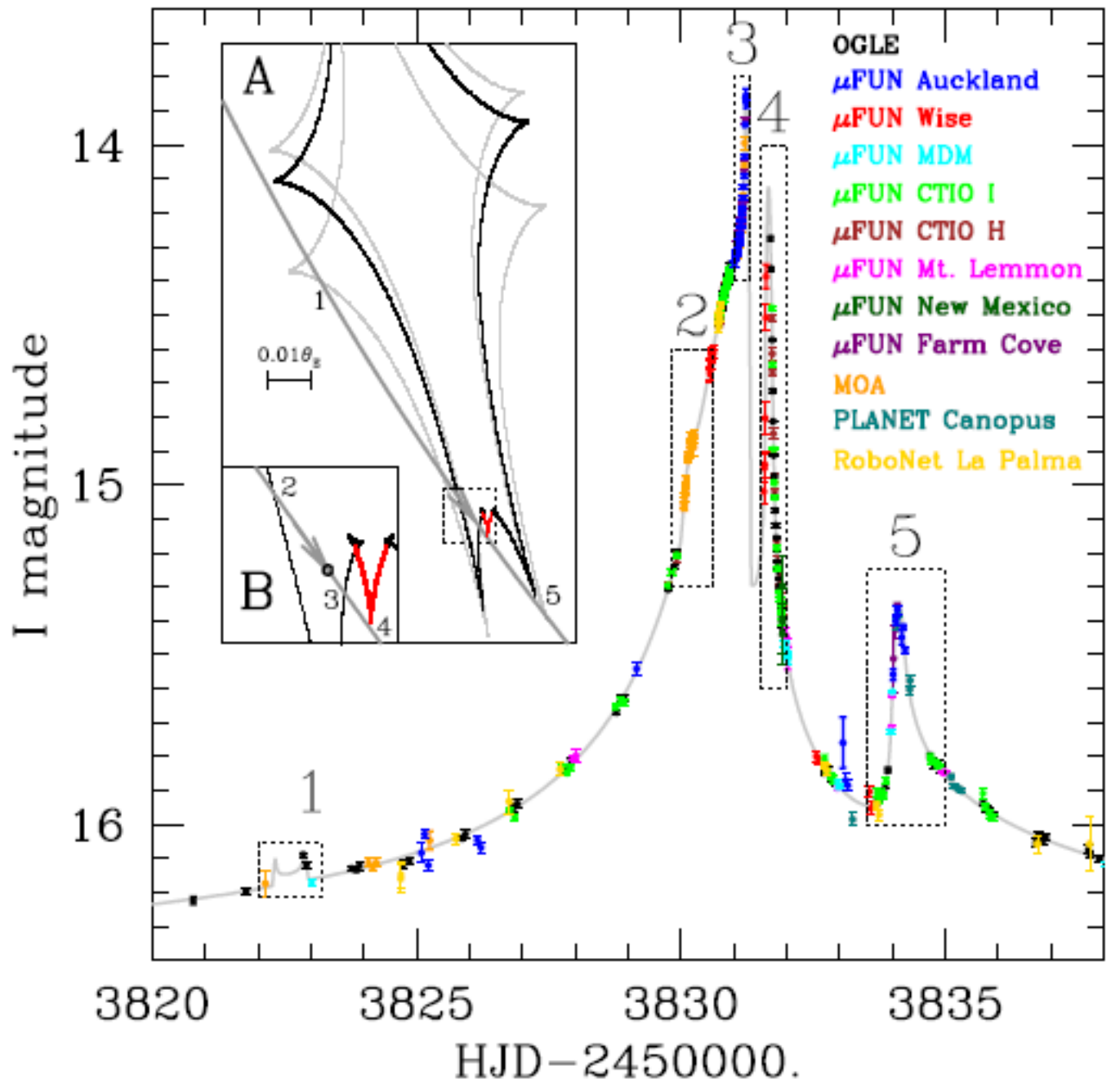

Figure 2.27: Data and best-fit model of the OGLE-2006-BLG-109Lb,c two-planet system [41].

The full potential of the microlensing technique can only be achieved from a space mission. Ground-based microlensing surveys are inherently limited in the mass and separations of the planets they can detect. Because of the high density of stars in the Galactic bulge, main-sequence source stars are not generally resolved in ground-based images. As a result, the photometry needed to detect Earth-mass planets can generally only be obtained when the source is significantly magnified. This, in turn, implies that ground-based microlensing is typically only sensitive to terrestrial planets located close to the Einstein ring (at ~1.5–5 AU). Furthermore, obtaining the accurate photometry of the smaller and fainter main-sequence stars that is required to detect planets with mass much smaller than the Earth is effectively impossible from the ground. Ground-based microlensing surveys also suffer losses in data coverage and quality due to poor weather and seeing. As a result, a significant fraction of the planetary deviations seen in a groundbased microlensing survey will have poor light-curve coverage and therefore poorly constrained parameters [43].

For all but a small fraction of planetary microlensing events, the detection of light from the host star is necessary to allow the star and planet masses and separation in physical units to be determined. This can be accomplished with HST or ground-based adaptive optics observations for a small number of planetary microlensing events [44], but space-based survey data will be needed for the detection of host stars for hundreds or thousands of planetary microlensing events.

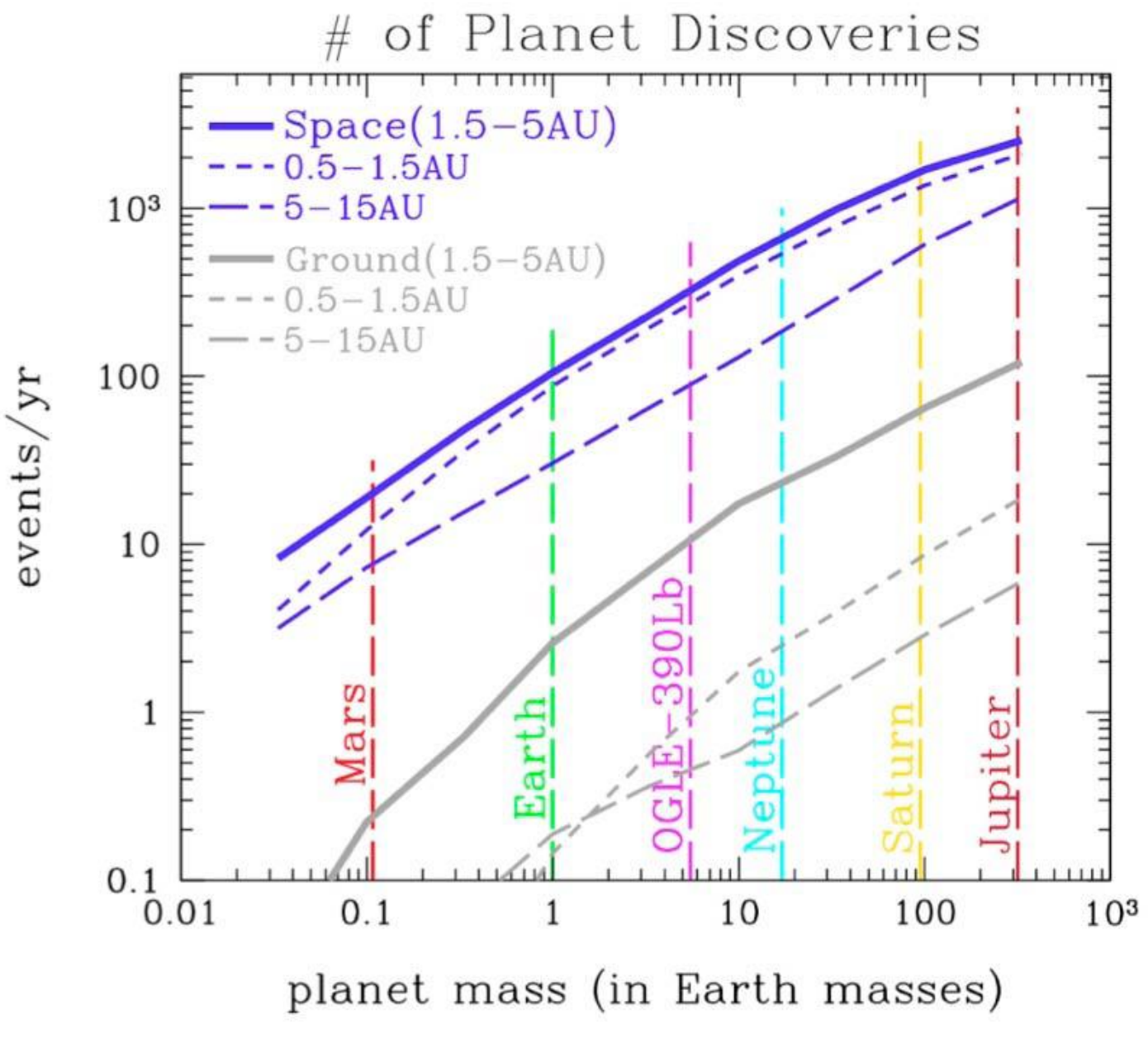

Figure 2.28: The expected number of planet discoveries as a function of the planet mass if every main-sequence star has a single planet within the specified range of separations for a next-generation ground-based survey, and a space-based survey [2].

## 2.4 Astrometry

A star in a planetary system would move around the barycentre in a circular or elliptical path projected on the plane of the sky. This motion can be observed and measured with astrometry. It is the oldest method used to detect extrasolar planets and it enables the study of regular wobble in a star's position which could have been caused by a companion planet.

First, one should determine the observed ellipse (Fig. 2.29). Next, data has to be fitted on a model of Keplerian orbital motion.

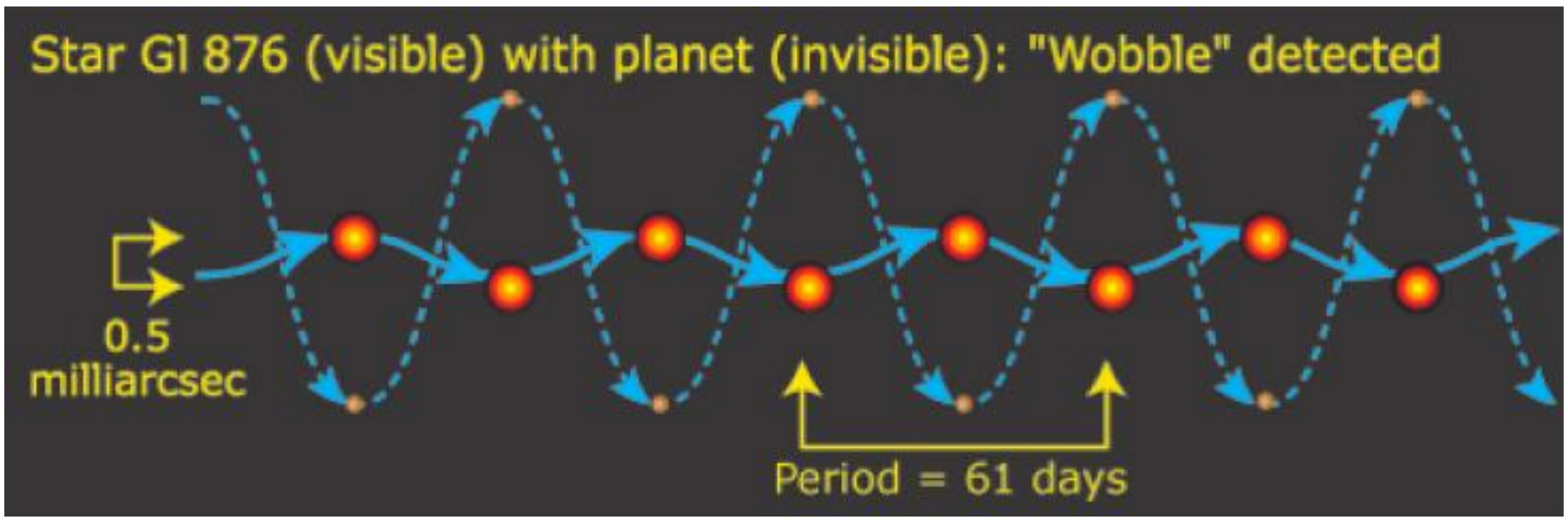

Figure 2.29: The perturbed motion of Gl 876 due to the gravitational pull from companion planet Gl 876b [45].

The Keplerian orbit of each companion is described by seven parameters: semi-major axis *a*, period P, eccentricity e, inclination i, position angle of the line of nodes Ω, argument of pericentre ω and epoch of pericentre passage τ.

Measuring the components of the orbital motion, it is possible to determine the inclination i, semi-major axis *a*, period P and eccentricity e [46]. Hence, combining it with radial velocities data it is possible to determine the mass of the exoplanet without ambiguity.

If the primary mass is $M_\star$ and the secondary is a planet of mass $M_p$, then, assuming a perfectly circular orbit, the apparent amplitude of the perturbation, i.e. the stellar orbital radius around the centre of mass of the system scaled by the distance from the observer, is the so-called *astrometric signature*:

$$\alpha = \frac{M_p}{M_*} \frac{a}{D}$$

(2.10)

If $M_p$ and $M_\star$ are given in solar mass units, a in AU, and D in parsec, then $\alpha$ is in arcsec [47].

Through astrometry, it is possible to survey a bigger sample of stars (i.e., more massive, young, pre main sequence) and to overcome some of the limits in the RV surveys, e.g., those due to stars with complex (few or broad) spectral features. It would be also possible to detect exoplanets around young stars, to probe the time scale of planet formation and migration processes [7].

Because the astrometric signal increases linearly with the semimajor axis a of the planetary orbit, systems with even rather small masses would be more easily detectable at large enough values of a. However, very long observation times will be required — years, and possibly decades, as planets far enough from their star to allow detection via astrometry also take a long time to complete an orbit.

The astrometric technique requires very accurate measurements of positions, in a well defined reference system, and at a number of epochs. An observer located at 10 pc from the Sun would observe an angular amplitude of 500 microarcsec due to the motion of Jupiter and an amplitude of 0.3 microarcsec in the case of the Earth. Thus, measurements require accuracy below 0.1 milliarcsec to detect objects smaller than Jupiter at distances of 50-200 pc. Table 2.3 shows some values of $\alpha$.

Measuring displacements of the order of a few milliarcsecs are impossible using standard imaging techniques from ground based observatories, due to effects of the atmosphere like turbulence and refraction that prevents the precise centering of images. Thus, interferometric techniques and space missions are ongoing to overcome these difficulties.

Table 2.3: Comparison between orders of magnitude of parallax, proper motion, and astrometric signatures induced by planets of various masses and different orbital radii. A 1-$M_\odot$ star at 10 pc is assumed [47].

| Source | α |
|---|---|
| Jupiter @ 1 AU | **100 μas** |
| Jupiter @ 5 AU | **500 μas** |
| Jupiter @ 0.05 AU | **5 μas** |
| Neptune @ 1 AU | **6 μas** |
| Earth @ 1 AU | **0.33 μas** |
| Parallax | **$10^5$ μas** |
| Proper Motion | **5 x $10^5$ μas/yr** |

All claims of a planetary companion made before 1996 using this method are likely spurious. In 2002 the Hubble Space Telescope did succeed in using astrometry to characterize a previously discovered planet around the star Gliese 876 [48]. In 2009 the discovery of VB 10b by astrometry was announced (Fig. 2.30 and [49]). This planetary object was reported to have a mass 7 times that of Jupiter and orbiting the nearby low mass red dwarf star VB 10. If confirmed, this will be the first exoplanet discovered by astrometry of the many that have been claimed through the years.

A number of smaller-scale interferometric astrometry programs have been developed on the ground and in space to demonstrate the technologies and science for a micro-arcsecond astrometry program. This includes The Hubble Space Telescope Fine Guidance Sensors (FGS) and The Palomar Testbed Interferometer (PTI). Moreover, there are projects under development carried out by ESO: PRIMA (Phase- Reference

Imaging and Micro-Arcsecond Astrometry); ESA: Gaia; and NASA: SIM (Space interferometry Mission).

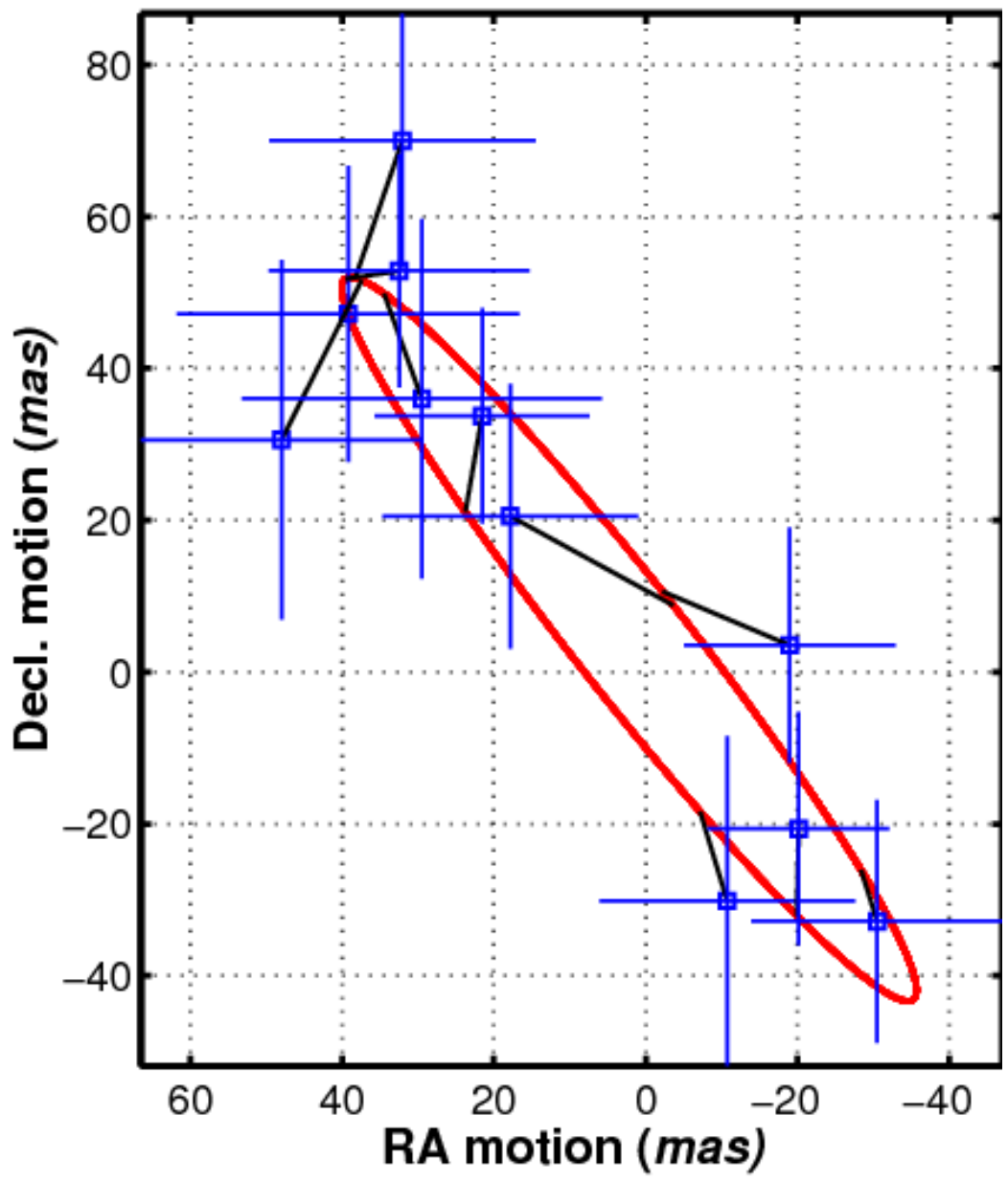

Figure 2.30: The Keplerian orbit of VB10b with the average data from each epoch shown [49].

Gaia is planned to be launched in 2011. It will achieve 100 μas per-measurement precisions (to V = 15) on 30 million stars (and reduced precision for up to a billion fainter stars), with each star revisited 1–250 times (typically ~90 one-dimensional measurements) over the course of the mission. It is a survey mission without capability to be pointed at a particular object, meaning the number of revisits cannot be increased for a high priority target. Gaia will saturate for very bright stars (V ~ 6), including the stars closest to Earth that are the highest priority targets when looking for Earth-like planets [2].

PRIMA narrow-angle differential astrometry instrument began commissioning by the European Very Large Telescope Interferometer (VLTI) in August 2008. The most optimistic estimates predict 10–20-μas performance. This will be an excellent tool for characterizing giant planets around nearby stars, and the expected long lifetime of this program will allow it to capitalize on astrometry's increasing sensitivity with planet orbital period. For ground-based observatories, there are no other facilities that would match or exceed the performance of VLTI within the 15-year time frame [50]. The VLTI will operate in the near-IR, and can study objects obscured at visible wavelengths including protostars and the galactic centre.

## 2.5 Circumstellar Disks

Disks of space dust (debris disks) surround many stars. The dust can be detected because it absorbs ordinary starlight and re-emits it as infrared radiation. Even if the dust particles have a total mass well less than that of Earth, they can still have a large enough total surface area that they outshine their parent star in infrared wavelengths [51].

The Hubble Space Telescope is capable of observing dust disks with its NICMOS (Near Infrared Camera and Multi-Object Spectrometer) instrument. Even better images have now been taken by its sister instrument, the Spitzer Space Telescope, which can see far deeper into infrared wavelengths than the Hubble can. Dust disks have now been found around more than 15% of nearby sunlike stars [52]. However, a small coronagraph or small interferometer in space is needed in order to reach the sensitivity required to detect the glow at the level of our own Solar System [2].

The dust is believed to be generated by collisions among comets and asteroids. Radiation pressure from the star will push the dust particles away into interstellar space over a relatively short timescale. Therefore, the detection of dust indicates continual replenishment by new collisions, and provides strong indirect evidence of the presence of small bodies like comets and asteroids that orbit the parent star [52]. For example, the dust disk around the star tau Ceti indicates that that star has a population of objects analogous to our own Solar System's Kuiper Belt, but at least ten times thicker [51].

More speculatively, features in dust disks sometimes suggest the presence of full-sized planets. Some disks have a central cavity, meaning that they are really ring-shaped. The central cavity may be caused by a planet "clearing out" the dust inside its orbit. Other disks contain clumps that may be caused by the gravitational influence of a planet. Both these kinds of features are present in the dust disk around epsilon Eridani, hinting at the presence of a planet with an orbital radius of around 40 AU (in addition to the inner planet detected through the radial-velocity method) [53].

## 2.6 Pulsar Timing

A pulsar is a rapidly-spinning neutron star with a strong magnetic field. Radiation produced by the neutron star is focused into two oppositely-directed beams by the magnetic field. As the star rotates, the beam is swept across the sky; if the beam intercepts the Earth once per rotation, then brief but regular pulses of radiation are seen.

When a planet is introduced, the mutual gravitational pull between it and the pulsar means that they both orbit about their common centre of mass. In the case of a pulsar and a planet, the centre of mass will lie very close to the pulsar, since it is much heavier than the planet. Therefore, during one orbit the pulsar will move a much lesser distance than the planet.

However, even thought the pulsar's 'wobble' is small, it has an effect on the timing of the pulses emitted by it. When the pulsar is moving away from the Earth, the time between each pulse becomes slightly longer; conversely, when the pulsar is moving toward the Earth, the time between pulses becomes slightly shorter.

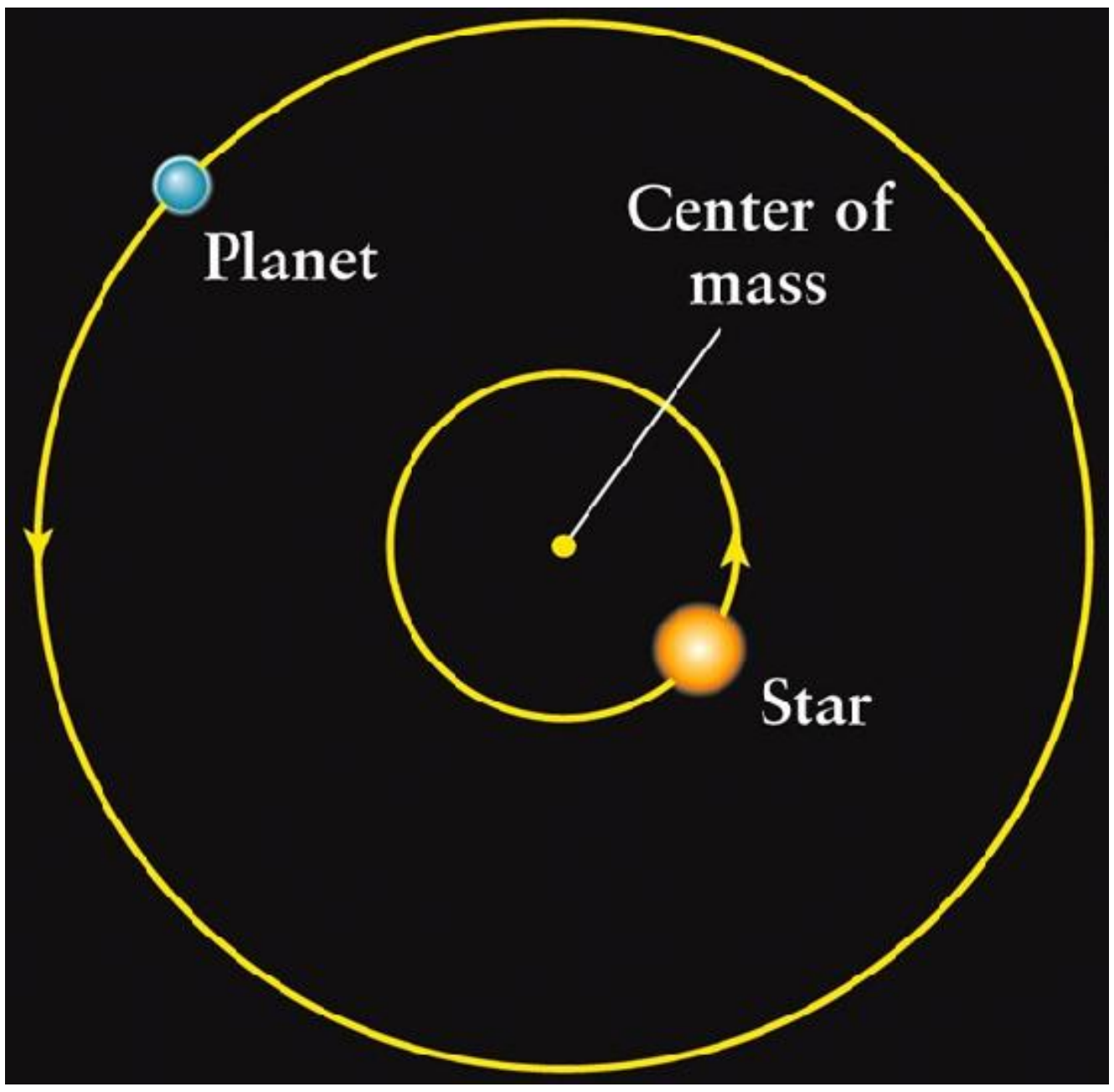

Figure 2.31: A star and planet orbiting around their common centre of mass.

By measuring these periodic changes in pulse timing, it is possible not only to deduce the existence of a planet orbiting around the pulsar, but also to estimate the semi-major axis of the planet's orbit, and place a lower limit on the mass of the planet [54]. This method was not originally designed for the detection of planets, but is so sensitive that it is capable of detecting planets far smaller than any other method can, down to less than a tenth the mass of Earth. It is also capable of detecting mutual gravitational perturbations between the various members of a planetary system, thereby revealing further information about those planets and their orbital parameters.

To date, seven planets have been detected using this method. The first three where discovered orbiting the pulsar PSR 1257+12 by Wolszczan and Frail in 1992 [55]. Detections of this sort are not of much interest in the search for extrasolar planets, since pulsars are created in supernovas, life as we know it could not survive on these planets since high-energy radiation there is extremely intense.

## 2.7 Magnetospheric Emission

Planets in the Solar System produce radio-frequency waves by the mechanism of radiation from electrons spiraling along the planet's magnetic field lines, and it is anticipated that exoplanets will broadcast similar radio waves. The radio luminosity is proportional to the solar wind power incident on a planet, so is expected to be largest for active stars. Since these stars are difficult to observe with astrometry or RV, magnetospheric emissions may offer a path to detecting planets in these systems.

In the case of the Earth, its magnetic field may contribute to its habitability by protecting its atmosphere from solar wind erosion and by preventing energetic particles from reaching its surface. Indirect evidence for at least some extrasolar giant planets also having magnetic fields includes the modulation of calcium emission lines of their host stars phased with the planetary orbits likely due to magnetic reconnection events between the stellar and planetary fields. If magnetic fields are a generic property of giant planets, then extrasolar giant planets should emit at radio wavelengths allowing for their direct detection. In the case of terrestrial-mass planets, if magnetized, they should also emit at radio wavelengths.

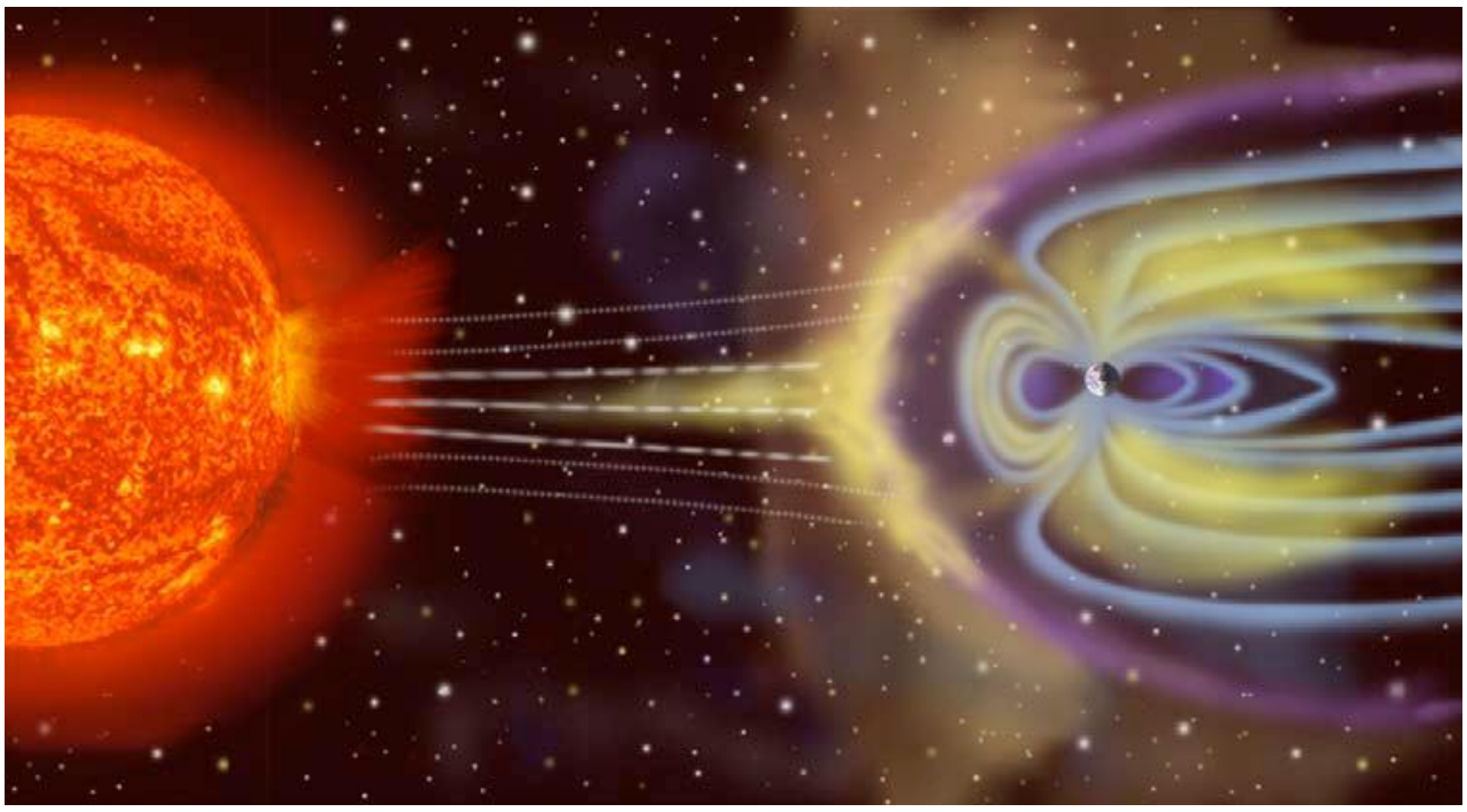

Figure 2.32: An illustration of the interaction between the Sun and Earth's magnetosphere (not to scale). In this example, an eruption from the Sun has reached the Earth and compressed the magnetosphere, injecting energetic particles into it [2].

There are a number of ground-based instruments under construction that promise significant improvements in sensitivity. The main technology needs are algorithmic, such as the development of improved radio frequency interference (RFI) avoidance and excision. Ground observations will ultimately be limited by a combination of the Earth's ionosphere and RFI, however antennas on the far side of the Moon could have large areas and very low RFI, and are therefore of special interest [2].

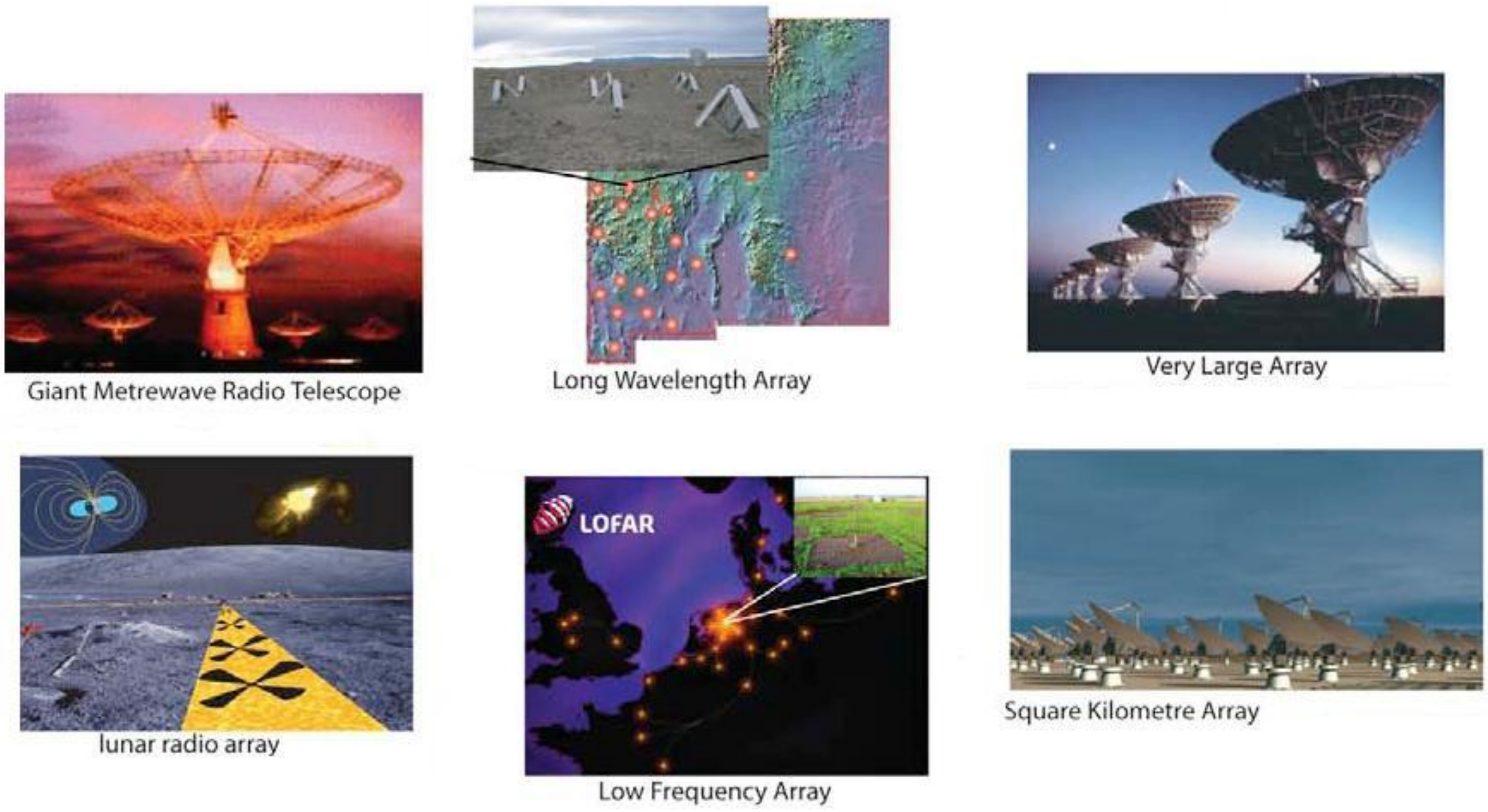

Figure 2.33: Images of present and future radio observatories [2].

*Chapter 3*

# EXOPLANET CHARACTERIZATION AND THE SEARCH FOR LIFE

In the first decade after the first discovery in 1995 [56], the task was to find more and more exoplanets. Now that more than 370 exoplanets are known, attention has switched from finding planets to characterizing them so that we can begin the long mission of understanding these astronomical bodies in the same way that we understand those in our own Solar System. This task includes looking for satellites and rings [81], probing their atmospheres for molecules, constraining their horizontal and vertical temperature profiles and estimating the contribution of clouds and hazes [4]. In addition to search missions such as CoRoT and Kepler, several missions have focused on follow-up of bright transiting systems. The MOST mission has observed transits of giant planets orbiting bright stars, and has been particularly successful at placing very stringent limits on the magnitude of reflected light from HD 209458b, a very low albedo planet (Rowe et al. [75]). Similarly, the EPOXI mission is concentrating on the properties of giant planets transiting bright stars (Christiansen et al. [82]; Ballard et al. [83]).

In fact, among the variety of exoplanets discovered so far, the best known are giant planets (EGPs) orbiting very close-in (hot-Jupiters). For the sixty two currently identified exoplanets [3] that transit their parent stars we know planetary and orbital parameters such as radius, eccentricity, inclination, mass (given by radial velocity combined measurements). This allows first order characterization on the bulk composition and temperature and offers a unique opportunity to estimate directly key physical properties of their atmospheres [57]. In particular, it is possible to exploit the wavelength dependence of the transit depth to identify key chemical components in the planet's atmosphere. This accomplishment permits enormous possibilities in terms of exoplanet characterization. To be able to do transit spectroscopy of terrestrial planets and/or colder atmospheres, we have to wait for JWST [58].

## 3.1 Characterization of Transiting Planet Atmospheres

Using transit techniques we can indirectly observe the thin atmospheric ring surrounding the optically thick disk of the planet -the annulus- while the planet is transiting in front of its parent star. This method was traditionally used to probe the atmospheres of planets in our Solar System and most recently, thanks to the Hubble Space Telescope and Spitzer, was successfully applied to a growing sample of "Hot-Jupiters". The idea was first theoretically proposed by Seager and Sasselov in 2000 [59] and permitted extraordinary accomplishments completely unexpected for the Spitzer and Hubble space telescopes.

- Primary transit

Since 2002, when Charbonneau et al. first detected the presence of Sodium in the atmosphere of a hot-Jupiter, the use of transmission spectroscopy to probe the upper layers of the transiting EGPs has been particularly successful in the UV and visible spectral ranges (Charbonneau et al. [60]; Vidal-Madjar et al. [61]; Knutson et al. [62], Pont et al. [63]) and in the Near and Middle Infrared spectral window (Deming et al. [64], Knutson et al. [65], Beaulieu et al. [66], Tinetti et al. [67], Swain et al. [68], Agol et al. [69]). Transmission spectra are sensitive to **atomic and molecular abundances** and less to temperature variation. Temperature influences the transmission spectrum by way of its influence on the atmospheric scale height [57] and the absorption coefficients.

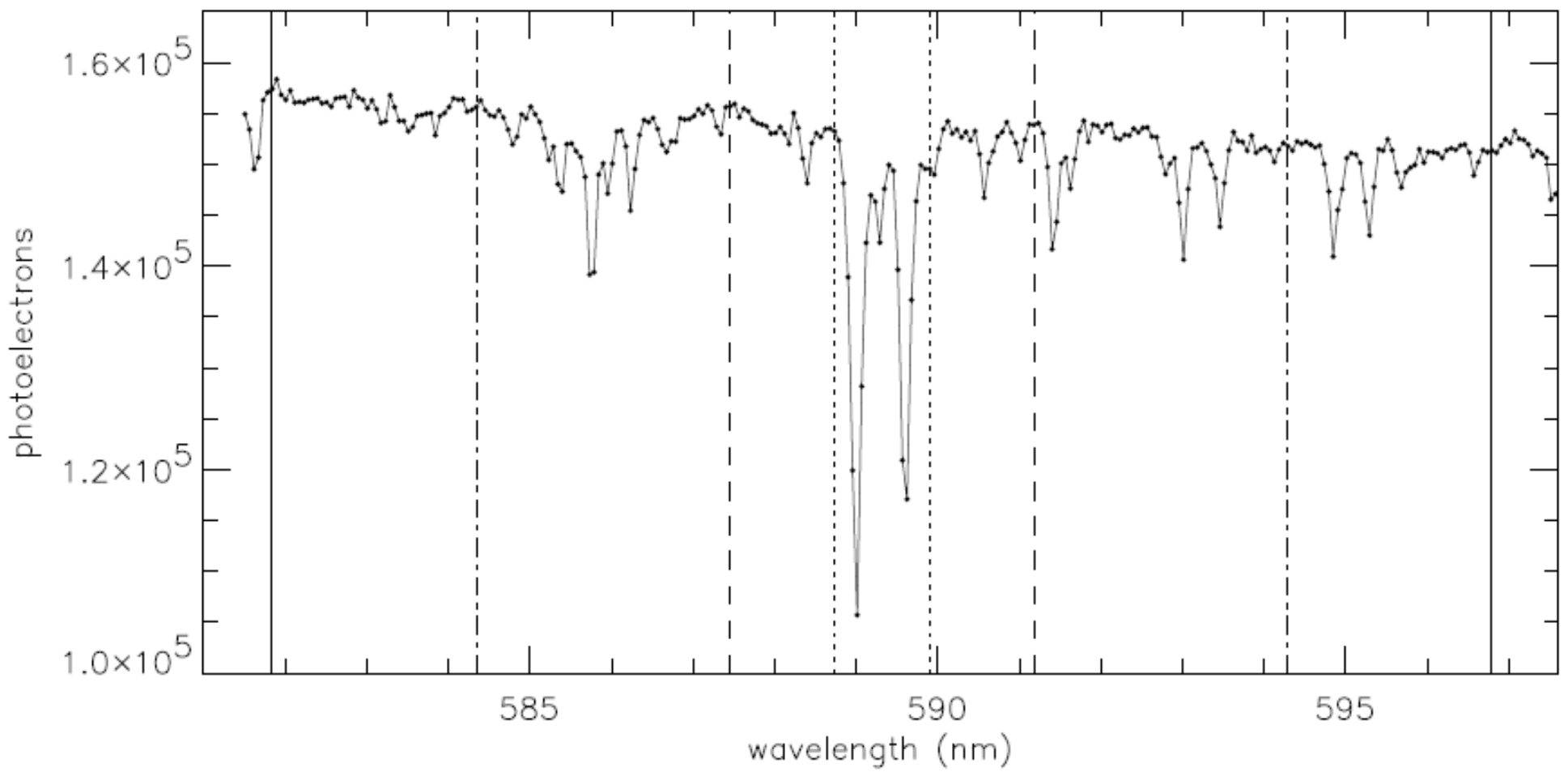

Figure 3.1: Shown is a portion of a STIS spectrum of HD 209458, centred on the Na D lines [60].

• Secondary transit

With this method, we observe first the combined spectrum of the star and the planet. Then, we take a second measurement of the star alone when the planet disappears behind it: the difference between the two measurements consists of the planet's contribution. This technique was pioneered by two different teams in 2005, using the Spitzer Space Telescope to probe two Hot-Jupiters in the Infrared (Deming et al. [70]; Charbonneau et al. [71]). So far, the focus has been on the brightest stars with transiting extrasolar planets, namely HD 209458b (Charbonneau et al. [14]), HD 189733b (Bouchy et al. [72]), GJ436b (Butler et al. [73]) and TRES-1 (Alonso et al. [74]).

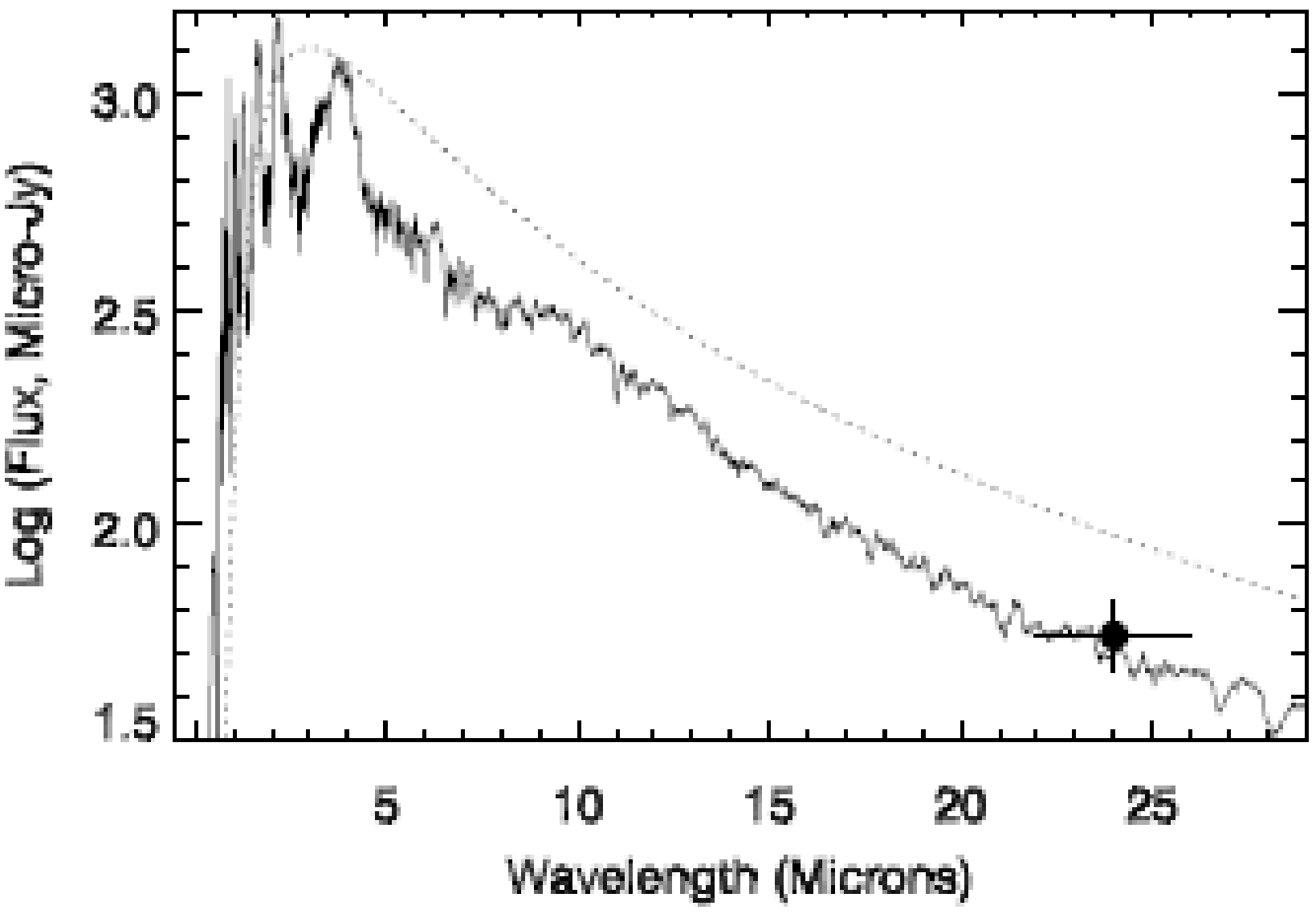

Figure 3.2: Measured infrared flux at 24 μm for HD 209458b. A theoretical spectrum (solid line) shows that planetary emission should be very different from a blackbody of $T_{eq}$ =1700 K. The suppressed flux at 24 μm is due to water vapour opacity [70].

In the infrared spectral range we measure directly the flux due to the planet and the star/planet contrast is lower, so with this technique we can not only detect the molecular species showing a noticeable rotational/vibrational signature, but also constrain the vertical thermal gradient [4] and the **bulk temperature**. This can be done by assuming that the depth of the secondary dip is the ratio of the planet emissivity over the star emissivity, modulated by the factor $(r/R)^2$ [16]. Compared to transmission spectroscopy, emission spectroscopy may scan different regions of the atmosphere for molecular signatures and cloud/hazes contributions [57]. Same considerations are valid in the UV-visible spectral range, except that the photons reflected by the planet do not bring any information about the planetary temperature and the thermal structure, but about the planetary **albedo** [75] and the presence of atomic/ionic/molecular species having electronic transitions.

Finally, especially if the planet is tidally locked, with primary and secondary transit techniques we can observe different phases of the planet along its orbit. During the primary transit we can sound the terminator, whereas during secondary we can above all observe the day-side.

- Light-curves

Monitoring the light-curve of the combined star-planet spectrum, can be a useful approach both for transiting (Knutson et al. [62], [65]; Kipping et al. [6]; Snellen [169]) and non-transiting planets (Harrington et al. [76]). In the latter case the planetary radius cannot be measured, but we can appreciate the temperature or albedo variations through time (depending if the observation is performed in the visible or infrared).

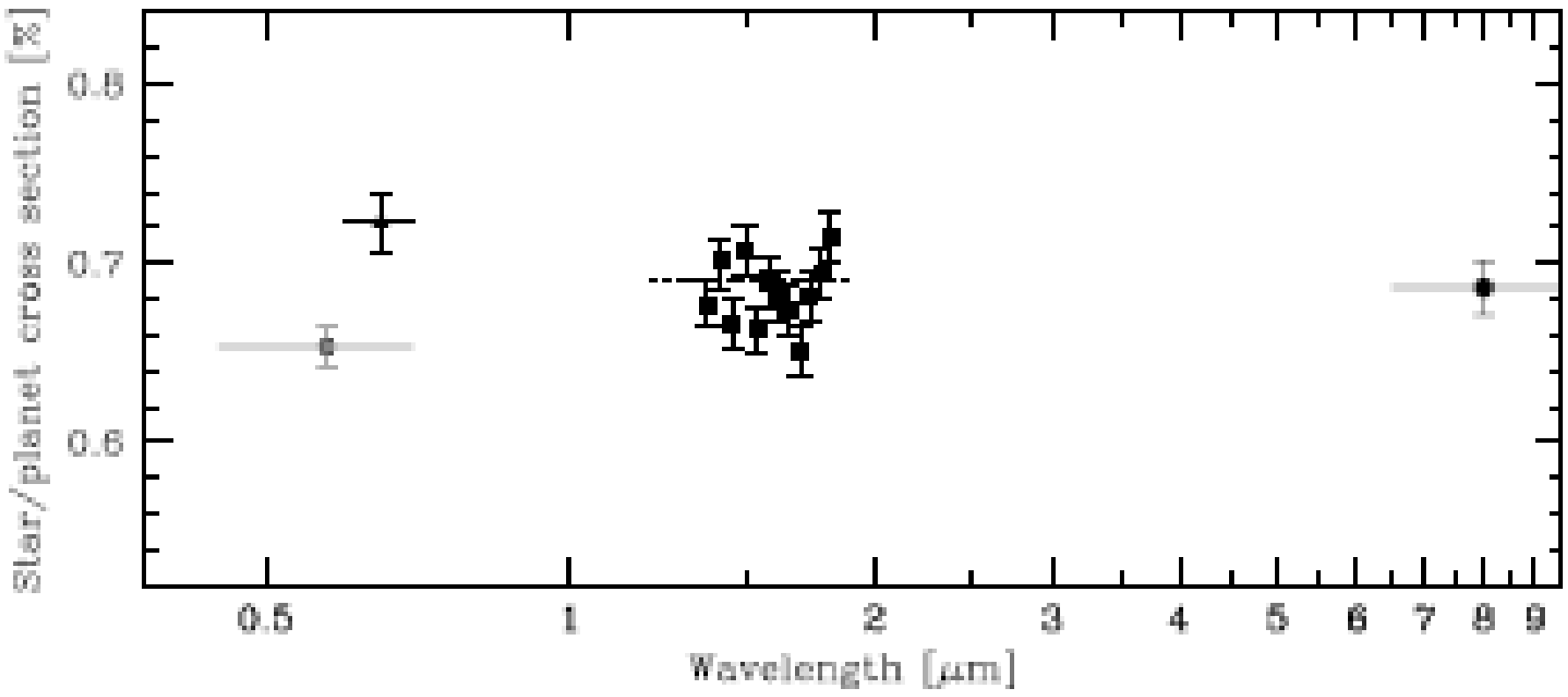

Figure 3.3: Planet cross-section (in percentage of the stellar disc covered) as a function of wavelength for GJ 436b [84].

The problems that we can face with current telescopes are:

• Detection of the main molecular species in the hot transiting planets' atmosphere.

• Constraint of the longitudinal and vertical thermal gradients in the hot exoplanet atmospheres.

• Presence of clouds or hazes in the atmospheres.

• Exploration of the differences between dayside and nightside atmospheric chemistry.

Today we can use two approaches to reach these objectives:

- Broad band or low resolution spectroscopy from a space based observatory. This can be accomplished by SPITZER or HST.

- High resolution spectroscopy from ground based observatories in the optical and NIR.

The next steps with these indirect techniques will be:

• Detection of minor atmospheric species and constraint of their abundance.

• More accurate spectral retrieval to map thermal and chemistry gradients in the atmospheres.

• Cloud microphysics: understanding the composition, location and optical parameters of cloud/haze particles.

• Cooler and smaller planets, possibly in the habitable zone.

Further into the future the James Webb Space Telescope or the JAXA/ESA SPICA mission concept (Nakagawa et al. [77]) will be the next generation of space telescopes to be online. They will guarantee high spectral resolution from space and the characterization of fainter targets, allowing us to expand the variety of "characterizable" extrasolar planets [4].

### *3.1.1 Temperature profiles*

With photometry, or low resolution spectroscopy in the Near and Mid Infrared we are today able to put some constraints on the thermal horizontal and vertical profiles of the planetary atmospheres. This knowledge would be localized by repeating the retrieval for different portions of the planet's orbit. For example, we are already in a position of appreciating the differences between HD189733b (Knutson et al. [78]) and Upsilon Andromedae (Harrington, et al. [76]): at 8 and 24 μm HD189733b shows a well mixed temperature distribution between the day and the night side, while we have the opposite for Upsilon Andromedae at 24 μm. HD209458b shows clear signs of a thermal inversion at relatively low altitude (Burrows et al. [79], Swain et al. [107]), the situation is different for HD189733b (Swain et al. [80]). However high resolution spectroscopy is needed to perform a more accurate spectral retrieval and better constrain the dynamics of planetary atmosphere.

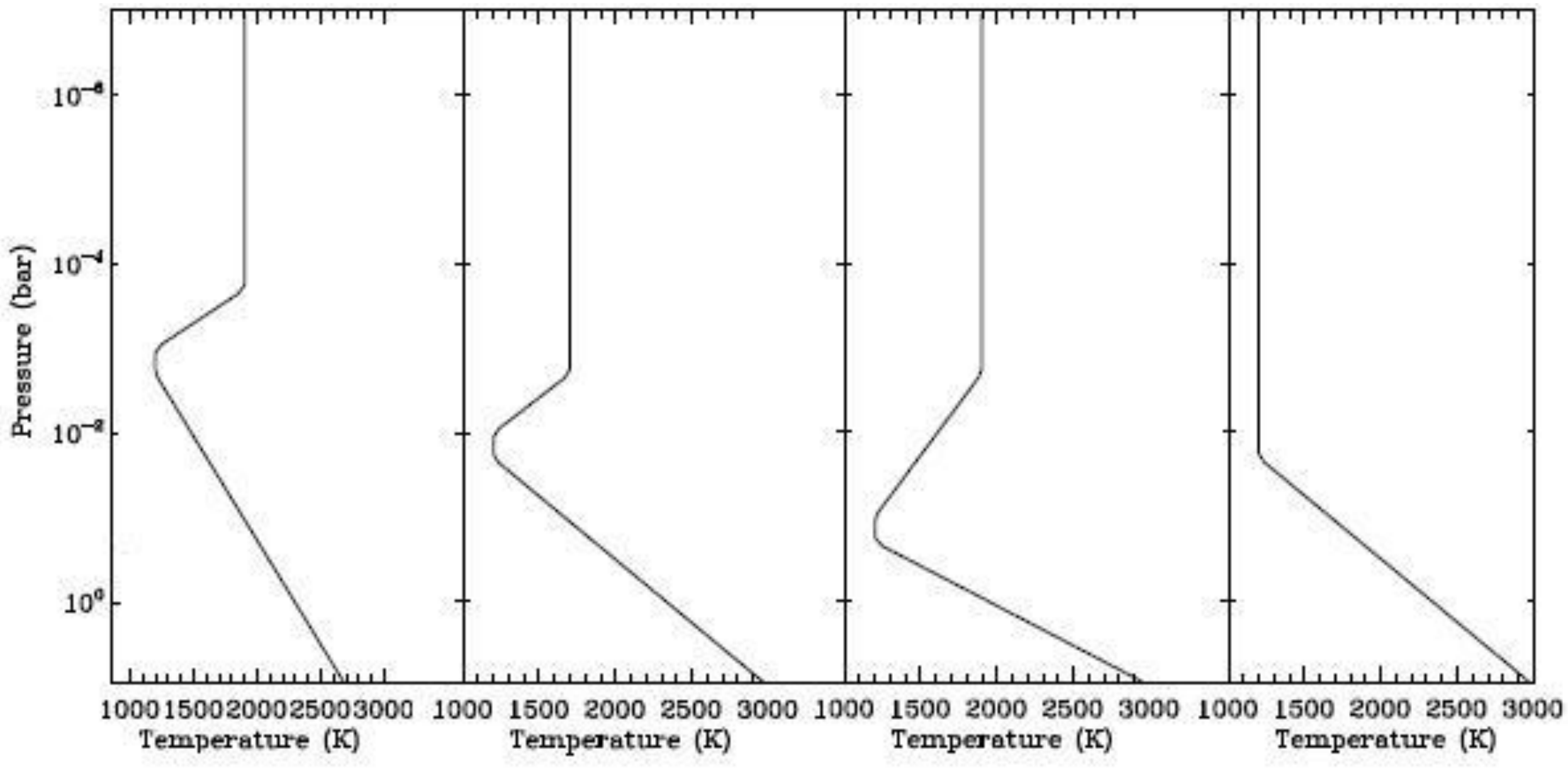

Figure 3.4: Dayside pressure-temperature profiles for four different models for HD 209458b. From left to right, the models correspond to a tropopause at 0.001 bar, 0.01 bar, 0.1 bar, and no tropopause respectively [107].

### *3.1.2 Molecules in the atmosphere of hot-Jupiters*

**Water**. In a star-planet system, a significant amount of water vapour can only exist in planetary atmospheres at orbital distances small enough (less than ~ 1AU for a solar like star). This closeness requirement is well met by hot Jupiters. According to photochemical models, $H_2O$ should be among the most abundant species (after $H_2$) in the lower atmosphere of giant planets orbiting close to their parent stars (Liang et al. [85]). Moreover, $H_2O$ should be the easiest of these species to detect in primary transit in the IR (Tinetti et al. [86]). A very accurate data list for water at hot Jupiter-like temperatures has been calculated by Barber et al. [87].

**Carbon bearing molecules**, such as CO, $CH_4$, $CO_2$, $C_2H_2$ are expected to be abundant as well, depending on the C/O ratio and the efficiency of the photochemistry in the upper atmosphere [85]. Each of these molecules probes the atmospheres in different ways. CO and $CH_4$ are primary reservoirs of carbon; the $CO/CH_4$ ratio is sensitive to temperature. $CO_2$ probes the vertical temperature profile and the vertical mixing ratio (potentially a diagnostic of photochemistry) using the V = 13–15 bands. $CO_2$ also serves as a proxy for CO, while CO and $CH_4$ abundances differing from thermochemical equilibrium might indicate photochemistry. For tidally locked, hot-Jovian planets, the chemistry of the atmosphere is expected to change, particularly the $[CO]/[CH_4]$ ratio, as a function of orbital phase and, in some cases, a dayside hot stratosphere might form [88]; the $[CO]/[CH_4]$ ratio at the terminator is also a potential diagnostic of variability, possibly associated with large-scale whirls generated by zonal winds[2]. CO, $CH_4$, $CO_2$ have already been detected (Charbonneau et al. [89], Swain et al. [68]). For less abundant species, or with spectral signatures which are harder to detect, we need to make the leap to high resolution spectroscopy.

Also, improved line lists at high temperatures are needed to better interpret the measurements. From preliminary results and models, it is not excluded that the chemistry might vary substantially from the highly irradiated day-side to the non-illuminated nightside of these planets (Swain et al. [68]).

**Nitrogen or sulphur-bearing molecules** are also likely to be present in the atmospheres of hot-Jupiters, but their weaker signatures may be difficult to be caught with a low resolving power (Sharp and Burrows [90], Zahnle [170]).

$H_3^+$ – the simple molecular ion formed by the photo-ionisation of $H_2$ – could be a crucial indicator of the escape processes in the upper atmosphere (Yelle [91], [95]). Now that HST/STIS instrument is no longer operative to observe the Lyman alpha line in the UV (Ben-Jaffel [92]), $H_3^+$ is the only molecular ion able to monitor the escape processes on hot Jupiters. So $H_3^+$ is a crucial detection target; even if detection is unsuccessful, such measurement provides at least an improved upper limit on its abundance (Shkolnik et al. [94]). Calculations of the $H_3^+$ abundance on a hot-Jupiter (Yelle [95]) show that the contrast between $H_3^+$ emission and stellar brightness places it just on the current limit of detectability with a large ground-based telescope. Even if it is a very challenging observation, it would be the best diagnostics to understand the properties of the upper atmospheres of Hot-Jupiters (Koskinen et al. [96]) whereas observations of the Lyman alpha line could be partially or totally contaminated by energetic neutral atoms from charge exchange between stellar wind protons and neutral hydrogen from the planetary exospheres (Holmstrom et al. [97]) or inadequately analyzed and understood (Ben-Jaffel [93]).

**Clouds and hazes**. At the spectral resolution we can obtain today from space, the best we can do is to assess their presence, as they are supposed to flatten the spectral signatures or modify the spectral shape. In the case of transmission spectroscopy, they cause the atmosphere to be opaque at higher altitudes (Brown [57]). Figure 3.5 shows the transmission spectrum of HD189733b in the wavelength range $0.5 - 25\mu m$. The HST observations from Pont et al. [98], show an almost featureless transmission spectrum in the range $0.5 - 1\mu m$ which may suggest the presence of hazes.

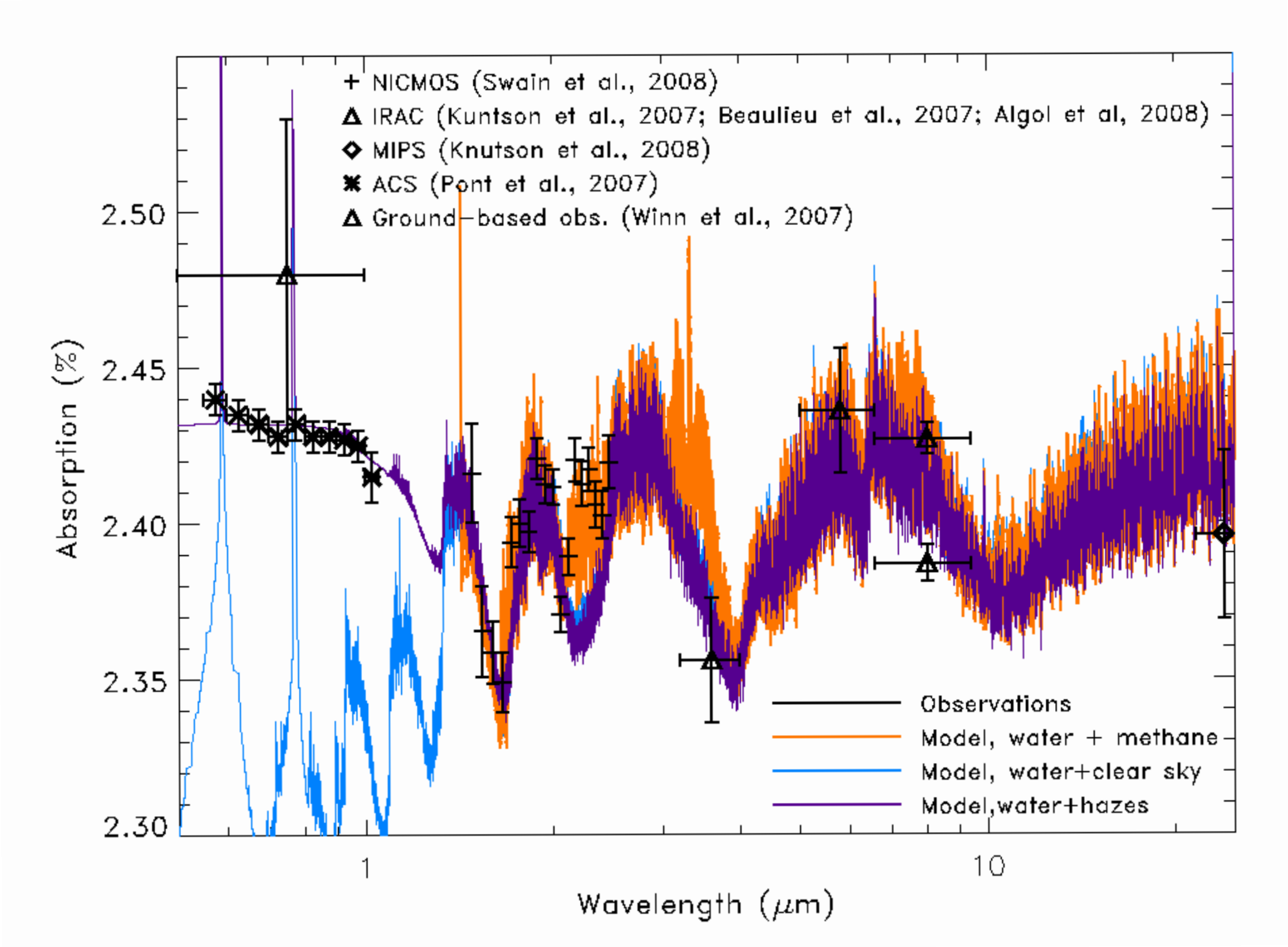

Figure 3.5: Simulated middle Infrared spectra of the transiting Hot Jupiter HD189733b in the wavelength range 0.5-25 μm. The overall transmission spectrum is shaped by the water absorption in the infrared (Tinetti et al. [67]) but methane is needed to explain the NIR (Swain, Vasisht, Tinetti [68]). At shorter wavelength, the increasing flatness of the spectrum could be explained by hazes [4].

The overall transmission spectrum is shaped by the water absorption in the infrared. $H_2$-$H_2$, methane and alkali metals absorptions are included, as well as a crude simulation of hazes opacity. It's remarkable that the different data collected by instruments over a wide wavelength range are giving consistent results. Most probably additional molecules are present, but it is not possible to appreciate their presence at this spectral resolution. For instance, in Figure 3.6, it is shown the additional contribution of a variety of plausible molecules as a function of wavelength: the contributions of Methane at 3.2 μm, CO at 4.5 μm and ammonia at 11 μm are quite noticeable. Although water can be detected with broad band photometry, it is clear that spectral resolution is needed in order to get the probe for the different species [4].

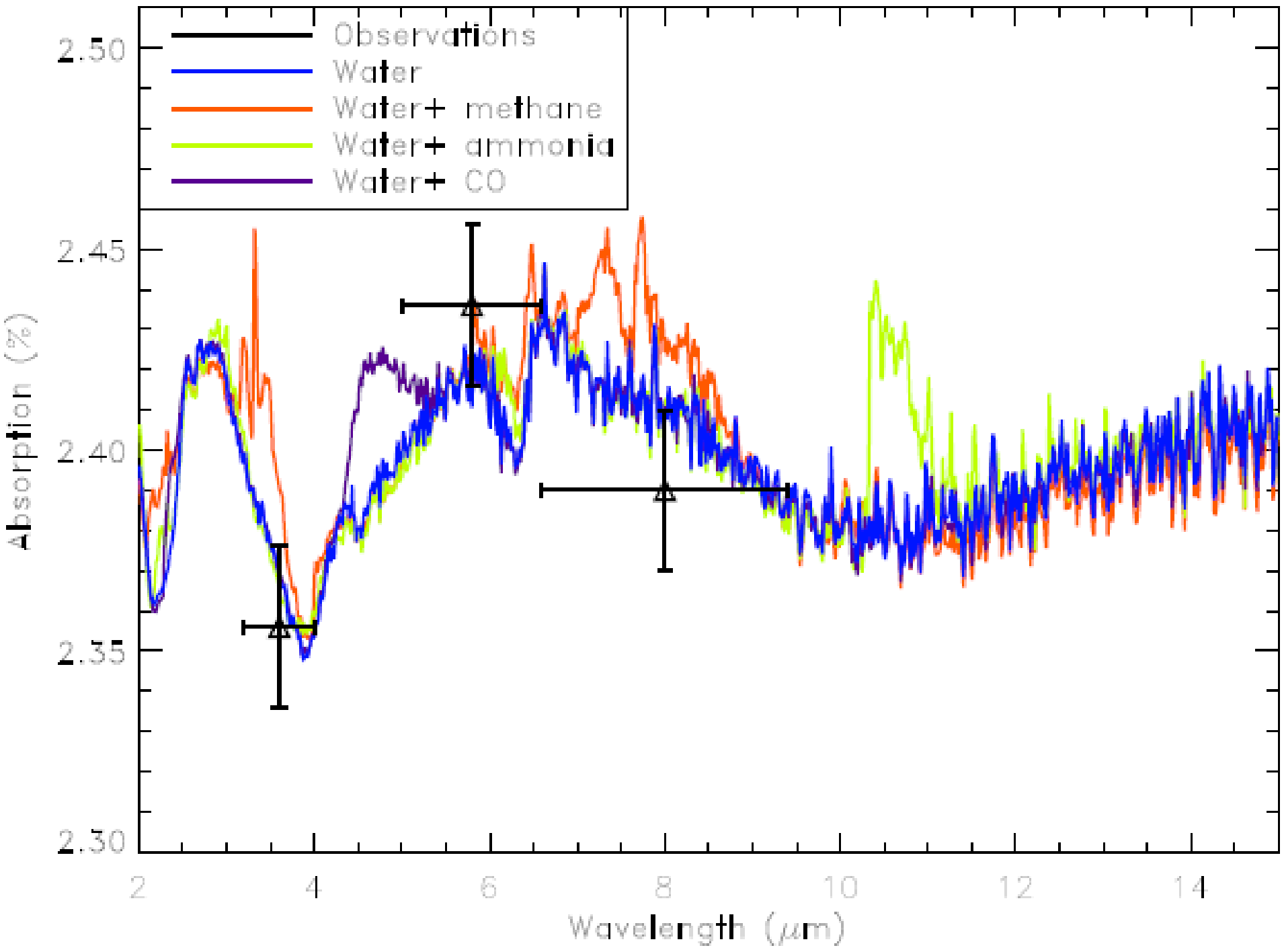

Figure 3.6: Simulated middle Infrared spectra of the transiting Hot Jupiter HD189733b in the wavelength range 2-15 μm together with IRAC measurements (Beaulieu et al. [66], Tinetti et al. [67]) [4].

### *3.1.3 Towards smaller mass planets*

For both primary and secondary transit methods, smaller size/colder planets will increase the challenge. Transmission spectroscopy can benefit from very extended atmospheres: this scenario can occur if the main atmospheric component has a light molecular weight and a high temperature. The lighter, the hotter and the smaller the core, the easier is the observation in transmission spectroscopy (i.e. the more detectable are the spectral features). In the case of secondary transits, the parameters playing the major role are the size of the planet compared to its parent star and the planetary temperature for observations in the IR or the albedo in the visible [4].

With current telescopes we can already approach the case of hot Neptunes transiting later type stars, e.g. Gliese 436b (Deming et al. [64]). For rocky planets within their star's habitable zones, we need to wait for JWST. They have the highest priority as these have the potential to harbor life. Our science goal is to find and characterize all nearby exoplanets so that within the decade we could answer long-standing questions about the evolution and nature of other planetary systems, and we could search for clues as to whether life exists elsewhere in our galactic neighborhood.

## 3.2 The Habitable Zone

The circumstellar *Habitable Zone* (HZ) is defined as the region around a star within which an Earth-like planet can sustain liquid water on its surface, a condition necessary for photosynthesis[1]. Within the HZ, starlight is sufficiently intense for a greenhouse atmosphere to maintain a surface temperature above 273 K, and low enough not to initiate runaway greenhouse conditions that can vaporize the whole water reservoir, allow photodissociation of water vapour into its constituent hydrogen and oxygen and the loss of hydrogen to space (This was the fate of Venus) [37]. The Continuously HZ is the region that remains habitable for durations longer than 1 Gyr. Figure 3.7 shows the limits of the Continuously HZ as a function of the stellar mass. Planets inside the HZ are not necessarily habitable. They can be too small, like Mars, to maintain active geology and to limit the gravitational escape of their atmospheres. They can be too massive, like HD69830d, and have accreted a thick $H_2$-He envelope below which water cannot be liquid. However, planet formation models predict abundant *Earth-like* planets with the right range of masses (0.5 – 8 M⊕) and water abundances (0.01-10% by mass) [101].

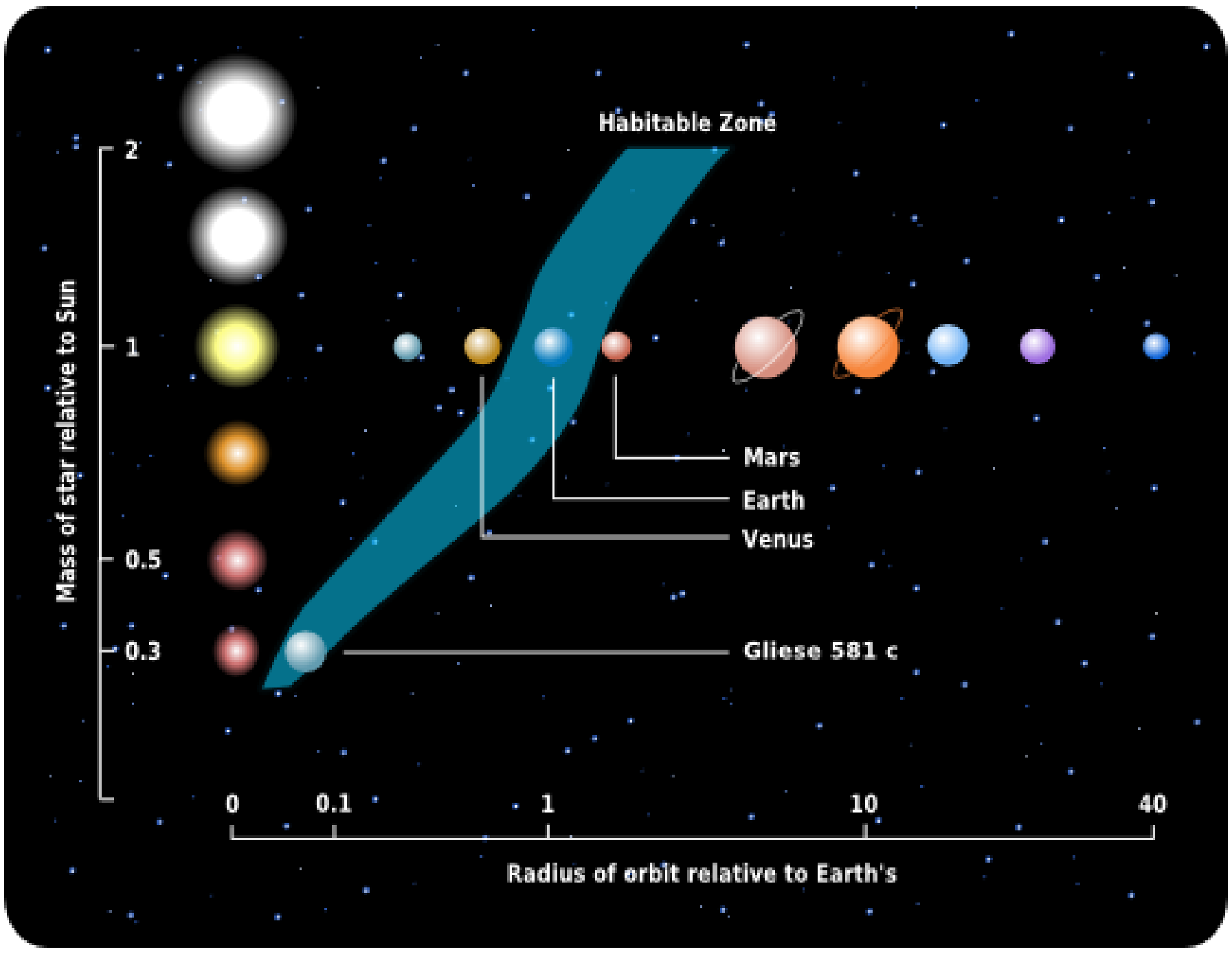

Figure 3.7: Continuous Habitable Zone (blue region) around M, K, G and F stars. The region around the Sun that remains habitable during at least 5 Gyrs extends from approximately 0.76 to 1.63 AU.

[1] As a matter of fact the equation for the oxygenic photosynthesis is $6\ CO_{2(g)} + 12\ H_2O_{(l)} + 686$ Kilocal/mol $\rightarrow C_6H_{12}O_{6(aq)} + 6\ O_{2(g)} + 6\ H_2O_{(l)}$

Again, we are interested in investigating what are the most suitable Stars in the contest of the development of life. Higher-mass stars tend to be larger and luminous than their lower-mass counterparts. Therefore, their habitable zones are situated further out. In addition, however, their HZs are much broader. As an illustration,

- a 0.2 solar-mass star's HZ extends from 0.1 to 0.2 AU

- a 1.0 solar-mass star's HZ extends from 1 to 2 AU

- a 40 solar-mass star's HZ extends from 350 to 600 AU

On these grounds, it would seem that high-mass starts are the best candidates for finding planets within a habitable zone. However, these stars emit most of their radiation in the far ultraviolet (FUV), which can be highly damaging to life, and also contributes to photodissociation and the loss of water. Furthermore, the lifetimes of these stars is so short (around 10 million years) that one can debate whether there is enough time for life to develop.

Then, we have the medium-mass stars like our own Sun. They make up about 15% percent of the stars in the galaxy, have reasonably-broad HZs, do not suffer from FUV irradiation, and have lifetimes of the order of 10 billion years. Therefore, they are among the best candidates for harboring planets where life might be able to begin.

After that, we should consider red dwarfs. They have the longest lifetimes of all, are the smallest and coolest type of star and by far the most common. Critical factors assumed to be impediments to habitability include relatively little energy output and thus reduced habitable zones and high stellar variation. In fact M dwarfs, especially during their early age, have strong magnetic activity, which makes them relatively bright in X-rays and UV radiation and causes them to flare frequently. As they age, red dwarfs become less magnetically active. Therefore, planets around red dwarfs in the HZ could be habitable if they can maintain a magnetic field for the 2 to 3 billion years that a red dwarf is active so that they can hold onto their atmosphere [104].

However, their HZs are very close in so the star's gravity would likely cause tidal lock. The daylight side of the planet would face the star all the time, while the night-time side would always face away from it. This implies they are rotating slowly around their axis so they could have a weak magnetic field that could shut down completely.

This is what happened to Mars. It had a magnetic field 3.5 billion years ago, but when its liquid iron core solidified, the field turned off. Without this protective shield, the solar wind stripped away most of the planet's atmosphere and liquid water. To avoid this fate around a red dwarf a planet might need to be more massive than Earth. The large liquid iron core inside a super Earth (see par. 3.2.1) could perhaps maintain a magnetic field in spite of the slower rotation rate.

Moreover, because of the tidal locking, potential life could be limited to a ring-like region between a hot area and a deep freeze. Joshi et al. [171] have shown that a planet's atmosphere needs to contain at least 100 mbars of greenhouse gas, e.g. $CO_2$, or 10% of Earth's atmosphere, for the star's heat to be effectively carried from the day side to the night side. This is well within the levels required for photosynthesis, though water may still remain frozen on the dark side.

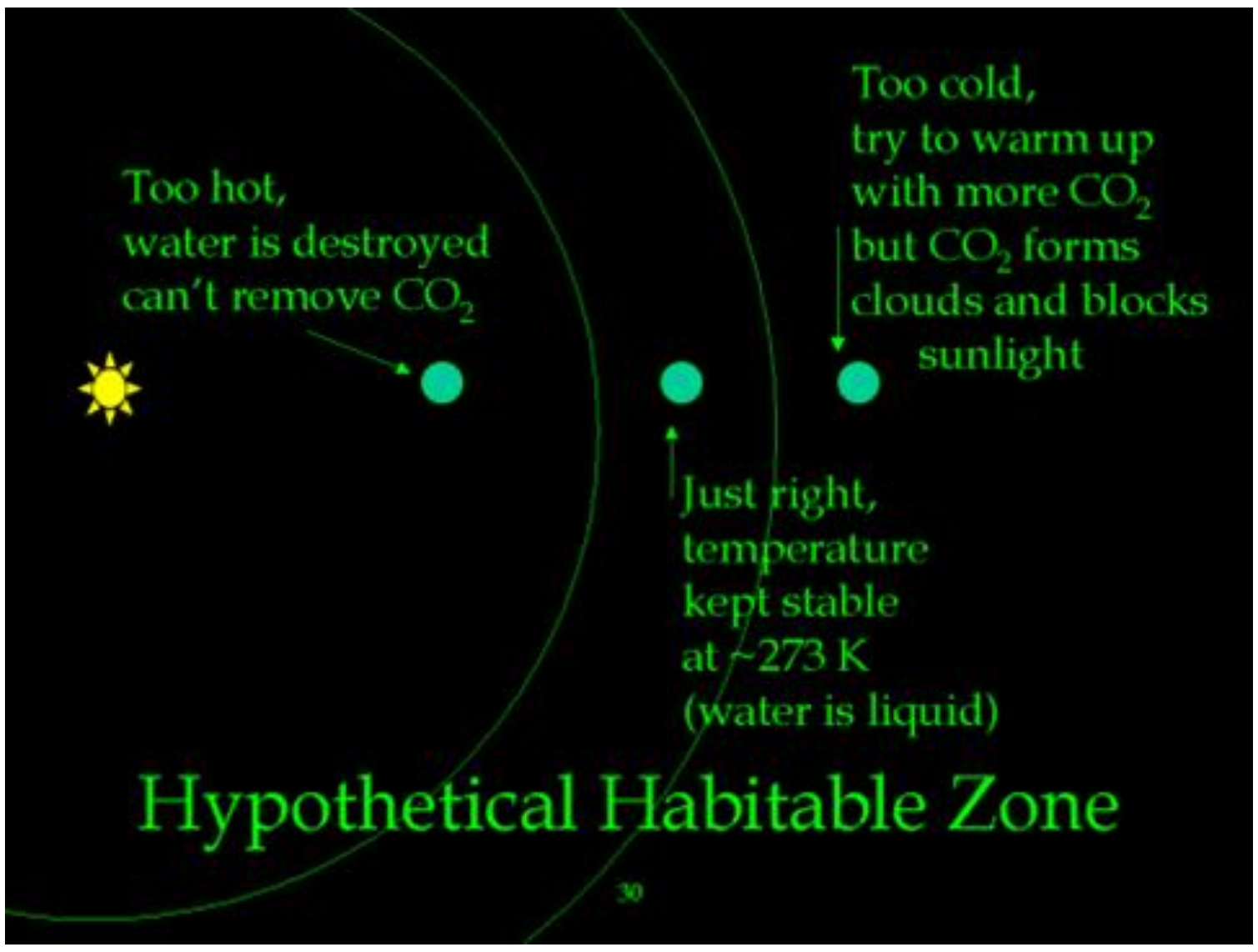

Figure 3.8: The boundaries of the HZ [54].

### *3.2.1 Possible Super-Earth Planets detected within the HZ*

Direct planetary observations led in 2007 to the discovery of the first possibly terrestrial planet within a star's habitable zone (Udry et al. [117]). Gliese 581 c is believed to be a rocky planet with a radius 1.5 times that of Earth and a minimum mass of roughly five times Earth around a M3V-type star. A direct measurement of the radius cannot be taken because the planet is not a transiting object but we know that it completes a full orbit in just under 13 days. It is notable as it is the planet with lowest minimum mass yet discovered in the habitable zone of another star, making it the most "earthlike" exoplanet found to date. According to Udry et al. [117], the *effective temperature* for Gliese 581 c, assuming an albedo such as Venus', would be −3 °C, and assuming an Earth-like albedo, then it would be 40 °C, a range of temperatures which overlaps with the range that water would be liquid at a pressure of 1 atmosphere. However, the effective temperature and actual surface temperature can be very different due to the greenhouse properties of the planetary atmosphere. For example, a speculative model by von Bloh et al. [121] estimates a temperature of around 500° due to a runaway greenhouse effect similar to that of Venus.

Subsequent observations revealed that also the outermost planet Gliese 581 d is within the habitable zone (M. Mayor et al. [122]). With a mass of roughly 7 Earths and an orbit of 66.8 Earth days [117], it is a potential candidate for being able to support life [121] [122].

However, the last word about these planets could be given only when we will be able to measure their spectra. So, starting with Kepler and JWST we now expect to detect and characterize terrestrial-size planets and determine if they are within the habitable zones of their stars. Such observational refinements may allow us to better gauge how common potentially habitable worlds are, and therefore allow us a much better idea of how common life in the universe might be.

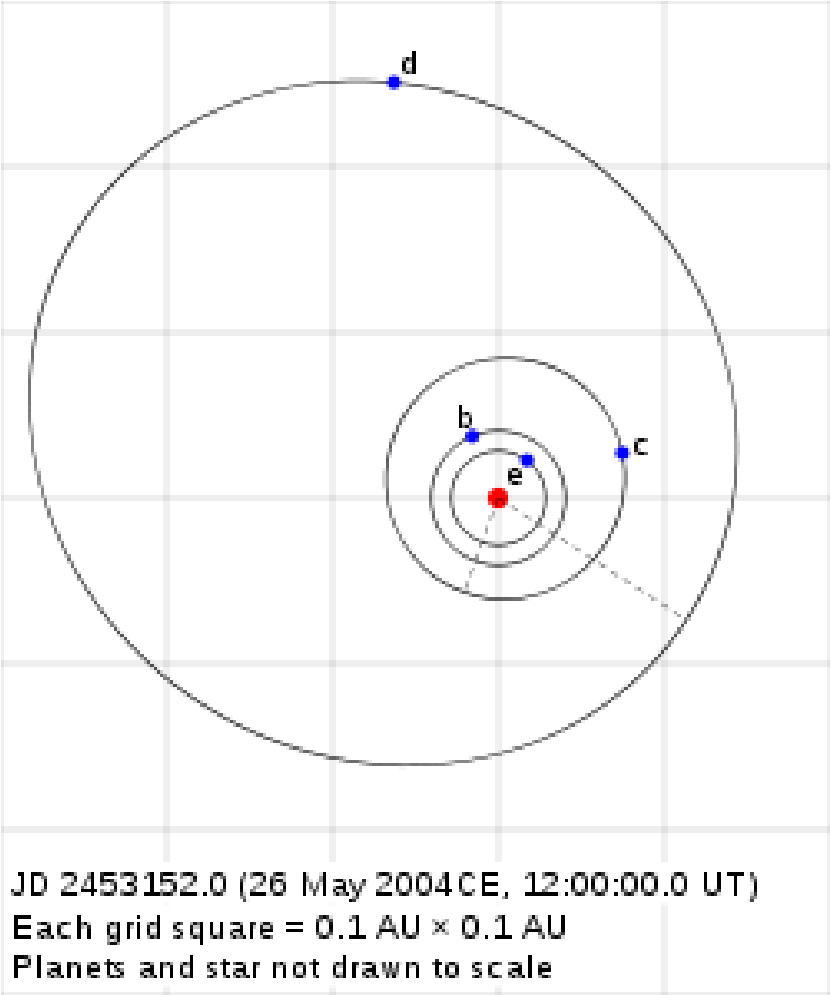

Figure 3.9: Gl 581’s planetary system.

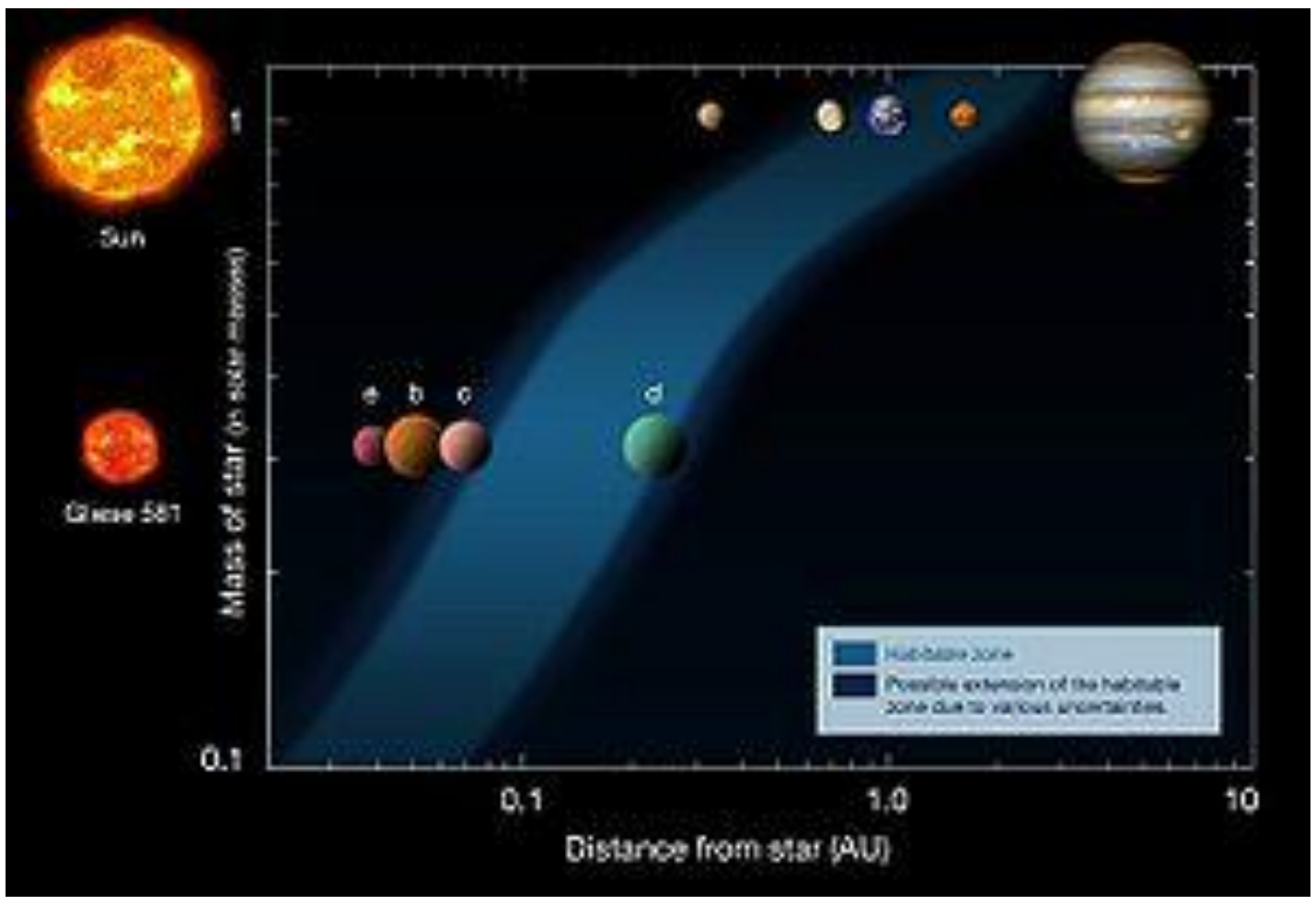

Figure 3.10: Comparison between the Solar System and the Gliese 581 System with the HZ highlighted.

### 3.3 Biosignatures

To know whether a planet in the HZ is *actually inhabited*, we need to search for biosignatures, features that are specific to biological activities and can be detected by remote sensing. There are at least two types of biosignatures: spectral and/or polarization features created by biological products and electromagnetic signals created by technology. Spectral biosignatures can arise from organic constituents (e.g., vegetation) and/or inorganic products (e.g., atmospheric $O_2$). The latter example of a biosignature requires SETI-like searches.

If we knew other examples of life forms, we could directly compare them and begin to discern general principles of the origins and evolution of life. We know the history of only one biosphere so, having few data, our task is more difficult. However, we can say that here on Earth most biological molecules are made from covalent combinations of six very common elements, whose chemical symbols are CHNOPS. Moreover, we can say that observations have led to some interesting discoveries: glycolaldehyde, the simplest possible sugar, was identified in a giant cloud of gas and dust near the center of the Milky Way Galaxy [172] and in a star-forming region [173]; copious amounts of nitrogen containing polycyclic aromatic hydrocarbons (PANHs), molecules critical to all known forms of life, have been found in a galaxy 12 million light-years away with Spitzer Space Telescope [174]; Glycine, an amino acid used by living organisms to make proteins, has been recognized in samples of comet Wild 2 returned by Nasa's Stardust spacecraft [175].

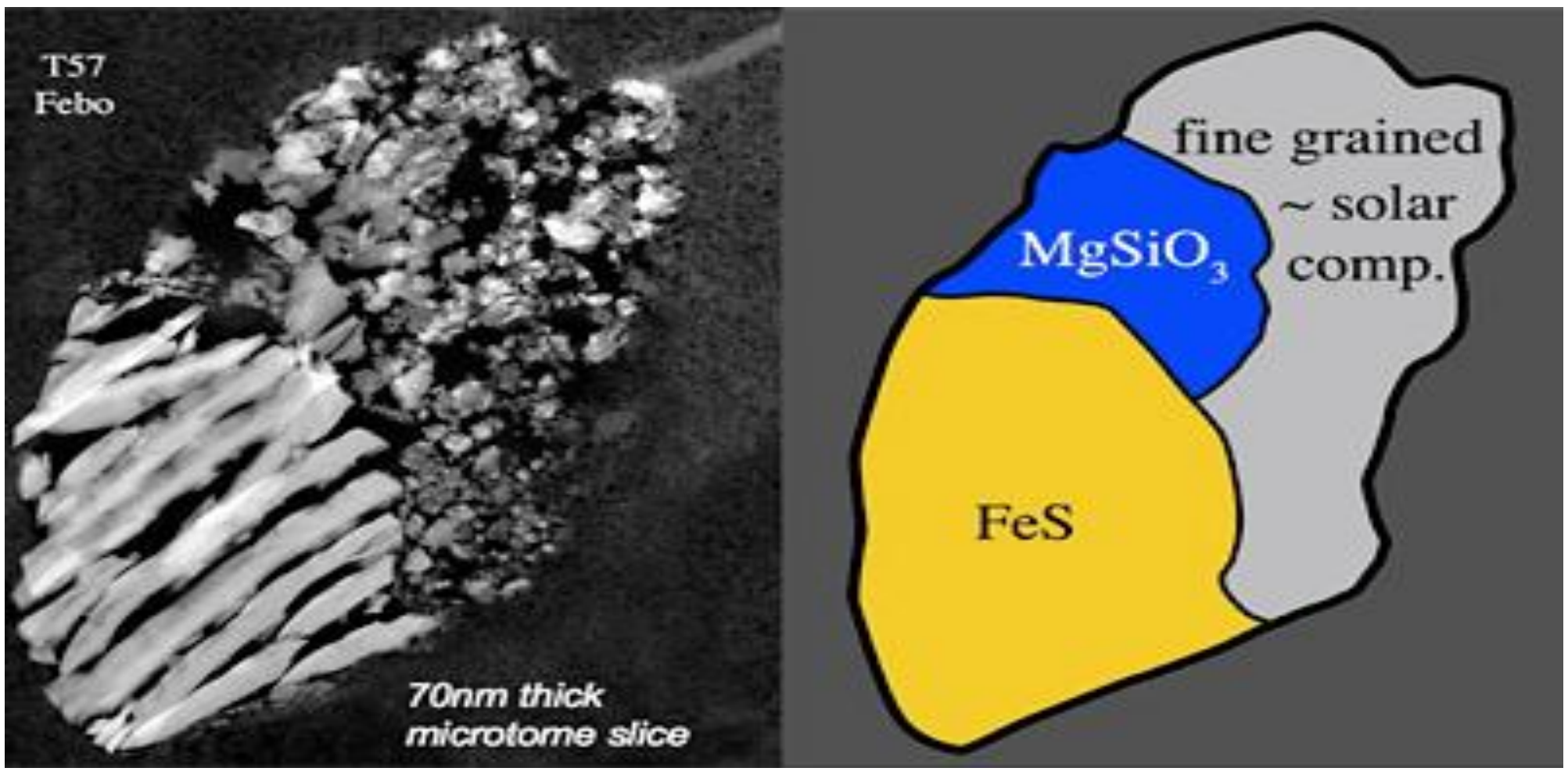

Figure 3.11: Example of one of the many organic particles collected and recovered by the Stardust mission.

The fact that complex, important biogenic compounds of CHNOPS elements have been found in space suggests that life around the universe can be based on these common elements and that the building blocks of life are prevalent in space. Herein we assume that all life requires complex organic compounds that interact in a liquid water solvent. These assumptions do not seem overly restrictive, given that life is an information-rich entity that depends fundamentally upon the strong polarity of its associated solvent. Carbon compounds and structures appear to be unrivaled in their potential for attaining high information contents. Other plausible solvents cannot match the strong polar–nonpolar dichotomy that water maintains with certain organic substances, and this dichotomy is essential for maintaining stable biomolecular and cellular structures [176]. However, our own biosphere utilizes only a small fraction of the number of potentially useful organic compounds. Alien life forms probably explored alternative possibilities and so their discovery will increase the known diversity of life. Hence, what signatures one would look for in the case of an habitable planet?

During the 1960s as a result of work for NASA concerned with detecting life on Mars [177], Lovelock formulated the Gaia hypothesis proposing that living and non-living parts of the earth, the biosphere, form a complex interacting system that has a regulatory effect on the Earth's environment that acts to sustain life [178][179]. Lovelock reasoned that many life forms on Mars would be obliged to make use of the Martian atmosphere and, thus, alter it. However, the atmosphere was found to be in a stable condition close to its chemical equilibrium, with very little oxygen, methane, or hydrogen, but with an overwhelming abundance of carbon dioxide. To Lovelock, the stark contrast between the Martian atmosphere and chemically-dynamic mixture of that of our Earth's biosphere was strongly indicative of the absence of life on the planet[177]. Therefore, one should conclude that a biosignature is a molecule which presence **alters the chemical equilibrium** of a planet atmosphere in a way that this disequilibrium cannot be explained solely by nonbiological processes.

### *3.3.1 The Earth in 1 dot*

In Fig. 3.12 we show a simple model (Traub and Jucks [100]) of spectra from the Sun and Solar System planets as seen from 10 pc, about the distance of a typical nearby star. This model assumes blackbody continuum spectra at the effective temperature of each object, plus reflectance spectra proportional to the visible albedo of each planet. The zodiacal light component is appropriate to a 0.1 AU diameter patch centered at 1.0 AU from the central star. The molecular band absorption features for the Earth's atmosphere have been superimposed for reference. We note that the contrast ratio (planet/star) of the Earth-twin is about $10^{-10}$ in the visible and $10^{-7}$ in the infrared.

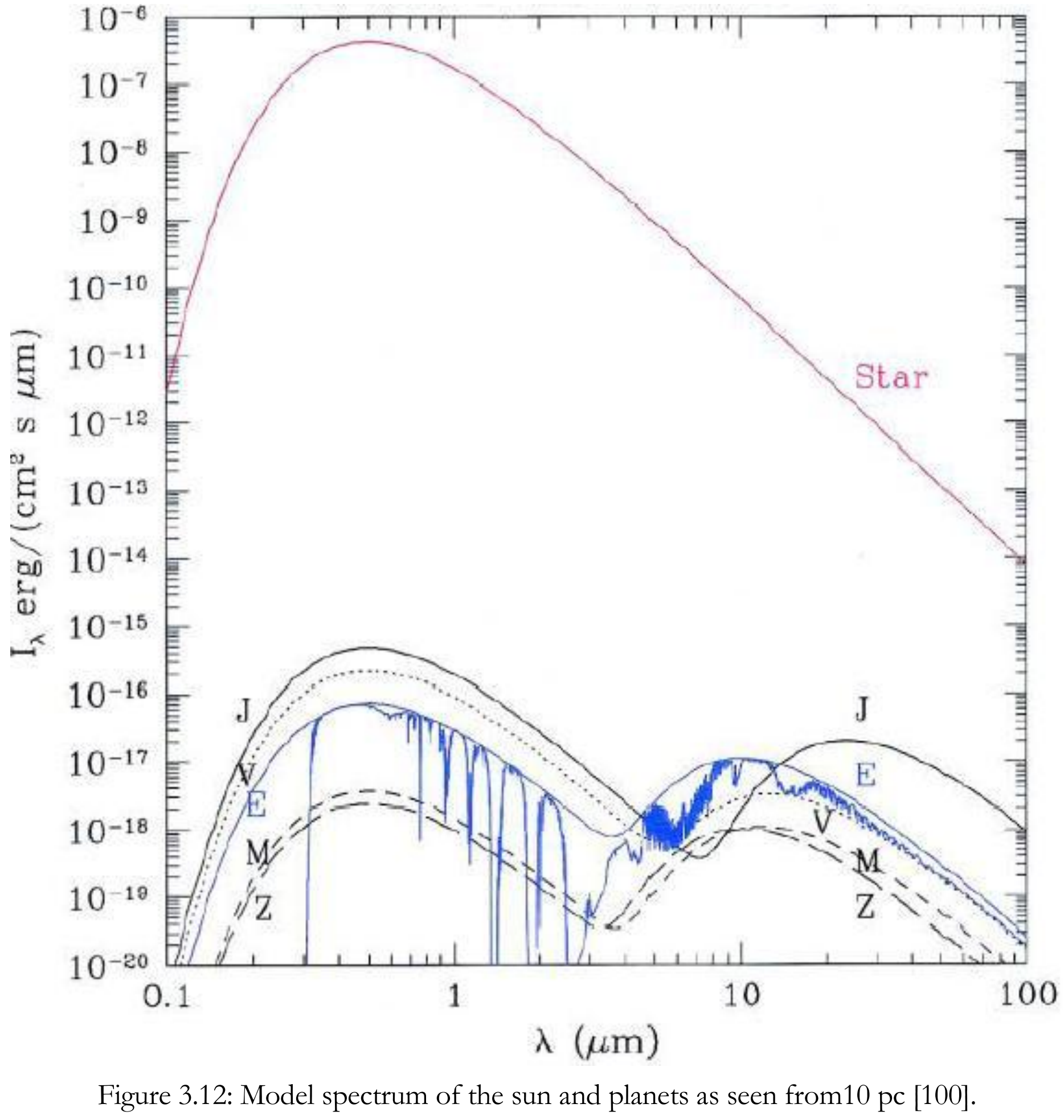

Figure 3.12: Model spectrum of the sun and planets as seen from10 pc [100].

The Earth's spectrum consists of two different parts. From the left side of the figure out to about 4 μm, the flux from Earth is proportional to that from the Sun because, at these wavelengths, one is seeing the Earth in reflected sunlight (the albedo of the Earth is 0.39, meaning that 39% of the Sun's radiation is reflected, and the remaining 61% is absorbed). Longward of 4 μm, one sees the Earth's own thermal-infrared radiation. Earth's spectrum is crudely similar to that of a blackbody with a temperature of ~255 K. Earth's surface is 33 degrees warmer than that due mostly to **greenhouse gases** in earth's atmosphere: water, carbon dioxide and sulphur dioxide are opaque to infrared radiation and their presence reduces the cooling of the planet by emission. Without the insulating effect provided by these greenhouse gasses all water would be frozen into ice [54].

Figure 3.13 shows that the mid-IR spectrum of Earth displays the 9.6 μm $O_3$ band, the 15 μm $CO_2$ band, the 6.3 μm $H_2O$ band and the $H_2O$ rotational band that extends longward of 12 μm [176]. The Earth's spectrum is clearly distinct from that of Mars and Venus, which display the $CO_2$ feature only.

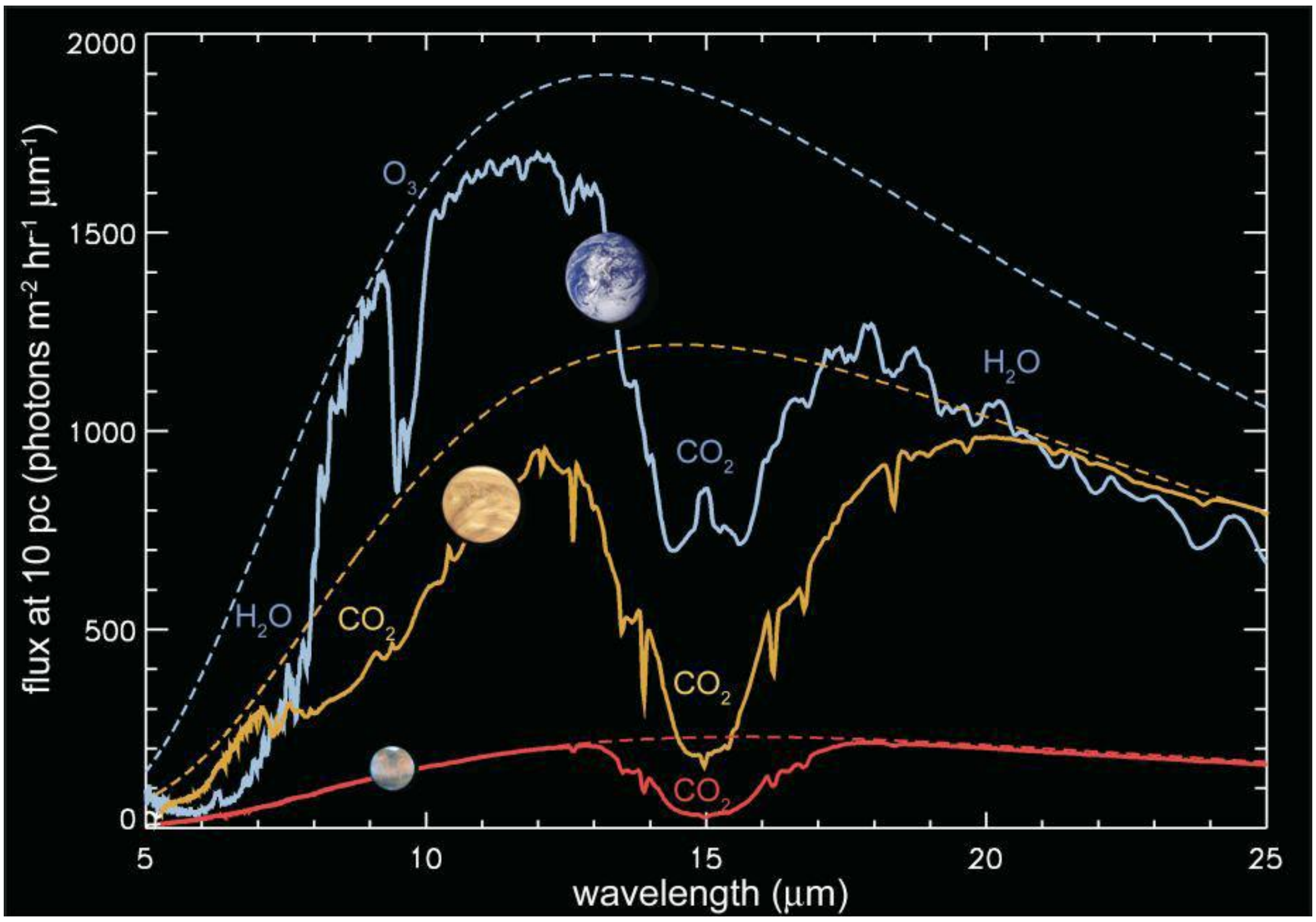

Figure 3.13: Mid-IR spectra of Venus, the Earth and Mars as seen from 10 pc.

A spectral feature of biological interest is free oxygen *producing photosynthesis* ($O_2$ band at 0.76 μm). For instance, the abundance of $O_2$ and its photolytic product $O_3$ have risen during the past billion years. If all oxygenic photosynthesis ceased today, then levels of free oxygen in Earth's atmosphere would decline rapidly (in geological terms), since the oxygen reacts readily with other compounds and is therefore quickly destroyed [106].

Other remarkable spectral features include methane ($CH_4$ band at 7.4 μm and 0.88 μm,) and species released as a consequence of biological fixation of nitrogen, such as ammonia ($NH_3$ at 6 and 9-11 μm), nitrous oxide ($N_2O$ at 7.8, 8.5 and 17 μm) and nitrogen dioxide ($NO_2$ at 6.2 μm). Methane and ammonia commonly appear in cold hydrogen-rich atmospheres, but they are not expected as abiotic constituents of Earth-size planetary atmospheres at habitable orbital distances. Known abiotic routes do not produce nitrous oxide and nitrogen dioxide.

According to some models of atmospheric evolution (Pavlov et al. [105]), methane was biologically maintained 2.5–3.0 billion years ago produced by methanogenic bacteria. But, $CH_4$ could have been produced abiotically as well from submarine outgassing. So, we would probably require additional information to decide whether a $CH_4$-rich atmosphere was really an indication of life [176].

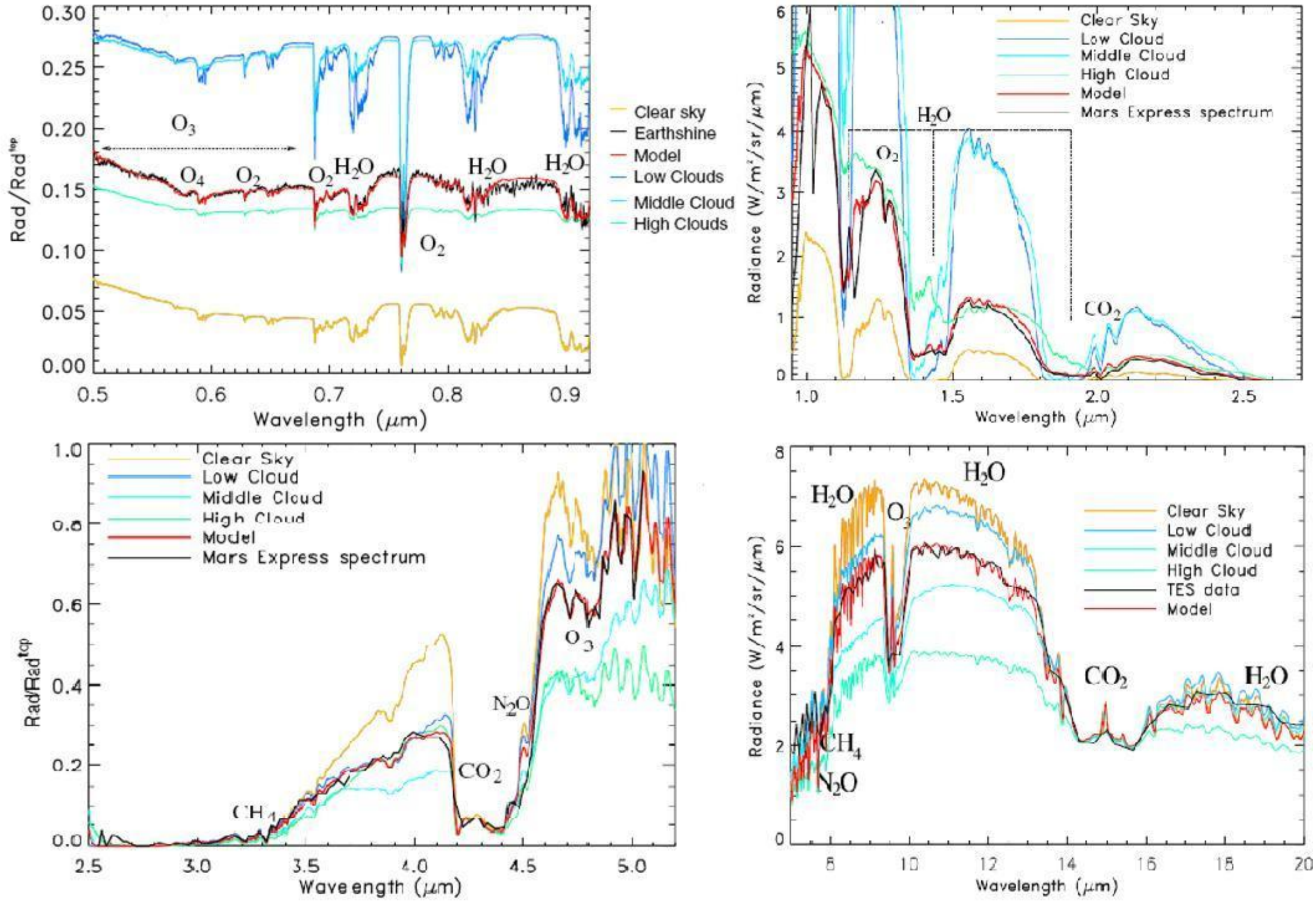

Figure 3.14: VIS, NIR and MIR Spectrum of the Earth observed with different instruments compared with synthetic models [185].

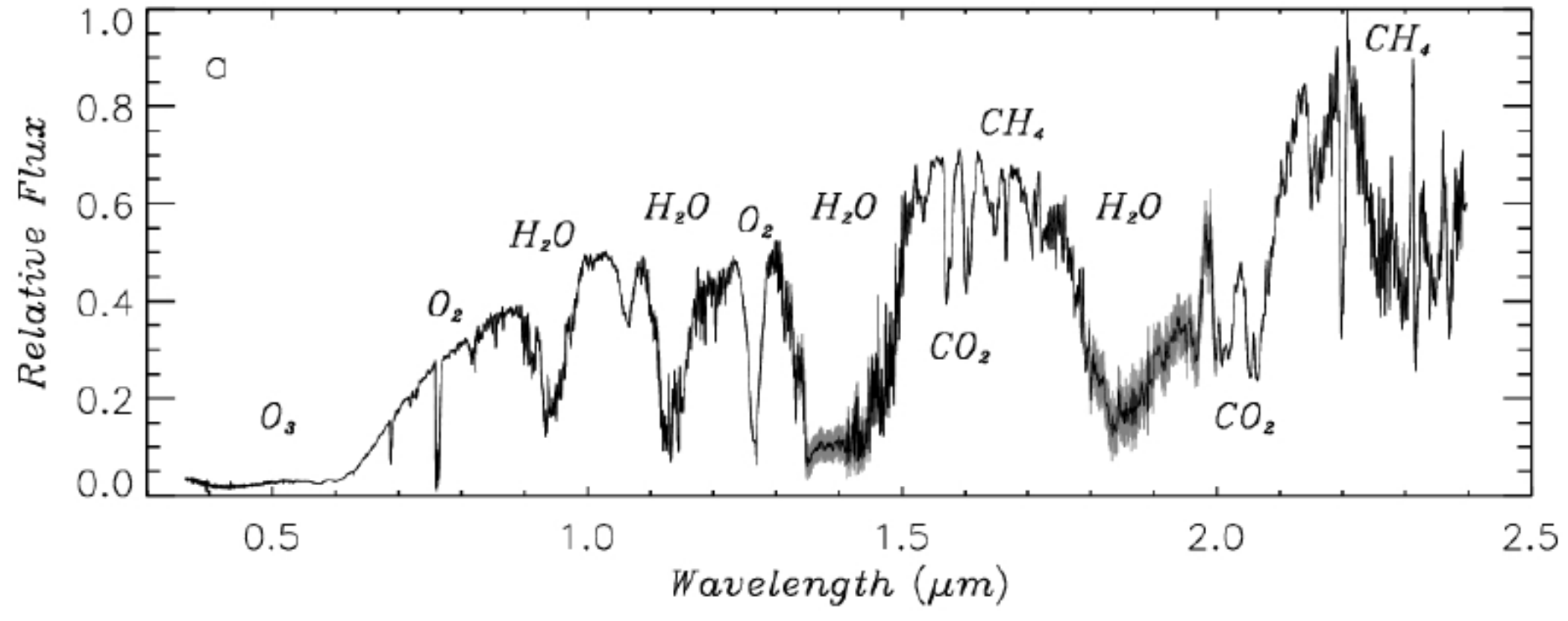

Figure 3.15: The Earth´s transmission spectrum from 0.36 to 2.40 μm. The major atmospheric features of the spectrum are marked [186].

### *3.3.2 Astrobiology of the Moons of the Outer Solar System*

The outer solar system provides a rich and rewarding assortment of planetary diversity of high interest to astrobiology. Spacecraft missions, particularly the Galileo mission and the Cassini-Huygens mission, indicate the potential habitability of planetary bodies in the outer solar system and in effect extended our concept of habitability (which was previously limited to Earth-like planets) to water worlds which are covered by ice, examples are Europa and Ganymede. Titan is a special case with its plethora of organic compounds closely related to prebiotic terrestrial chemistry for which habitability may be linked to an organic solvent that forms large liquid pools on the moon's surface. Enceladus is another unusual case, possibly containing subsurface reservoirs of liquid water and organic compounds that could have led to prebiotic chemistry or even life.

Europa is primarily made of silicate rock and likely has an iron core. It has a tenuous atmosphere composed primarily of oxygen. This young surface is striated by cracks and streaks, while craters are relatively infrequent. The apparent youth and smoothness of the surface have led to the hypothesis that a water ocean exists beneath it, which could conceivably serve as an abode for extraterrestrial life [181]. Heat energy from tidal flexing ensures that the ocean remains liquid and drives geological activity [182].

Enceladus is a relatively small satellite, with a mean diameter of 505 km. Cassini performed several close flybys of Enceladus and revealed a plume venting from the moon's south polar region containing salt-water and organic compounds such as carbonates and dust grains [102]. This discovery, along with the presence of escaping internal heat, extensive linear cracks [183] and very few impact craters in the south polar region, shows that Enceladus is geologically active today. Moreover, it gives strengthen evidence of the presence of sub-surface liquid water, which might also make it possible for Enceladus to support life (Parkinson et al. [184]).

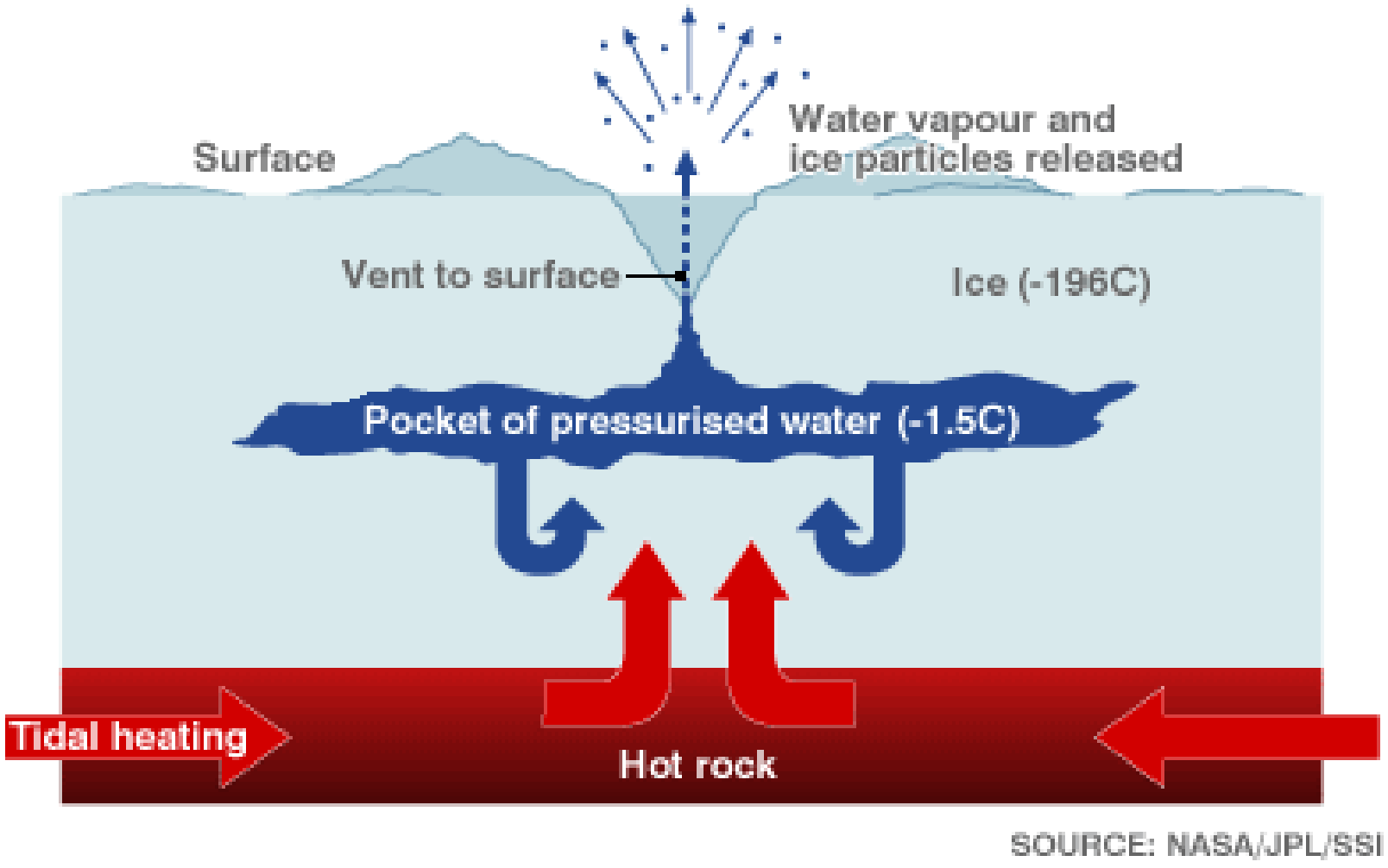

Figure 3.16: Illustration the hydrological, geochemical and chemical cycles on Enceladus.

There are two major Outer Planet flagship missions currently in preparation by NASA and ESA. They are the Europa Jupiter System Mission (EJSM), proposed for a launch in 2020, and the Titan Saturn System Mission (TSSM).

The EJSM concept consists of two primary flight elements operating in the Jovian system: the NASA-led Jupiter Europa Orbiter (JEO), and the ESA-led Jupiter-Ganymede Orbiter (JGO). JEO and JGO are designed to execute a choreographed exploration of the Jupiter System before settling into orbit around Europa and Ganymede, respectively. JEO and JGO are anticipated to determine abundances and distributions of surface materials, and characterize water oceans beneath the ice shells of Europa and Ganymede.

At arrival, the TSSM mission would perform an orbit insertion burn to capture into Saturn orbit. The hot-air balloon, targeted for Titan, would be dropped off just prior to the first Titan flyby following Saturn orbit insertion. Data relay from the balloon would continue through its six-month mission via the orbiter telecommunications system. The lander element, targeted for Kraken Mare (a northern lake) would be dropped off at the second Titan flyby and the orbiter would perform dedicated science data capture and relay for the nine-hour length of the lander's mission. During a two-year Saturn tour phase, the orbiter would perform seven close flybys of Enceladus sampling compounds ejected into its atmosphere. Finally, the Titan orbit phase would commence with a Titan orbit insertion burn, placing the orbiter in an elliptical orbit that would be used for concurrent aerobraking and aerosampling. The orbit would be circularized over two months, beginning a 20-month Titan orbit phase [180].

The primary significance of any life detection in the outer solar system would be that this life would most likely constitute a separate origin from life on Earth, thus allowing an assessment of how different (or exotic) life can be and insights into whether life is common in the universe or not.

*Chapter 4*

# PROBING THE ATMOSPHERE OF HD 209458B WITH TRANSMISSION PHOTOMETRY IN THE INFRARED

As an example of characterization of exoplanet atmospheres here we present a work [5] in which the author has been involved during the thesis preparation. It concerned the observation of the primary transit of HD 209458b at 3.6, 4.5, 5.8 and 8.0 μm using the Infrared Array Camera on the Spitzer Space Telescope.

The extrasolar planet HD 209458b orbits a main sequence G star at 0.046 AU (period **3.52 days**). It was the first exoplanet for which repeated transits across the stellar disk were observed (~ 1.5% absorption; Charbonneau et al. [14]). Using radial velocity measurements (Mazeh et al. [109]), the planet's mass and radius were able to be determined ($M_p \sim 0.69 M_{Jup}$, $R_p \sim 1.4 R_{Jup}$), confirming the planet is a gas giant with one of the **lowest densities** so far discovered. Consequently it must have a very extended atmosphere making it one of the optimum candidates for observation using primary transit techniques, and it was indeed the first exo-atmosphere probed successfully using this method in the visible (Charbonneau et al. [60]) and more recently in the IR (Richardson et al. [110]).

Following the work on HD 189733b where the first detections of water vapour (Tinetti et al. [67]; Beaulieu et al. [66]) and methane (Swain et al. [68]) have been achieved, the group has been awarded 20 hours as a Director Discretionary Time on Spitzer (PI Tinetti) to probe the atmosphere of HD 209458b in primary transit in the four IRAC bands. Water vapour was proposed to be present in the atmosphere of HD 209458b by Barman [111], to fit the data recorded by Hubble-STIS in the visible (Knutson et al. [78]). Also, water vapour in emission is a reasonable suggestion to fit the secondary transit photometric data observed in the Mid-IR (Deming et al. [70]; Knutson et al. [78]; Burrows et al. [79]). Our understanding of the thermal profile and composition has improved thanks to more recent secondary transit spectroscopic data in the Near and Mid-IR, indicating the additional presence of methane and carbon dioxide in the atmosphere of HD 209458b (Swain et al. [107]).

The mid-infrared primary transit observations described here allow us to probe the terminator region of HD 209458b (see par. 3.1) between the bar and millibar level.

First of all we desire to introduce three important equipments that have allowed these observations and the subsequent data reduction: Spitzer Observatory, the Infrared Array Camera and SExtractor package.

### 4.1 Spitzer

The Spitzer Space Telescope (formerly the Space Infrared Telescope Facility, SIRTF) has revolutionized the observational characterization of exoplanets by detecting infrared light emitted from these objects. Broadband, photometric measurements have been reported for HD 209458b, TrES-1, HD 189733b and υ Andromeda b (see par. 3.1). The impact of these measurements on our understanding of exoplanets is difficult to overstate.

This $800 million telescope was launched from Cape Canaveral Air Force Station by a Delta II rocket on 25 August 2003. The mission is operated and managed by the Jet Propulsion Laboratory and the Spitzer Science Center, located on the Caltech campus in Pasadena, California. It is the fourth and final of NASA's Great Observatories - a family of four orbiting observatories, each observing the Universe in a different wavelength range (visible, gamma rays, X-rays and infrared). Other missions in this program include the Hubble Space Telescope, Compton Gamma-Ray Observatory and the Chandra X-Ray Observatory.

Earlier infrared observations had been made by both space-based and ground-based observatories. Ground-based observatories have the drawback that at infrared wavelengths or frequencies, both the Earth's atmosphere and the telescope itself will radiate strongly. Additionally, the atmosphere is opaque at most infrared wavelengths. Previous space-based satellites (such as IRAS, the Infrared Astronomical Satellite, and ISO, the Infrared Space Observatory) were operational during the 1980s and 1990s and great advances in astronomical technology have been made since then.

In keeping with NASA tradition, the telescope was renamed after successful demonstration of operation, on December 18 2003. Unlike most telescopes which are named after famous deceased astronomers by a board of scientists, the name for SIRTF was obtained from a contest open to the general public. The result was it being named in honor of Lyman Spitzer, one of the 20th century's great scientists.

Spitzer follows an "earth-trailing" orbit (period 372 days) drifting away from Earth's orbit at approximately 0.1 astronomical unit per year. Cryogenic satellites that require liquid helium (LHe, T ≈ 4 K) temperatures in near-Earth orbit are typically exposed to a large heat load from the Earth, and consequently entail large usage of LHe coolant, which then tends to dominate the total payload mass and limits mission life. Placing the satellite in solar orbit far from Earth allowed innovative passive cooling (such as the sun shield, against the single remaining major heat source) to drastically reduce the total mass of helium needed, resulting in an overall smaller lighter payload, with major cost savings. This orbit also simplifies telescope pointing, but does require the Deep Space Network for communications.

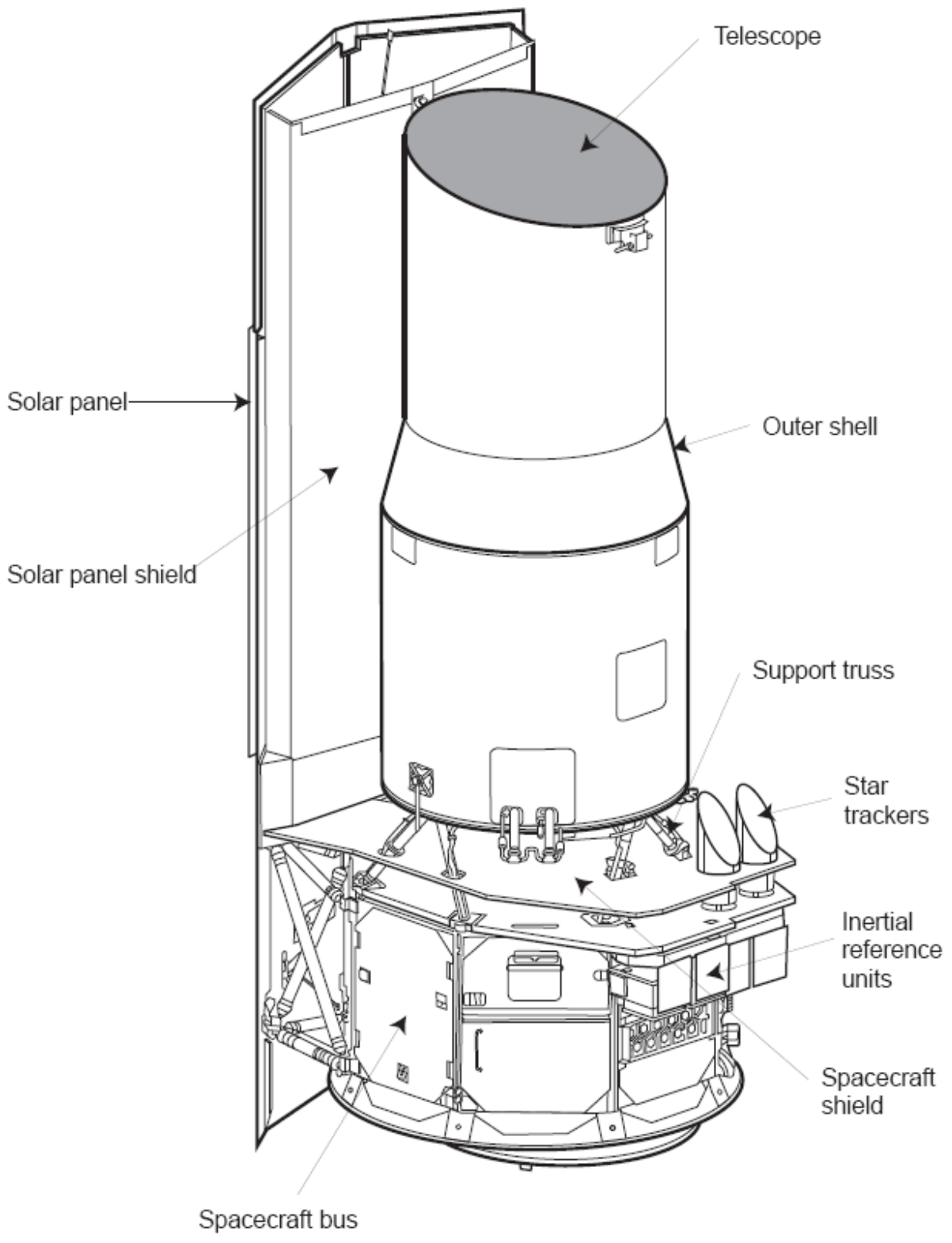

Figure 4.1: Observatory exterior.

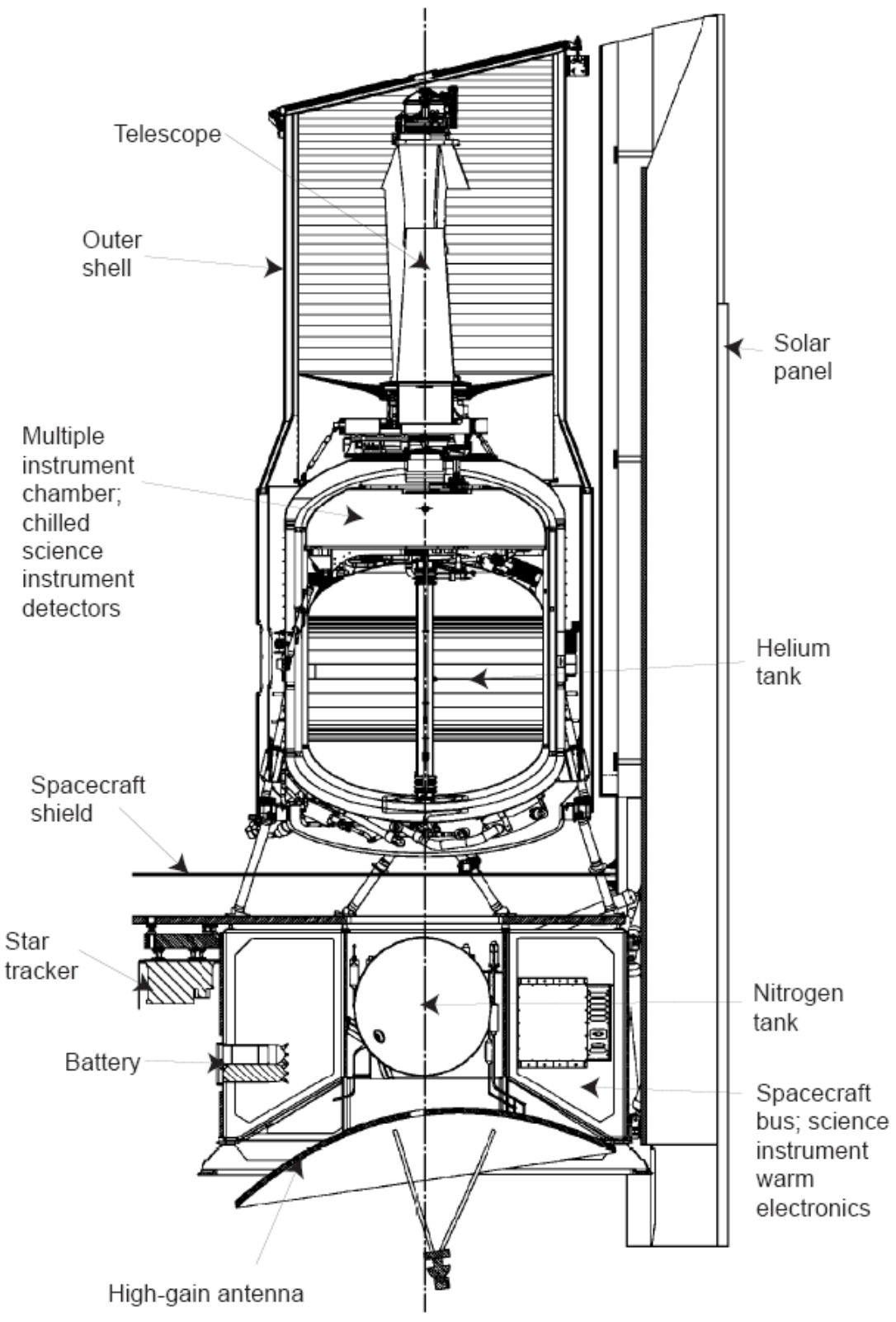

Figure 4.2: Observatory cutaway.

Once a week, a new sequence of commands is uplinked, telling the observatory what scientific targets to observe and when. The mission's telescope has a Ritchey-Chrétien design, the primary mirror is 85 cm in diameter, f/12 and made of beryllium. The telescope is attached to the top of the vapor-cooled cryostat shell. Only the portions of the three science instruments that need to be chilled are contained in the multiple instrument chamber, which is mounted directly to the helium tank in the cryostat shell. These three instruments have allowed performing imaging and photometry from 3 to 180 micrometers, spectroscopy from 5 to 40 micrometers, and spectrophotometry from 5 to 100 micrometers. They are:

- IRAC (Infrared Array Camera): an infrared camera which operates simultaneously on four wavelengths (3.6 μm, 4.5 μm, 5.8 μm and 8 μm). Each module uses a 256 × 256 pixel detector -- the short wavelength pair use indium antimonide technology, the long wavelength pair use arsenic-doped silicon impurity band conduction technology (see par. 4.2).

- IRS (Infrared Spectrograph): an infrared spectrometer with four sub-modules which operate at the wavelengths 5.3-14 μm (low resolution), 10-19.5 μm (high resolution), 14-40 μm (low resolution), and 19-37 μm (high resolution). Each module uses a 128x128 pixel detector -- the short wavelength pair use arsenic-doped silicon blocked impurity band technology, the long wavelength pair use antimony-doped silicon blocked impurity band technology.

- MIPS (Multiband Imaging Photometer for Spitzer): three detector arrays in the far infrared (128 × 128 pixels at 24 μm, 32 × 32 pixels at 70 μm, 2 × 20 pixels at 160 μm). The 24 μm detector is identical to one of the IRS short wavelength modules. The 70 μm detector uses gallium-doped germanium technology, and the 160 μm detector also uses gallium-doped germanium, but with mechanical stress added to each pixel to lower the bandgap and extend sensitivity to this long wavelength [123].

The planned nominal mission period was to be 2.5 years with a pre-launch expectation that the mission could extend to five or slightly more years until the onboard liquid helium supply (360 liters at launch) was exhausted. This occurred on 15

May 2009. The two shortest wavelength modules of the IRAC camera are still operable in a "warm" (~ 30K) telescope so surveys will continue at reduced sensitivity in these wavebands only in the Spitzer Warm Mission.

Spitzer has several systematic errors that include pointing, background contamination, charge trapping, and detector fringing. One difficulty with Spitzer is that no on-board calibration exists to monitor the changes in the instrument systematic. Thus, calibrated instrument stability for spectroscopy at the level of one part in $10^4$ may be difficult to achieve [124].

## 4.2 IRAC

IRAC is an imaging camera operating at four “channels". The short wavelength detectors are InSb devices, while the two longer wavelength detectors are Si:As devices. The instrument observes two fields of view (FOVs) simultaneously, with their centres about $6'.5$ apart, leaving a gap of about $1'.5$ between the FOVs. Each FOV is seen simultaneously by two detectors (one InSb and one Si:As detector), fed through lenses and filters and through a beamsplitter. The arrays are read out via four readout amplifiers. Data are obtained simultaneously in all four arrays for all the frame times, regardless of whether only one or two FOVs were specified in the observing request, although the various channels may observe nearby, but separate, parts of the sky.

There are two components to IRAC: the cold assembly and the warm electronics. The cold assembly primarily contains the detector arrays. It is temperature-controlled to milli-Kelvin levels and is contained within the cryogenically cooled multi-instrument chamber. The warm electronics is connected to the cold assembly by a cable.

### *4.2.1 IRAC Data Products*

The Basic Calibrated Data (BCD) are the calibrated, individual images. These are in array orientation and have a size of 256 x 256 pixels for the full array images, and 64 planes times 32 x 32 pixels for the subarray images. These data are fully calibrated and have detailed file headers. The Post-BCD pipeline combines the BCD images into mosaics (per wavelength and per frame time). Calibration observations designated as darks or flats go through a similar but separate pipeline.

Here we describe some of the important header keywords. An almost complete IRAC image header description is included in Appendix A.

AORLABEL is the name of the Astronomical Observation Request as defined by the user when requesting the observations. The P.I. will be listed as the OBSRVR of each project. AORKEY is a unique digit sequence for each observation; it is also part of the filename for each BCD. EXPID is an exposure counter incremented within a given AOR for each data-taking command. The DCENUM is a counter of individual frames (per wavelength) from an individual command; it can be used to separate frames generated with internal repeats.

DATE OBS is the time at the start of the AOR. Other times in the header include the time since IRAC was turned on (both for the beginning and end of the frame). EXPTIME is the effective integration time, UTCS OBS is the start of IRAC data taking sequence.

BUNIT gives the units (MJy/sr) of the images. FLUXCONV is the calibration factor derived from standard star observations; its units are (MJy/sr)/(DN/s). The raw files are in "data numbers" (DN). To convert from MJy/sr back to DN, one should divide by FLUXCONV and multiply by EXPTIME. To convert DN to electrons, one should multiply by GAIN. CRVAL1 and CRVAL2 give the coordinates of the image centre.

For one of our AOR (AORKEY 24740096), we have 2860 files of each type for each channel. The final fits images from this AOR in channel 3 are as follows:

SPITZER_I3_24740096_0000_0000_1_bcd.fits
…
SPITZER_I3_24740096_2859_0000_1_bcd.fits

After the name of the telescope, the first partition gives the instrument ("I" = IRAC), and the number after the "I" gives the channel (in this case, 3). The next part gives the AORKEY, then we have the EXPID, DCENUM, and the version number (how many times these data have been processed through the pipeline). One should generally use only the data from the highest version number, in case multiple versions have been downloaded from the archive. Finally, there is a group of letters that specify what kind of data are in the file (in this case, BCD data) and the file type (here "fits").

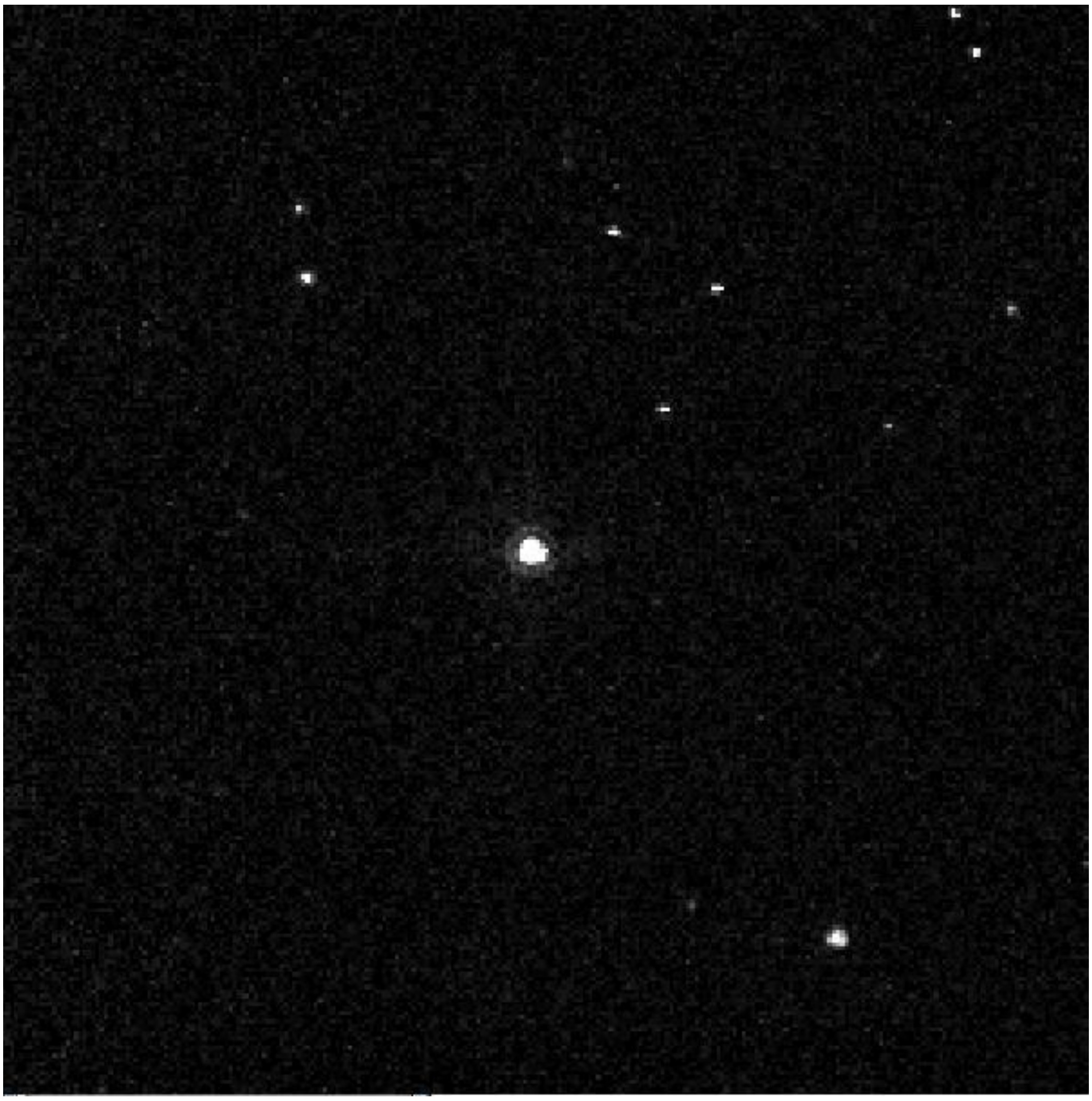

Figure 4.3: SPITZER_I3_24740096_2417_0000_1_bcd.fits with HD 209458b in the centroid.

### *4.2.2 IRAC Science BCD Pipeline and Detector Artifacts*

The IRAC BCD pipeline is designed to take a single raw image from a single IRAC detector and produce a flux-calibrated image which has had all well-understood instrumental signatures removed. It consists of several modules, each of these modules corrects a single instrument signature.

• Dark Subtraction

The true dark current in the IRAC detectors is actually very low - the most notable dark current features are the electronic glows seen in the Si:As arrays (channels 3 & 4). There are two steps for the dark subtraction, one using a dark from the ground-based laboratory measurements (the so-called "lab dark"), and another using a delta dark which is the difference between the lab dark and the sky dark measured at a low zodiacal background light region. The skydarks are taken in pre-determined regions which have been specifically chosen to be the darkest parts of the sky and are in regions that are relatively free of bright objects such as stars.

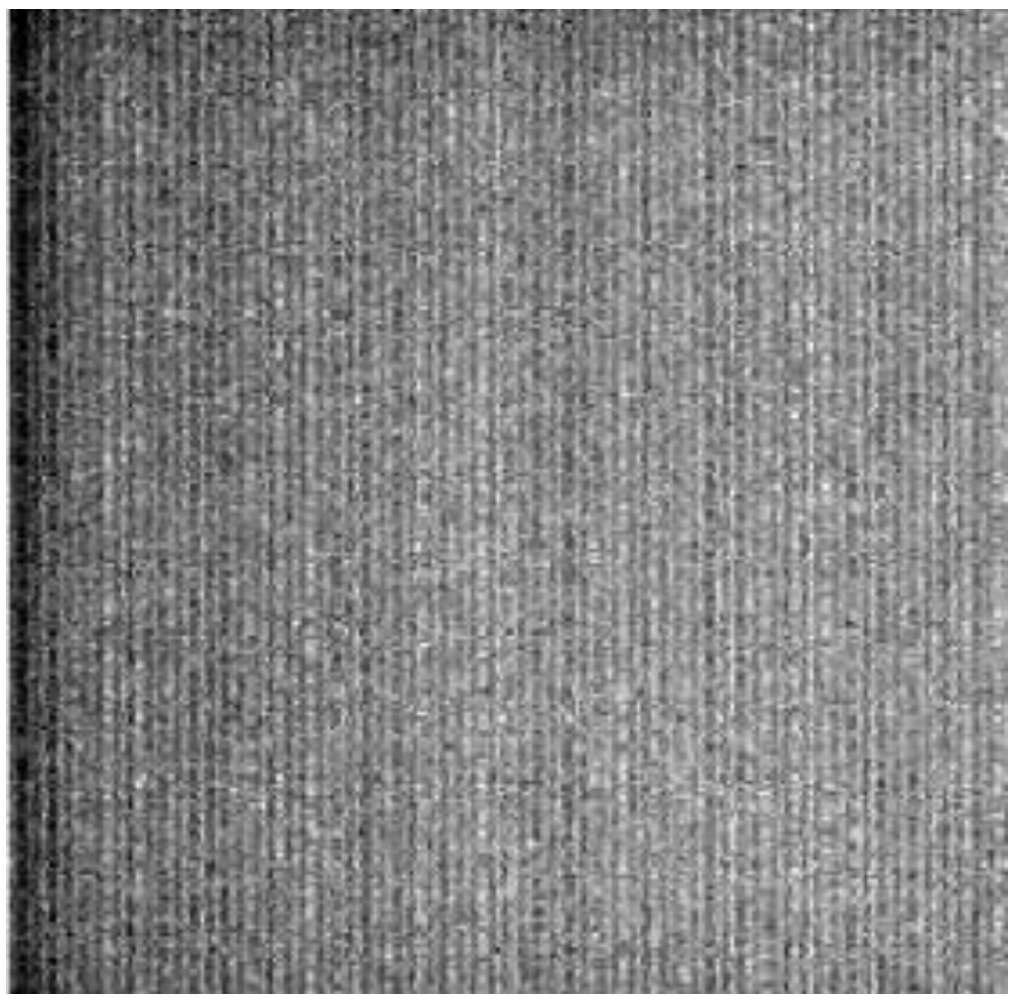

Figure 4.4: IRAC skydark with a frametime of 12 seconds for channel 3. Data from AORKEY = 9552128.

• Flat-Fielding

Similarly to all imaging detectors, each of the IRAC pixels has an individual response function (i.e. DN/incident photon conversion). To account for this pixel-to-pixel responsivity variation, each IRAC image is corrected by dividing it by a map of these variations, called a flat field. Observations are taken of approximately 20 pre-selected regions of high-zodiacal background chosen to be as free as possible of stars and Galactic cirrus. Between 80 and 100 dithered frames of 100 seconds in each channel are taken. Additionally, objects are detected during the averaging and rejected. The flat fields are then normalized to one. Analysis of data from the first two years of operations has showed that the flat field response of IRAC is unchanging at the limit of our ability to measure. As a result, so-called "super skyflats" were generated.

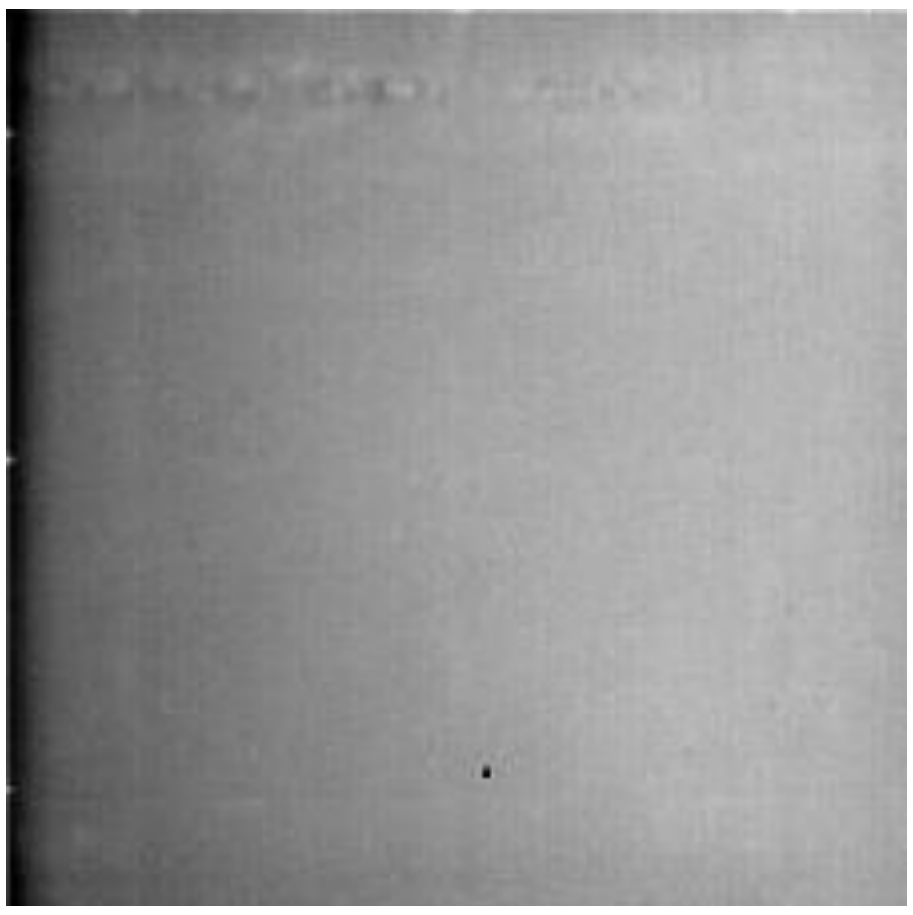

Figure 4.5: Superskyflat for IRAC's channel 3. This was made by combining the flat fields from the first two years of operations.

- Cosmic Ray Detection

Individual frames are analyzed for probable radiation hits (cosmic rays), and the results appear as a flag in a radhit image. This is computed by a median filtering technique. Input images are read in, and a median filter is applied. The difference between the input image and the median-filtered image is then computed. Pixels above a specified threshold are then flagged. Each IRAC array receives approximately 1.5 cosmic ray hits per second, with ~ 2 pixels affected in channels 1 and 2, and ~ 6 pixels per hit affected in channels 3 and 4. Radiation hits do increase suddenly and dramatically during some major solar proton events.

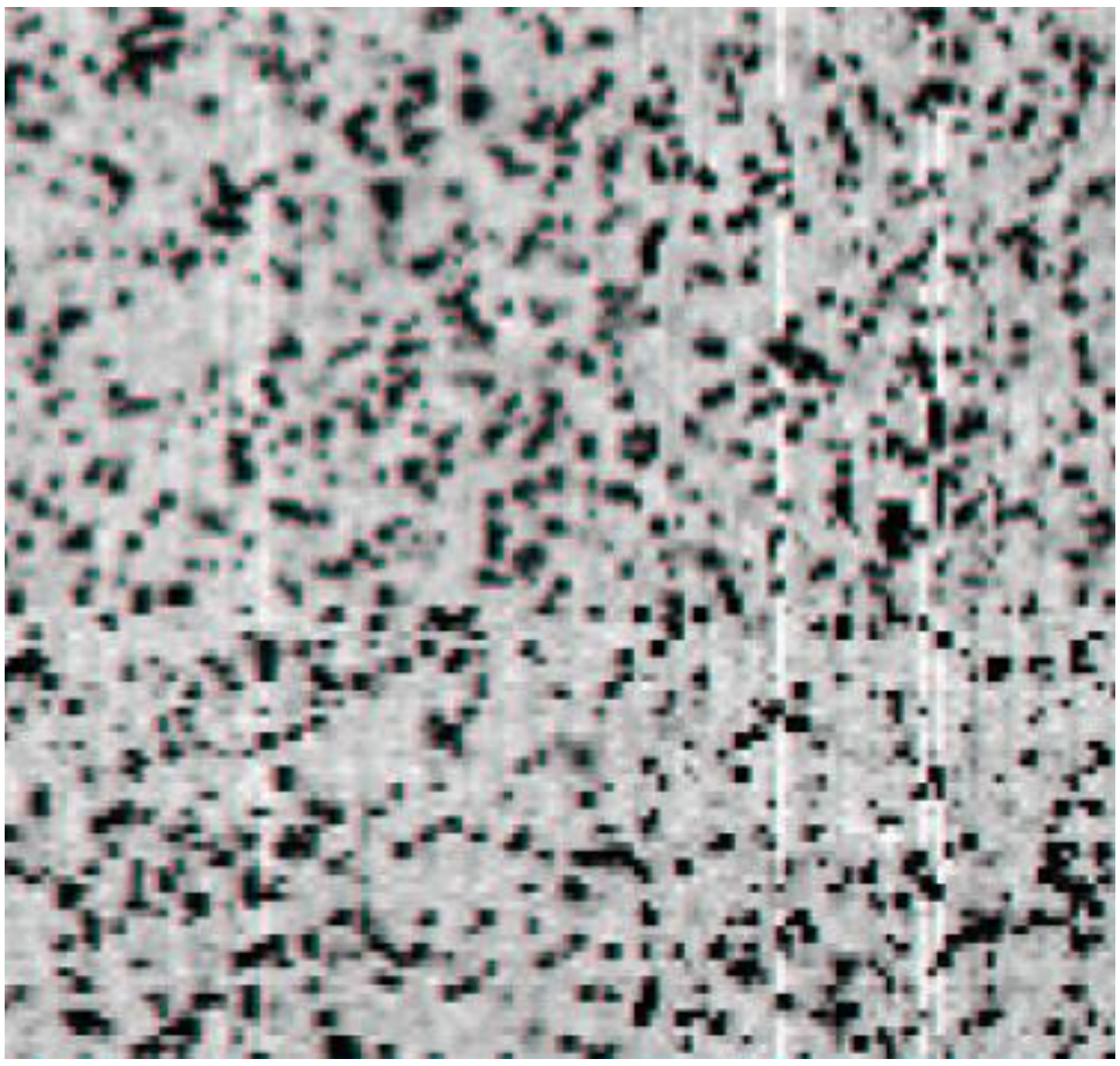

Figure 4.6: The central 128 x 128 pixels of channel 2 12-second image taken on January 20, 2005 during a major solar proton event.

• Bad Pixels and Optical Ghosts

Hot pixels usually appear bright. These pixels have high dark currents and are usually isolated but sometimes in a clump. "Dead" pixels are really just very hot pixels, so hot that they saturate before the first pedestal sample. In a BCD image, hot pixels do not appear bright because they have been canceled by the labdark or skydark subtraction. Most hot pixels appeared after launch and are the result of hits by energetic nuclei. By annealing the arrays, we restore most pixels that get activated. Some of them cannot be restored, and thus they become "permanent" hot pixels. Some pixels jump randomly from normal to high dark current and back, dwelling in one state for anywhere from a few minutes to weeks, so they may not be canceled by a skydark subtraction. These are IRAC's "rogue pixels".

There are three types of known or potential optical ghosts visible in IRAC images. The brightest and most common ghosts are produced by internal reflections within the filters. Similar ghosts are created by internal reflections within the beamsplitters. These only affect channels 3 and 4 which are transmitted through the beamsplitters. The faintest identified ghosts appear as images of the Spitzer entrance pupil, i.e., the primary mirror shadowed by the secondary and supports. These pupil ghosts are only found in channels 2 and 4, and require an extremely bright source (e.g., a first magnitude star in a 12-second frame) to be seen due to their low surface brightnesses.

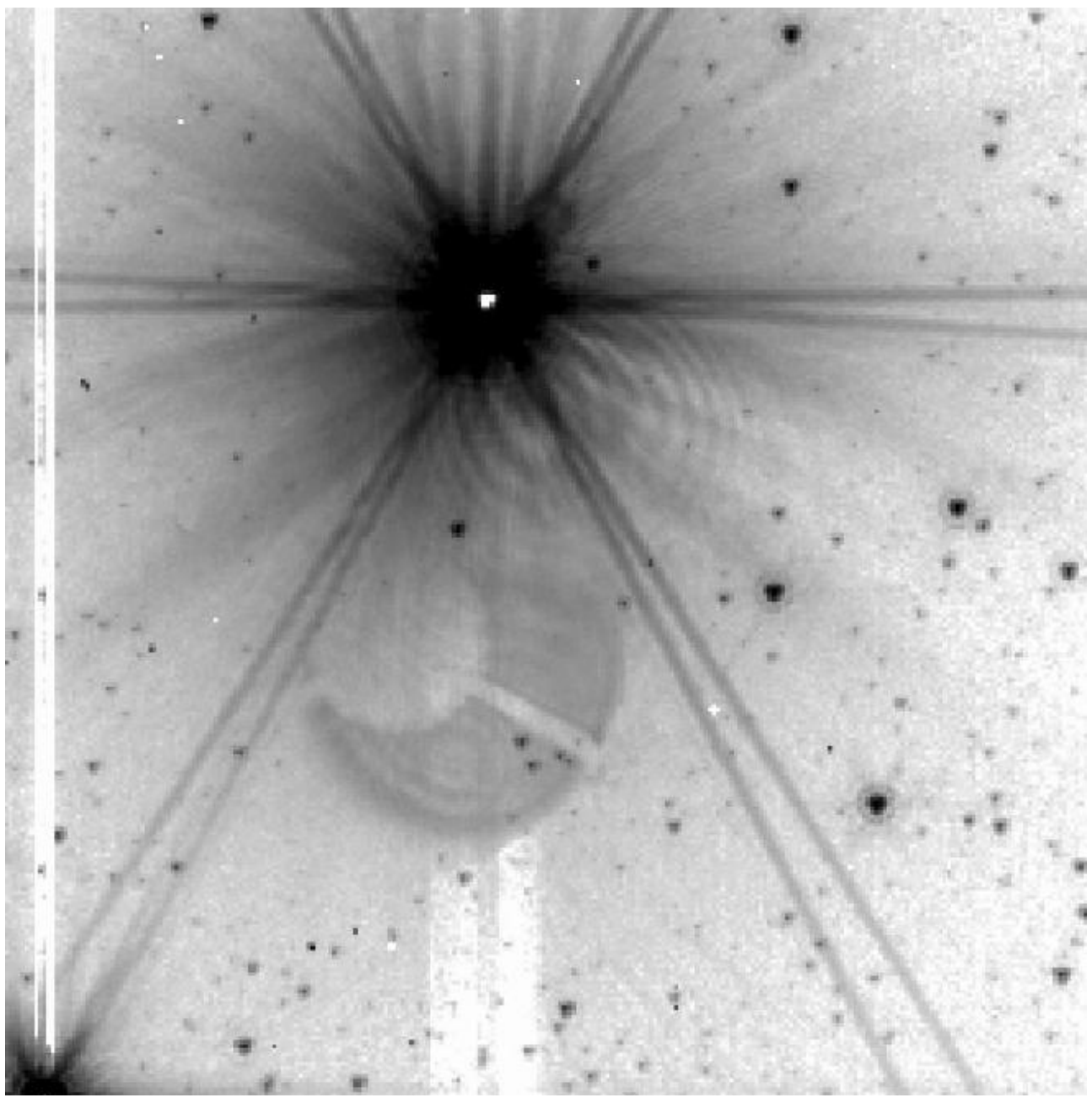

Figure 4.7: Pupil ghost in channel 2 from V416 Lac.

### *4.2.3 Photometry and Calibration*

IRAC is calibrated using both so-called primary and secondary calibrator stars. The primary stars are used to monitor long-term variations in the absolute calibration. They currently number 11 stars, are located in the continuous viewing zone (CVZ), and are thus observable year-round. They are observed once at the beginning, and once at the end of each campaign, i.e., about every 10 days whenever the instrument is switched on. The secondary calibrator stars are used to monitor short-term variations in the absolute calibration.

- Pixel Phase

The flux density of a point source measured from IRAC images depends on the exact location where the peak of the Point Response Function (PRF) falls on a pixel. This effect is due to the variations in the quantum efficiency of a pixel, and combined with the undersampling of the PRF, it is most severe in channel 1. The correction can be as much as **4%** peak to peak. The effect is graphically shown in Figure 4.8 where the normalized measured flux density (y-axis) is plotted against the distance of the source centroid from the centre of a pixel. The correction for channel 1 can be calculated from:

$$\text{Correction} = 1 + 0.0535 \times \left( \frac{1}{\sqrt{2\pi}} - p \right) \tag{4.1}$$

where *p* is the pixel phase ( $p = \sqrt{(x - x_0)^2 + (y - y_0)^2}$ ), where (x,y) is the centroid of the point source and $x_0$ and $y_0$ are the integer pixel numbers containing the source centroid. The correction was derived from photometry of a sample of stars, each star observed at many positions on the array. The "ratio" on the vertical axis in Figure 4.8 is the ratio of the measured flux density to the mean value for the star. To correct the flux of a point source, one should calculate the correction from Equation 4.1 and divide the source flux by that correction.

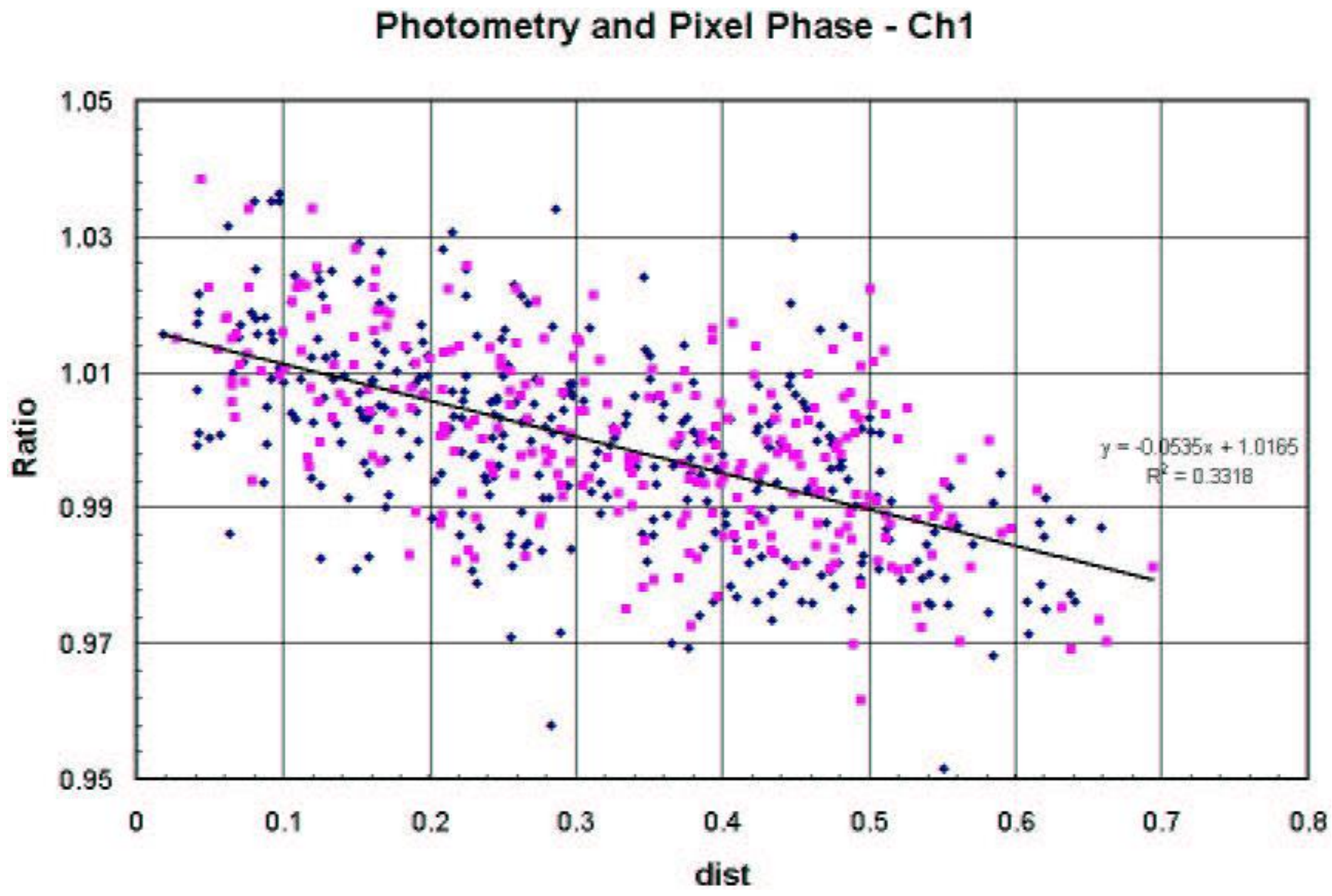

Figure 4.8: Dependence of point source photometry on the distance of the centroid of a point source from the nearest pixel centre in channel 1.

- Point-Spread Function

Point Source Photometry using IRAC data is no different from that with any other high-quality astronomical data. Both aperture photometry and PSF-fitting can be tried. Aperture photometry is most commonly used, so we will discuss it briefly. The radius of the on-source aperture should be chosen in such a way that it includes as much of the flux from the star (thus, greater than 2 arcseconds) as possible, but it should be small enough that a nearby background annulus can be used to accurately subtract unrelated diffuse emission, and that other point sources are not contributing to the aperture. For calibration stars, an annulus of 12 arcseconds is used; such a wide aperture will often not be possible for crowded fields. The dominant background in regions of low interstellar medium is zodiacal light, which is very smooth. In regions of significant interstellar emission, it is important to use a small aperture, especially in IRAC channels 3 and 4. The flux of a source can then be calculated in the standard way, taking the average over the background annulus, subtracting from the pixels in the on-source region, and then summing over the on-source region. Then, the measured brightness should be multiplied by the aperture corrections in Table 4.1, ranging from 12% to 23% [125].

Table 4.1: IRAC aperture corrections.

| radius on source | background annulus | Aperture correction | | | |
|---|---|---|---|---|---|
| | | 3.6 μm | 4.5 μm | 5.8 μm | 8.0 μm |
| **infinite** | N/A | 0.944 | 0.937 | 0.772 | 0.737 |
| **10** | 10-20 | 1.000 | 1.000 | 1.000 | 1.000 |
| **5** | 10-20 | 1.049 | 1.050 | 1.058 | 1.068 |
| **5** | 5-10 | 1.061 | 1.064 | 1.067 | 1.089 |
| **3** | 10-20 | 1.112 | 1.113 | 1.125 | 1.218 |
| **3** | 3-7 | 1.124 | 1.127 | 1.143 | 1.234 |
| **2** | 10-20 | 1.205 | 1.221 | 1.363 | 1.571 |
| **2** | 2-6 | 1.213 | 1.234 | 1.379 | 1.584 |

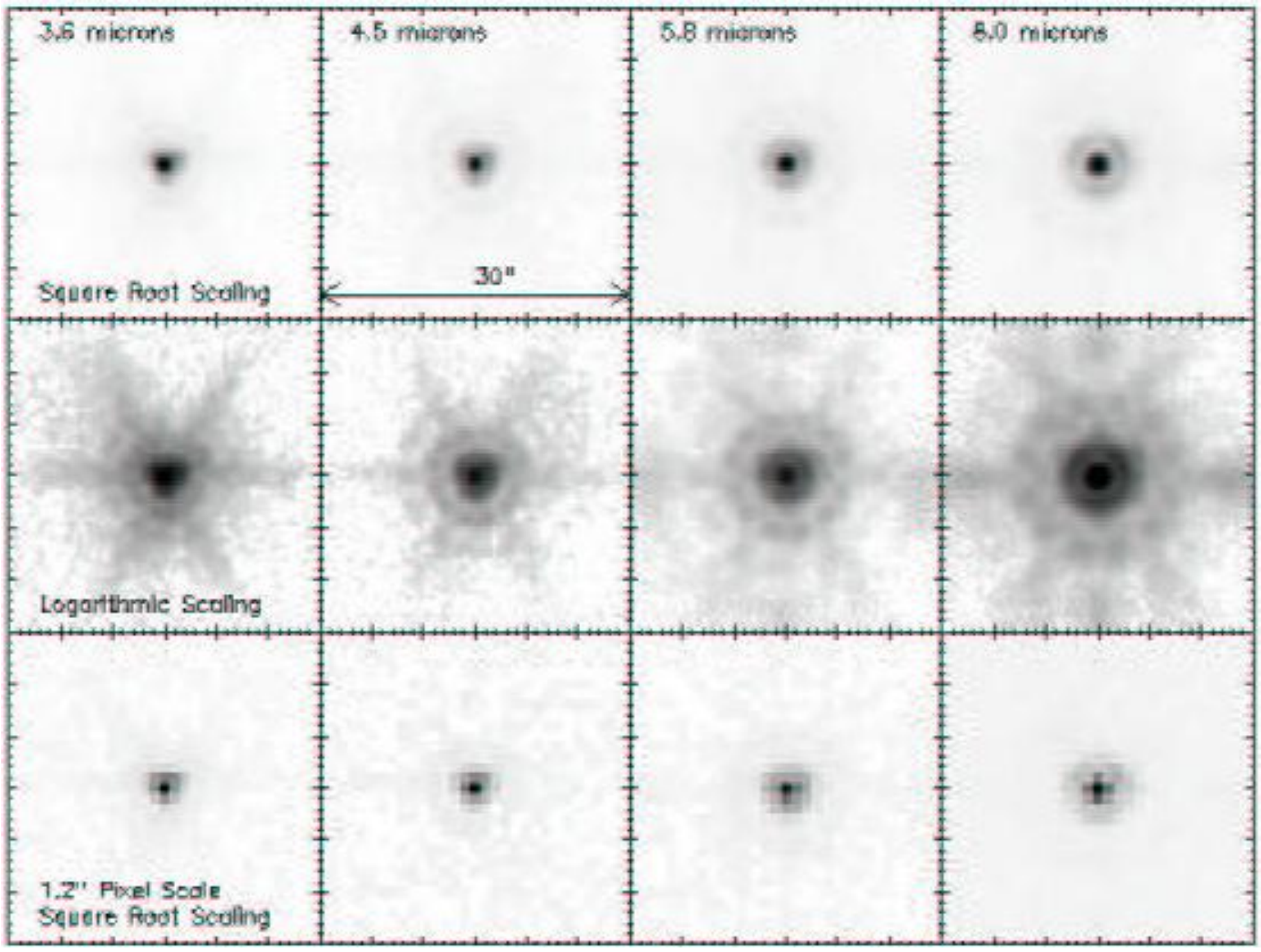

Figure 4.9: The IRAC PSF with both a square root and logarithmic scaling, emphasizing the structure in the core and wings of the PSF, respectively. There is also shown the PSF as it appears at the IRAC pixel scale of 1".22 [125].

## 4.3 SExtractor

SExtractor (Source-Extractor) is a program that builds a catalogue of objects from an astronomical image. Since the beginning in 1993, the development of SExtractor was always made on Unix systems.

### *4.3.1 Using SExtractor*

SExtractor is run from the shell with the following syntax:

% sex image [-c configuration-file] [ -Parameter1 Value1 ] [ -Parameter2 Value2 ] ...

The part enclosed within brackets is optional. Any "-Parameter Value" statement in the command-line overrides the corresponding definition in the configuration-file or any default value (see below).

If no configuration file-name is specified in the command line, SExtractor tries to load a file called "default.sex" from the local directory. If default.sex is not found, it loads default values defined internally.

A number of configuration parameters can be used in SExtractor. Here we report just the ones used during our analysis for this work.

Table 4.2: Some SExtractor parameters [126].

| Parameter | type | Description |
|---|---|---|
| CATALOG_NAME | string | Name of the output catalogue where data from sources extracted are written |
| PHOT_APERTURES | floats ($n \leq 32$) | Aperture diameters in pixels |

In the output catalogue SExtractor returns MAG_APER, an estimation of the flux above the background within a fixed circular aperture. The diameter of the aperture in pixels (PHOT_APERTURES) is supplied by the user (in fact it does not need to be an integer since each "normal" pixel is subdivided in 5 × 5 sub-pixels before measuring the flux within the aperture) otherwise, it is set to 10 by default [127].

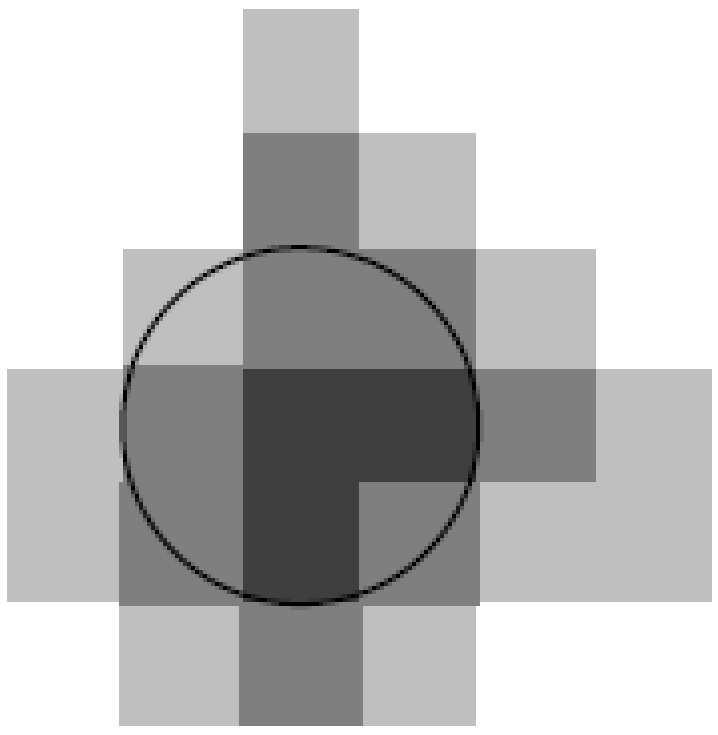

Figure 4.10: Illustration of MAG_APER [128].

### *4.3.2 SExtractor analysis*

SExtractor accepts images stored in FITS[2] format. The complete analysis of an image is done in two steps through the data. During the first step, a model of the sky background is built and a couple of global statistics are estimated. During the second step, the image is background-subtracted, filtered and thresholded "on-the-fly". Detections are then deblended, photometered, classified and finally written to the output catalog (see Fig. 4.11).

The detection of sources is part of a process called *segmentation* in the image-processing vocabulary. Segmentation normally consists of identifying and separating image regions which have different properties (brightness, colour, texture...) or are delineated by edges. In the astronomical context, the segmentation process consists of separating objects from the sky background. This is however a somewhat imprecise definition, as astronomical sources have, on the images — and even often physically — no clear boundaries, and may overlap. We shall therefore use the following working definition of an **object** in SExtractor: a group of pixels selected through some detection process and for which the flux contribution of an astronomical source is believed to be dominant over that of other objects. Note that this means that a simple x, y position vector alone cannot be handled by SExtractor as a detection.

[2] Flexible Image Transport System

Segmentation in SExtractor is achieved through a very simple thresholding process: a group of connected pixels that exceed some threshold above the background is identified as a detection. Each time an object extraction is completed, the connected set of pixels passes through a sort of filter that tries to split it into eventual overlapping components [126]. This case appears when the field is crowded, so does not occur for HD 209458.

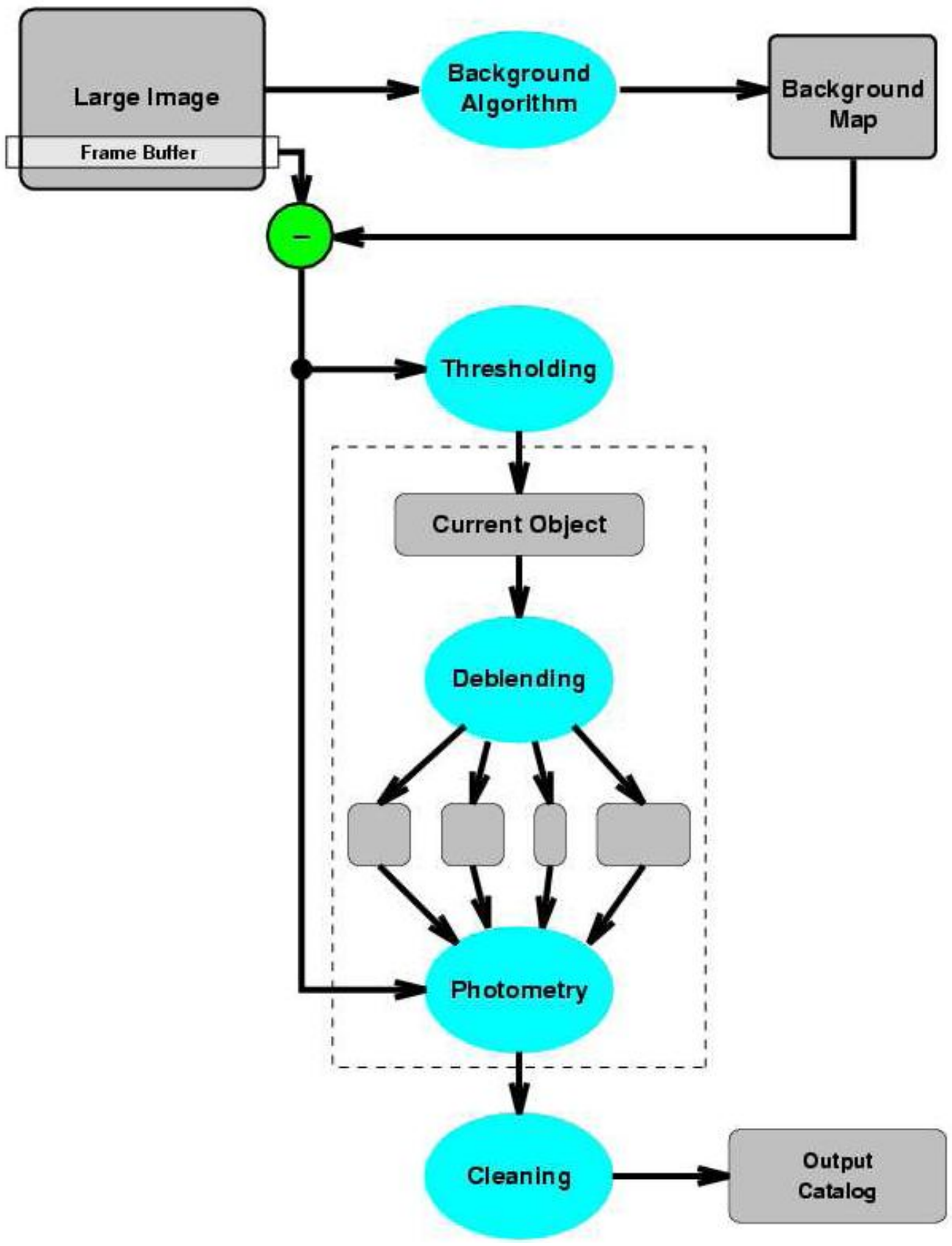

Figure 4.11: SExtractor flow diagram [128].

## 4.4 Observations and Data Reduction

Three HD209458 primary transits were observed with the IRAC camera on board the Spitzer Space Telescope. Channels 1 and 3 (3.6 and 5.8 μm) observed at two epochs, on December 30 2007 and July 18 2008, and channels 2 and 4 (4.5 and 8 μm) observed on July 20, 2008. Since HD 209458 is G0V star with a 2MASS Ks[3] magnitude of 6.3, the IRAC predicted fluxes are 878, 556, 351 and 189 mJy in channels 1-4, respectively. For our observations we required extremely high signal to noise as the modelled contribution to the absorption due to $H_2O$ and CO was predicted to be ~0.05% of the stellar flux.

Observing a transiting planet with Spitzer requires monitoring the target continuously without dithering, in order to be able to quantify the systematic effects described below:

- The amount of light detected in channels 1 and 2 will show variability that depends on the relative position of the source with respect to the pixel centre (see par. 4.2.3). The time scale of this variation is of the order of 50 minutes. These effects have to be estimated from the data in order to evaluate the proper corrections to apply.

- In channels 3 and 4 there are only very minor pixel-scale effects, but a variation of the response of the pixels to a long period of illumination and latent build-up effect, affect the 5.8 and 8.0 μm observations, respectively.

- From previous observations with Spitzer (Beaulieu et al. [66]), it was noticed that the first 20 minutes were affected by a systematic effect that was different in nature to the ones affecting the subsequent hours of observation. There was an initial sharp decrease in signal, followed by a shallower one that was present in all the remaining observations. These pre and post transit data have been critical in the understanding of systematic effects. Accordingly, we obtained a slightly longer "pre-transit" data set, to let the satellite to settle in a "repeatable" jitter pattern and a shorter post-transit data set. The time scale of the pixel-phase being of the order of 50 minutes, we chose a 120 min of pre- and 80 min of post-transit data baseline.

[3] Ks is a near-infrared band centered on 2.17μm. This is the K magnitude from the Two Micron All-Sky Survey (2MASS).

It is important to note that the ~ 184 minute transit of HD 209458b means that our data contain three full cycles of the pixel-scale phase variation in the transit itself, giving an excellent opportunity to have a full control on the behavior of the systematic effects by evaluating them both in and outside the transit.

Our observations employed the IRAC 0.4/2 second stellar photometry mode[4]. Using the regular Astronomical Observation Templates (AOTs), a total of two transits per field of view is required to achieve the desired sensitivity at 4.5 (but we have just one epoch for it) and 5.8 μm (the arrays with the limiting sensitivity). Unfortunately, the AOTs as designed was not the most efficient way to perform this observation. Each stellar mode frame effectively incurs 8 seconds of overheads due to data transfer from the instrument to the spacecraft. As our observations only required the data in the field-of-view with the star, it was possible to save both data volume by collecting data in only two channels and with a cadence of 4 seconds. Consequently, we designed a special engineering template (Instrument Engineering Request; IER) to optimize the observations. IERs can typically double the efficiency (so that 1 epoch at 4.5 μm can be enough) and our IER enabled us to reduce the total required observing time for all four channels to only 13.4 hours (so at least 2 observations instead of 4).

We used the BCD files (see par. 4.2.2) provided by the Spitzer pipeline. Each channel has been treated separately (for a detailed discussion of the data reduction carried out by G. Campanella see [127]). We measured the flux of the target on each image using the SExtractor package (see par. 4.3), with a standard set of parameters for Spitzer. The centroid determination was done by PSF fitting. We performed both aperture photometry, and PSF fitting photometry. In Fig. 4.12, for all the six observed transits, we give the raw magnitude measurements, the variation of the centroid in X and Y axis, and the distance of the centroid from the lower left corner of the pixel (the pixel phase). A rapid inspection shows that all observations contain correlated noise of different nature, as expected when using the IRAC camera.

---

[4] 0.4 and 2 s is the integration time for the exposures of channels 1&2 and channels 3&4, respectively. The short exposure times in IRAC "stellar mode" avoid the saturation of the detector.

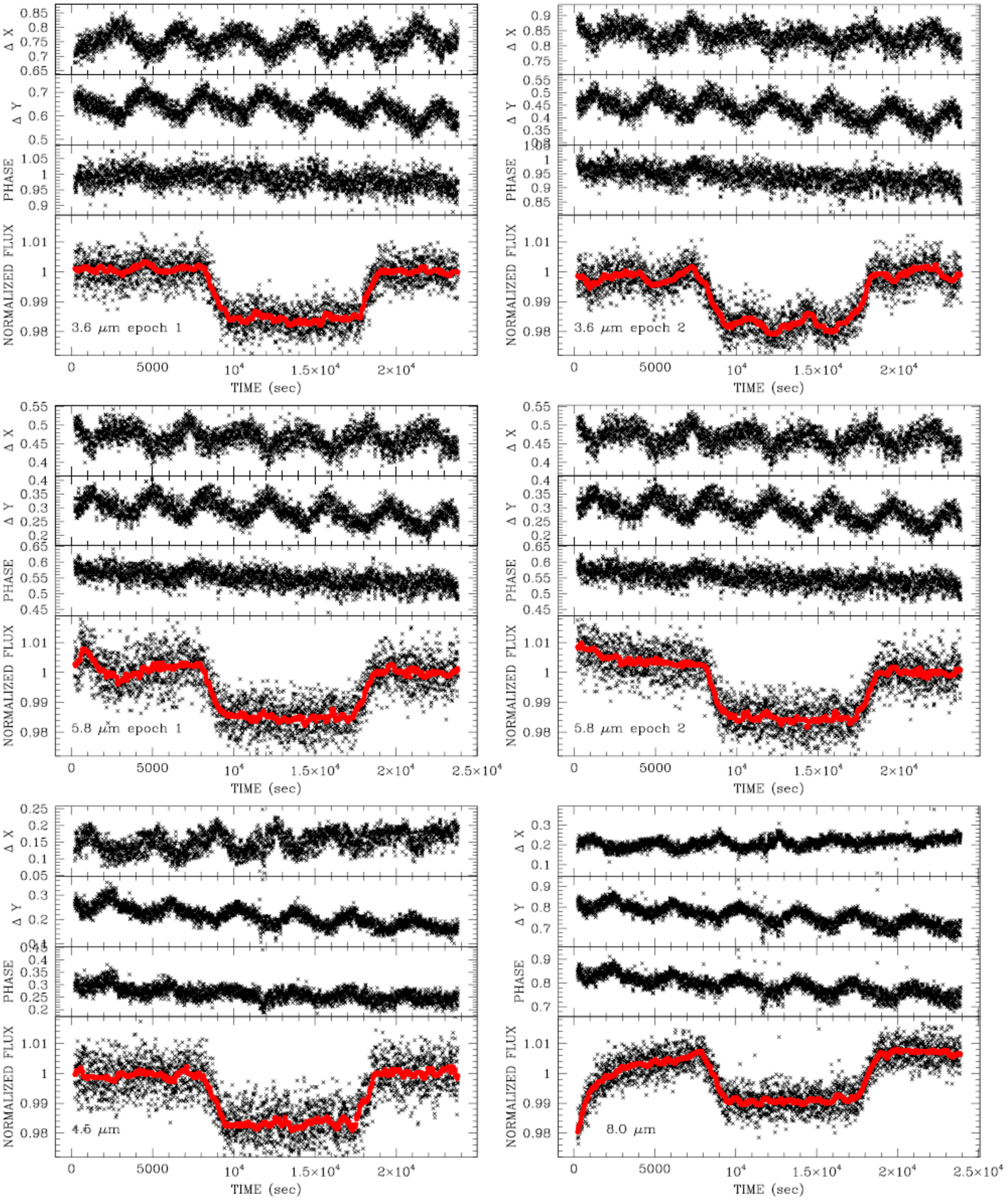

Figure 4.12: Six panels showing the raw photometric data for 3.6 μm (epoch 1 and 2), 5.8 μm (epoch 1 and 2), 4.5 and 8 μm obtained with IRAC on Spitzer. The lower panel of each plot is the primary transit, and over-plotted the 50-point median-stack smoothing. They provide a clear view of the systematic trends at work in IRAC Spitzer data.

## 4.5 Estimating and Attenuation of Correlated Noise

Pixel-phase information is retrieved by using SExtractor's PSF fitting to obtain estimates of the X and Y pixel-phase for each exposure. In contrast, the flux for each exposure is obtained through aperture photometry since this offers substantially larger signal-to-noise compared to the PSF flux estimates. The PSF fitted estimates of X and Y exhibit significant scatter due to photon noise and therefore only the average trends in X and Y represent the true physical motion of the spacecraft, which are ultimately responsible for this effect. It is this spacecraft motion that we wish to correlate to the flux to and not the photon-noise-induced randomisation of X and Y. For example, in channel 1, we find a sinusoidal-like variation in the X-phase with r.m.s. amplitude of 0.023 pixels whereas the uncertainty on each pixel phase determination is 0.026.

We find that there is a dominant ~1 hour period sinusoidal-like variation in X and Y, characteristic of small elliptical motion in Spitzer's pointing, with a more complex time trend overlaid. For each channel, we apply a non-linear regression of a sinusoidal wave to the phases, in order to determine the best-fit dominant period, $P_{phase}$ (typically close to one hour, e.g. Fig 4.13).

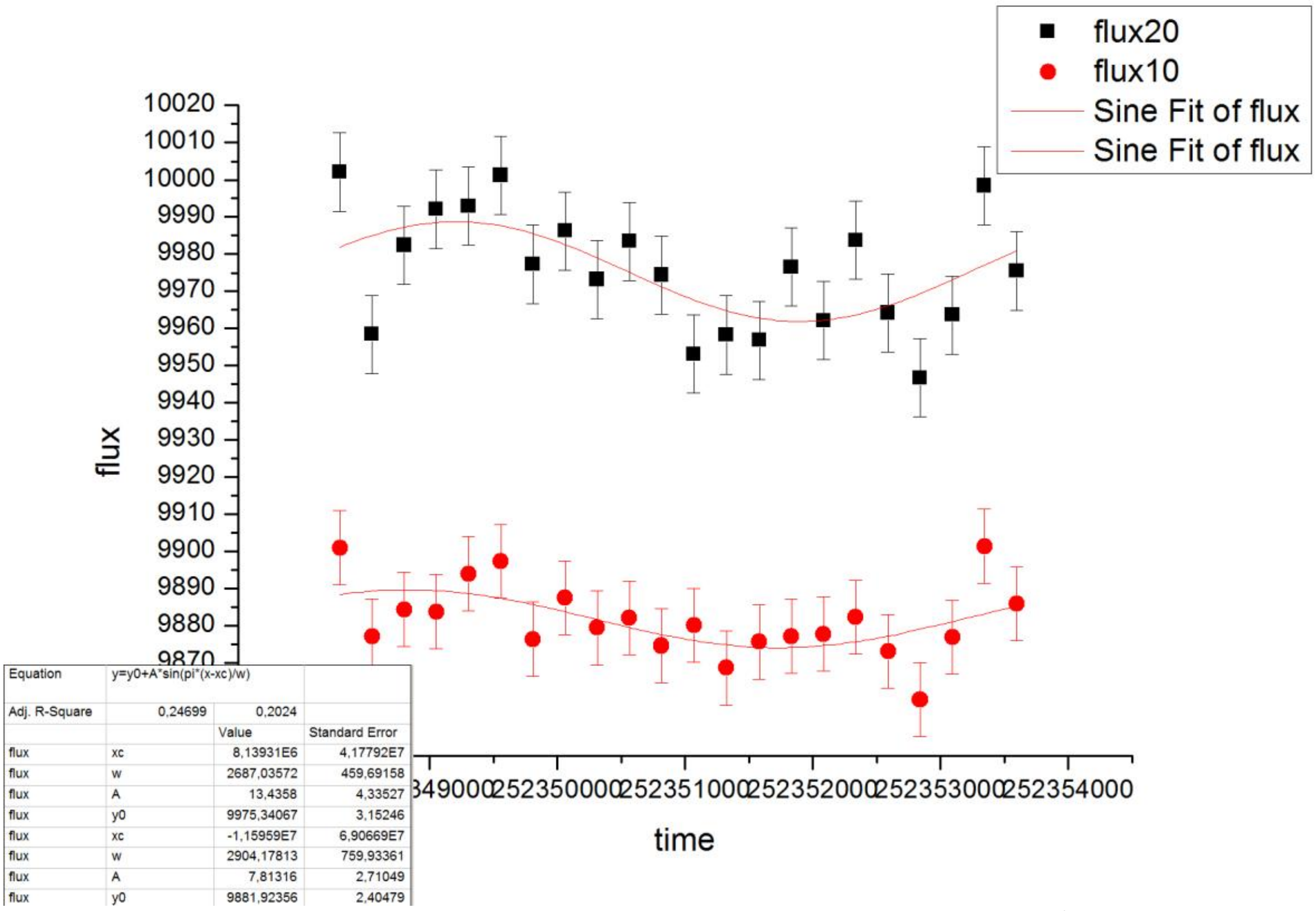

Figure 4.13: Sinusoidal fittings on the post transit data of the first epoch of 5.8 μm (20/10-pixel-aperture). We find $P_{phase}$~ 50 min [127].

Having removed the pixel phase variation of X and Y with respect to time, as induced by spacecraft motion we obtain a robust determination of the second-order phase variations, which may then be fitted using a polynomial, of orders varying from 2 to 4 depending on the degree of curvature in the resultant phase trends. For channel 1 (epochs 1 and 2) and channel 2, we removed pixel-phase effects of r.m.s. amplitude 0.49, 1.51 and 0.57 mmag respectively. Channel 1 epoch 2, is particularly polluted by pixel phase response possibly due to a large inhomogeneity in response close to the PSF centroid position (pixel 131,128 of the detector).

In identifying the systematic trends present in the 5.8 μm data, we use 50-point median-stack smoothing of the out-of-transit data points, which provides a clearer view. The first epoch of 5.8 μm exhibits a discontinuity between the photometry at ~ 2800 seconds (see Fig. 4.14). The behaviour of the photometry in the regime $0 \leq t \leq 5500$ seconds does not match either a "linear drift" usually associated with this channel, or a "ramp" style-effect. The origin of the observed behaviour is unclear and is present in many different trial aperture sizes, between 2.5 to 20 pixels radius, suggesting an instrumental effect located very close to the centroid position [127].

The second epoch of 5.8 μm also, even if more subtly, exhibit a discontinuity between the photometry in the regime $0 \leq t \lesssim 2800$ seconds, visible in the median-smoothed lightcurve. Channel 3 seems to require a certain amount of time to settle into a stable regime. As a result, we can again treat this data by simply excluding the first ~ 30 minutes and then de-trend the remaining data.

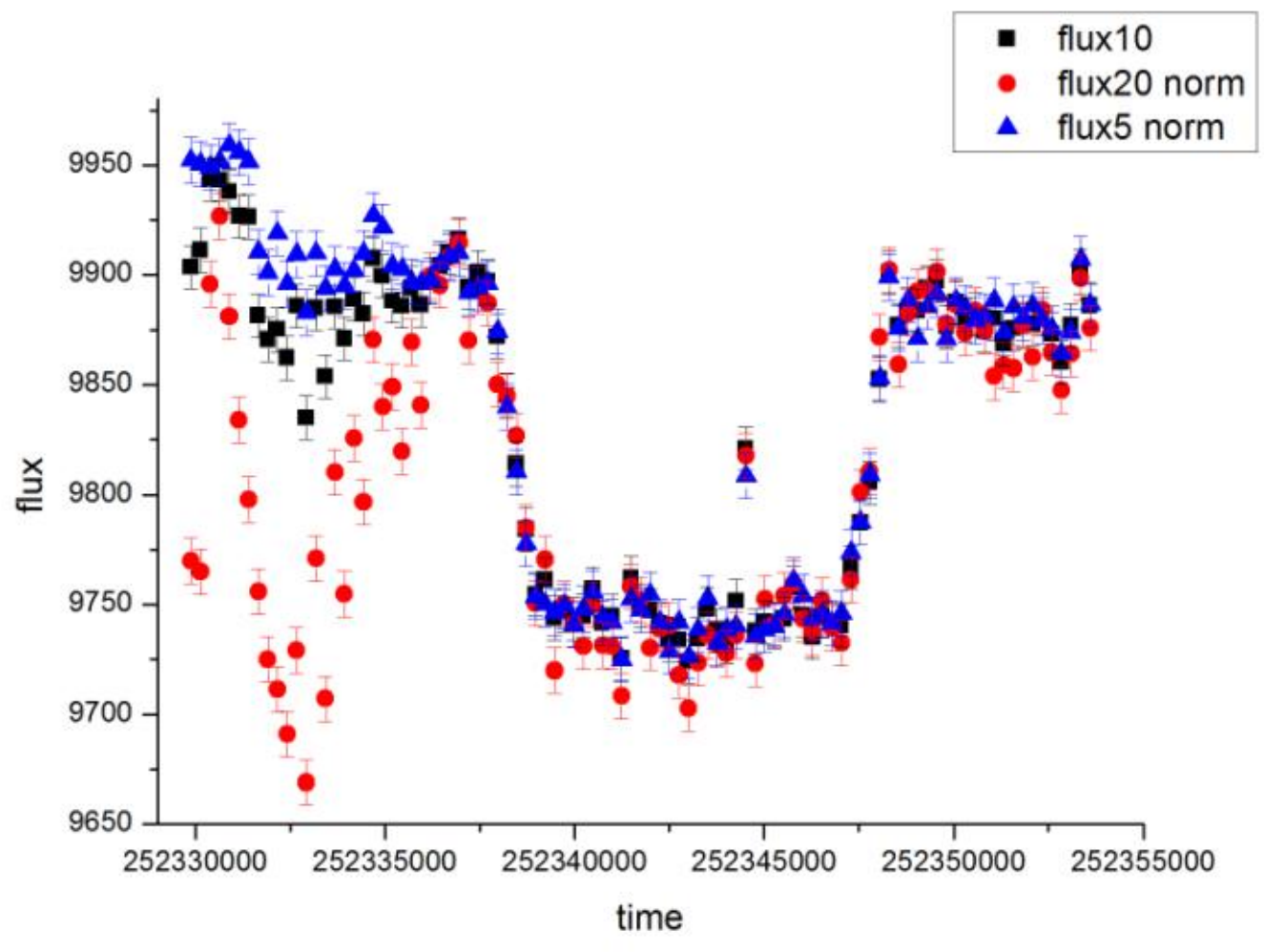

Figure 4.14: Photometric data for 5.8 μm (epoch 1) with 20/10/5-pixel-aperture. The discontinuity is clearly visible [127].

The ramp effect at 8 μm is well documented and so too is the methodology for correcting this phenomenon (Knutson et al. [78], Agol et al. [69]). We fit a time trend to the out-of-transit data of the form $a+bt+c\log(t-t_0)+d(\log(t-t_0))^2$ where $t_0$ is chosen to be 32.7 seconds before the observations begin, which provides the best correction.

### 4.6 Fitting the Transit Light Curves

Among the 6 transit light curves, we have four of high quality with well understood and corrected systematic effects: the first epoch at 3.6 μm, 4.5 μm, the second epoch at 5.8 μm and 8 μm. The second epoch at 3.6 μm and the first epoch at 5.8 μm will be treated separately.

Accurate limb darkening coefficients were calculated for each of the four IRAC bands following the procedure in Claret [34] (see Appendix B).We adopted the following stellar properties: $T_{eff}$ = 6100 K, log g = 4.38 and [Fe/H] = 0. The coefficients are given in Table 4.3.

Table 4.3: Limb darkening coefficients.

| channel | c1 | c2 | c3 | c4 |
|---|---|---|---|---|
| 3.6 μm | 0.2670569 | 0.1396675 | -0.1900802 | 0.064018 |
| 4.5 μm | 0.3325055 | -0.1999922 | 0.1858255 | -0.0703259 |
| 5.8 μm | 0.3269256 | -0.2715499 | 0.2258883 | -0.0684003 |
| 8 μm | 0.2800222 | -0.2278080 | 0.1451840 | 0.0273881 |

We adopt the physical model of a transit light curve through the expressions of Mandel & Agol [35] and orbital eccentricity using the equations of Kipping [23].We fit the data with Markov Chain Monte Carlo and prayer-bead Monte Carlo techniques [5].

As some physical parameters should be the same for all bands, we made a simultaneous fit to the best observations, namely 3.6 μm (epoch 1), 4.5 μm , 5.8 μm (epoch 2) and 8 μm, in which four parameters are shared by all channels: e, i, ω and a/R*. We decided to fit separately 3.6 μm (epoch 2), forcing the four shared parameters to be equal to the values derived from the best fit with the four other channels. We obtain extremely similar results concerning the transit depth for the different wavelengths with these techniques. The final results are listed in table 4.4.

Table 4.4: Best-fit inclination *i*, a/R*, ratio of radii *k*, transit depths *d*.

| Wavelength (μm) | i (°) | a/R*. | k | $d = \left(R_P/R_*\right)^2$ |
|---|---|---|---|---|
| 3.6 (epoch 1) | 86.76 ± 0.10 | 8.77 ± 0.07 | 0.121215 ± 0.00054 | 1.469 ± 0.013% |
| 3.6 (epoch 2) | 86.76 ± 0.10 | 8.77 ± 0.07 | 0.120343 ± 0.00053 | 1.448 ± 0.013% |
| 4.5 | 86.76 ± 0.10 | 8.77 ± 0.07 | 0.121568 ± 0.00072 | 1.478 ± 0.017% |
| 5.8 (epoch 2) | 86.76 ± 0.10 | 8.77 ± 0.07 | 0.1244 ± 0.00059 | 1.549 ± 0.015% |
| 8 | 86.76 ± 0.10 | 8.77 ± 0.07 | 0.12390 ± 0.00046 | 1.535 ± 0.011% |

We asked for two epochs for HD 209458b at 3.6 and 5.8 μm with the prime intention of demonstrating the possibility of co-add multiple epoch observations, and/or be able to check for the variability in the system. The two epochs are separated by 7 months, and the observing setups are identical.

Firstly, at 3.6 μm the data are affected by systematic trends of the same nature due to the pixel scale effect. We notice a factor 3 in the amplitude of the systematic trends between the two epochs. The results are compatible between the two channels.

Secondly, at 5.8 μm the situation is more complex. The second epoch showed the expected behaviour, and we have been able to correct for systematics, and to fit it. In the first epoch systematic trends affect the measurements up to the mid transit. We took a conservative approach here, excluding the data up to the mid-transit. We made a fit adopting the physical parameters (i, a/R*) reported in Table 4.4. In Fig. 4.15 we compare epoch 1 data and the model fitted on the data from the mid-transit. Inspection of the residuals after the mid-transit indicates a good fit to the data. The data from the first half of the observations show the uncorrected systematic trends at work. We measure the two transit depth to be 1.552 ± 0.032% and 1.549 ± 0.015% for epoch 1 and epoch 2 respectively. The results are compatible between the two channels.

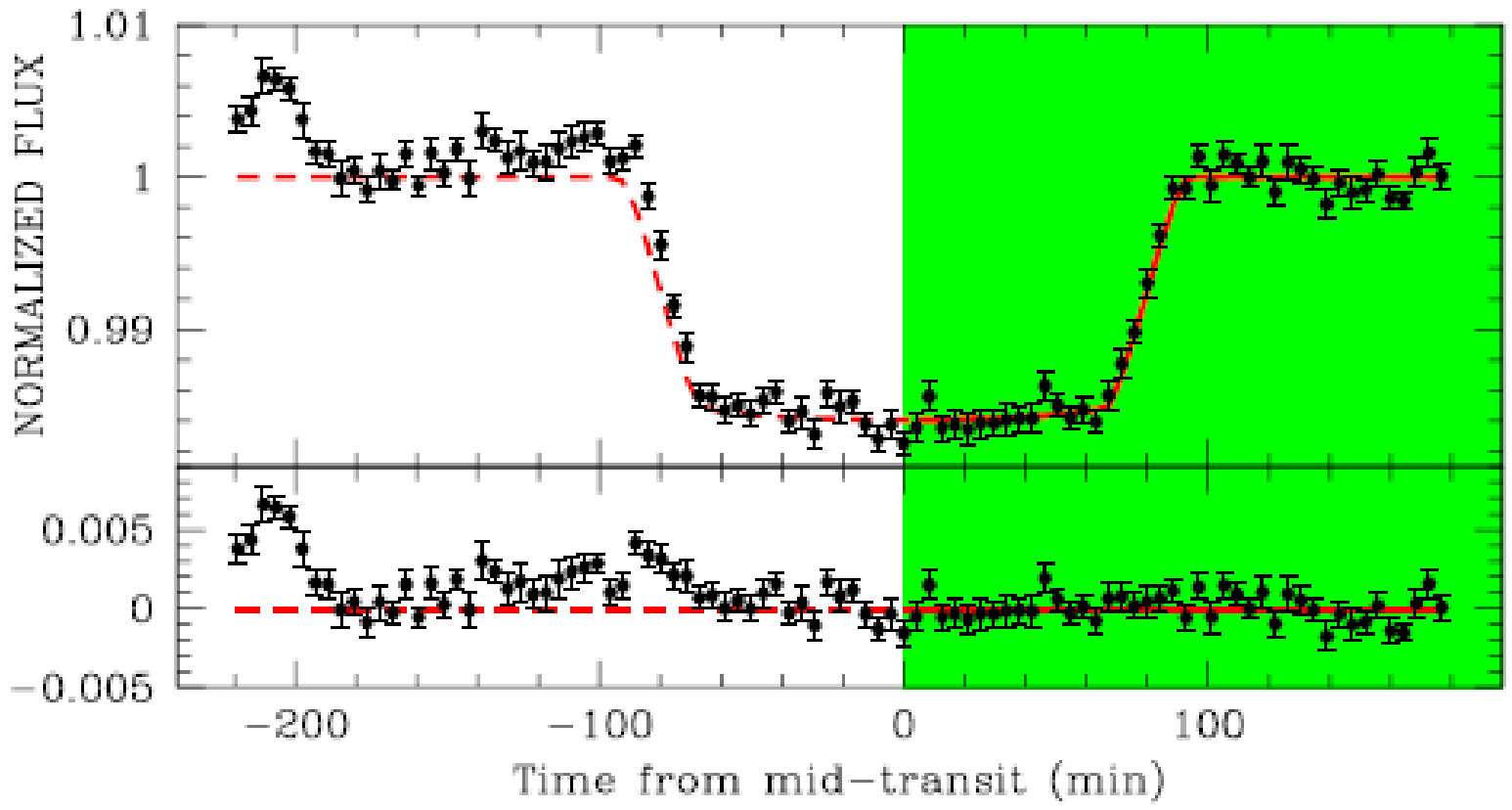

Figure 4.15: The uncorrected and binned data from first epoch at 5.8 μm and the underlined model computed for the second epoch (corrected for systematics). The shaded area marks the second half of the transit and the post transit observations used in the fit.

## 4.7 Data Interpretation

Our analysis includes the effects of water, methane, carbon dioxide, carbon monoxide, pressure-induced absorption of $H_2{-}H_2$. We do not consider the presence of particulates, because there is no indication of particles large enough (~ 3μm ) to affect the planet's middle-infrared spectrum (see Beaulieu et al. [5] for more details). In Fig. 4.16, we show the contribution of the different molecules combined to water.

The 3.6 μm (and to a lesser degree the one at 8 μm) IRAC channel measurement can be affected by the presence of methane. By contrast, $CO_2$ and CO may contribute in the passband at 4.5 μm. We find absorption by water alone can explain the spectral characteristics of the photometric measurements, which probe pressure levels from 1 to 0.001 bars (see Fig. 4.17). The determined water abundance depends on the assumed temperature profile and planetary radius. We find that the data can be interpreted, e.g., with a thermochemical equilibrium water abundance of 4.5 $10^{-4}$ (Liang et al. [85]), assuming temperature profiles from Swain et al. [107] (see par. 3.1.1). However, ~ 1% difference in the estimate of the planetary radius, is compatible with water abundances 10 times smaller or larger. The assumed temperature profiles affects the derived water abundance to a lesser degree. While the contribution of other constituents is not necessary to interpret the measurements, mixing ratios of $10^{-7}$, $10^{-6}$ and $10^{-4}$ of $CO_2$, $CH_4$ and CO, respectively, are allowed in our nominal model. Spectroscopic data are needed to further investigate the composition and thermal structure of this planetary atmosphere.

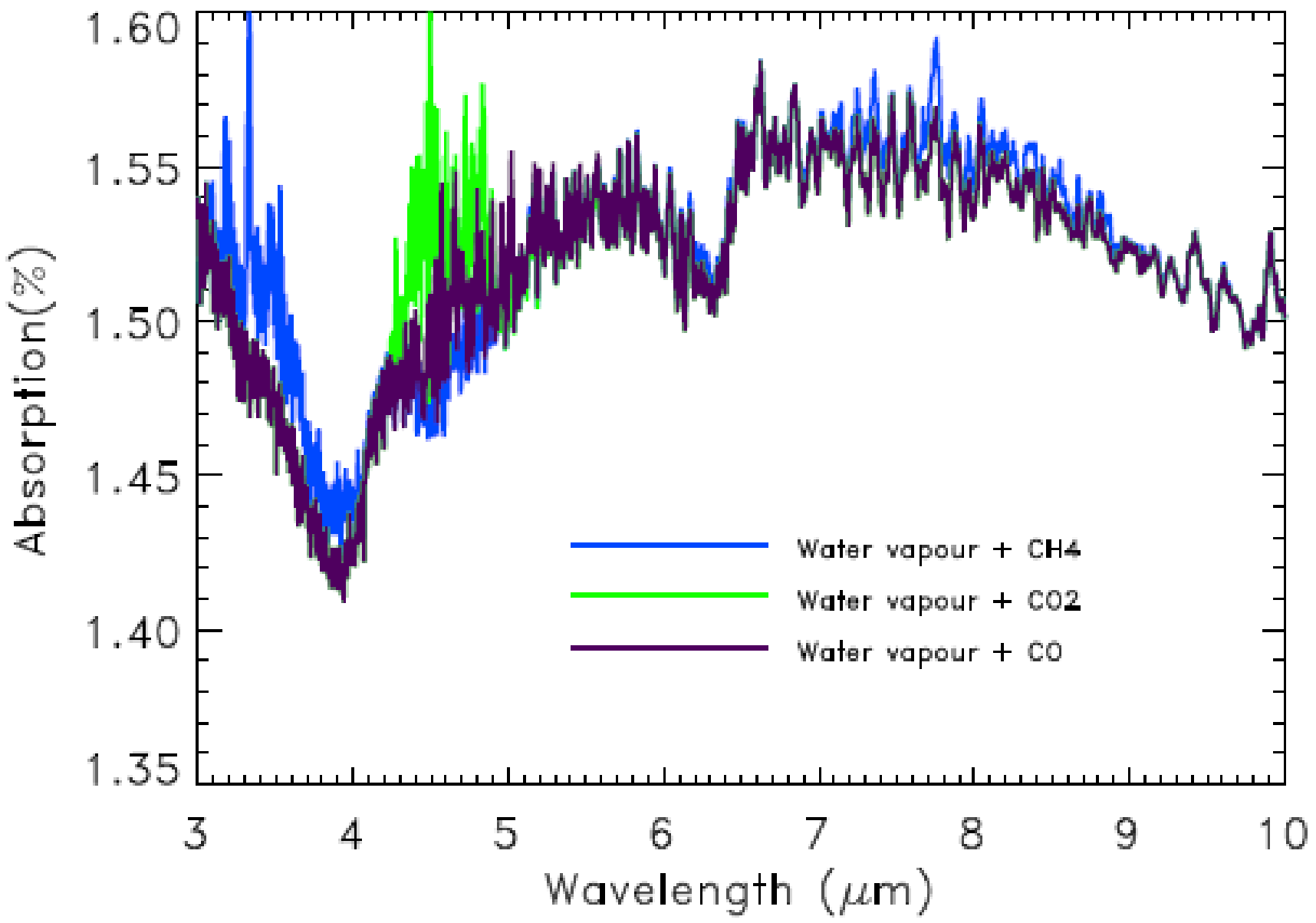

Figure 4.16: Simulated middle Infrared spectra of the transiting Hot Jupiter HD209458b in the wavelength range 3-10 μm. Water absorption is responsible for the main pattern of the spectra. The additional presence of $CH_4$, CO and $CO_2$ are simulated in the blue, violet and green spectra respectively.

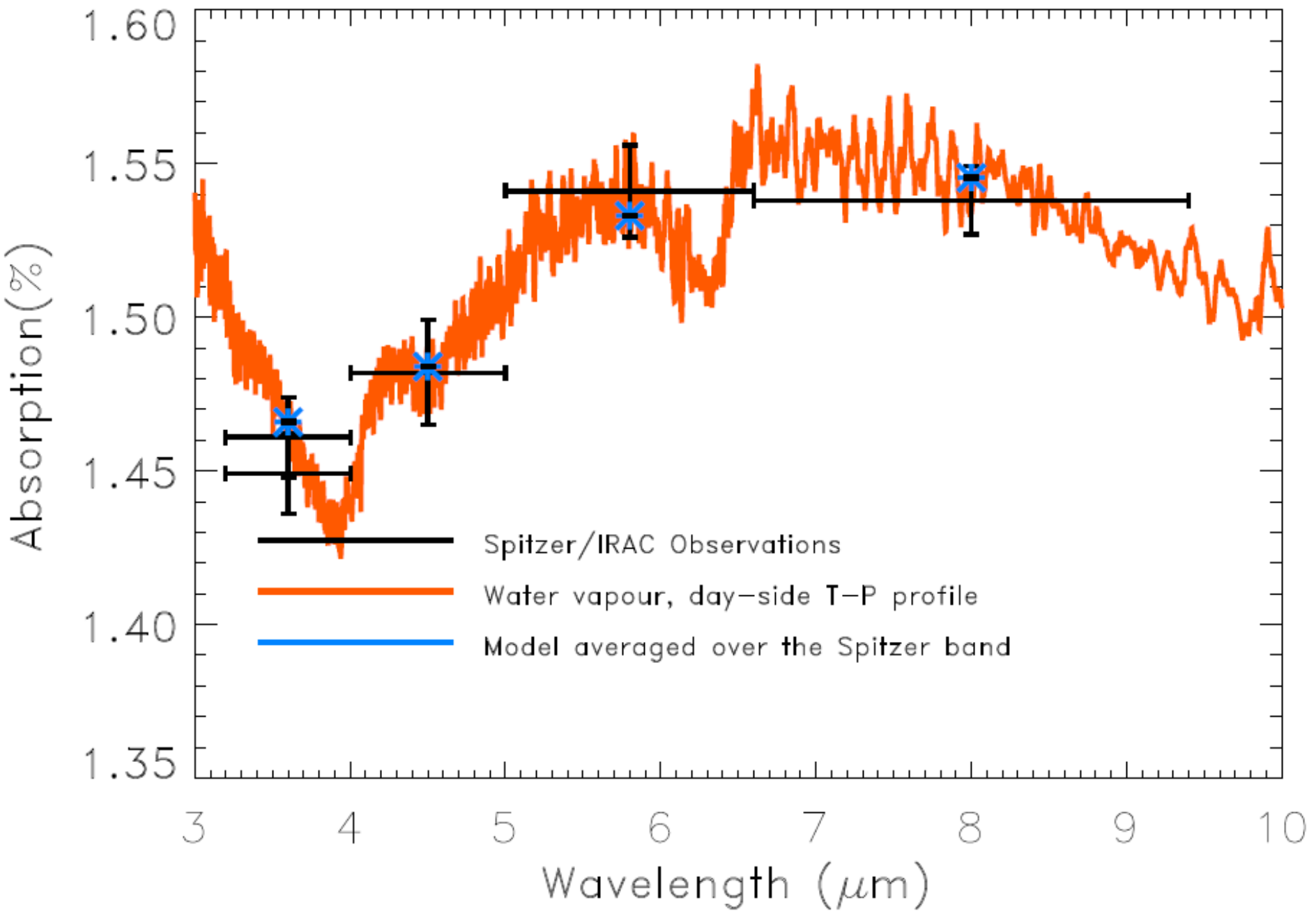

Figure 4.17: Observations and spectral simulations of the atmosphere of HD 209458b. Black: Spitzer measurements where the horizontal bar is the IRAC bandwidth. Orange spectrum: water vapour and a thermal profile compatible with the day-side spectroscopy and photometry data (Swain et al. [107]; Griffith and Tinetti, in prep.). Blue stars: simulated spectrum integrated over the Spitzer bands.

*Chapter 5*

# THE SEARCH FOR EXOMOONS

An exomoon is a smaller, natural satellite that orbits an extrasolar planet. As the number of known exoplanets continues to grow, the question as to whether such bodies frequently host satellites has become one of increasing interest.

At the time of writing, no detection of an exomoon has ever been made but this is most probably due to a selection effect of the insensitivity of current methods. As many of the planets of own solar system host satellites and since some of the most promising candidates for extraterrestrial habitable locations within our own solar system are the moons of gas giants (see par. 3.3.2), for example Europa (Reynolds et al. [129], Marion et al. [130]) and Enceladus (Spencer et al. [102]), we think that the search for exomoons will offer exciting developments and almost certainly it will take a more prominent role in future astronomy.

The theory of exomoon detection begun with Sartoretti & Schneider [131] proposing the use of transit timing. This theory was developed further by Szabo et al. [132], Simon et al. [133] and most recently by Kipping ([134], [135]).

The detection of exomoons would have a number of implications:

1. It could help to understand how important the Moon was to life on the Earth (influence on geological activities, rotation axis and tidal effects). In fact, several authors (Waltham [136], Lathe [137]) suggest that complex life may not form on planets without large moons. So even if we detect an Earth-like planet in the future, the search for a moon around such a body may be critical in assessing its habitability.

2. It could have a role in the field of planetary formation theory. It may allow several theories of the stability of exomoons to be tested, for example that of Barnes & O'Brien [138] and Domingos et al. [139]. It would also allow a test of whether most satellite systems agree with the mass-scaling relation proposed by Canup & Ward [140].

3. It could expand the number of rocky and icy bodies we know.

4. There may be more habitable exomoons than exoplanets. In support of this view, Scharf [141] suggested exomoons could support habitable environments even outside of the conventional habitable zone due to tidal heating from the host planet.

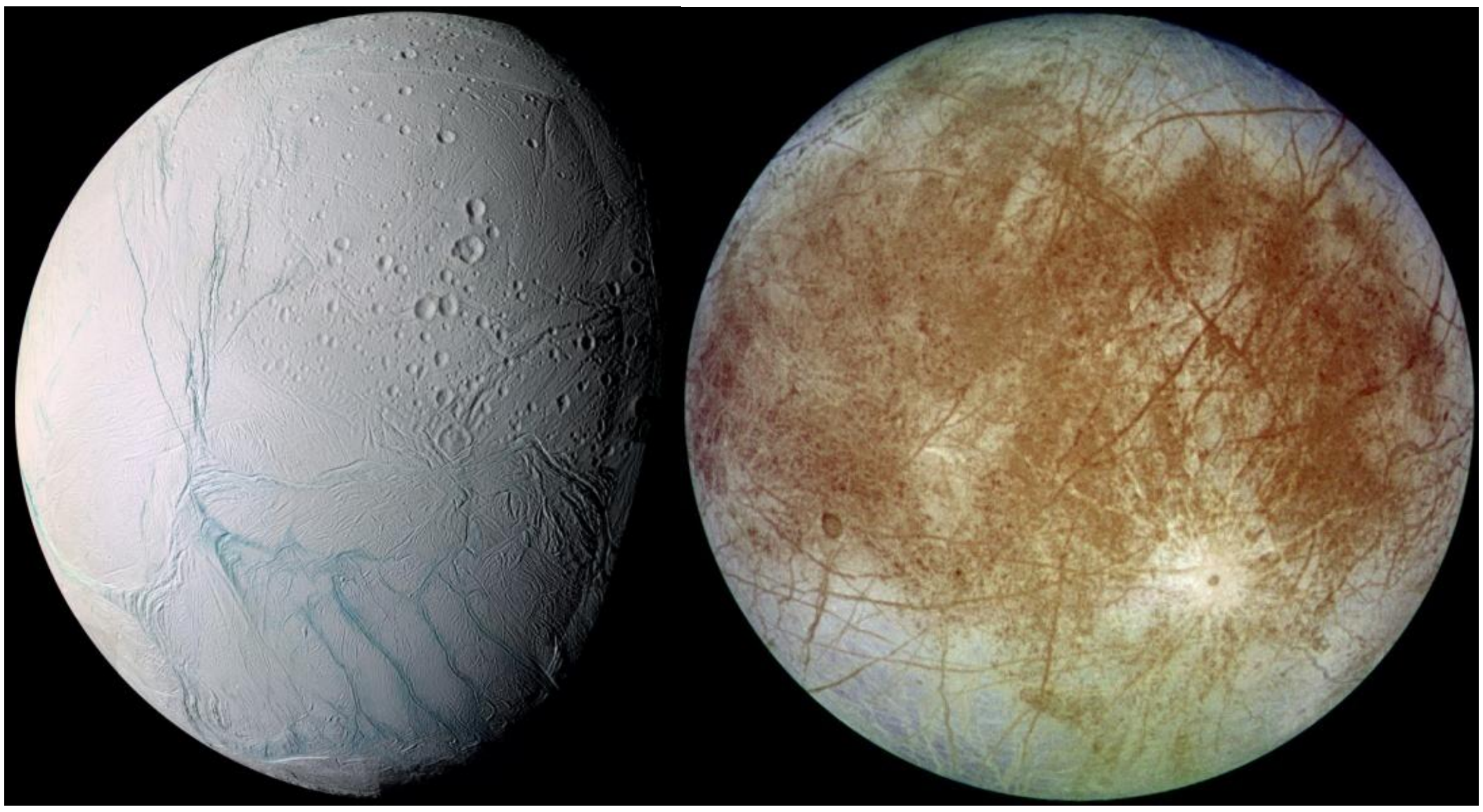

Figure 5.1: Enceladus and Europa, the most promising candidates for hosting extraterrestrial life within the solar system.

## 5.1 Proposed Detection Methods

Despite the great successes of planet hunters with Doppler spectroscopy of the host star [3], exomoons cannot be found with this technique. This is because the resultant shifted stellar spectra due to the presence of a planet plus additional satellites would behave identically to a single point-mass moving in orbit of the host star. Cabrera & Schneider [144] argue that if the emitted light of an exoplanet could be measured, then Doppler spectroscopy of the planet itself could be achieved which would allow for satellite detection. However, achieving high spectral resolutions of light from a planet alone remains a challenging prospect.

In recognition of this, there have been several other methods proposed for detecting exomoons; we are now going to present some of them.

### 5.1.1 Direct imaging

Direct imaging even an exoplanet is extremely challenging due to the large difference in brightness between the objects and the small angular size of the planet. These problems are exacerbated for small exomoons. Even if a coronographic or interferometric technique can be utilised to extinguish the star light, the moon light must still compete with the planetary light. The most significant difficulty is therefore likely to be achieving an angular resolution sufficient to resolve the two objects. At a distance of 10pc, the Moon-Earth separation is 0.5 mas whereas the current best interferometric precision in exoplanet astronomy is ~ 25 mas [142]. Therefore, one would need at least two coronographic telescopes combined with an interferometer in order to image such systems [143] and there exists no such design in planning. Ergo, direct imaging is not a feasible detection method at present.

### 5.1.2 Pulsar timing

In 2008, Lewis et al. [145] proposed using pulsar timing to detect the moons of pulsar planets. The authors applied their method to the case of PSR B1620-26 b and found that a stable moon orbiting this planet could be detected, if the moon had a separation of about one fiftieth of that of the orbit of the planet around the pulsar, and a mass ratio to the planet of 5% or larger. Perhaps it would not be surprising if the first exomoon was found through this method given that the first exoplanet was discovered around a pulsar star by Wolszczan & Frail [55].

### *5.1.3 Microlensing*

In 2002, Han & Han [146] proposed that microlensing could be used to detect the moons of extrasolar planets. While a space mission concept like Euclid could detect exomoons [103], microlensed planets and moons could not be followed up due to their great distances from the Earth and thus characterising these moons would be unlikely.

### *5.1.4 Occultation*

If a planet and a moon passed in front of a host star, both objects should produce a dip in the observed light [133]. Using this technique, Brown et al. [19] were able to set an upper limit of 1.2 $R_\oplus$ for an orbiting exomoon around HD209458b.

To date, the smallest transiting planet discovered is still Neptune-sized, Gliese 436b. An exomoon is likely to be sub-Earth sized, based upon our Solar System, and so even a specialized transit space-based telescope like COROT or Kepler will struggle to spot the signature (Sartoretti & Schneider [131]).

Moreover, the occultation method does suffer from the problem that a "lucky" detection is required. The exomoon could well be hiding in front of behind the planet the moment of transit. Another way of putting this is that the modal position of the moon's lightcurve perfectly coincides with the planetary lightcurve, as pointed out by Cabrera & Schneider [144]. This means one would likely require multiple transits in order to see just one clean moon transit. After this one moon transit had been observed, the whole process would need to repeat several times in order to show that the event was not a statistical fluke or some form of noise.

As an example, in figure 5.2 there is shown the transit lightcurve of a planet with $r_p = r_{Jup}$ and $T_P = 50$ days and a satellite with $r_s = 2.5\ r_\oplus$ and $T_S = 0.5$days. The stellar flux shows a marked but moderate decrease at $t = -5$ hr, as the satellite first transits in front of the star. When the planet also starts to transit, at $t = -3.5$ hr, the stellar flux drops more abruptly. Soon after, at $t = -1$ hr, the relative flux maximum produced by the planet-satellite transit appears clearly. Then, the stellar flux sharply increases as the planet leaves the star, and the satellite, which now follows the planet, continues to moderately occult the star from about t = 3.5 to 4.5 hr. Also shown by crosses are the results of simulated 10 min exposure observations with a Poisson noise of $10^{-4}$.

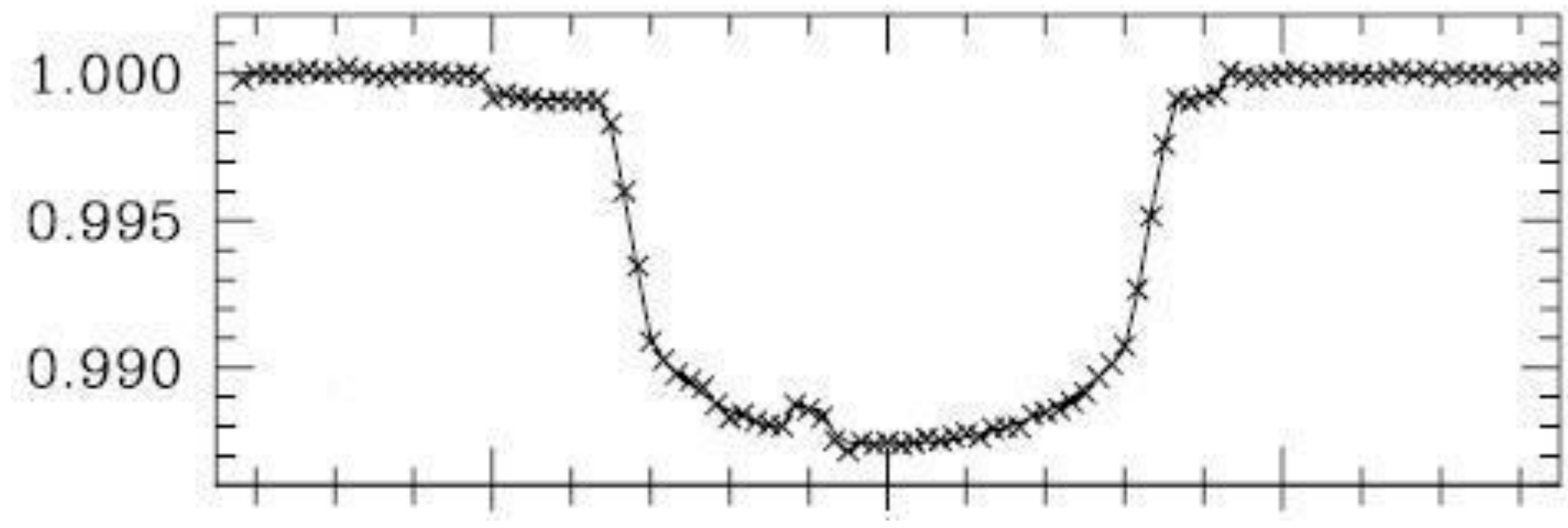

Figure 5.2: Transit lightcurve for a planet with $r_p = r_{Jup}$ and $T_P$ = 50 days and a satellite with $r_s$ = 2.5 $r_\oplus$ and $T_S$ = 0.5days [131].

## 5.2 Transit timing effects

If a satellite does not produce a detectable signal in the stellar lightcurve, it may still be detected indirectly through the associated rotation of the planet around the barycentre of the planet-satellite system. This requires that a planetary transit be observed at least 3 times, as the effect of the rotation will be a periodical time shift of the lightcurve minima induced by the planet transits.

### *5.2.1 Transit time variation (TTV)*

We consider the simple case where:

- the planet-moon orbital plane is co-aligned with the planet-star orbital plane at $i$ = 90°.
- both the planet and the moon are on a circular orbit.

In this case, the projected diameter of the planet orbit around the barycentre of the planet-satellite system is $2a_s M_s M_p^{-1}$ where $M_s$ is the satellite mass. Therefore, the expected time shift between transits will be

$$\Delta t \approx 2a_s M_s M_p^{-1} \times T_P \left(2\pi a_p\right)^{-1} \tag{5.1}$$

Measurements of $\Delta t$ provide also an estimate of the product of its mass and orbital radius [131]. Introducing the root-mean-square amplitude of the TTV signal, $\delta_{TTV}$, we have [134]:

$$\delta_{TTV} \propto a_s M_s \tag{5.2}$$

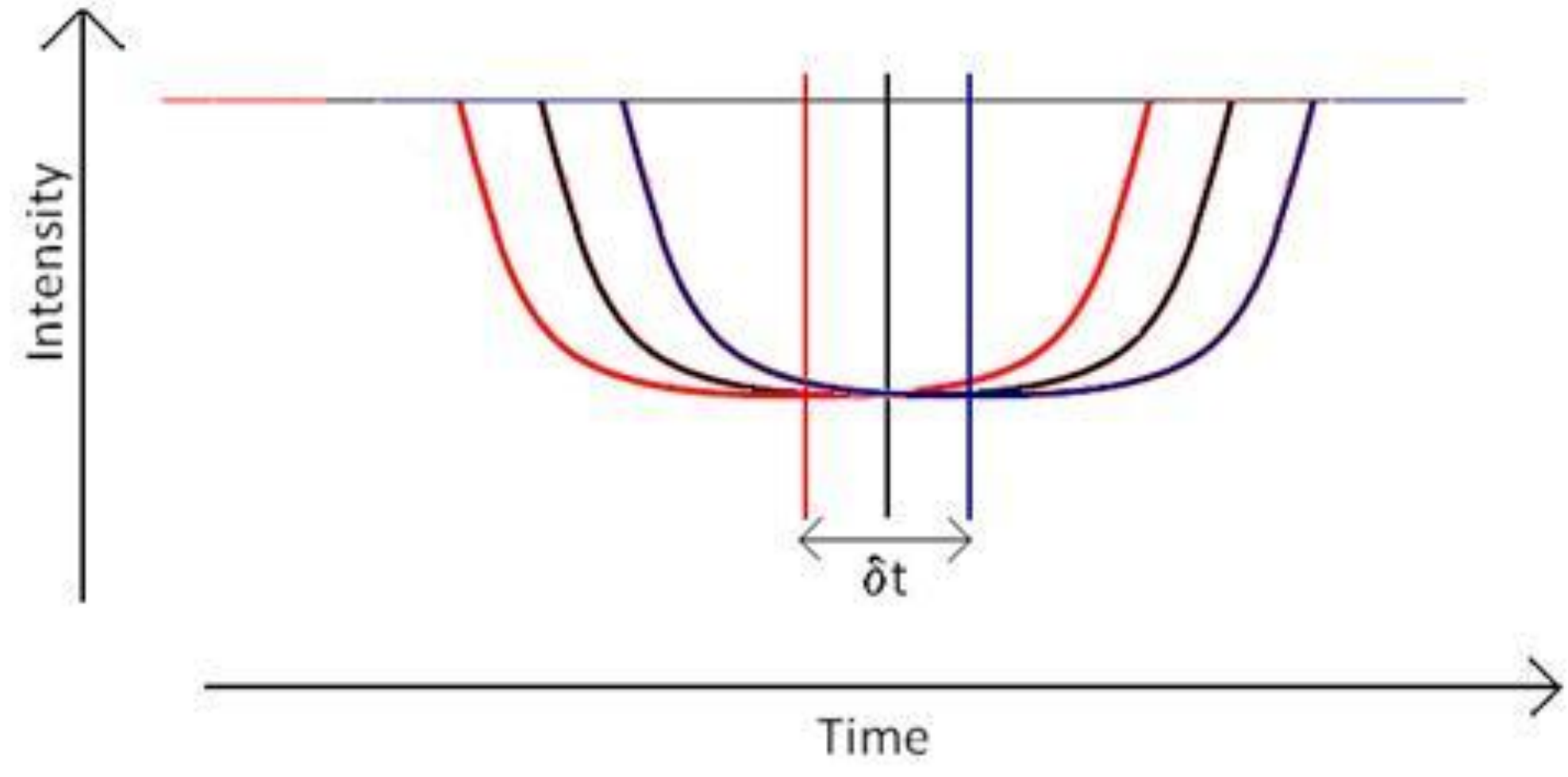

Figure 5.3: TTV due to an exomoon.

Unfortunately, transit timing variation can also be provoked by a multitude of phenomenon, including gravitational influence of other planets in the system or companion stars, general relativistic precession of the orbit, tidal deformations of both the star and the planet, torques due to spin-induced quadrupole moment of the star, stellar peculiar motion and parallax effects (Kipping [143]). Due to the rich plethora of phenomenon that could cause TTV, claiming any observed signal was indeed due to a moon would be extremely challenging.

Another critical problem with TTV from an exomoon is that the period of an exomoon must always be much smaller than the period of the host exoplanet, due to stability arguments outlined by Kipping [134]. As a result, the observed signal will always be under-sampled and have a frequency above the Nyquist frequency. This means that it would be impossible to deduce an accurate period for the exomoon; only a list of possible harmonic frequencies could be derived. The subsidiary effect is that since the TTV amplitude is proportional to the mass and to the orbital distance of the moon, then it is also impossible to derive the mass of the exomoon through TTV.

### *5.2.2 Transit duration variation (TDV)*

Transit duration variation (TDV) is the periodic change in the duration of a transit (τ) over many measurements.

We have that the duration of a transit is inversely proportional to the projected velocity of the planet across the star, $v_{P\perp}$ :

$$\tau \propto 1/v_{P\perp} \tag{5.3}$$

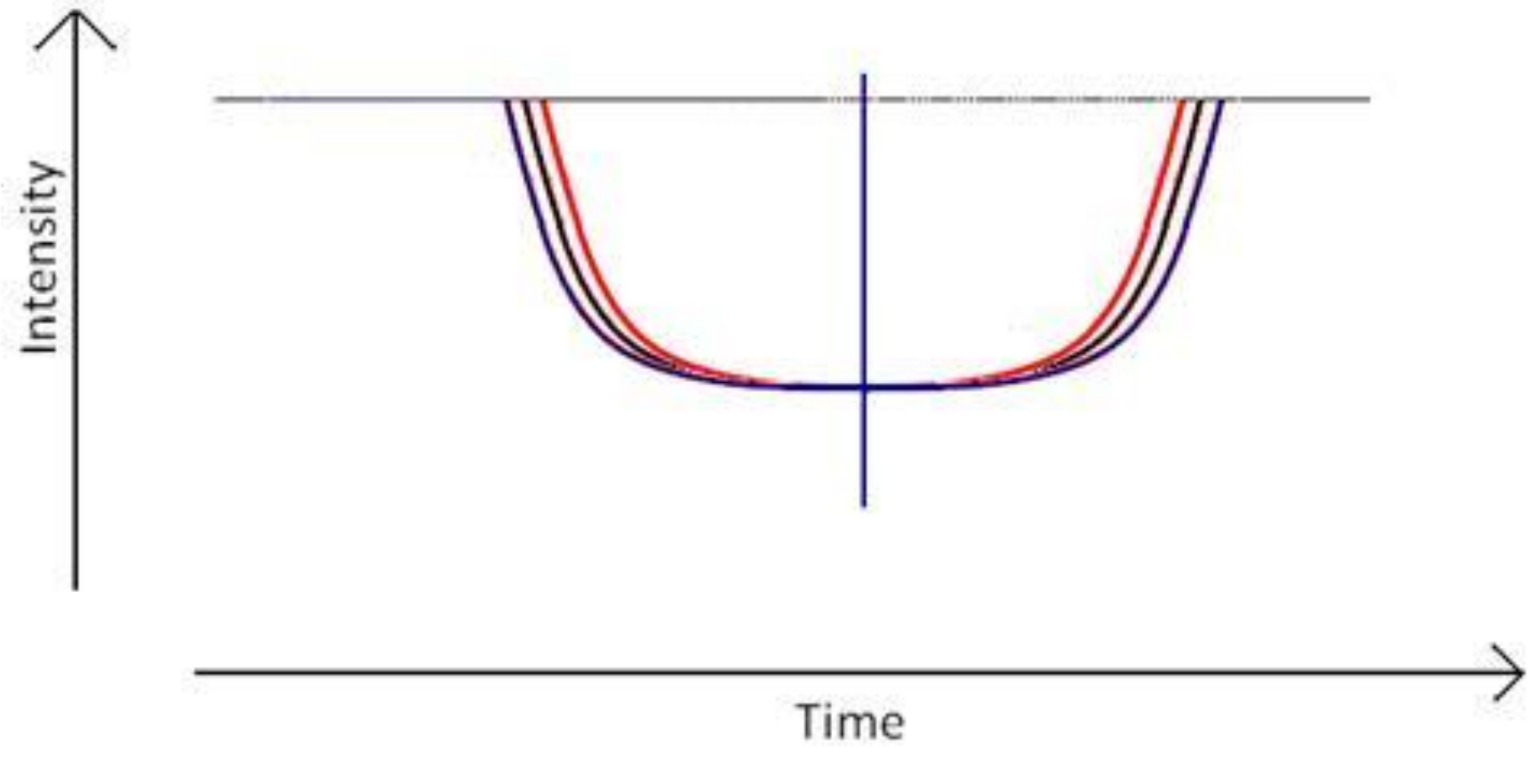

Figure 5.4: TDV due to an exomoon.

Furthermore, making the following assumptions:

- The projected velocity of the planet-moon barycentre during the transit, relative to the star, does not vary significantly over the time-scale of the transit duration.

- The projected velocity of the planet during the transit, relative to planet-moon barycentre, does not vary significantly over the time-scale of the transit duration.

- The orbital inclinations of the planet (i = 90°) and exomoon do not vary from orbit to orbit.

- We do not consider additional perturbing bodies in the system.

We obtain that any transit duration variation is solely due to the variation of the velocity. Moreover, it can be shown [134] that the rms amplitude of the TDV signal, $\delta_{TDV}$, has the following proportionality:

$$\delta_{TDV} \propto M_s a_s^{-1/2} \tag{5.4}$$

In the more general case of $i \leq 90°$ and an exomoon inclined from the planet-star plane, the TDV effect due to an exomoon has two major constituents: one due to the velocity variation effect (V) already described and one new component due to transit impact parameter variation (TIP). The TIP-component is due to the planet moving between higher and lower impact parameters as a result of the wobbling (see Fig. 5.5). Moreover, it is $\propto a_S M_S$, which is the same as TTV's proportionality. A summary of the properties of the three known transit timing effects due to an exomoon can be seen in Table 5.1.

This additional TDV component acts constructively with the V-component in the case of a prograde exomoon orbit and destructively for a retrograde orbit. So, this asymmetry could allow for future determination of the orbital sense of motion [135].

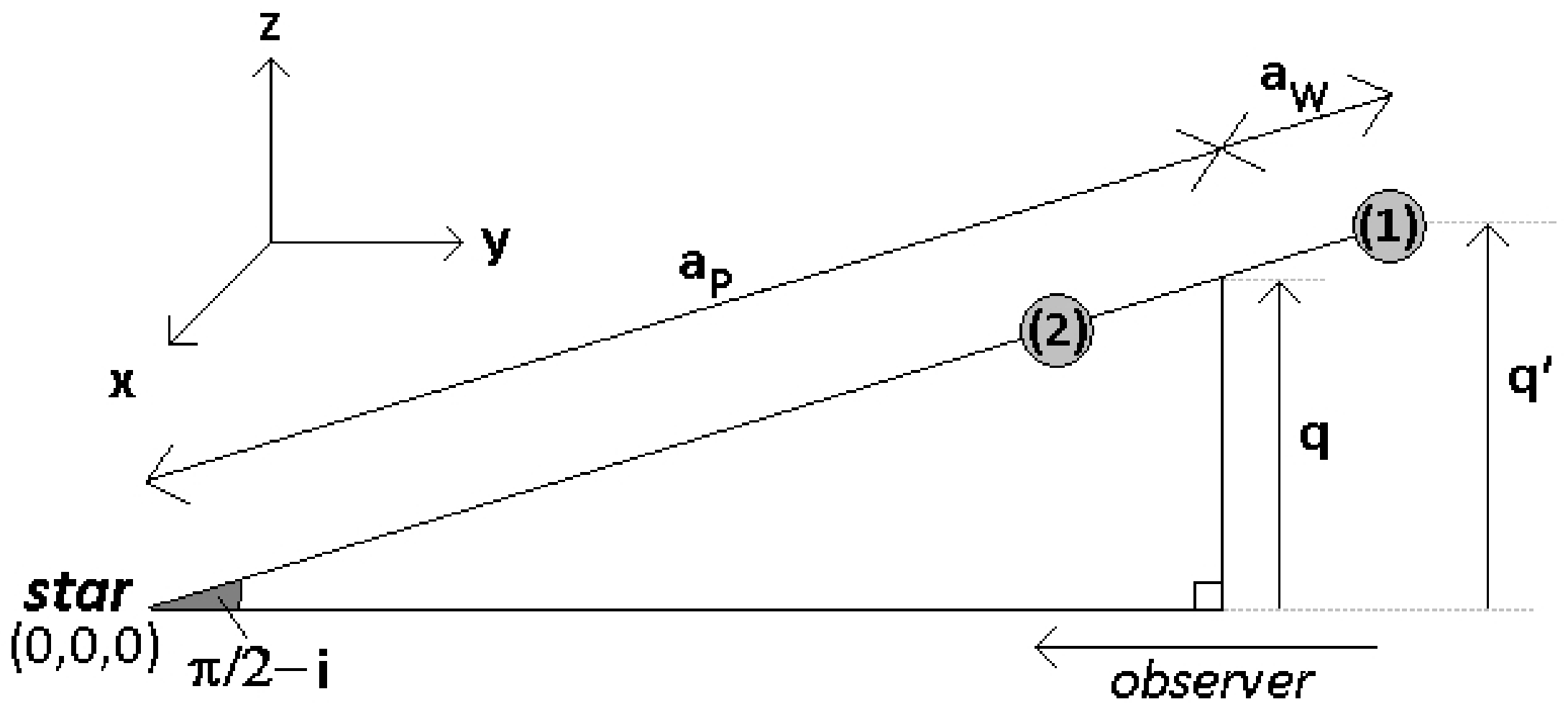

Figure 5.5: Sketch of the side-on view of the star-planet-moon system. In this schematic, the star lies in the bottom left, the observer lies at y = +∞ and the exomoon is not shown. The wobble of the planet is represented by the two grey spheres, (1) and (2), being the planet's maximal positions. The presence of a moon causes distance a perturbation in the distance q [135].

Table 5.1: Summary of key properties of the three known transit timing effects due to an exomoon.

| | TTV | TDV-V | TDV-TIP |
|---|---|---|---|
| Type of effect | Positional | Velocity | Positional |
| Direction | $\hat{x}$ | $\hat{x}$ | $\hat{z}$ |
| Proportionality | $a_s M_s$ | $M_s a_s^{-1/2}$ | $a_s M_s$ |
| Relative phase | 0 | $\pi/2$ | $\pm\pi/2$ |

### *5.2.3 TTV & TDV as complementary methods*

The introduction of TDV allows for the precise measurement of $M_S$ without any assumption on the orbital separation. In fact, using equations (5.2) and (5.4) we can directly measure the orbital separation of the exomoon $a_S$ and hence $M_S$.

Another major advantage of TDV is that the signal should lag TTV by a $\pi/2$ phase difference, originating from the fact TTV is a spatial effect whereas TDV is a velocity effect. Unfortunately, in the case i = 90° the direction and value of the phase shift remains unchanged between prograde and retrograde satellites [134].

Combining TTV and TDV should allow for a much more significant confirmation of a potential exomoon than just using TTV alone. The phase difference can be seen in the example waveforms in figure 5.6, where there are plotted the TTV and TDV waveforms due to a hypothetical exomoon of 1M⊕ around a GJ436b for $e_S$ = 0, 0.3 and 0.6 (i = 90°).

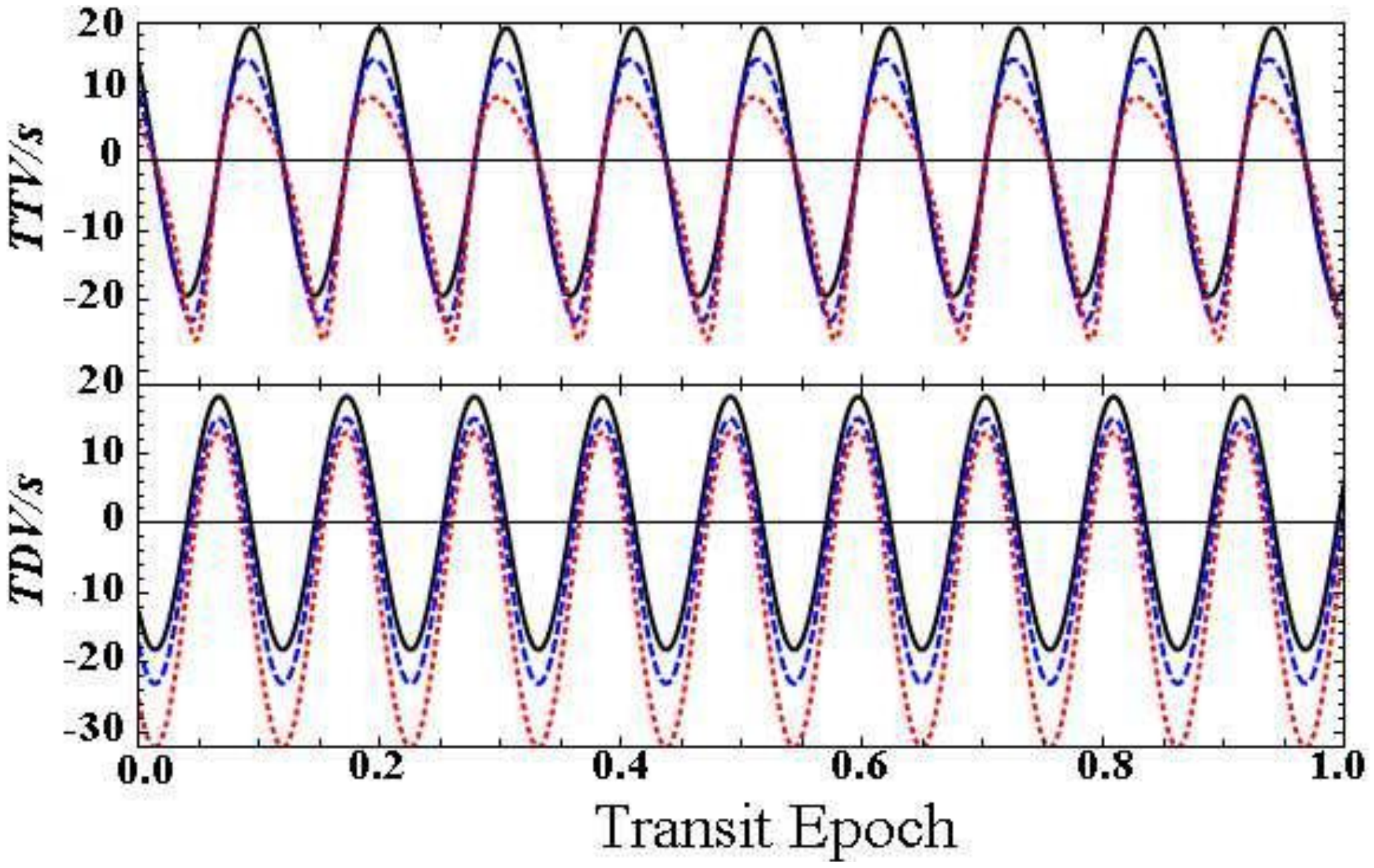

Figure 5.6: The TTV and TDV waveforms due to a hypothetical 1M⊕ exomoon around GJ436b. The solid line gives $e_S$ = 0, the dashed line $e_S$ = 0.3 and the dotted line $e_S$ = 0.6 [134].

The orbital radius of any satellite around a planet must lie somewhere between the Hill radius $R_H$ and the Roche limit $d_{min}$ to maintain stability. So, $a_S$ can be expressed by assuming that it is equal to some fraction χ of the Hill radius: $a_s = \chi \cdot R_H$

In Table 5.2, there are evaluated the TTV and TDV (both V- & TIP- components) rms amplitudes for a variety of known transiting planets in the case of a 1M⊕ exomoon with $\chi = 1/3$ and $e_S = 0$. We can deduce that the TDV-V signal is of a similar order of magnitude to the TTV signal while TDV-TIP effect is often an order of magnitude less than the V-component. Moreover, the timing signal of Neptune-mass host planets should be between 10 and 20 seconds. With some current ground based instruments reaching 7 seconds, for example Johnson et al. [147], and the Kepler mission expected to achieve second accuracy then large exomoons can be already detectable.

Table 5.2: Predicted TTV & TDV rms amplitudes due to a 1M⊕ exomoon, for a selection of transiting planets [135].

| Planet | $\delta_{TTV}$ (s) | V-part of $\delta_{TDV}$ (s) | TIP-part of $\delta_{TDV}$ (s) |
|---|---|---|---|
| GJ436b | 14.12 | 13.68 | 1.30 |
| CoRoT-Exo-4b | 7.58 | 9.15 | 0.00 |
| OGLE-TR-111b | 4.63 | 7.32 | 0.11 |
| HAT-P-1b | 4.58 | 6.82 | 0.47 |
| WASP-7b | 3.26 | 5.88 | 0.00 |
| TrES-1b | 3.04 | 5.95 | 0.05 |
| HD209458b | 2.97 | 5.95 | 0.07 |

### *5.2.4 An Earth-like habitable exomoon*

One interesting hypothetical example to reflect on is an Earth-like exomoon around a Neptune-like planet. Consider a system like GJ436 but with a transiting planet which has a period of $T_P$ = 35.7 days and is on a circular orbit (i = 90°). This would put the planet into the habitable zone of GJ436 with Teq ≃ 300K. Although the Neptune-like planet itself would not be an ideal place to search for life, an Earth mass exomoon would be. Suppose there is a 1M⊕ exomoon with this planet on a circular orbit, χ=0.25 and hence $P_S$ ≃ 2.5 days and the planet-moon orbital plane is coaligned to the star-planet orbital plane, then the predicted rms TTV amplitude would be 138s and the TDV amplitude would be 60s [134]. For GJ436b, Alonso et al. [148] used the 1.52m Carlos Sanchez telescope and achieved a timing accurateness of ∼ 13 seconds and a duration error of ∼ 50 seconds. This suggests that the detection of the exomoon should be presently possible through TTV from the ground and feasible with TDV in the near future. This illustrates that even ground-based instruments could detect an Earth-like body in the habitable zone using timing effects.

*Chapter 6*

# EVALUATING THE DETECTABILITY OF HABITABLE EXOMOONS WITH KEPLER-CLASS PHOTOMETRY

Moons could potentially be a perfect environment for life to flourish (see par. 3.3.2) but the detection of a moon outside our own solar system remains a challenge to current astronomical instrumentation. The Kepler Mission, while designed to detect Earth-like planets, is expected to detect many more gas giants. If such bodies harboured extrasolar moons, such planets should exhibit timing deviations in each transit.

As part of my work at UCL, we have estimated the maximum detectable range of exomoons that the Kepler Mission or Kepler-class photometry (KCP) could detect through transit timing effects, with particular attention to the exciting prospect of habitable-zone moons around gas giant planets [6].

The question as to whether current telescopes can detect exomoons could have far-reaching implications for the future of exoplanet science. Probably, exomoons could be common habitable environments in the galaxy and would therefore be of great interest to astrobiologists. Our work shows that exomoons are indeed already quite detectable with KCP. We emphasise the use of Kepler-class photometry due to the increasingly remarkable results being achieved from the ground which are matching space-based photometry, for example Johnson et al. [150]. In addition, groundbased observations are often more ideally suited for transit-timing studies due to the fewer constrictions placed on the system, such as telemetry-limited data-download speeds.

## 6.1 Modelling the detectability of exomoons

In order to investigate a large range of parameter space, it is more opportune and efficient to utilize analytic expressions rather than repeated individual simulations for thousands of different scenarios. Our hypotheses are here described:

- Only one exomoon exists around the gas giant exoplanet of interest.
- The moon and planet are both on circular orbits and the moon's orbit is prograde.
- The moon's orbital plane is coaligned to that of the planet-star plane which is itself perpendicular to the line-of-sight of the observer, i.e. $i = 90°$.
- If a planet is within the habitable zone, then any moon around that planet may also be considered to be **habitable**.
- A transiting planet must be detected to 8-$\sigma$ confidence to be accepted as genuine.
- An exomoon must be detected through either 1) TTV to 8-$\sigma$ and TDV to 3-$\sigma$ confidence or 2) TTV to 3-$\sigma$ and TDV to 8-$\sigma$ confidence, in order to be accepted as authentic.
- The Kepler Mission or KCP will be used in high cadence mode for the transit timing of a target of interest for $\simeq$ 4 years.
- $n = M/P_P$ where $M$ is the mission duration and $P_P$ is the period of transiting planet.
- At least three transits are needed to detect both a planet and a moon.

In most of the cases we will consider, many more than three transits will be detected and three can be seen to be the limiting case for G0V stars, where the habitable zone is sufficiently distant to only permit three transits in a 4-year timespan. Although statistically speaking three transits is sufficient, there is a risk of an outlier producing a false positive. We then consider detections of habitable exomoons in early G-type star systems to be described as "tentative", whereas once four transits are detected, for stars of spectral type G5V and later, this risk can be considered to be reduced.

The nominal mission length of Kepler is 3.5 years and it may be extended to up to 6 years, which justifies our choice of 4 years of transit timing observations. A ground-based search achieving KCP may easily be operational for 4 years or more. We choose 8-$\sigma$ as the signal detection threshold since this is the same as that used by Kepler [36]. The second signal may be detected to lower significance since it is only used to confirm the phase difference between the two and also derive the exomoon period.

The analytic expressions we need are:

1. TTV & TDV signal amplitudes

2. Transit mid-time and transit-duration errors

3. Confidence of detection, based on signal-to-noise

1. The TTV and TDV root mean square amplitudes are derived from the equations presented in Kipping [134] [135] which we have been described in par. 5.2.1 and 5.2.2.

2. For the purposes of TDV measurements, the primary requirement is to use a measure of transit duration which has the lowest possible uncertainty. Carter et al. [151] derived analytic expressions for the uncertainty on the mid-transit time ($t_c$) and duration ($T$) which consider the transit duration to be defined as the time taken for the planet to move between contact points 1.5 to 3.5 (see Fig. 6.1). This is in contrast to the definition used by Seager & Mallén-Ornelas [20], for example, who consider $t_T$ as being the duration between contact points 1 & 4 and $t_F$ as the duration between contact points 2 & 3 (see Fig. 2.6). By calculating the covariances of the lightcurve, it can be shown that $T$ can be calculated more precisely than either $t_T$ or $t_F$ and so we select $T$ as a robust duration parameter to investigate the TDV effect.

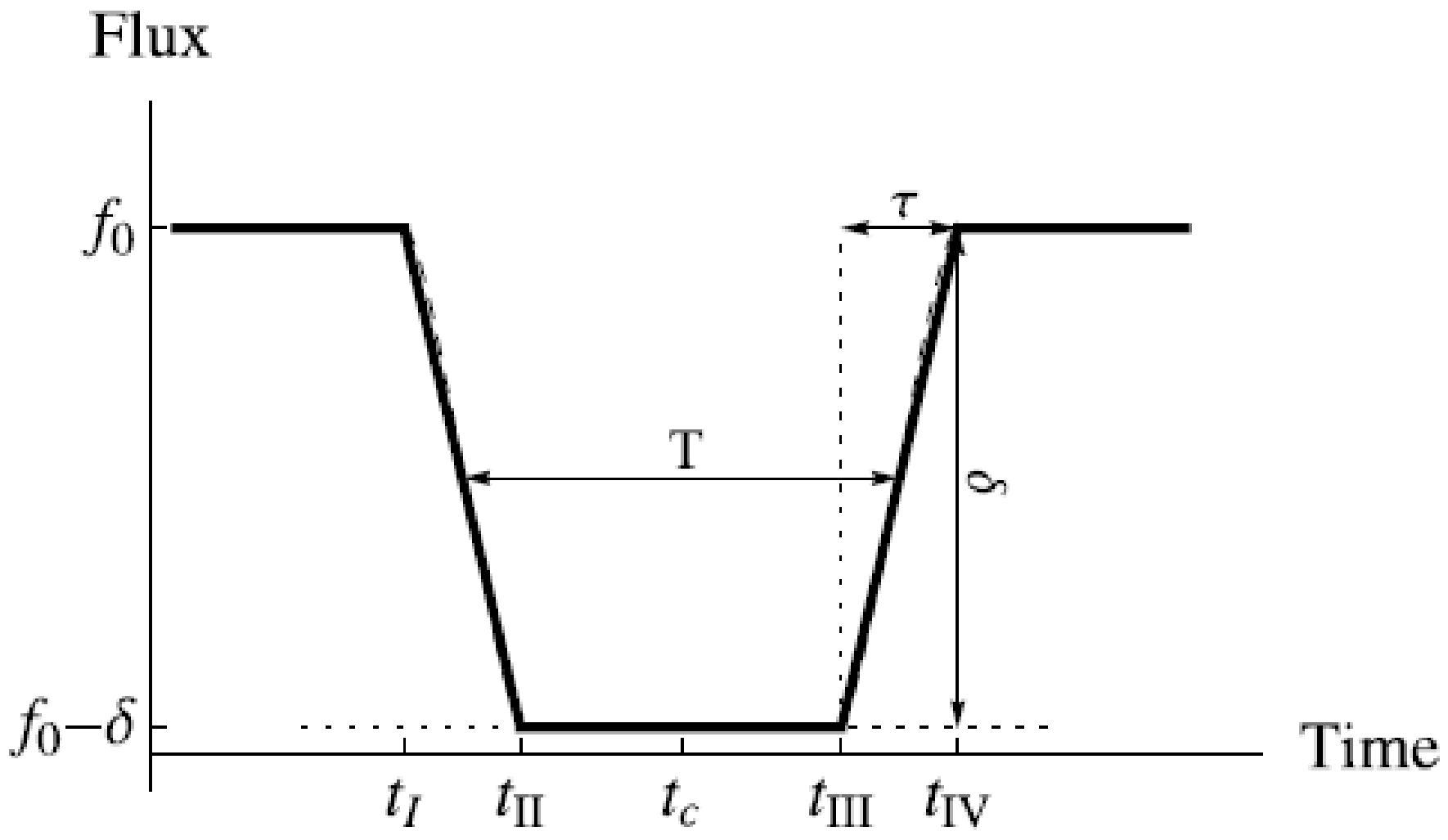

Figure 6.1: Lightcurve model with the transit duration *T* defined [151].

So the equations that we use for the uncertainties on transit depth *d*, transit duration *T* and mid-transit time $t_C$ are [151]:

$$\sigma_d = W^{-1}d \tag{6.1}$$

$$\sigma_T = W^{-1}\sqrt{2T\tau} \tag{6.2}$$

$$\sigma_{t_c} = W^{-1}\sqrt{T\tau/2} \tag{6.3}$$

$$\text{with } W = d\sqrt{\Gamma_{ph}T} \tag{6.4}$$

where τ is the ingress/egress duration and $\Gamma_{ph}$ is the photon collection rate.

These expressions do not hold for a poorly sampled ingress or egress and consequently we assume a cadence of 1 minute, corresponding to Kepler's transit-timing/asteroseismology mode. Moreover, they require as inputs the ingress duration τ and the transit duration *T*. In order to calculate these values we use the expressions of Seager & Mallén-Ornelas [20] modified for the time between contact points 1.5 & 3.5. For the photon collection rate, we use the same estimate made for the Kepler Mission (Borucki et al. [36], Yee & Gaudi [152]):

$$\Gamma_{ph} = 6.3\times10^{8}\,hr^{-1}10^{-0.4(m-12)} \tag{6.5}$$

where m is the apparent magnitude.

3. For a normal transit depth observed *n* times, the confidence *C* to which the transit is detected, in terms of the number of standard deviations, is defined as:

$$C_{photometric} = \frac{d}{\sigma_d}\sqrt{n} \tag{6.6}$$

In contrast, for TTV and TDV signals, which are periodic in nature with a frequency always much higher than the sampling frequency, we search for statistically significant excess variance and the $\chi^2$-distribution to calculate the confidence in a frequentist formulation:

$$C_{time} = \sqrt{2}erf^{-1}\left\{1 - Q\left[\frac{n}{2}, \frac{n}{2}\left(1 + \frac{\delta^2}{\Delta^2}\right)\right]\right\} \tag{6.7}$$

where *erf* is the error function, Q[a, b] is the incomplete upper regularized Gamma function, δ is the r.m.s. amplitude of the transit timing signal and Δ is the uncertainty on the mid-transit time $t_C$/transit duration *T*.

- The total noise

The expressions (6.1)-(6.4) only consider shot noise through the $\Gamma_{ph}$ parameter. In order to account for red noise, we may assume that the impact on $\sigma_d$, $\sigma_T$ and $\sigma_{t_c}$ are approximately equivalent for additional uncorrelated noise and for correlated noise, so we can simply modify W to absorb the effects of red noise. For uncorrelated noise we add the additional sources of noise in quadrature as done in the technical documents for Kepler (see for example [36]).

In general, there are expected to be three major types of noise present in the Kepler data in the form of **shot noise**, **instrument noise** *I* and **stellar variability** *S*. Instrument noise is due to a variety of effects and has been modelled in depth [36] to quantify its effect as a function of magnitude. With all three noise sources, we modify W to W', given by:

$$\frac{1}{W'} = \frac{1}{d}\sqrt{\frac{1}{\Gamma_{ph}T} + I^2 + S^2} \tag{6.8}$$

We illustrate all three noise sources plotted as a function of magnitude in Figure 6.2. We assume a constant value for stellar variability of 10 ppm across all spectral types, a realistic supposition, given that 65–70% of F7-K9 main-sequence stars in the Kepler field are likely to have analogous or lower intrinsic variability than the Sun ([153], [36]) on timescales significant to transit detections. We also note that (6.8) is equivalent to the formulation used in the original technical design papers for Kepler (e.g. equation (1) of [36]).

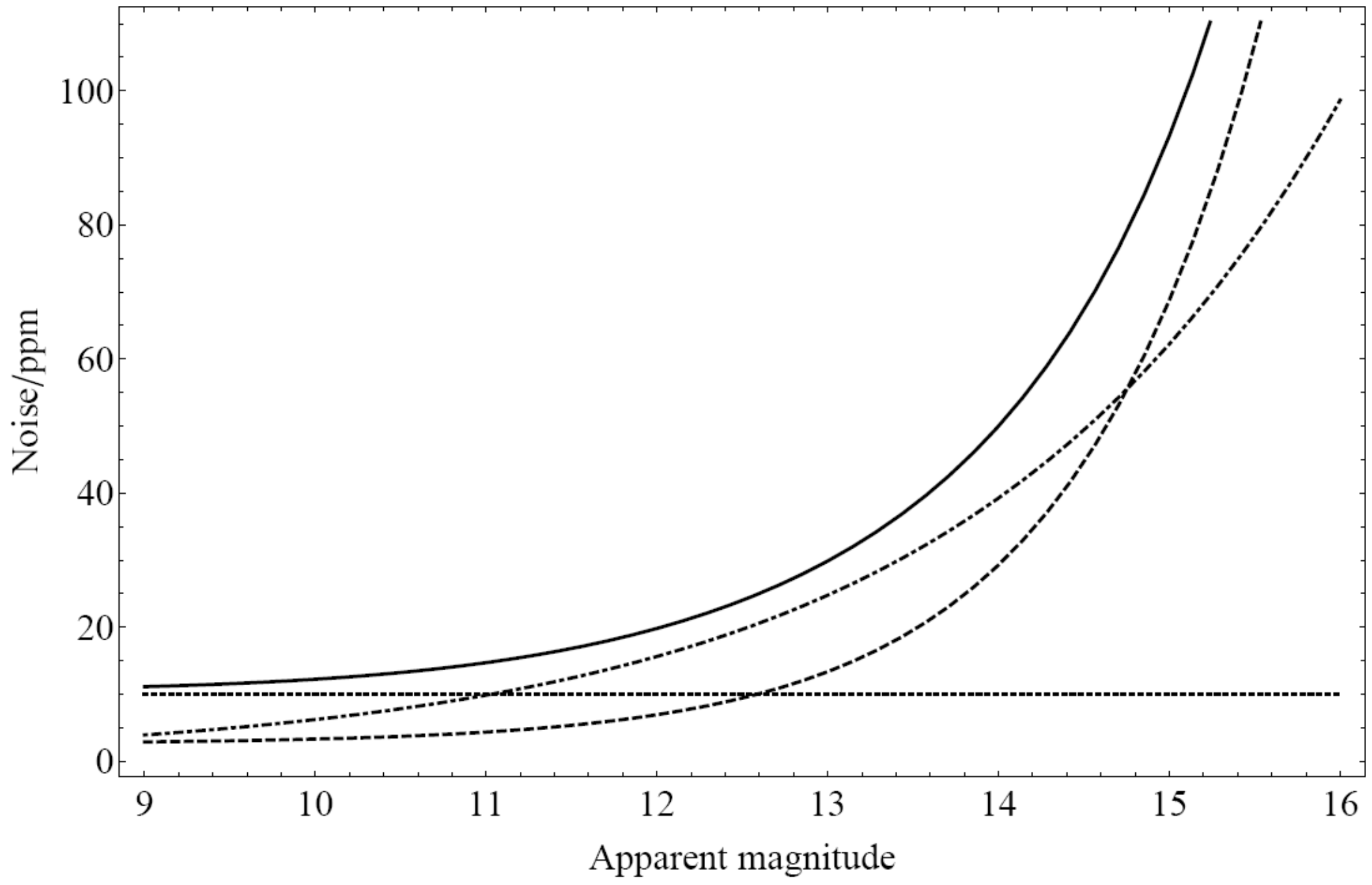

Figure 6.2: Noise sources predicted to affect Kepler photometry as a function of magnitude. Instrument noise is dashed, photon noise is dot-dashed, stellar variability is dotted and the total is solid [6]. Note as I is lower than the other sources till m=12.6 reflecting the good performances expected with Kepler.

- Properties of host star

In our study, we just consider single main-sequence stars because they offer the best potential for hosting habitable environments. We consider spectral types from M5V to F0V and assume for each an approximate mass, radius and effective temperature as given by Cox [21], and a luminosity derived from data therein. We use the Kepler bandpass (see Fig. 6.3) to calculate the *absolute* magnitude $M_{Kep}$ of these stars.

For each stellar type, we assume the stars are not young and may be considered to be slow rotators. Since stellar variability is correlated to rotational period (see Dorren et al.

[154]), we therefore limit ourselves to quiet stars. This is the same assumption used for Kepler's ability to detect Earth-like planets since very active stars will be too variable for the detection of such bodies. In Table 6.1 we list the different star properties.

Table 6.1: Properties of stars used in our calculations [21].

| Star type | $M_*/M_\odot$ | $R_*/R_\odot$ | $L_*/L_\odot$ | $T_{eff}$ (K) | $M_{Kep}$ |
|---|---|---|---|---|---|
| **M5V** | 0.21 | 0.27 | 0.0066 | 3170 | 11.84 |
| **M2V** | 0.40 | 0.50 | 0.0345 | 3520 | 9.49 |
| **M0V** | 0.51 | 0.60 | 0.0703 | 3840 | 8.42 |
| **K5V** | 0.67 | 0.72 | 0.1760 | 4410 | 7.06 |
| **K0V** | 0.79 | 0.85 | 0.4563 | 5150 | 5.78 |
| **G5V** | 0.92 | 0.92 | 0.7262 | 5560 | 5.02 |
| **G2V** | 1.00 | 1.00 | 1.0091 | 5790 | 4.63 |
| **G0V** | 1.05 | 1.10 | 1.3525 | 5940 | 4.34 |
| **F5V** | 1.4 | 1.3 | 2.9674 | 6650 | 3.47 |
| **F0V** | 1.6 | 1.5 | 5.7369 | 7300 | 2.71 |

- The habitable zone

We choose to consider a moon-hosting gas-giant exoplanet around the variety of main-sequence stars shown in Table 6.1. For each star we calculate the habitable zone orbital distance $a_{hab}$, to be defined as the distance where a planet would receive the same insolation as the Earth. This straightforwardly permits a reasonable estimation of the habitable zone for each star type.

$$a_{hab} = \sqrt{L_*/L_\odot}\,\mathrm{AU} \tag{6.9}$$

For each planet-moon system we consider, the period of the transiting planet is calculated using Kepler's Third Law:

$$P_{hab} = 2\pi\sqrt{\frac{a_{hab}^3}{G(M_* + M_P + M_S)}} \tag{6.10}$$

We choose to work in the time domain, rather than the orbital-distance formulation, since a major limiting factor in our study is the *Kepler Mission* duration.

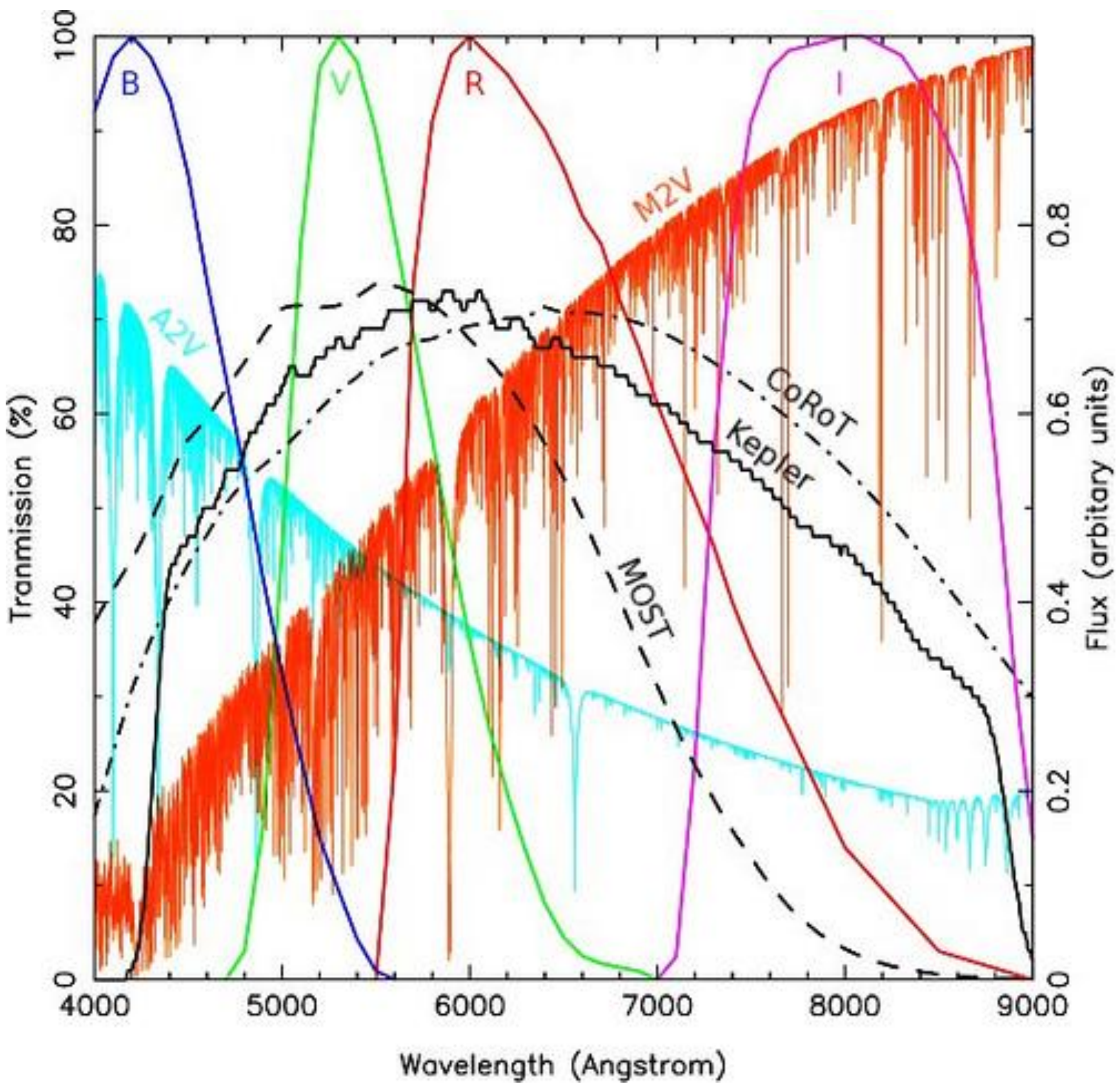

Figure 6.3: The transmission functions for the Johnson B,V,R,I filters. The Kepler bandpass is shown in black which peaks at approximately 70% throughput. The CoRoT bandpass is shown by the dot-dashed line. The MOST bandpass is shown by the dashed line. The spectrum for an A2V star which peaks in the UV and the spectrum for a M2V star which peaks in the infrared are shown as well. The two spectra have been scaled to have equal flux in the Johnson V filter.

• Properties of exomoons

Barnes & O'Brien [138] and Domingos et al. [139] have developed a set of analytic expressions for the stability of exomoons around exoplanets which provide excellent agreement to numerical simulations. In particular, using their equations we are able to estimate the range of allowed values for the planet-moon separation, in units of Hill radii $R_H$, which we label as ξ.

An estimate of the **maximum distance $a_E$** at which an exomoon is stable for prograde orbits is [139]:

$$a_E \approx 0.4895\,(1.0000-1.0305e_P-0.2738e_s) \tag{6.11}$$

for $e_P = e_s = 0$ we have $a_E \approx \xi/2$ with $\xi = a_P \left(\frac{M_P}{3M_*}\right)^{1/3}$.

For the **minimum distance $a_I$**, we calculate the Roche limit $r_R$ of the planet in all cases [134]:

$$r_R = R_P \cdot \left(2\frac{\rho_P}{\rho_S}\right)^{1/3} \tag{6.12}$$

If this value is greater than $2R_P$ we use the Roche limit as the minimum distance, otherwise $2R_P$ is adopted (i. e. $a_I \geq 2R_P$ in every scenario considered). We assume that in the range of allowed values, there is no reason for a moon to exist at any particular value of ξ and thus the prior **distribution** is **flat**.

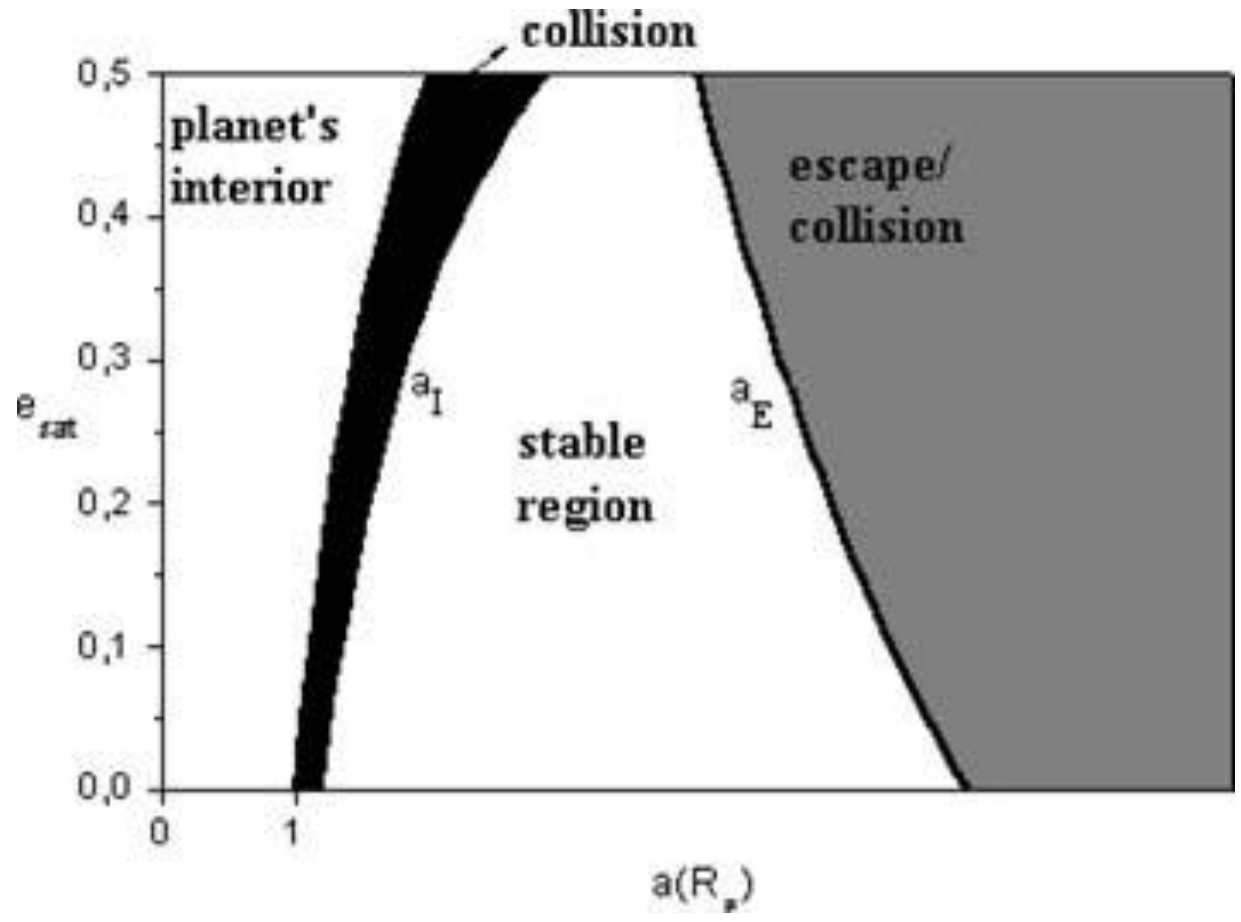

Figure 6.4: Sketch illustrating the stability and instability regions of prograde satellites in the initial conditions space, $a_S \times e_S$. The stable region is delimited by the internal ($a_I$) and external ($a_E$) critical semimajor axes. The left-hand border of the black region is determined by $a(1 - e_S) = R_P$ [139].

## 6.2 Lightcurve simulations

Having modified W to W' into the equation of Carter et al. [151] in order to account for both the effects of instrument and stellar noise, we choose to test this modified formulation through generation of synthetic lightcurves.

We are now going to present the programmes that have been written in FORTRAN 77/90 [155] in order to: generate the synthetic lightcurves, add noise to them and fit the noisy lightcurves.

### *6.2.1 The synthetic lightcurves generator: JASMINE*

Jasmine is a 3222-line Fortran 90 module which can be used with the Mandel-Agol code [35] to generate limb darkened lightcurves in the case of both circular and eccentric orbits.

As input, Jasmine needs the following parameters: the eccentricity *e*, argument of periastron *Ω* (redundant for circular orbits), $\frac{a_P}{R_*}$ where $a_P$ can be found using eq. (6.9) and $R_*$ should be known, the orbital period of the planet $P_P$ which can be found using eq. (6.10), the semi-amplitude of the stellar radial velocity signal $K_*$ which can be calculated using eq. (2.1) (see par. 2.2.5 on the necessity for spectroscopic follow up observations), the planet-star radii ratio *k*, the orbital inclination *i*, the specific 4 coefficients of the nonlinear limb-darkening law (see par 2.2.3) from Claret [156] (see Appendix B for an example about the Sun).

Jasmines uses the Seager & Mallén-Ornelas [20] expressions to compute the motion of the planet since we are dealing with circular orbits only. It computes the geometry of the system for the given inputs and then finds the intersection points of the orbit with the stellar limb, in "true anomaly" space. These true anomalies are then converted into times.

Jasmine produces two key sets of data:

1) Defining $z = \frac{\sqrt{x^2 + y^2}}{R_*}$ as the impact parameter (i.e. the apparent distant between the centre of the star and the centre of the planet), Jasmine computes z at different values of time. Then, the z-table is used in the Mandel-Agol code.

2) Additionally, Jasmine produces an output file which contains a list of durations of interest. T(i,j) means the time between the i-th and j-th contact point for the primary transit (1, 1.5, 2, 3, 3.5, 4 as shown in Fig. 6.1). S(i,j) refers to the secondary eclipse (see Appendix D for an example).

### *6.2.2 The noise generator: NOISY*

Noisy is a 184-line Fortran 90 program written by G. Campanella to account for all the contributions from the different noise sources (the code is in Appendix C). Noisy needs both the flux vs. time data as generated by Jasmine, as well as the characteristics of the noise to be generated.

The first noise source we add is **shot noise** generated by taking a random real number from a normal distribution of mean zero and standard deviation given by:

$$\sigma_{shot} = \frac{1}{\sqrt{t_{\exp}\Gamma_{ph}}} \tag{6.13}$$

where $t_{exp}$ is the time between each consecutive measurement, which is assumed to be one minute in our calculations. In order to generate a normal distribution of true-random numbers the Box-Muller algorithm has been used [157].

For the **stellar noise**, Noisy overlays a sinusoidal signal of period ~ 5 minutes and r.m.s. amplitude 10 ppm, which matches the amplitude prediction of Borucki et al. [36] for a G2V star.

For the **instrument noise**, in Noisy we make the assumption that it can be modelled approximately by a sinusoidal signal of period ~ 1 hour. This is consistent with the dominant source of instrument noise with other space telescopes, for example *Spitzer*, which suffers slight pointing variations which introduce intrapixel pixel phase response noise. We calculate the r.m.s. amplitude of this noise using the Borucki et al. [36] model for *Kepler*.

The periods of the instrument noise and stellar noise are slightly offset by a random number between -1 and +1 seconds in order to avoid any kind of resonance between the two waveforms. We also give the two correlated noises a random phase difference between 0 and $2\pi$.

Finally, Noisy returns the lightcurve in an output file which lists: 1) time in days from mid-transit and 2) relative flux+noise.

### *6.2.3 Lightcurve fitting with JASPER*

Jasper is a 3459-line Fortran 77 module used to obtain best-fit values for T and $t_c$. In all cases we fit for $a_P/R_*$, i, $k$ and $t_c$ and therefore assume that the out-of-transit baseline is well known and its errors are essentially negligible. This is a reasonable assumption for high quality photometry with large amounts of out-of-transit data.

The fitting code finds the best-fit to the lightcurve by utilising a genetic algorithm, **PIKAIA** (Metcalfe & Charbonneau [159]), to get close to a minimum in $\chi^2$.

In fact, genetic algorithms (GA) are a class of heuristic search techniques that apply basic evolutionary operators in a computational setting, this makes them well suited to search for the region of parameter-space that contains the global minimum. The central idea is to solve an optimization problem by evolving the global solution, starting with an initial set of purely random guesses. The GA evaluates the model for each set of parameters, and the predictions are compared to observations. Each point in the "population" of trials is subsequently assigned a fitness based on the relative quality of the match. A new generation of sample points is then created by selecting from the current population of points according to their computed fitness, and then modifying their defining parameter values with two genetic operators in order to explore new regions of parameter-space.

The two basic genetic operators are crossover, which emulates sexual reproduction, and mutation which emulates somatic defects. The crossover procedure pairs up the strings, chooses a random position for each pair, and swaps the two strings from that position to the end. The mutation operator spontaneously replaces a digit in the string with a new randomly chosen value.

Under the "pressure" of fitness-based selection, crossover acting on successive generations of strings modifies the frequency of a given substring in the population at a rate proportional to the difference between the mean fitness of the subset of strings incorporating that substring, and the mean fitness of all strings making up the current population. So, the GA can be thought of as a classifier system that continuously sorts out and combines the most advantageous substrings that happen to be present across the whole population at a given time. In this context the role of mutation is to inject "novelty" continuously, by producing new digit values at specific string positions, which might not otherwise have been present in the population or may have been selected against during earlier evolutionary phases.

The approximate solution from PIKAIA is then used as a starting point for a $\chi^2$-minimisation performed with the **AMOEBA routine** (Press et al. [157]).

The AMOEBA algorithm, originally published in 1965 by Nelder and Mead [160], is designed to solve the classical unconstrained optimization problem of minimizing a given nonlinear function. The Nelder-Mead method is simplex-based, a simplex *S* in $\Re^n$ is defined as the convex hull of n + 1 vertices $x_0,\ldots,x_n \in \Re^n$. For example, a simplex in $\Re^2$ is a triangle, in our case n = 4.

The method begins with a set of n + 1 points $x_0,\ldots,x_n \in \Re^n$ that are considered as the vertices of a working simplex *S*, and the corresponding set of function values at the vertices $f_j := f\left(x_j\right)$, for $j = 0,\ldots,n$.

The method then performs a sequence of transformations of the working simplex *S*, aimed at decreasing the function values at its vertices. At each step, the transformation is determined by computing one or more test points, together with their function values, and by comparison of these function values with those at the vertices. This process is terminated when the working simplex becomes sufficiently small in some sense, completing the minimization of the function.

Subsequently, the AMOEBA solution is tested by randomly perturbing it and refitting in 20 trials. Then, Jasper calculates the quantities of interest and writes 2 files: one with the AMOEBA model for the lightcurve and another with a report.

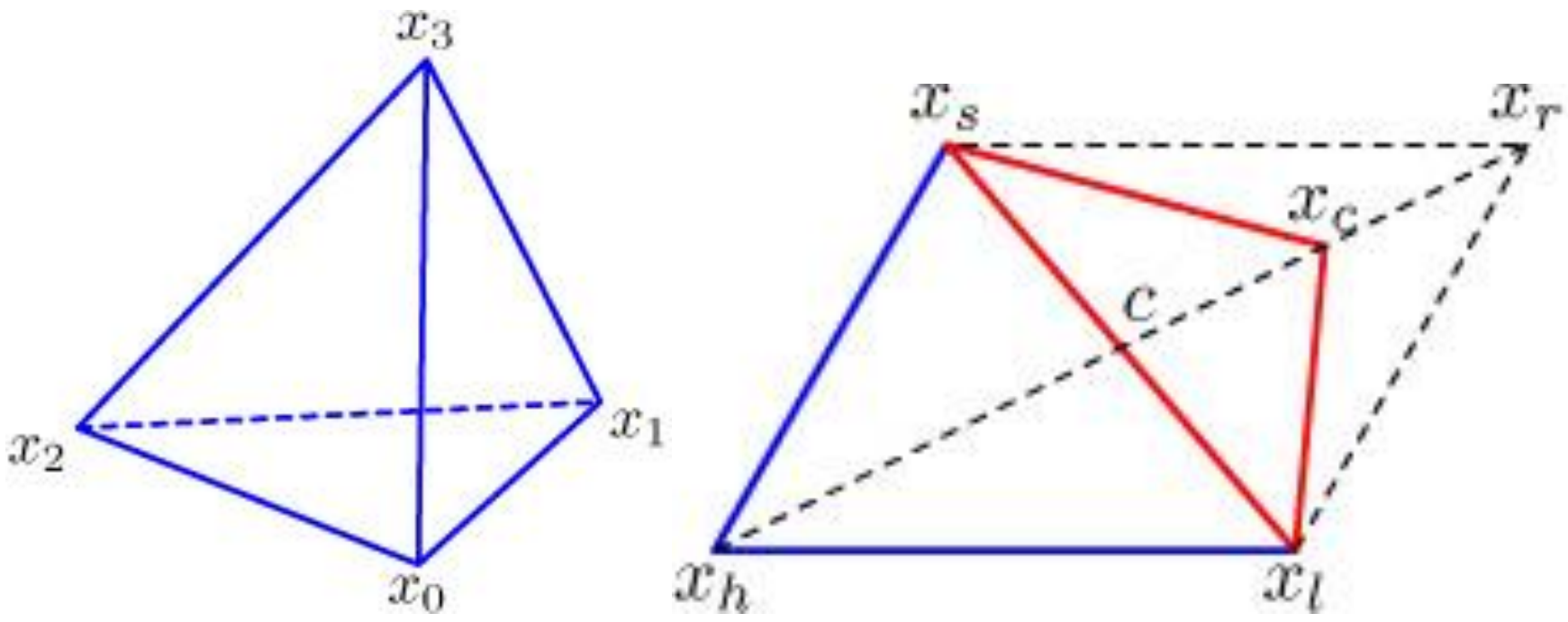

Figure 6.5: Left: A simplex in $\Re^3$. Right: The simplex "contract" transformation in $\Re^2$, the new simplex is shown in red [161].

### *6.2.4 Lightcurves generation and fitting*

We generate the lightcurves for a Neptune, Saturn and Jupiter-like planet in the habitable zone of a $m_{Kep} = 12$, G2V Sun-like star $(a_P = 1AU)$ with i = 90° and e = 0. In Table 6.2 there are summarized the inputs entered in Jasmine. Each lightcurve is generated to have 1441 data points evenly spaced with a 1-minute cadence (so to simulate 60 x 24 = 1 day of observations) centred on the mid-transit time.

Table 6.2: The values for the planets used during the simulations. Note that to calculate $K_*$, we have assumed $M_p$ = mass of planet + mass of Earth [158]. This is because we assume an earth mass moon around each planet so the total barycentric mass has this number added on.

| Planet | $a_P/R_*$ | $P_P$ (days) | $K_*$ (m/s) | $k$ |
|---|---|---|---|---|
| Jupiter | 215.094 | 365.071 | 28.510 | 0.10279 |
| Saturn | 215.094 | 365.193 | 8.602 | 0.08665 |
| Neptune | 215.094 | 365.236 | 1.624 | 0.03561 |

Noisy was used to estimate the noise. As we want to mimic the non-stochastic nature of real stellar variability, this time for the stellar noise we generate 1000 sinusoidal waveforms with varying periods randomly selected between one minute up to 24 hours. For the amplitude, we have used a 1/f frequency dependency with maximum amplitude at the low frequency being 120% of <A> and minimum amplitude being 80% of <A> where <A> = 0.8 ppm. Using this setup and adding all them gives a single synthetic stellar signal (see Fig 6.6) with a standard deviation of around 10ppm over 6.5 hour integration.

Moreover, now we consider that the instrument noise is composed of hundreds of different noise sources periodic in nature varying on timescales from one minute to one day. There is no benefit of including timescales longer than this since transit events will not last longer than ~ 1 day. For a star of *visual* magnitude $m_{Kep}$ = 12, we take an r.m.s. instrument noise of 7.47 ppm.

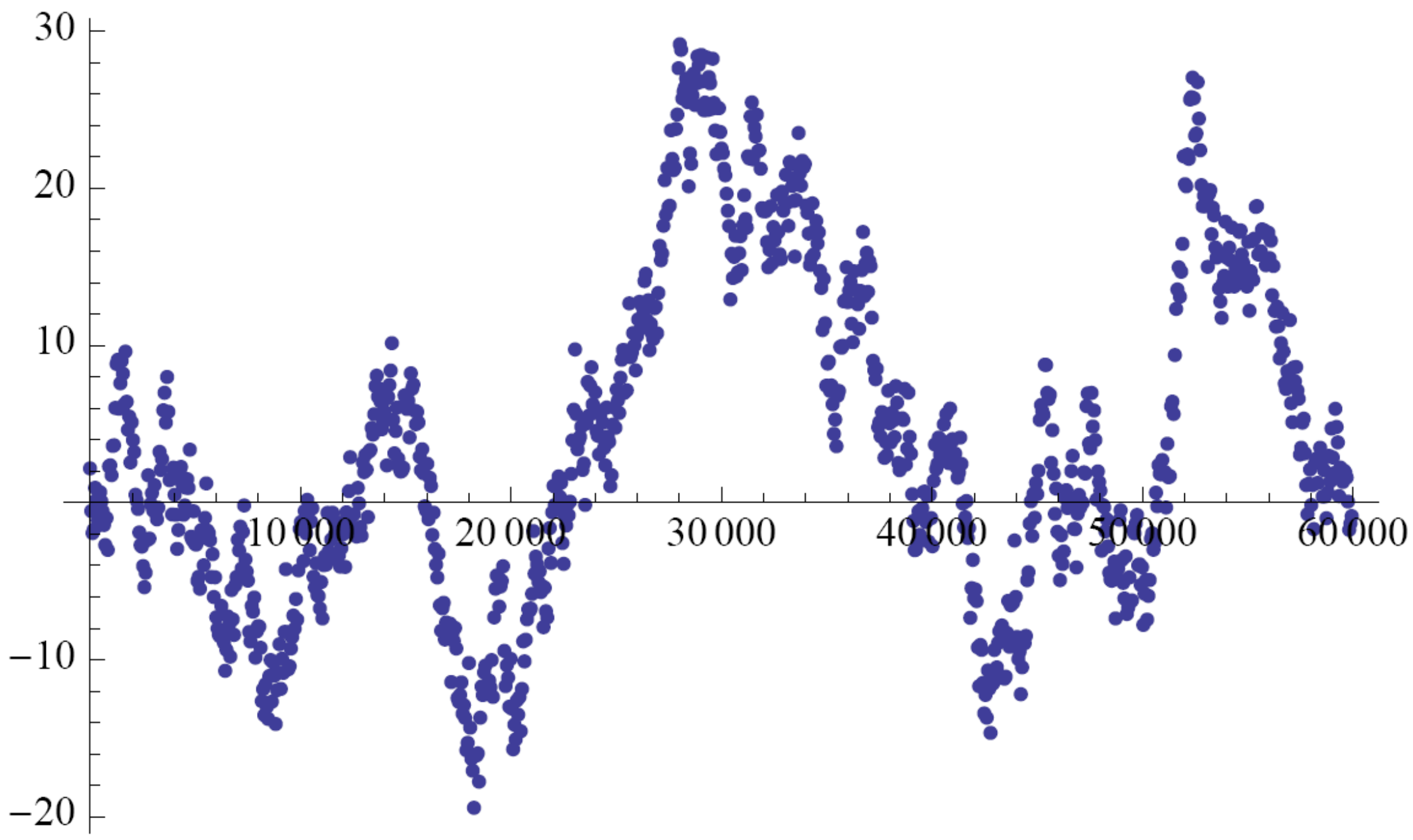

Figure 6.6: An example of stellar noise.

The noisy lightcurves are generated 10,000 times with the correlated noises and photon noise being randomly generated in all cases. In Figs. 6.7 and 6.8 we show some examples of what we have obtained.

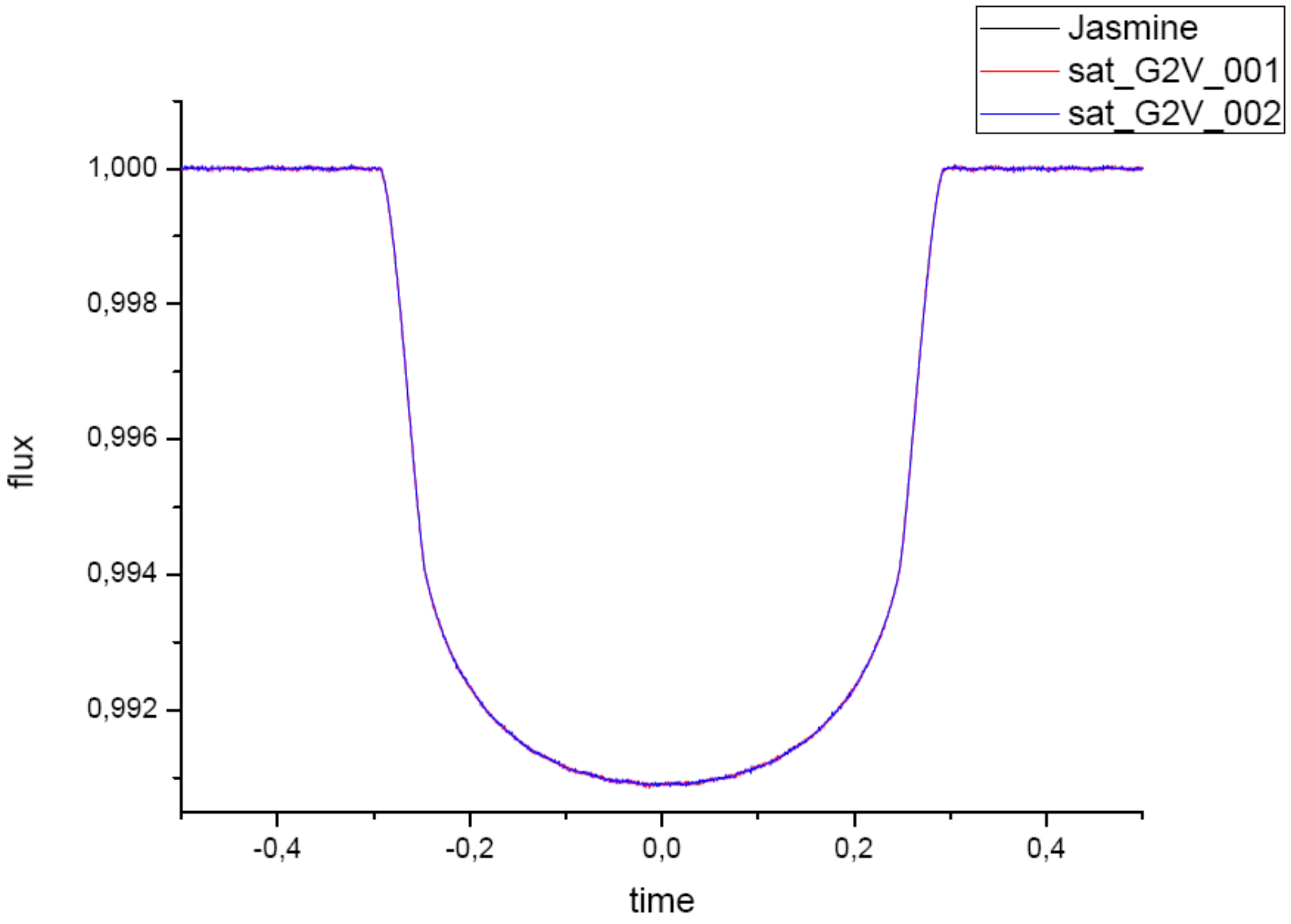

Figure 6.7: The lightcurve for the Saturn-like planet generated by Jasmine. Moreover, there are shown 2 noisy lightcurves.

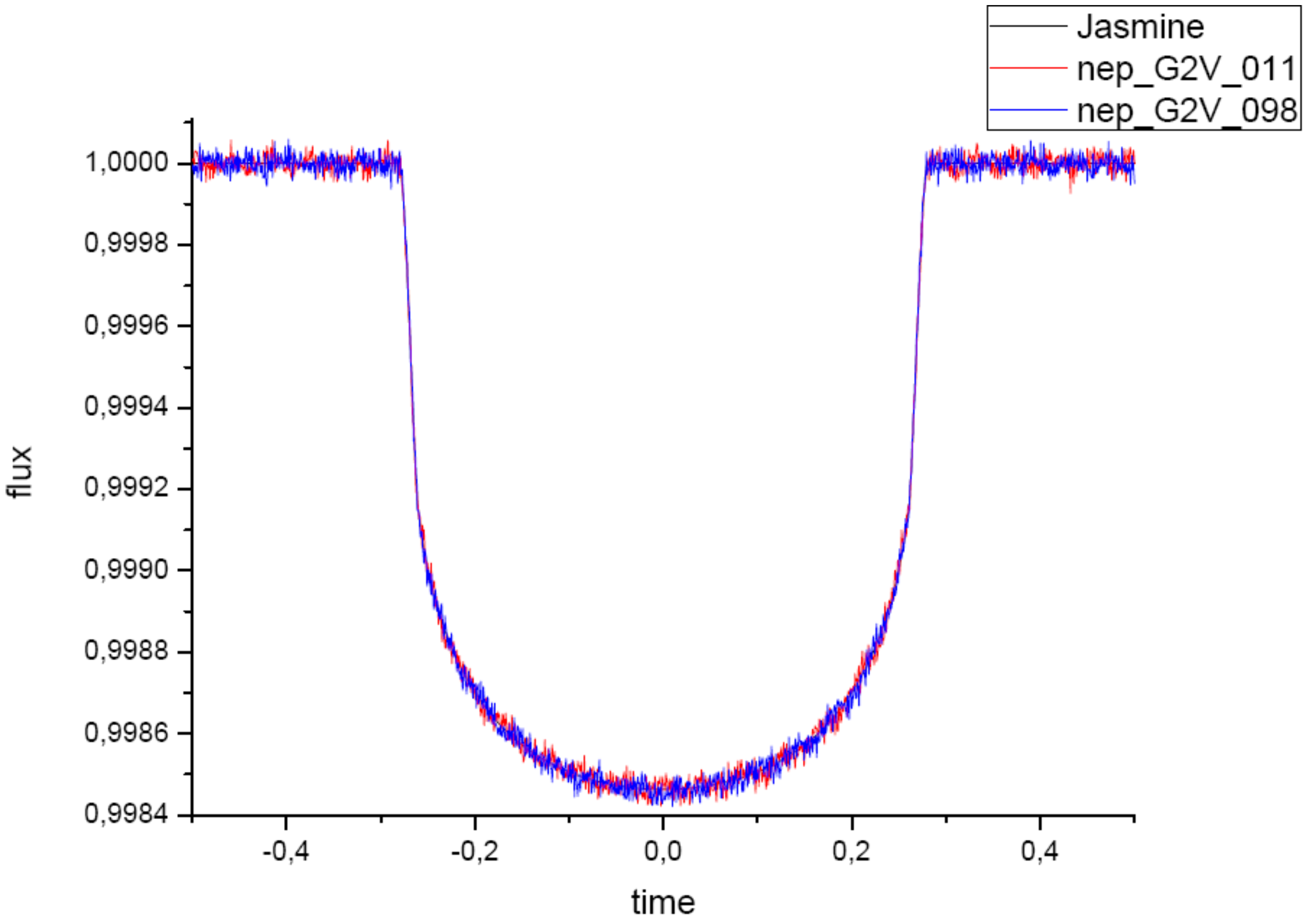

Figure 6.8: The lightcurve for the Neptune-like planet generated by Jasmine. Moreover, there are shown 2 noisy lightcurves. Note as here the noise is more relevant.

The lightcurves are then passed onto Jasper. In each subsequent fitting run of the 10,000 lightcurves, the AMOEBA routine starts from the original best-fit. In Appendix D, we show in detail an example of simulation.

### *6.2.5 Comparison to the analytic expressions*

For the three cases of a Neptune, a Saturn and a Jupiter, we obtain 10,000 estimates of T and $t_c$ in each case. These values are binned and plotted as a histogram and then compared to the distribution expected from the modified expressions 6.1-6.4. In all cases, we find excellent agreement between the predicted uncertainties and the theoretical values. In Figure 6.9, we plot a histogram of the results for the Saturn-case T values, and overlay the predicted value of $\sigma_T$ for comparison where the quality of the agreement is evident from the plot. Note that this overlaid Gaussian is not a fit but a theoretical prediction.

We find excellent agreement for both $\sigma_T$ and $\sigma_{t_C}$ and the theoretical expressions slightly overestimate the uncertainties. Also a very nice result is that $\sigma_T \approx 2\sigma_{t_C}$, almost precisely what eqs. (6.2)-(6.3) predict. Thus any results from this study can in fact be considered to be slightly conservative. We prefer to adopt a conservative approach in our analysis since there may be unexpected sources of noise which increase the timing uncertainties.

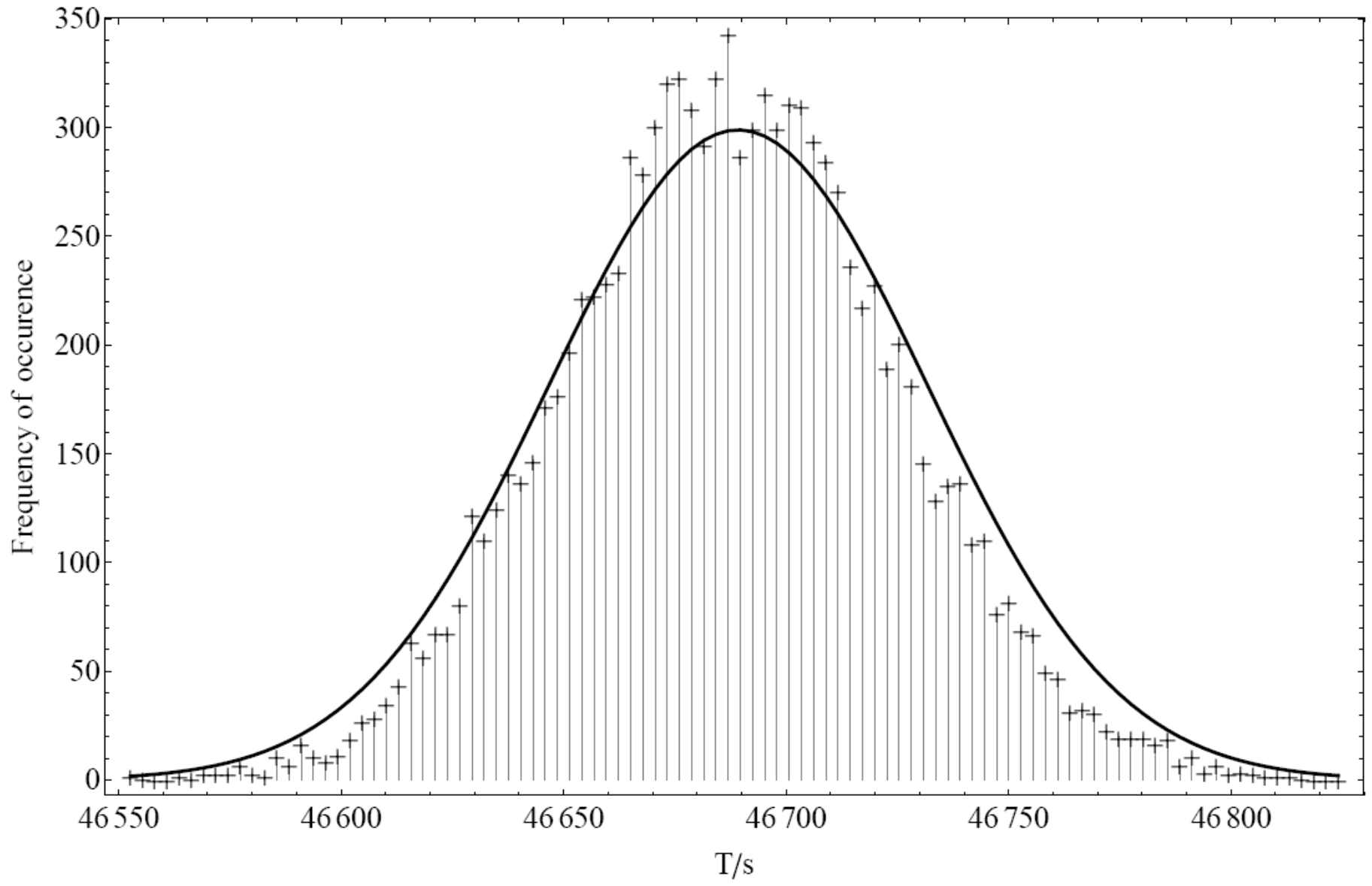

Figure 6.9: Comparison of the distribution in T found using simulations (scattered points) of a transit of a Saturn-like planet versus that from theoretical prediction (smooth line).

## 6.3 Results

Kepler is designed to look at ∼ 100,000 stars between $6^{th}$ to $16^{th}$ visual magnitude. Having implemented our model, we would like now to answer some questions about the potentiality expected with KCP.

### *6.3.1 Earth-transit magnitude limit*

The first question that we would like to address is about the magnitude limit in detecting an Earth-like transit with KCP. For each spectral type in Table 6.1, we are able to compute the faintest visual magnitude to which a habitable Earth-like transit can be detected. *Kepler* was conceived with the goal of detecting an Earth-Sun transit of 6.5 hours in duration, which represents about half the duration if the Earth has an impact parameter of zero. However, the maximum magnitude to which an Earth-like transit could be detected will be for the cases of equatorial transits (b = 0, i = 90°) as these maximise the transit duration and hence integration time. We establish the following criteria for a reliable transit detection:

• Each transit must be detected to ≥ 1-σ confidence.

• The fold lightcurve must have a significance of ≥ 8-σ.

• The time between contact points 2 & 3 ≥ 1 hour, in order to be detected with Kepler's survey-mode cadence of 30 minutes.

• At least 3 transits must be detected over the mission duration.

With these criteria, we calculate the magnitudes of the faintest stars hosting a habitable, transiting Earth which can be detected with KCP, for both full transit durations (i.e. equatorial transits) and half-durations (i.e. impact parameter chosen to be such that the transit duration is half the full transit). The results are given in Table 6.3. As expected, smaller host stars can be fainter due to their smaller radius and hence deeper transit depths. We also collect more transits towards smaller stars since the orbital period of the habitable Earth decreases.

Even if Kepler was extended to a mission length of 6 years, an F5V star would have a habitable zone so far out that detecting three transits within this region would be impossible. Ergo, we do not consider F type stars in our analysis. Moreover, considering 6th-magnitude stars as the bright limit, we can calculate $d_{min}$, the minimum distance for each spectral type which avoid the saturation of the CCDs.

Table 6.3: Faintest stars for which a habitable transiting Earth could be detected for different star types. Final column gives maximum distance of such a star with a full transit duration $D_{full}$, based on absolute magnitude in the Kepler bandpass. A blank indicates that no magnitude can satisfy the detection criteria.

| Star type | $D_{half}$ $m_{Kep,max}$ | $D_{full}$ $m_{Kep,max}$ | $d_{min}$ (pc) | $d_{max}$ (pc) |
|---|---|---|---|---|
| **M5V** | - | 18.111 | 0.68 | 179.56 |
| **M2V** | 15.992 | 16.190 | 2.00 | 218.78 |
| **M0V** | 15.206 | 15.465 | 3.28 | 256.45 |
| **K5V** | 14.278 | 14.618 | 6.14 | 324.79 |
| **K0V** | 13.286 | 13.712 | 11.07 | 385.83 |
| **G5V** | 12.762 | 13.240 | 15.70 | 440.56 |
| **G2V** | 12.236 | 12.763 | 18.79 | 423.25 |
| **G0V** | 11.520 | 12.106 | 21.48 | 357.44 |
| **F5V** | - | - | - | - |
| **F0V** | - | - | - | - |

### *6.3.2 Jupiters vs Saturns vs Neptunes*

We now use the approximate analytic expressions to get a feel on the general trends in exomoon detection. The first question we would like to ask is what is the optimum planet to search for moons around, out of the three classes of Neptunes, Saturns and Jupiters? We take each of these planets and simulate the detectability of the TTV signal as a function of planetary orbital period $P_P$, for an 0.2M⊕ exomoon with $a_S$ = 0.4895 $R_H$. We fix the star to be a G2V, $m_{Kep}$ = 12 object and work with i = 90° for simplicity.

We calculate the signal amplitude and mid-transit time uncertainty in all cases and hence find the $\chi^2$ value for a range of orbital periods. In Figure 6.10, we plot $\chi^2$ as a function of host planet's orbital period. The plot reveals that Jupiters are the hardest to search for exomoons around whilst Saturns are the easiest. This is due to Saturn's low density meaning a large transit depth but a low enough mass such that an exomoon still affects it significantly. We find the same order of detectability consistently in many different orbital configurations and for TDV as well.

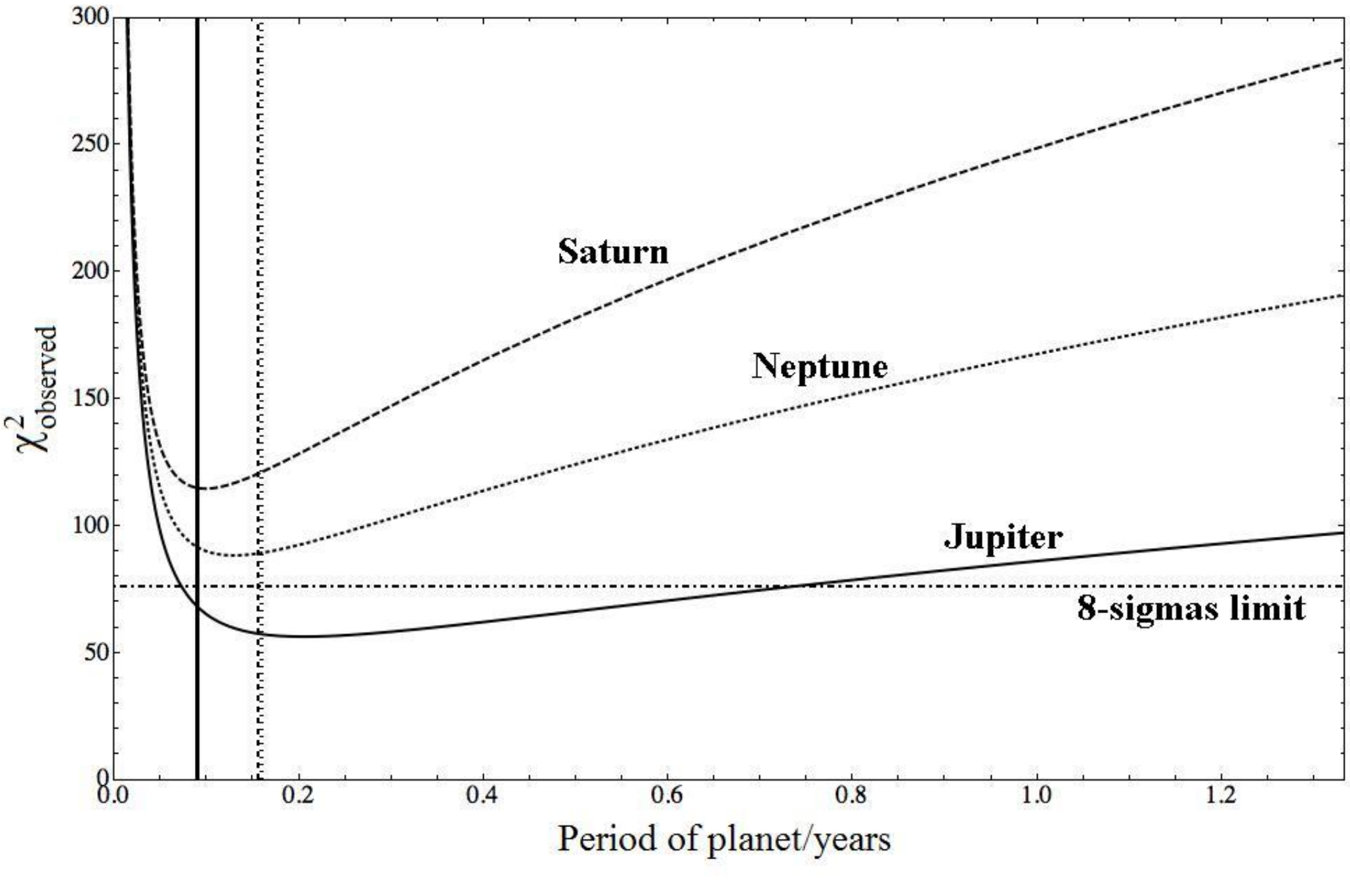

Figure 6.10: Plot showing the TTV detectability of a 0.2M⊕ (ξ = 0.4895) exomoon around a Jupiter, Saturn and Neptune-like exoplanet for a G2V, $m_{Kep}$ = 12 star. Values of $\chi^2 \gg 76$ are detectable at ≥ 8-σ confidence. Saturns present the strongest signal. The vertical lines represent the stability limit of such a moon, calculated using the model of Barnes & O'Brien [138].

### *6.3.3 ξ-$m_{Kep}$ parameter space*

We now consider an exomoon in the optimum condition of a Saturn hosting a single satellite in the habitable zone of the host star. The detectability of the timing signals depends on the planet-moon separation $a_S = \xi R_H$, the apparent magnitude of the host star in the Kepler bandpass $m_{Kep}$, mass of the exomoon $M_S$ and finally the host star's mass and radius (i.e. spectral type).

TDV increases with lower values of $\xi$ and TTV increases with higher values of $\xi$ (see eqs. 5.2, 5.4), but we must maintain a balance such that at least one of the effects is detected to 8-σ and the other to 3-σ confidence (dictated by our previous detection criteria). To find the limits of interest, we assume a zero-impact-parameter transit of our optimum planet-moon arrangement. The orbital period of the host planet is always given by $P_{hab}$. We prefer to calculate this best-case scenario since we may then consider it unfeasible to detect habitable moons beyond this limit.

For a given star type and exomoon mass, we plot the contours of $m_{Kep}$ and ξ which provide TTVs of 3- and 8-σ confidence and then repeat for TDVs of the same confidences. There are two possible acceptance criteria (see par. 6.1): 1) TTV confidence is ≥ 3-σ and TDV confidence is ≥ 8-σ; 2) TTV confidence is ≥ 8-σ and TDV confidence is ≥ 3-σ. The loci of points below these lines represents the potential detection parameter space. For each star type and exomoon mass, these loci will be different. In Table 6.4 we show the data of interest in the case of $M_S = 1/3 M_\oplus$ and in Figure 6.11 we plot an example for an M0V-type star.

There are certain cases which constrain the stars of interest. The Barnes & O'Brien [138] limit may be calculated for a Saturn harbouring an exomoon in a system of 5 Gyr age. This suggests that the maximum moon mass that a habitable zone Saturn could hold around a M5V star would be ~ 0.3 Ganymede[5] masses, assuming the maximum prograde orbital distance is 0.4895 $R_H$, as calculated by Domingos et al. [139]. In contrast, an M2V star allows for a habitable Saturn to hold onto a 0.4-Earth-mass moon for over 5 Gyr.

[5] M(Ganymede)=1.4819 x $10^{23}$ Kg ≈ 0.025 $M_\oplus$

A second lower limit exists from tidal forces and the Roche limit. An upper limit is given by the fact we require three transits in a 4 year observation duration and thus the most distant habitable zone assumed here corresponds to a period of 1.33 years, which excludes the F0V and F5V stellar types.

Table 6.4: Values of interest for a zero-impact-parameter transit (i = 90°) of a 1/3-Earth-mass moon orbiting a Saturn-like planet in the habitable zone of different stellar type. Note as the moon would not be stable in the case of a M5V star.

| Star type | $a_P/R_*$ | $P_{hab}$ (days) | $K_*$ (m/s) | $k$ |
|---|---|---|---|---|
| M5V | 64.71969789 | 18.4414602 | 65.366817438 | 0.320941502 |
| M2V | 79.90387125 | 46.2085195 | 31.333501490 | 0.173308411 |
| M0V | 95.05056217 | 69.7995617 | 23.227546659 | 0.144423676 |
| K5V | 125.329156 | 121.2125361 | 16.111680482 | 0.120353063 |
| K0V | 170.936413 | 228.0808689 | 11.693495705 | 0.101946124 |
| G5V | 199.2363481 | 299.4842301 | 9.647697652 | 0.094189354 |
| G2V | 216.0704522 | 367.6476191 | 8.523216538 | 0.086654206 |
| G0V | 227.4072097 | 446.9329498 | 7.730562053 | 0.078776551 |
| F5V | 285.0184484 | 697.7774733 | 5.501065587 | 0.066657081 |
| F0V | 343.459599 | 1070.1691359 | 4.363963528 | 0.057769470 |

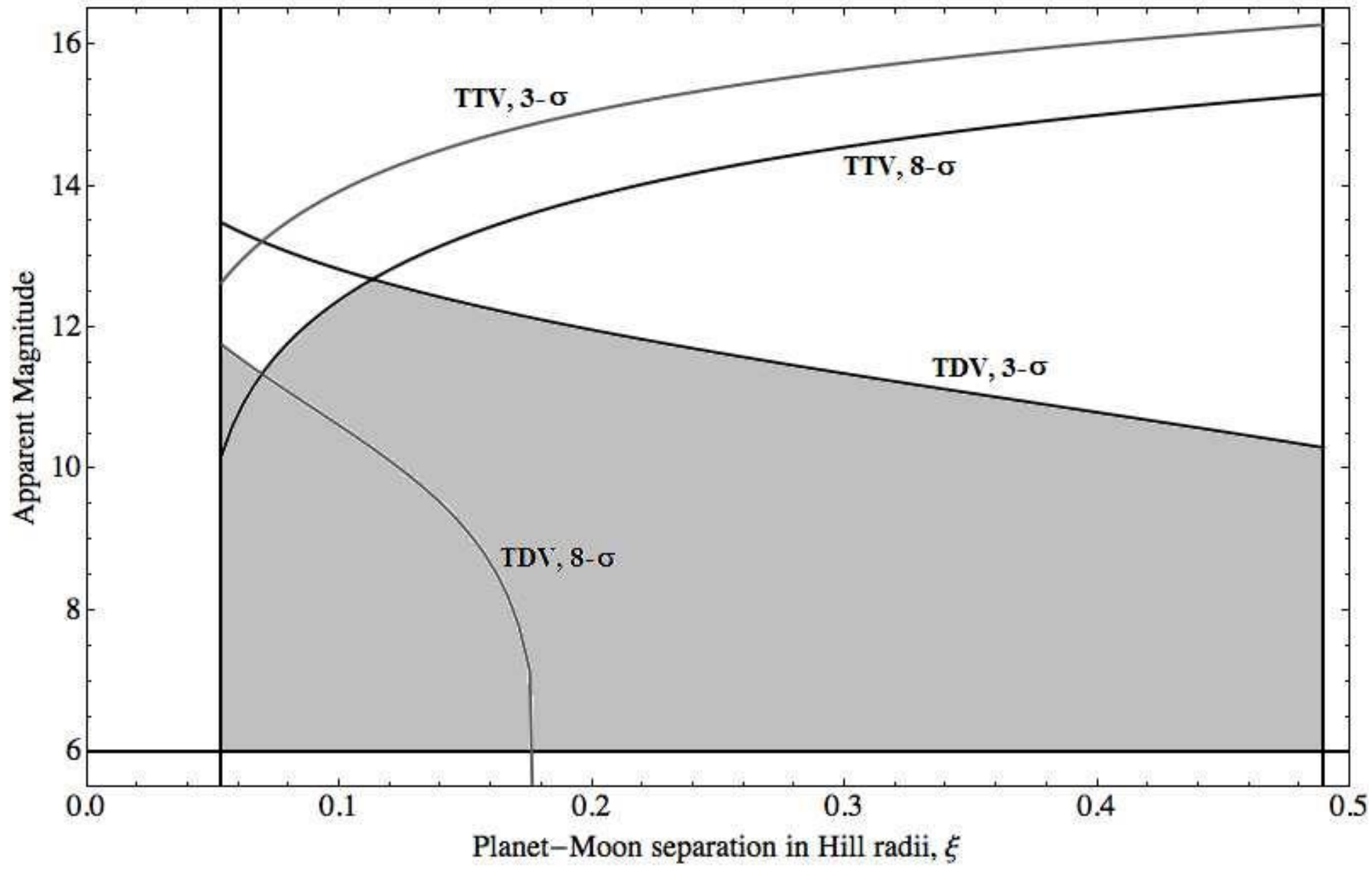

Figure 6.11: Detectable range of a habitable 1/3 M⊕ exomoon for an M0V star with respect to apparent magnitude and ξ parameter space. The grey area represents the area satisfying our detection criteria shown by the four curves. Additional constraints are the lower Roche limit, the upper dynamical stability limit and the bright-star magnitude limit.

### *6.3.4 Minimum detectable habitable exomoon masses*

Assuming that the probability distribution of exomoons with respect to $\xi$ is approximately flat between $\xi_{min}$ and $\xi_{max}$, we can easily calculate the magnitude limits to **detect** 25%, 50% or 75% of the exomoons in the given sample. For the case shown in Figure 6.11, the quartile values are $m_{Kep}$ = 12.0, 11.5 and 10.9 respectively (see Fig. 6.12). This means that for a M0V star of $m_{Kep}$ = 12.0 there is a 25% probability of detecting a 1/3 M⊕ exomoon orbiting a Saturn-like planet in the habitable zone.

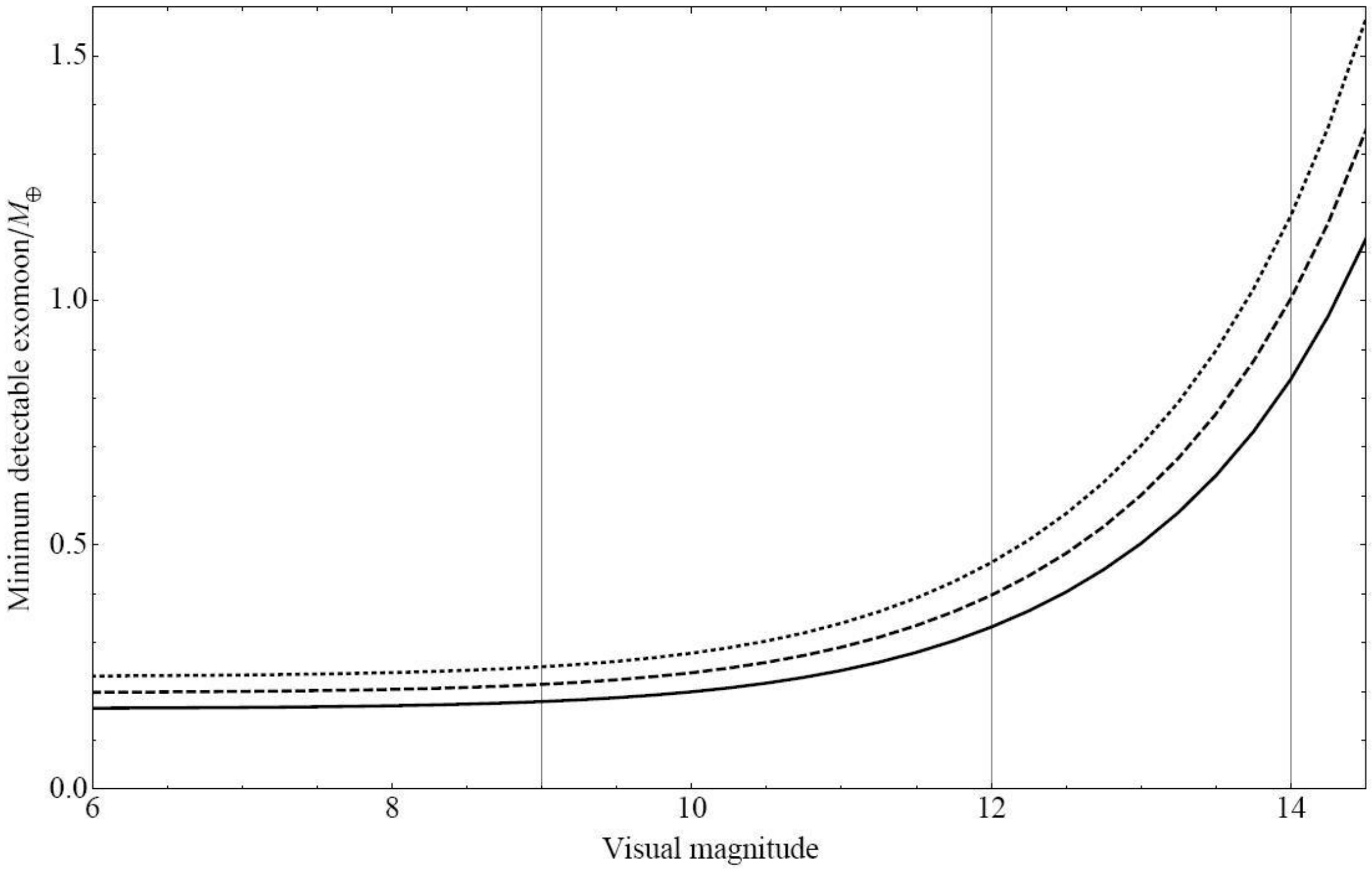

Figure 6.12: Minimum detectable exomoon mass around a Saturn-like planet host in the habitable zone of an M0V star. The top line is the 75% catch-rate, the middle is 50% and the lowest is for a 25% exomoon catch-rate.

We may then calculate these quartile values for different exomoon masses and various star types. Effectively, we are able to determine the minimum detectable exomoon mass as a function of magnitude for three different detection yields: i) 25%, ii) 50%, and iii) 75%. In Figure 6.13, we plot the minimum detectable exomoon mass, for several contours of visual magnitude, as a function of stellar mass, in the 25% detection-yield case. We also show an estimate of the upper limit on moon mass calculated from Barnes & O'Brien [138].

The $m_{Kep}$ = 12.5 limit almost intersects the Barnes & O'Brien [138] limit at 0.4M☉ and ~ 0.4M⊕ for the exomoon corresponding to an M2V star ~ 40 pc away.

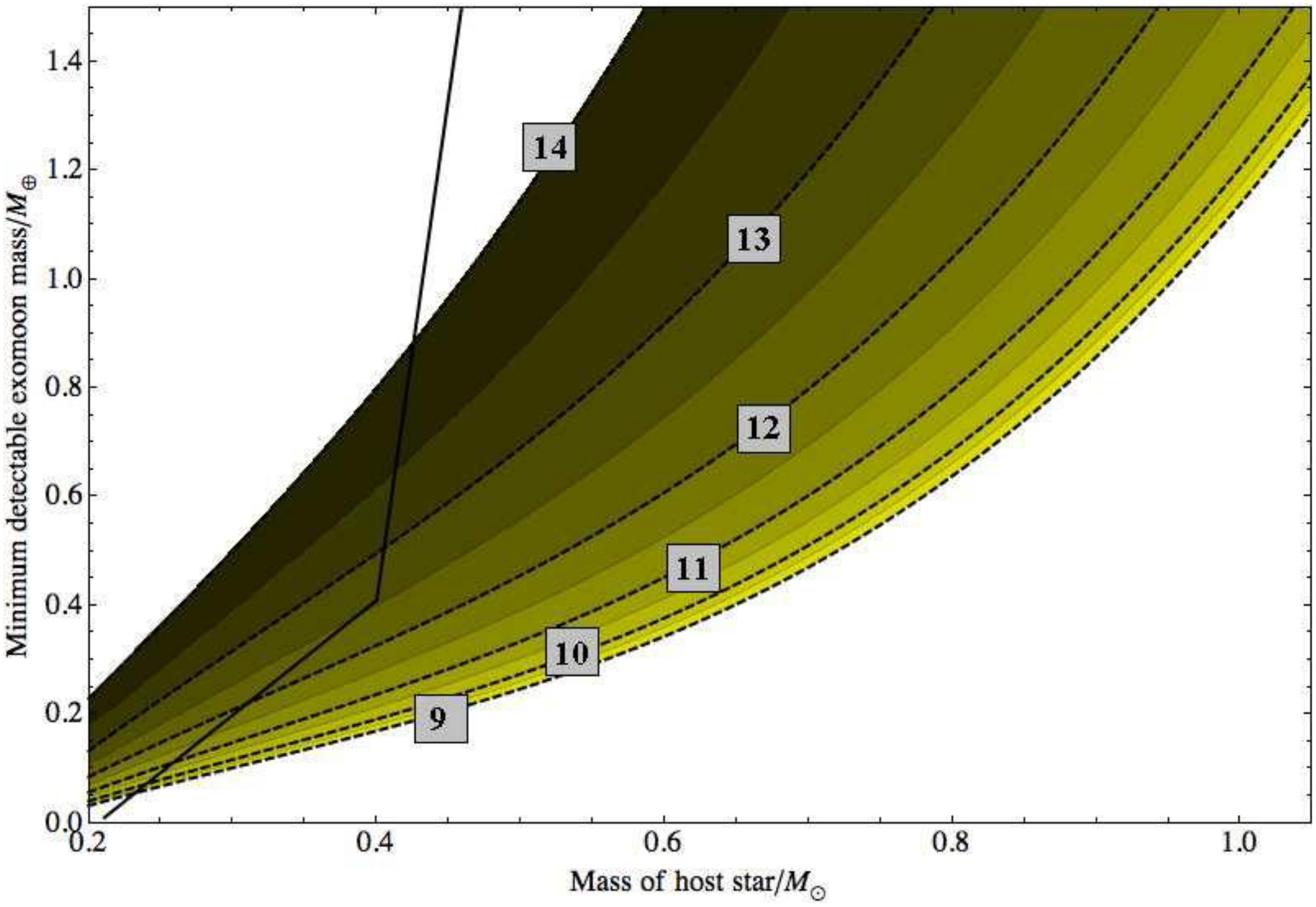

Figure 6.13: Minimum detectable exomoon mass around a Saturn-like planet host in the habitable zone of various star types. Contours show the different magnitude limits for a 25% exomoon catch-rate and overlaid is the mass stability limit (black, solid).

If we set a lower limit of 10 pc for our target star, then the lowest-mass habitable-zone exomoon which could be detected with KCP around a M2V star will be 0.18M⊕ which is below the stability limit of 0.41 M⊕. We therefore conclude that the minimum detectable habitable-zone exomoon mass with KCP is ~ 0.2 M⊕.

### *6.3.5 Detectability of a 1 M⊕ habitable exomoon and of an Earth-Moon analogue*

For the 25% detection yield case, we can convert the magnitude limit for detecting a 1 M⊕ habitable exomoon into a distance limit by making use of the absolute magnitudes for each star type. The distance limit for detecting a 1 M⊕ habitable exomoon may then be compared to the limit found for a transiting 1 R⊕ habitable exoplanet (see par. 6.3.1). In Figure 6.14, we compare the two distance limits, which reveals that the distance limit for moons takes the same shape as the transit limit but is ~ 1.6 times lower. This translates to a volume space diminished by a factor of ~ 4.

So, assuming that there are about $10^5$ useable stars in Kepler's field of view, we estimate roughly that 25,000 stars would be within range for habitable-exomoon detection.

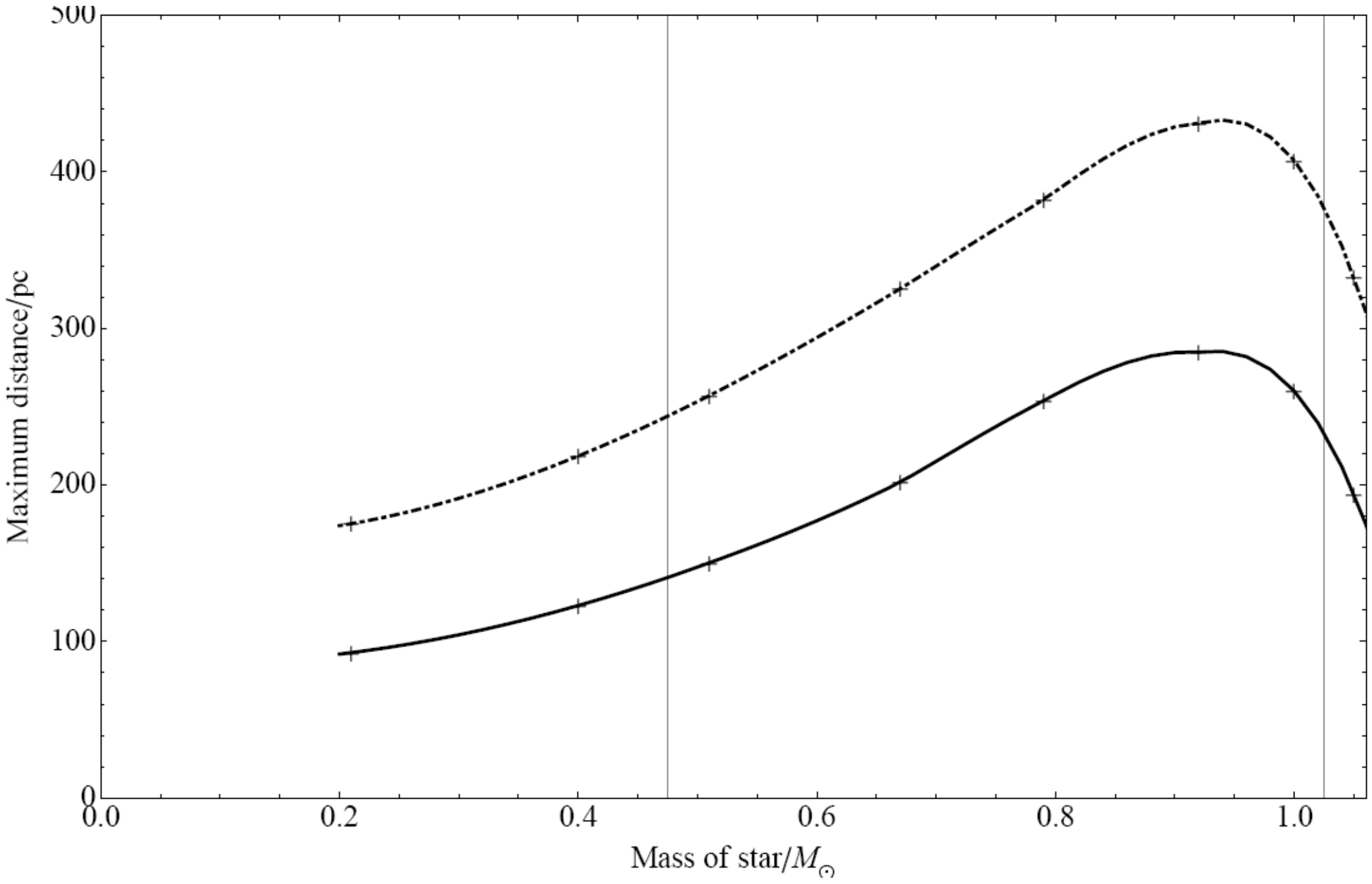

Figure 6.14: Distance limit for detecting a 1 $M_\oplus$ habitable exomoon (solid) and a 1 $R_\oplus$ exoplanet (dot-dashed) as a function of stellar mass, for main sequence stars. We assume a Saturn-like planet as the host for the exomoon case. We also mark the lower stellar mass stability limit of ~ 0.475 $M_\odot$ and the upper mass limit of ~ 1 $M_\odot$ imposed by the requirement to observe ≥ 3 transits. The drop at 0.9 $M_\odot$ is due to $P_{hab}$ becoming longer.

To conclude, we consider the detectability of an Earth-Moon system. Using the orbital parameters of the Sun-Earth-Moon system and assuming the most favourable configuration of i = 90°, we find the Earth would exhibit a TTV r.m.s. amplitude of 112.4 seconds and a TDV of 13.7 seconds. The TDV is particularly low due to the large value of $\xi = 0.26$ for the Moon.

In comparison, KCP of a V = 6 star would yield timing errors of $\sigma_{tc} = 389.6$ seconds and $\sigma_T = 779.2$ seconds. So, even for the most favourable case the TTV has a signal-to-noise of 0.3, which makes it undetectable. We therefore conclude that KCP cannot detect a Sun-Earth-Moon analogue through transit timing effects.

*Chapter 7*

# CONCLUSIONS

Since 1995, the search for exoplanets has imposed itself like one of the most dynamic research fields. As a matter of fact, we can count more than 80 ground-based and 20 space-based ongoing programs and future projects. More than 370 exoplanets are now known thanks mostly to indirect detection techniques. This number is due to increase exponentially in the near future thanks to the improved capabilities of the groundbased telescopes and space-missions devoted to the detection of exoplanets. One of them, already operational, is the Kepler mission. The purpose of a mission like Kepler is to monitor ~100,000 solar-type stars for four years in order to get statistical data on the population and distribution of occurrence of Earth analogs and of larger-size planets as well.

In the first decade, the task was to find more and more of these astronomical bodies. In recent years, attention has switched from finding planets to characterizing them. Among the variety of exoplanets discovered, particular attention is being devoted to those planets that transit their parent star. For these objects planetary and orbital parameters such as eccentricity, inclination, radius, mass, are known, allowing first order characterization on the bulk composition and temperature. Most significantly, it is possible to use the wavelength dependence of this extinction to identify key chemical components in the planetary atmosphere, which opens up enormous possibilities in terms of exoplanet characterization.

As an example of atmospheric characterization we have presented the IRAC photometry data recording the primary transit of HD209458b in four infrared bands. We have corrected the systematics and performed Monte Carlo fits. Our results are consistent with the presence of **water vapour** in the atmosphere of HD209458b. It is possible that additional molecules, such as $CH_4$, CO and/or $CO_2$ are also present, but the lack of spectral resolution of our data have prevented these from being detected.

Additional data in transmission at different wavelengths and/or higher will be required to gain information about these other molecules.

Whilst significant advances have been made in the field of understanding the atmospheres of hot Jupiters, we have to wait for results from Kepler and then from JWST in order to begin the characterization of terrestrial planets and investigate the possibility for them to harbour life. In fact, while within our own Solar System the search for extraterrestrial life and evidence about the origin of earthly life will likely be confined to Europa, Enceladus and Titan; the advantages of studies beyond the Solar System are the greater diversities of both planetary environments and their stages of development that are available for investigation.

Moreover, many other planetary attributes remain a mystery for us. One example is the question as to whether exoplanets frequently host satellites, as many of the planets of own solar system do. We have presented a model for the TTV and TDV signals, by employing transit time variation (TTV) and transit duration variation (TDV) in combination, it is possible to not only identify exomoons but also derive their mass and orbital separation from the host planet. Once the period, mass and ephemeris of an exomoon can be determined using these techniques, it should be possible to scrutinise the lightcurve and search for the dip in light due to the moon itself which would ultimately reveal the exomoon radius.

We have also shown that current ground-based telescopes could detect a $1M_\oplus$ exomoon in the habitable zone around a Neptune-like exoplanet. So, the natural subsequent question was about the possibilities that the Kepler Mission would offer in this search.

Our analysis finds that habitable exomoons are detectable up to ~ 100-200 pc away around early-M, K and later-G dwarf stars with the Kepler Mission or photometry of equal quality (KCP). This photometric quality should be sensitive down to $0.2M_\oplus$ habitable exomoons in the idealised cases and could survey ~ $10^6$ stars for $1M_\oplus$ habitable exomoons, with a full Galactic Plane survey. This number is around 25,000 stars for Kepler's field-of-view.

We find that Saturn-like planets are the ideal host candidates for detection due to their large radius to mass ratio. Additionally, we find lower-mass exomoons may be found around M-dwarfs due to the closer habitable zone permitting a larger number of transits in a 4-year window.

The exciting prospect of discovering a habitable exomoon is well within the grasp of KCP and whether such worlds exist or whether they will be classed as truly habitable worlds are questions we can hope to answer in the coming years. It is interesting to note that $\sim 0.2M_\oplus$ is the minimum habitable mass for a planet or moon proposed by several authors. Moreover, it has been proposed that the habitable zone could be extended for exomoons due to tidal heating and so our calculations may infact be an underestimate.

We find that Kepler will be incapable of finding Ganymede-mass moons around Saturn-like planets. If the maximum moon mass which could form from a planetary debris disk is $2 \times 10^{-4}$ of that of the primary's, as happen in the Solar System, unlikely KCP would detect any moons which formed around a planet in such a way. Nevertheless, moons above this mass limit are still dynamically stable and could have been captured by a planet or formed through an impact, for example like the formation of Triton and the Moon respectively. Whether or not such objects are common is unknown but KCP could make the first in-depth search.

Our results suggest it is easier to detect an Earth-like exoplanet than an Earth-like exomoon around a gas giant. However, we have no statistics to draw upon to estimate which of these scenarios is more common. If a roughly equal number of both are discovered, it would indicate that the latter is more common due to the detection bias.

All these results highlight the promising opportunity of making the **first exomoon detection** using the Kepler telescope, or photometry of equivalent quality, especially the feasibility of detecting habitable-zone exomoons.

SAMPLE IRAC BCD HEADER

```
ORIGIN    = 'Spitzer Science Center' / Organization generating this FITS file
CREATOR   = 'S17.0.4 '           / SW version used to create this FITS file
TELESCOP  = 'Spitzer '           / SPITZER Space Telescope
INSTRUME  = 'IRAC    '           / SPITZER Space Telescope instrument ID
CHNLNUM   = 3 / 1 digit instrument channel number
EXPTYPE   = 'sci     '           / Exposure Type
REQTYPE   = 'IER     '           / Request type (AOR, IER, or  SER)
AOT_TYPE  = 'IRAC    '           / Observation template type
AORLABEL  = 'p461_IRAC_HD209458_primary1_fpa13_rb' / AOR Label
FOVID     = 67 / Field of View ID
FOVNAME   = 'IRAC_Center_of_3.6&5.8umArray' / Field of View Name

/ PROPOSAL INFORMATION

OBSRVR    = 'Giovanna Tinetti'   / Observer Name (Last, First)
OBSRVRID  = 34318 / Observer ID of Principal Investigator
PROCYCL   = 7 / Proposal Cycle
PROGID    = 461 / Program ID
PROTITLE  = 'Probing water vapor and CO in the atmosphere of HD 209458b' / Program
PROGCAT   = 31 / Program Category

/ TIME AND EXPOSURE INFORMATION

DATE_OBS  = '2007-12-31T05:10:15.049' / Date & time at DCE start
MJD_OBS   = 54465.215452 / [days] MJD at DCE start (,JD-2400000.5)
HMJD_OBS  = 54465.2157446 / [days] Corresponding Heliocen. Mod. Julian Date
UTCS_OBS  = 252349815.049 / [sec] J2000 ephem. time at DCE start
SCLK_OBS  = 883545201.808 / [sec] SCLK time (since 1/1/1980) at DCE start
SPTZR_X   = 57584784.137474 / [km] Heliocentric J2000 x position
SPTZR_Y   = 127475885.053542 / [km] Heliocentric J2000 y position
SPTZR_Z   = 58316007.936835 / [km] Heliocentric J2000 z position
SPTZR_VX  = -27.510977 / [km/s] Heliocentric J2000 x velocity
SPTZR_VY  = 10.133945 / [km/s] Heliocentric J2000 y velocity
SPTZR_VZ  = 4.154083 / [km/s] Heliocentric J2000 z velocity
SPTZR_LT  = 505.510476 / [sec] One-way light time to Sun's center
ET_OBS    = 252349880.233 / [sec] Ephemeris time (seconds past J2000 epoch)
AORTIME   = 2. / [sec] Duration of AOR
SAMPTIME  = 0.2 / [sec] Sample integration time
FRAMTIME  = 2. / [sec] Time spent integrating (whole array)
COMMENT   Photons in Well = Flux[photons/sec/pixel] * FRAMTIME
EXPTIME   = 1.2 / [sec] Effective integration time per pixel
COMMENT   DN per pixel = Flux[photons/sec/pixel] / GAIN * EXPTIME
FRAMEDLY  = 6.4 / [sec] Frame Delay Time
FRDLYDET  = 'T       '           / Frame Delay Time Determinable (T or F)
```

| | |
|---|---|
| INTRFDLY | = 6.4 / [sec] Inter Frame Delay Time |
| IMDLYDET | = 'T ' / Immediate Delay Time Determinable (T or F) |
| AINTBEG | = 779414.65 / [Secs since IRAC turn-on] Time of integ. start |
| ATIMEEND | = 779416.61 / [Secs since IRAC turn-on] Time of integ. end |
| AFOWLNUM | = 4 / Fowler number |
| AWAITPER | = 2 / [0.2 sec] Wait period |
| ANUMREPS | = 1 / Number of repeat integrations |
| AREADMOD | = 0 / Full (0) or subarray (1) |
| HDRMODE | = F / DCE taken in High Dynamic Range mode |
| ABARREL | = 2 / Barrel shift |
| APEDSIG | = 0 / 0=Normal, 1=Pedestal, 2=Signal |

/ TARGET AND POINTING INFORMATION

| | |
|---|---|
| OBJECT | = 'HD 209458' / Target Name |
| OBJTYPE | = 'TargetFixedSingle' / Object Type |
| CRVAL1 | = 330.794331876722 / [deg] RA at CRPIX1,CRPIX2 (using Pointing Recon |
| CRVAL2 | = 18.8853674330823 / [deg] DEC at CRPIX1,CRPIX2 (using Pointing Reco |
| RA_HMS | = '22h03m10.6s' / [hh:mm:ss.s] CRVAL1 as sexagesimal |
| DEC_DMS | = '+18d53m07s' / [dd:mm:ss] CRVAL2 as sexagesimal |
| RADESYS | = 'ICRS ' / International Celestial Reference System |
| EQUINOX | = 2000. / Equinox for ICRS celestial coord. system |
| RA_REF | = 330.795068055556 / [deg] Commanded RA (J2000) of ref. position |
| DEC_REF | = 18.8844044444445 / [deg] Commanded Dec (J2000) of ref. position |
| RECONFOV | = 'IRAC_Center_of_5.8umArray' / Reconstructed Field of View |

/ PHOTOMETRY

| | |
|---|---|
| BUNIT | = 'MJy/sr ' / Units of image data |
| FLUXCONV | = 0.5952 / Flux Conv. factor (MJy/sr per DN/sec) |
| GAIN | = 3.8 / e/DN conversion |
| RONOISE | = 9.1 / [Electrons] Readout Noise from Array |
| ZODY_EST | = 2.49335 / [MJy/sr] Zodiacal Background Estimate |
| ISM_EST | = 0.2164939 / [MJy/sr] Interstellar Medium Estimate |
| CIB_EST | = 0. / [MJy/sr] Cosmic Infrared Background Estimate |
| SKYDRKZB | = 1.612603 / [MJy/sr] Zodiacal Background Est of Subtracted |
| SKYDKMED | = -0.948284 / [MJy/sr] Median of Subtracted Skydark |
| SKDKRKEY | = 24621568 / Skydark AOR Reqkey |
| SKDKTIME | = 2. / [sec] Skydark AOR Duration Time |
| SKDKFDLY | = 13.39 / [sec] Average Frame Delay Time of Skydark |
| SKDKIDLY | = 13.39 / [sec] Average Immediate Delay Time of Skydark |

/ DATA FLOW KEYWORDS

| | |
|---|---|
| DATE | = '2008-01-04T01:45:20' / [YYYY-MM-DDThh:mm:ss UTC] file creation date |
| AORKEY | = 24740096 / AOR or EIR key. Astrnmy Obs Req/Instr Eng Req |
| EXPID | = 2395 / Exposure ID (0-9999) |
| DCENUM | = 0 / DCE number (0-9999) |

*Appendix B*

## CLARET'S COEFFICIENTS FOR LIMB-DARKENING FUNCTIONS

Claret [34] has found that the most accurate functions that can describe the limb-darkening of main-sequence stars are the quadratic and the "nonlinear" law. In [156] Claret presents 2 models: ATLAS and PHOENIX. Given a value for: the rotation velocity of the star *VT*, its surface gravity *g*, its effective temperature $T_{eff}$ and metallicity *M,* each model returns the 2, 4 coefficients to be used in the quadratic and the "nonlinear" law respectively. The coefficients to be used are different depending on the band of observation considered (u', g', r', i', z').

In the following Table we shown the coefficients valid for the Sun and the Kepler Mission (the filter r' = 625nm is a good approximation) which we have used during our analysis.

Table B.1. The coefficients from the ATLAS model which describe the limb-darkening of the Sun. The coefficients can be read in the r'-column, the ones to be used in the quadratic law are u1 and u2, while the ones for the "nonlinear" law are named c1, c2, c3 and c4.

| Coefficient | $VT/kms^{-1}$ | $log(g/cms^{-2})$ | $T_{eff}/K$ | log(M/H) | r' |
|---|---|---|---|---|---|
| u1 | 0 | 4.5 | 5750 | 0 | 0.3702 |
| u2 | 0 | 4.5 | 5750 | 0 | 0.3115 |
| | | | | | |
| c1 | 0 | 4.5 | 5750 | 0 | 0.5596 |
| c2 | 0 | 4.5 | 5750 | 0 | -0.1245 |
| c3 | 0 | 4.5 | 5750 | 0 | 0.7238 |
| c4 | 0 | 4.5 | 5750 | 0 | -0.4005 |

*Appendix C*

## NOISY

```
program noisy

! Author: Giammarco Campanella (g.campanella@ucl.ac.uk)
!
! Generates noise and adds it to a transit curve created with jasmine

! INPUT:
! The input data (time, flux, z, mu0) should be listed in oqMDL.dat
! The input data for the noise should be listed in noise.txt following this order:
! r.m.s. amplitude of the shot noise (in ppm)
! r.m.s. amplitude of the instrument noise (in ppm)
! period of the instrument noise (in mins)
! r.m.s. amplitude of the stellar noise (in ppm)
! period of the stellar noise (in mins)

! OUTPUT:
!  Output is written to filenames
!  The output file is a two-column listing:
!   col 1: time in days from mid-transit
!   col 2: relative flux+noise
!
! COMPILING: g95 -o noise noise.f90
!          And run with ./noise
!
! Version history:
!   1.0: 2009 May 8
!   1.1: 2009 May 15 version for jasmine
!   1.2: 2009 May 18 reads from a file the characteristics of the noise to be generated
and gives 'true random numbers'
!                i.e. different numbers every time it is run
…
!   1.3.3: 2009 Jun 3 if ndatasets=1 data are saved in transitdata.dat
!   1.4: 2009 Jun 4 periods of the instrument noise and stellar signal are offset by a
random number between -1 and +1 seconds

implicit none

integer i, n, j, k, ndatasets
parameter (n=1441)
double precision r(n), time(n), flux(n), fn(n), instrum(n), stellar(n), z(n), mu0(n)
double precision instrrmsampl, stellarrmsampl, instrperiod, starperiod, sigma
character(len=7) filename
character(len=25), dimension(:), allocatable :: filenames
character(len=3) number
```

```
double precision, parameter :: pi=3.14159265358979323846D00

! read data from oqMDL
open (1, file='LC_quad.dat')
  do j=1, n
  read (1,*) time(j), flux(j), z(j), mu0(j)
  end do
close (1)

! read the characteristics of the noise to be generated
open (3, file='noise.txt')
  read (3,*) sigma
  read (3,*) instrrmsampl
  read (3,*) instrperiod
  read (3,*) stellarrmsampl
  read (3,*) starperiod
close (3)

  sigma=sigma/1.0d6
  instrrmsampl=instrrmsampl/1.0d6
  stellarrmsampl=stellarrmsampl/1.0d6

  write(*,*) 'Insert the number of datasets to be generated: '
  read (*,*) ndatasets

  if (ndatasets>1) then
    write(*,*) 'Insert the first part of the name of the files where data will be written (7
characters e.g. nep_G2V): '
    read (*,*) filename
  end if

  allocate(filenames(ndatasets))

     do k=1, ndatasets
      if (ndatasets==1) then
      filenames(k)='transitdata.dat'
    else
       write(number,999) k
   999  format(i3.3)
       filenames(k)=filename//'_'//number//'.dat'
     end if

       call shot (n, r, sigma)
      call instrument (n, instrum, time, instrrmsampl, instrperiod)
      call star (n, stellar, time, stellarrmsampl, starperiod)

      ! add noise to flux
       open(unit=k+30,file=filenames(k))
        do i=1, n
```

```
        fn(i)=flux(i)+r(i)+instrum(i)+stellar(i)
        write(k+30,*) time(i), fn(i)
       end do
      close (k+30)
    end do

! report of the characteristics of the noise generated
! open (4, file='noisereport.txt')
! write(4,*) 'characteristics of the noise generated:'
! write(4,*) ''
! write(4,*) 'r.m.s. amplitude of the shot noise       =', sigma*1.0d6, 'ppm'
! write(4,*) 'r.m.s. amplitude of the instrument noise =', instrrmsampl*1.0d6, 'ppm'
! write(4,*) 'period of the instrument noise          =', instrperiod, 'mins'
! write(4,*) 'r.m.s. amplitude of the stellar noise    =', stellarrmsampl*1.0d6, 'ppm'
! write(4,*) 'period of the stellar noise             =', starperiod, 'mins'
! close (4)

stop

contains

! generate shot noise: standard deviation 15.63 ppm
subroutine shot (n, r, sigma)
  integer i, n
  double precision seeda, seedb, sigma, r(n)

  call random_seed()

  do i=1, n
  call random_number(seeda)
  call random_number(seedb)
  r(i)=sqrt(-2.0d0*log(seeda))*dcos(2.0d0*pi*seedb) ! box-muller algorithm from
Numerical Recipes
  r(i)=sigma*r(i)
  end do
end subroutine shot

! generate instrument noise: period 1 hour, r.m.s. amplitude 7.47 ppm
subroutine instrument (n, instrum, time, instrrmsampl, instrperiod)
  integer i, n
  double precision amplitudei, instrum(n), time(n), periodi, seed, seed2, random2,
phasei, instrrmsampl, instrperiod

  amplitudei=sqrt(2.0d0)*instrrmsampl !instrument noise amplitude

  periodi=instrperiod/(dble(n-1)) ! convert period from minutes to days
  ! write (*,*) periodi
  call random_seed()
  call random_number(seed2)
  ! write (*,*) seed2
```

```
  random2=2.0D0*seed2-1.0D0       !random numbers with uniform distribution in the
range (-1,+1)
  random2=random2/(dble(n-1)*60.0d0)   !convert random in seconds
  ! write (*,*) random2
  periodi=periodi+random2           ! an offset of about +/- 1 sec is added to the period

  call random_seed()
  call random_number(seed)
  phasei=2.0D0*pi*seed          ! the phase created is a random number with uniform
distribution in the range (0,2PI)

  do i=1, n
  instrum(i)=amplitudei*dsin((2.0d0*pi*time(i)/periodi)+phasei)
  ! a phase is added so that instrument noise and stellar signal are never resonant
  end do
end subroutine instrument

! generate stellar signal: r.m.s. amplitude 10ppm, period 5 minutes
subroutine star (n, stellar, time, stellarrmsampl, starperiod)
  integer i, n
  double precision amplitudes, stellar(n), time(n), periods, seed1, seed3, random3,
phases, stellarrmsampl, starperiod

  amplitudes=sqrt(2.0d0)*stellarrmsampl !stellar signal amplitude

  periods=starperiod/(dble(n-1)) ! convert period from minutes to phase
  ! write (*,*) periods
  call random_seed()
  call random_number(seed3)
  ! write (*,*) seed3
  random3=2.0D0*seed3-1.0D0       !random numbers with uniform distribution in the
range (-1,+1)
  random3=random3/(dble(n-1)*60.0d0)   !convert random in seconds
  ! write (*,*) random3
  periods=periods+random3           ! an offset of about +/- 1 sec is added to the period

  call random_seed()
  call random_number(seed1)
  phases=2.0D0*pi*seed1         ! the phase created is a random number with uniform
distribution in the range (0,2PI)

  do i=1, n
  stellar(i)=amplitudes*dsin((2.0d0*pi*time(i)/periods)+phases)
  ! a phase is added so that instrument noise and stellar signal are never resonant
  end do
end subroutine star

end program noisy
```

## LIGHTCURVE GENERATION AND FITTING

Here we report an example of simulation carried out for an hot-Jupiter.

- The following values are entered in Jasmine (see par. 6.2.1):

0.0D0 = e
94.8D0 = Omega
8.1D0 = aoverR
3.25D0 = P
70.0D0 = Kstar
0.10D0 = k
85.2D0 = i
0.7164D0 = u1
0.1378D0 = u2

- Jasmine returns the lightcurve and a file with the following quantities calculated:

```
----------------------------------
        PRIMARY TRANSIT
----------------------------------

Impact parameter =  0.677722989601881
D/hrs = T(1.5,3.5) =  2.253708435480541
tT/hrs = T(1,4) =  2.655608132944167
tF/hrs = T(2,3) =  1.8149541766066593
ingress/hrs = T(1,2) =  0.420326978168754
egress/hrs = T(3,4) =  0.420326978168754

T(mid,true)/hrs = T(1,mid) =  1.3278040664720836
T(mid,app)/hrs = [0.5*T(1,4)] =  1.3278040664720836
Delta T(mid)/mins =  0.

----------------------------------
      SECONDARY TRANSIT
----------------------------------

Impact parameter =  0.6778580722867117
D/hrs = S(1.5,3.5) =  2.253708435480541
sT/hrs = S(1,4) =  2.655608132944167
sF/hrs = S(2,3) =  1.8229230378558192
ingress/hrs = S(1,2) =  0.420326978168754
egress/hrs = S(3,4) =  0.420326978168754

S(mid,true)/hrs = S(1,mid) =  1.3278040664720836
S(mid,app)/hrs = [0.5*S(1,4)] =  1.3278040664720836
Delta S(mid)/mins =  0.
```

```
----------------------------------
       PRI TO SEC TIMES
----------------------------------

LT/mins =  0.
dyn/days =  0.
T(sec)-T(pri)/days =  0.

----------------------------------
     OTHER CARRIED PARAMETERS
----------------------------------

 Planetary surface gravity =  10.312801286097992  m/s^2
  (Biased) stellar density =  951.7602329446462  kg/m^3
```

- The inputs for Noisy are (see par. 6.2.2):

```
15.63d0  ppm  = r.m.s. amplitude of the shot noise
0.0d0   ppm  = r.m.s. amplitude of the instrument noise
0.0d0   mins = period of the instrument noise
0.0d0   ppm  = r.m.s. amplitude of the stellar noise
0.0d0    mins = period of the stellar noise
```

Noisy is asked to generate 100 noisy lightcurves with the photon noise being randomly generated in all cases. Results are written in "dat" files.

- The inputs for Jasper are (see par. 6.2.3):

```
0.7164D0 = u1
0.1378D0 = u2
1441  = nobs                  // number of data
0    = wtflag                 // weight flag: 0=unweighted, 1=weighted
1.0  = wtscale                // scaling factor for errors
0    = mctrials               // number of Monte Carlo trials - 0 = no MC trials at all
0    = mcstyle, 0=shuffle, 1=string      // MC style - shuffle (0) or string of beads (1)
```

- Jasper returns a lightcurve model in aq.dat and a report which we show here:

```
JASPER running        :
  Number of data read :  1441

 Pikaia output:
      seed:  82   // random seed for the initialization
    status:  0
         x:  0.189768 0.269962 0.500118 0.461002 // the array with the 4 scaled solutions
found by Pikaia
  fitness,f:  16674.965          // goodness-of-fit measure for the chi squared

 PIKAIA Fitted parameters:          // the quantities calculated by Pikaia using the 4
solutions
  Planet/star ratio   :  0.0948840007185936
```

Orb. incl. (degrees): 86.64284789893657
Impact parameter, b : 0.5399240255355835
a_semi/star radius : 9.22003984451294
Mean HJD (observed): 5.816921468369141E-18
Transit centre (HJD): 0.000029504299163824173
Transit depth (%): 1.0738672540965277
durationP : 8159.7876646951445
durationS : 8159.7876646951445
tT : 9233.76132243135
tF : 7042.307477253868
ingress : 1095.7269225887412
egress : 1095.7269225887412
rhostar : 14.815160305859592
surfgrav : 1403.6889846672043

AMOEBA initialization // Amoeba begins with a set of n + 1 points whit n=4
Initial values and rms at:
point 1 : 0.1850238 0.26321295 0.48761508 0.44947696 0.00030874563
point 2 : 0.189768 0.269962 0.500118 0.461002 0.000059970145
point 3 : 0.1945122 0.27671108 0.512621 0.47252703 0.00032285327
point 4 : 0.1992564 0.2834601 0.5251239 0.4840521 0.0006302125
point 5 : 0.2040006 0.29020917 0.53762686 0.49557716 0.0009230944

Amoeba output:
Number of iterations: -1
Best points and rms: 0.18976223 0.27017277 0.500051 0.46108234 0.00005990534
Best points and rms: 0.18976223 0.27017277 0.500051 0.46108234 0.00005990534
Best points and rms: 0.18976223 0.27017277 0.500051 0.46108234 0.00005990534
Best points and rms: 0.18976223 0.27017277 0.500051 0.4610823 0.000059905327
Best points and rms: 0.18976223 0.27017277 0.500051 0.46108234 0.00005990534
mean point: 0.18976223 0.27017277 0.500051 0.46108237 // Amoeba solutions
rms at mean: 0.000059905353
equiv pikaia fitness: 16693. // note as the fitness is improved with respect to pikaia
reduced chi^2: 3.5886514E-9

AMOEBA Fitted parameters: // the quantities calculated by Amoeba using the 4 solutions
Planet/star ratio : 0.09488111734390259
Orb. incl. (degrees): 86.64081028528403
Impact parameter, b : 0.5403455495834351
a_semi/star radius : 9.221647381782532
Mean HJD (observed): 5.816921468369141E-18
Transit centre (HJD): 0.000012755393981939408
Transit depth (%): 1.07354866183218
durationP : 8155.743016255804
durationS : 8155.743016255804
tT : 9229.802393682961

tF : 7038.076203291475
ingress : 1095.863095195743
egress : 1095.863095195743
rhostar : 14.821258579916663
surfgrav : 1404.423322861648

// AMOEBA is set to "robustness mode": the solution is tested by randomly perturbing it and refitting in 100 trials

initial chisq : 3.5886512031650077E-9
point 1 : 0.19958363 0.27979428 0.48754975 0.39578763 0.0007538731
point 2 : 0.20470117 0.2869685 0.500051 0.40593603 0.0006794137
…
point 5 : 0.21057022 0.3015926 0.5375347 0.36357817 0.0012991638
AMOEBA improved in trial : 48
new chisq : 2.2617008217698092E-10 // AMOEBA has found a better solution
point 1 : 0.17665711 0.27117956 0.48753145 0.33747244 0.0010253706
point 2 : 0.18118678 0.2781329 0.50003225 0.34612557 0.00093369116
….
point 5 : 0.23435625 0.35322848 0.5375347 0.4127448 0.0012175601

No. of improved solutions: 5
Best trial no. : 48

Number of MC trials : 0 // we do not run MC trials since the estimate on the uncertainties of the quantities is obtained statistically by considering the results from all the 100 noisy lightcurves

AMOEBA Fitted parameters: // this is the final solution of the simulation, we can compare it with the inputs and output of Jasmine and note the good result obtained

Planet/star ratio : 0.0999637022614479
+/- : 0.
Orb. incl. (degrees): 85.20900791450762
+/- : 0.
Impact parameter, b : 0.6771102547645569
+/- : 0.
a_semi/star radius : 8.107048869132996
+/- : 0.
Mean HJD (observed): 5.816921468369141E-18
Transit centre (HJD): 0.000008061528205877396
+/- : 0.
Transit depth (%): 1.078842459689855
+/- : 0.
tT : 9557.232006563052
+/- : 0.
tF : 6537.3369315887585
+/- : 0.
durationP : 8113.203572476131
+/- : 0.
durationS : 8113.203572476131

+/- : 0.
ingress : 1509.9475374871467
+/- : 0.
egress : 1509.9475374871467
+/- : 0.
rhostar : 10.3381255225885
+/- : 0.
surfgrav : 954.2471489648906
+/- : 0.

In Figure D.1 we show the lightcurves obtained during the simulation.

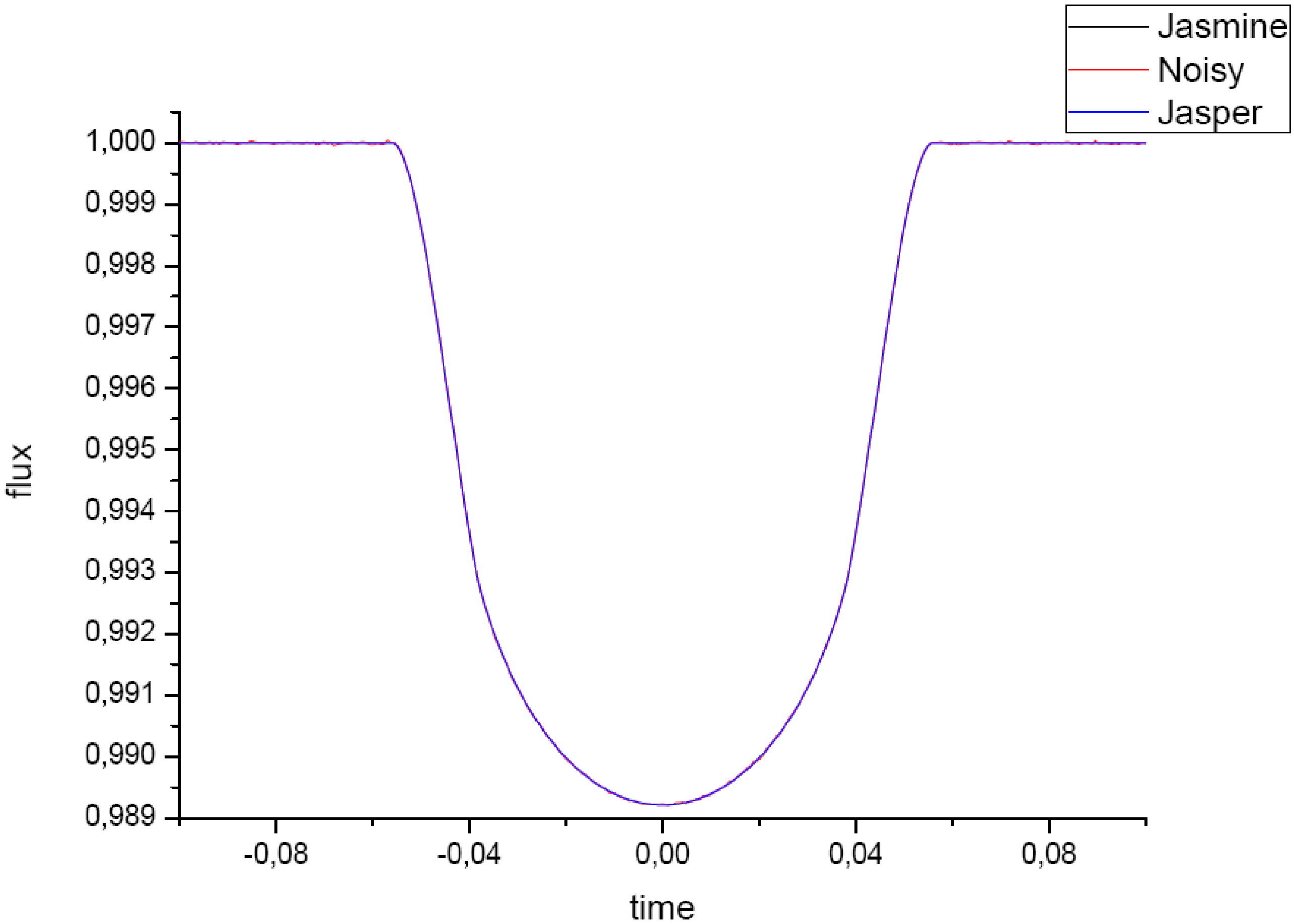

Figure D.1: Zoomed on the transit, we show the lightcurves generated by Jasmine, Noisy and Jasper. Note the good agreement of the results.

# PUBLICATIONS

## REFEREED ARTICLES

D. Kipping, S. J. Fossey and **G. Campanella**. On the detectability of habitable exomoons with *Kepler*-class photometry. MNRAS in press, 2009. (arXiv:0907.3909)

J.P. Beaulieu, D. Kipping, V. Batista, G. Tinetti, I. Ribas, S. Carey, J A. Noriega-Crespo, C. Griffith, **G. Campanella**, S. Dong, J. Tennysson, R. Barber, P. Deroo, S. Fossey, D. Liang, M. Swain, Y. Yung, N. Allard. Water in HD209458b's atmosphere from 3.6-8 μm IRAC photometric observations in primary transit. Submitted to MNRAS, 2009.